\documentclass[]{aa}   
\usepackage{amsmath,amsfonts,amsbsy,mathrsfs,amssymb}
\usepackage{graphicx}
\usepackage[T1]{fontenc}
\usepackage{lmodern}
\usepackage[utf8]{inputenc}
\usepackage[english]{babel}
\usepackage{supertabular,booktabs}
\usepackage{subcaption}
\usepackage{float}
\usepackage[flushleft]{threeparttable}
\usepackage{array}
\usepackage{longtable}
\usepackage{adjustbox}
\usepackage[pdftex=true, colorlinks=true, linkcolor=blue, citecolor=blue, pdfborder={0 0 0}, bookmarks=false, urlcolor=blue, breaklinks=true, pdftitle={HERSCHEL observations of extraordinary sources: Full HERSCHEL/HIFI molecular line survey of SAGITTARIUS B2$($M$)$}, pdfsubject={}, pdfauthor={}, pdfkeywords={}]{hyperref}

\usepackage{natbib,twoopt}
\bibpunct{(}{)}{;}{a}{}{,}          

\makeatletter
\newcommandtwoopt{\citeads}[3][][]{\href{http://adsabs.harvard.edu/abs/#3}%
{\def\hyper@linkstart##1##2{}%
\let\hyper@linkend\@empty\citealp[#1][#2]{#3}}}
\newcommandtwoopt{\citepads}[3][][]{\href{http://adsabs.harvard.edu/abs/#3}%
{\def\hyper@linkstart##1##2{}%
\let\hyper@linkend\@empty\citep[#1][#2]{#3}}}
\newcommandtwoopt{\citetads}[3][][]{\href{http://adsabs.harvard.edu/abs/#3}%
{\def\hyper@linkstart##1##2{}%
\let\hyper@linkend\@empty\citet[#1][#2]{#3}}}
\newcommandtwoopt{\citeyearads}[3][][]%
{\href{http://adsabs.harvard.edu/abs/#3}
{\def\hyper@linkstart##1##2{}%
\let\hyper@linkend\@empty\citeyear[#1][#2]{#3}}}
\makeatother

\newcolumntype{C}[1]{>{\centering\arraybackslash}p{#1}}
\newcolumntype{R}[1]{>{\flushright\arraybackslash}p{#1}}

\begin{document}
    \title{Herschel observations of extraordinary sources: Full Herschel/HIFI molecular line survey of Sagittarius B2(M)}


    \author{T.~M\"{o}ller\inst{1}
            \and
            P.~Schilke\inst{1}
            \and
            A.~Schmiedeke\inst{2}
            \and
            E.~A.~Bergin\inst{3}
            \and
            D.~C.~Lis\inst{4}
            \and
            \'{A}.~S\'{a}nchez-Monge\inst{1}
            \and
            A.~Schw\"{o}rer\inst{1}
            \and
            C.~Comito\inst{1,5}
            }

    \institute{I. Physikalisches Institut, Universit\"{a}t zu K\"{o}ln,
               Z\"{u}lpicher Str. 77, D-50937 K\"{o}ln, Germany\\
               \email{moeller@ph1.uni-koeln.de}
               \and
               Max-Planck-Institut für extraterrestrische Physik, Gießenbachstraße 1,
               85748 Garching bei München, Germany
               \and
               Department of Astronomy, University of Michigan,
               500 Church Street, Ann Arbor, MI 48109, USA
               \and
               Jet Propulsion Laboratory, California Institute of Technology,
               4800 Oak Grove Drive, Pasadena, CA 91109, USA
               \and
               Forschungszentrum Jülich GmbH
               Institute for Advanced Simulation (IAS)
               Jülich Supercomputing Centre (JSC)
               Wilhelm-Johnen-Straße, 52425 Jülich, Germany
              }

    \date{Received January 01, 2021 / Accepted January 31, 2021}

    \abstract
    {We present a full analysis of a broadband spectral line survey of Sagittarius~B2 (Main), one of the most chemically rich regions in the Galaxy located within the giant molecular cloud complex Sgr B2 in the Central Molecular Zone.}
    {Our goal is to derive the molecular abundances and temperatures of the high-mass star-forming region Sgr~B2(M) and thus its physical and astrochemical conditions.}
    {Sgr~B2(M) was observed using the Heterodyne Instrument for the Far-Infrared (HIFI) on board the \emph{Herschel} Space Observatory in a spectral line survey from 480 to 1907~GHz at a spectral resolution of 1.1~MHz, which provides one of the largest spectral coverages ever obtained toward this high-mass star-forming region in the submillimeter with high spectral resolution and includes frequencies $>$ 1~THz unobservable from the ground. We model the molecular emission from the submillimeter to the far-IR using the XCLASS program, which assumes local thermodynamic equilibrium (LTE). For each molecule, a quantitative description was determined taking all emission and absorption features of that species across the entire spectral range into account. Because of the wide frequency coverage, our models are constrained by transitions over an unprecedented range in excitation energy. Additionally, we derive velocity resolved \emph{ortho} / \emph{para} ratios for those molecules for which \emph{ortho} and \emph{para} resolved molecular parameters are available. Finally, the temperature and velocity distributions are analyzed and the derived abundances are compared with those obtained for Sgr~B2(N) from a similar HIFI survey.}
    {A total of 92 isotopologues were identified, arising from 49 different molecules, ranging from free ions to complex organic compounds and originating from a variety of environments from the cold envelope to hot and dense gas within the cores. Sulfur dioxide, methanol, and water are the dominant contributors. Vibrationally excited HCN~(v$_2$=1) and HNC~(v$_2$=1) are detected as well. For the \emph{ortho}/\emph{para} ratios we find deviations from the high temperature values between 13 and 27~\%. In total 14~\% of all lines remain unidentified.}
    {Compared to Sgr~B2(N), we found less complex molecules like CH$_3$OCH$_3$, CH$_3$NH$_2$, or NH$_2$CHO, but more simple molecules like CN, CCH, SO, and SO$_2$. However some sulfur bearing molecules like H$_2$CS, CS, NS and OCS are more abundant in N than in M. The derived molecular abundances can be used for comparison to other sources and for providing further constraints for astrochemical models.}


    \titlerunning{Molecular line survey of Sagittarius B2(M)}
    \authorrunning{T.~M\"{o}ller \textit{et al.}}

    \maketitle

    \section{Introduction}\label{sec:Introduction}

The Sagittarius B2 complex (Sgr~B2) is one of the most massive molecular clouds in the Galaxy with a mass of 10$^7 \, M_\odot$ and H$_2$ densities of $10^3$ -- $10^5$~cm$^{-3}$ \citepads{2016A&A...588A.143S, 1995A&A...294..667H, 1989ApJ...337..704L}. Located at a projected distance of 107~pc from Sgr~A$^{*}$, the compact radio source associated with the supermassive black hole in the Galactic center at a distance of $8.178 \pm 0.013_{\rm stat.} \pm 0.022_{\rm sys.}$~kpc \citepads{2019A&A...625L..10G}, Sgr~B2 is part of the central molecular zone (CMZ) and has a diameter of 36~pc \citepads[][]{2016A&A...588A.143S}\footnote{In order to be consistent with previous models we use in the rest of the paper a distance of 8.5~kpc.}.

The complex Sgr~B2 harbors two main sites of active high-mass star formation, Sgr~B2 Main (M) and North (N), which are separated by $\sim$48$\arcsec$ ($\sim$1.9~pc in projection). They have comparable luminosities of $2 - 10 \times 10^6 \, L_\odot$, masses of $5 \times 10^4 \, M_\odot$ and sizes of $\sim$0.5~pc \citepads[see][]{2016A&A...588A.143S}. These two sites of active star formation are located at the center of the envelope, occupy an area of around 2~pc in radius, contain at least $\sim$50 high-mass stars with spectral types in the range from O5 to B0, and constitute one of the best laboratories for the search of new chemical species in the Galaxy. Their masses ($3 \times 10^4 \, M_\odot$ and $6 \times 10^4 \, M_\odot$), however, correspond to only a small fraction ($\sim$1\%) of the total mass of the cloud. It is still not clear if all the available mass will form new stars only in the central cores (resulting in the formation of one or two rich star clusters), or if the envelope will fragment and form stars spread over the whole region. Understanding the structure of the Sgr~B2 molecular cloud complex is necessary to comprehend the most massive star forming region in our Galaxy, which at the same time provides a unique opportunity to study in detail the nearest counterpart of the extreme environments that dominate star formation in the Universe. Sgr~B2(M) shows a higher degree of fragmentation, has a higher luminosity, and more ultracompact \hbox{H\,{\sc ii}} regions than N \citepads[see e.g.][]{1992ApJ...389..338G, 2011A&A...530L...9Q, 2017A&A...604A...6S, 2019A&A...628A...6S, 2019A&A...630A..73M}. The larger number of \hbox{H\,{\sc ii}} regions found in Sgr~B2(M) suggests a more evolved stage and larger amount of feedback compared to Sgr~B2(N). Furthermore they have different chemical compositions: M is very rich in sulphur-bearing molecules, whereas organics dominate in N \citepads{1991ApJS...77..255S, 1998ApJS..117..427N, 2004ApJ...600..234F, 2013A&A...559A..47B, 2014ApJ...789....8N}. All this has always been interpreted as being caused by M being more dynamically evolved than N, but their age difference cannot be very large, because both of them have a high luminosity, indicating ongoing high-mass star formation, and the development of \hbox{H\,{\sc ii}} regions and their resulting feedback, which is more prevalent in M than in N, is fast at these densities and accretion rates. Thus it appears that we are observing a very crucial stage in the rapid development of massive clusters, i.e.\ M just after and N just before cluster fragmentation \citepads[see discussion of super stellar cluster in][]{2019A&A...628A...6S}. In general, high mass proto-clusters such as Sgr~B2 have complex, multi-layered structures which are difficult to study. Spectral line surveys give insights into their thermal excitation conditions and dynamics by studying line intensities and profiles, which allows one to separate different physical components and to identify chemical patterns.

Although the molecular content of Sgr~B2 was analyzed in many line surveys before \citepads[see e.g.][]{1986ApJS...60..819C, 1989ApJS...70..539T, 1991ApJS...77..255S, 1998ApJS..117..427N, 2004ApJ...600..234F, 2013A&A...559A..47B, 2014ApJ...789....8N}, many transitions of smaller molecules are located in the submillimeter and the far infrared regime, where Earth's atmosphere inhibits investigations with ground-based telescopes across much of the spectral bandwidth. Additionally, for many other molecules that are observable from the ground (e.g.\, CO, HCN), high-excitation transitions tracing the most energetic gas in star-forming regions are located in this frequency range.

The Heterodyne Instrument for the Far-Infrared \citepads[HIFI, ][]{2010A&A...518L...6D} on board of the \emph{Herschel} Space Observatory \citepads{2010A&A...518L...1P} was an ideal instrument for making these observations. It was a high-resolution ($R > 5 \times 10^5$) instrument with nearly continuous spectral coverage, which enabled the first high spectral resolution investigations of star-forming regions in this wavelength region. The complete spectral surveys of Sgr~B2(M) presented in this study were obtained as part of the Herschel Observations of EXtraOrdinary Sources (HEXOS) guaranteed time key program and span a frequency range from 480 to 1907~GHz, providing extraordinary frequency coverage in the submillimeter and far-IR.

Previously, some small sections of the spectrum were studied to find transitions of molecules rarely observed before in the interstellar medium (ISM): Oxidaniumyl (H$_2$O$^{+}$) by \citetads{2010A&A...521L..11S}, hydronium (H$_3$O$^+$) by \citetads{2014ApJ...785..135L}, water (H$_2$O) partly by \citetads{2010A&A...521L..26L}, deuterated water (HDO) by \citetads{2010A&A...521L..38C}, CH by \citetads{2010A&A...521L..14Q}, HCN by \citetads{2010A&A...521L..46R}, HF by \citetads{2011ApJ...734L..23M}, and argonium (ArH$^+$) by \citetads{2014A&A...566A..29S}. This article now presents the complete analysis of the full survey, with the key purpose of providing a reliable identification of the detected lines. Therefore, our model parameters might be slightly different from those reported in previous studies, because the contribution of all identified species is now fully taken into account. Additionally, we apply an improved fitting procedure \citepads[compared to e.g.][]{2014ApJ...789....8N}, use a different dust description, and take local line overlap effects into account.

This paper is organized in the following way. In Section~\ref{sec:DataRed}, we present the observations and outline the data reduction procedure, whereas Section~\ref{sec:DataAnalysis} describes the modeling methodology used to analyze the data set. This includes integrated line intensities and unidentified (U) line statistics. Descriptions of individual molecular fits are given in Section~\ref{sec:Results}. We give a discussion of our results in Section~\ref{sec:Discussion} and summarize our conclusions in Section~\ref{sec:Conclusions}.

    \section{Observations and data reduction}\label{sec:DataRed}

\begin{figure}[!tb]
   \centering
   \includegraphics[width=0.95\columnwidth]{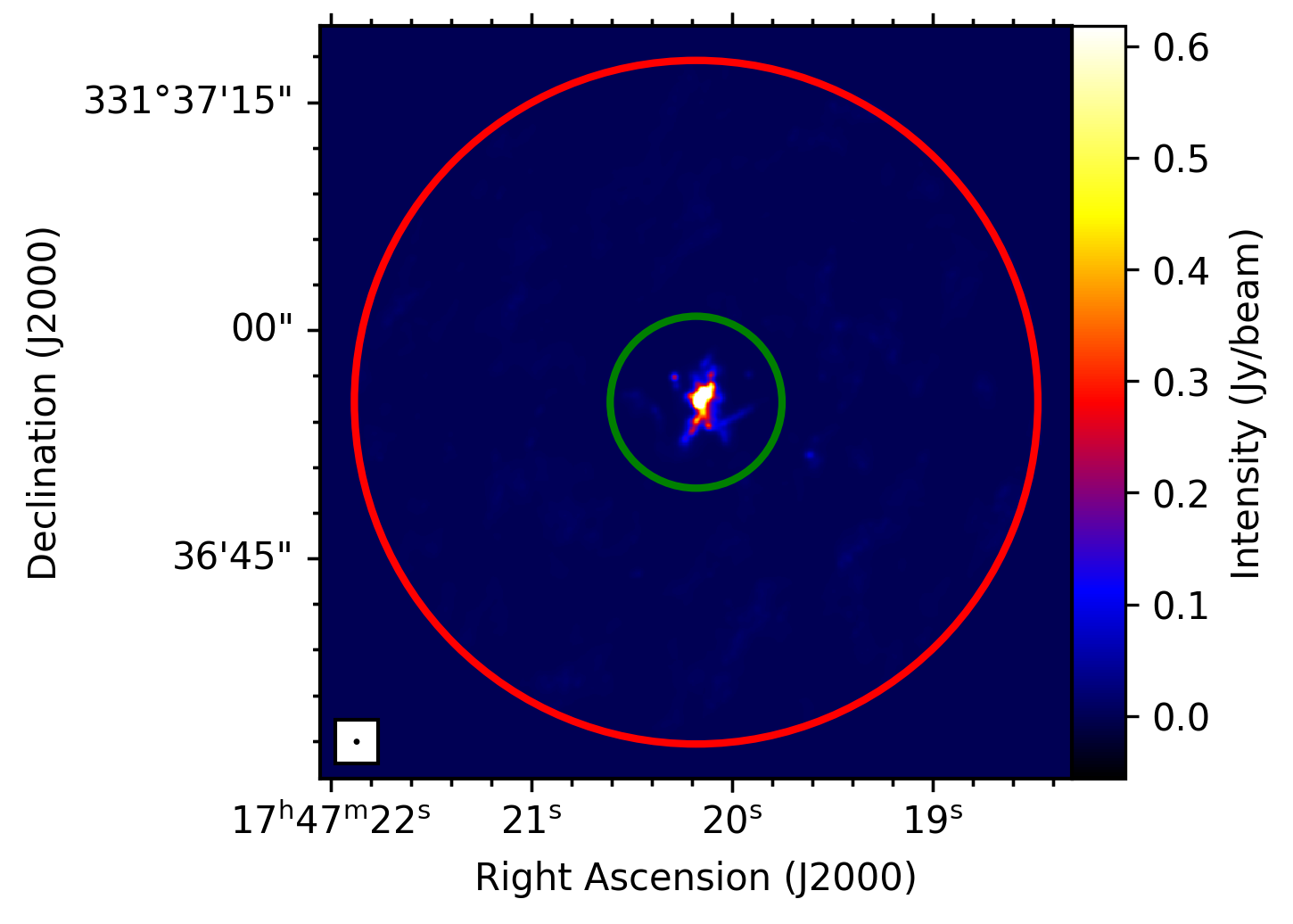}\\
   \caption{ALMA 235.5~GHz continuum image of Sgr~B2(M) at 0.4$\arcsec$ angular resolution \citepads[from]{2017A&A...604A...6S} with the FWHM HIFI beam at 480~GHz (band~1a) and 1907~GHz (band~7b) shown as red and green circles, respectively.}
   \label{fig:hifibeam}
\end{figure}

The Herschel/HIFI guaranteed time key project HEXOS \citepads[Herschel/HIFI observations of EXtraOrdinary Sources,][]{2010A&A...521L..20B} includes full line surveys of Sgr~B2(N) towards $\alpha_{\rm J2000} = 17^{\rm h} 47^{\rm m} 19 \fs 88$, $\delta_{\rm J2000} = 28^\circ 22' 18 \farcs 4$ and Sgr~B2(M) towards $\alpha_{\rm J2000} = 17^{\rm h} 47^{\rm m} 20 \fs 35$, $\delta_{\rm J2000} = 28^\circ 23' 3 \farcs 0$, covering frequency ranges from 479.581 -- 1280.148~GHz and 1444.999 -- 1907.119~GHz at a spectral resolution of 1.1~MHz with corresponding half-power beam widths of 44.9 -- 16.8$\arcsec$ and 14.9 -- 11.3$\arcsec$, see Fig.~\ref{fig:hifibeam}.

\begin{figure*}[!tb]
   \centering
   \includegraphics[width=0.95\textwidth]{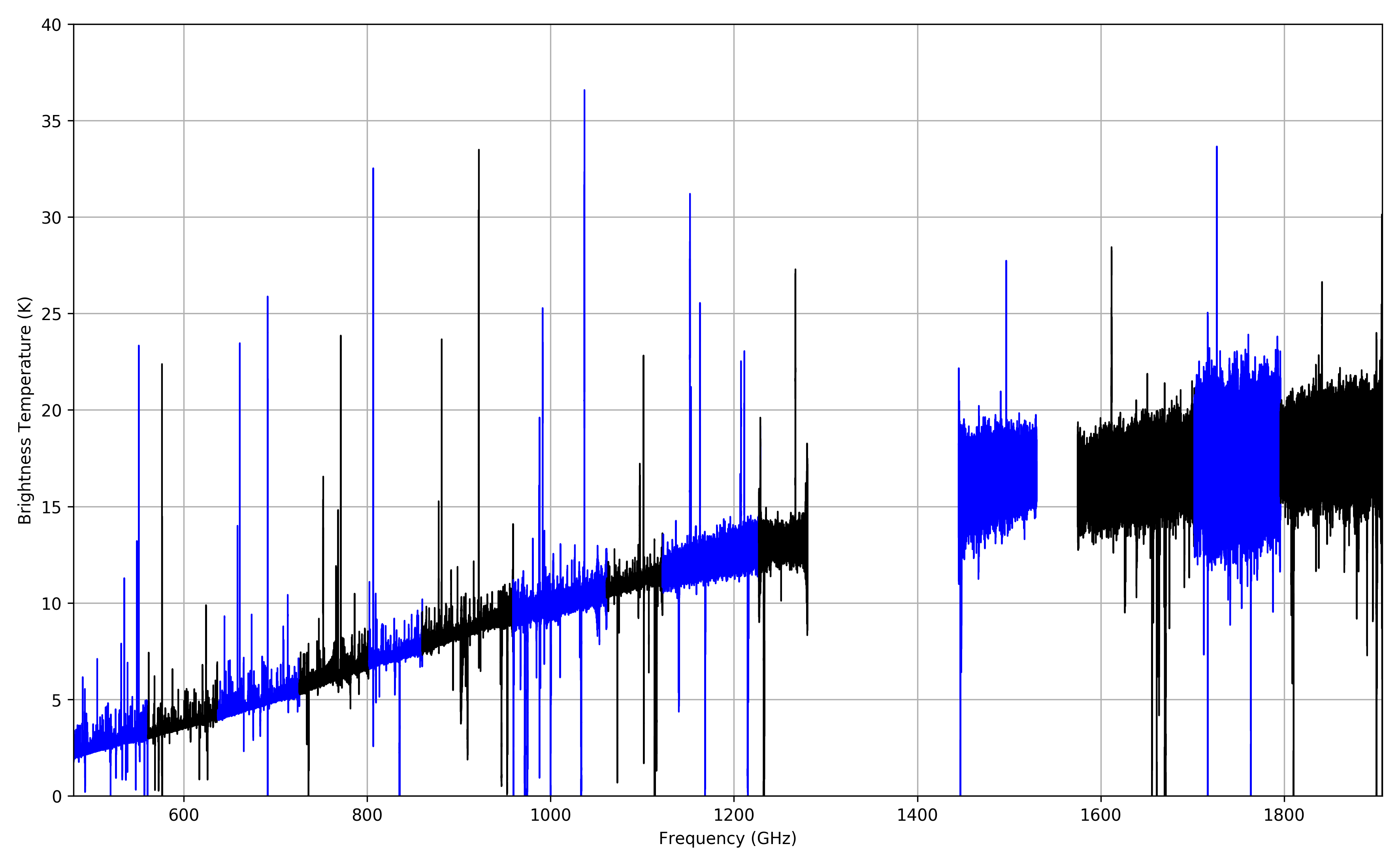}\\
   \caption{Full view of the (uncorrected) Sgr~B2(M) HIFI survey, where the HIFI bands are divided into "a" (blue) and "b" (black) bands).}
   \label{fig:fulldata}
\end{figure*}

Full spectral scans of HIFI bands 1a up to 7b towards Sgr~B2(M) were carried out respectively on March 1, 2, and 5 2010, providing coverage of the frequency range 479 through 1907~GHz. HIFI spectral scans were carried out in dual beam switch (DBS) mode, where the DBS reference beams lie approximately 180$\arcsec$ east and west.

The spectral scans have been calibrated with HIPE version 10.0 \citepads{2012A&A...537A..17R}. The resulting double-sideband (DSB) spectra were reduced and deconvolved with the GILDAS CLASS\footnote{\url{https://www.iram.fr/IRAMFR/GILDAS/}} package. For a detailed description of the calibration process see \citetads[][sec.2.1]{2016A&A...588A.143S}.

    \section{Data analysis}\label{sec:DataAnalysis}

The survey was modeled using the eXtended CASA Line Analysis Software Suite \citepads[XCLASS\footnote{\url{https://xclass.astro.uni-koeln.de/}},][]{2017A&A...598A...7M} with additional extensions (M\"{o}ller in prep.). XCLASS enables the modeling and fitting of molecular lines by solving the 1D radiative transfer equation assuming local thermal equilibrium (LTE) conditions and an isothermal source
\begin{align}\label{myXCLASS:modelFirstDist}
  T_{\rm mb}(\nu) = \sum_{m,c \in i} &\Bigg[\eta \left(\theta_{\rm source}^{m,c}\right) \left[S^{m,c}(\nu) \left(1 - e^{-\tau_{\rm total}^{m,c}(\nu)}\right)\right. \\
  &\left.+ I_{\rm bg} (\nu) \left(e^{-\tau_{\rm total}^{m,c}(\nu)} - 1\right) \right] \Bigg]\nonumber\\
  &+ \left(I_{\rm bg}(\nu) - J_\mathrm{CMB} \right),\nonumber
\end{align}
where the sums go over the indices $m$ for molecule, and $c$ for component, respectively. Here, $\eta(\theta^{m,c})$ indicates the beam filling (dilution) factor, $S^{m,c}(\nu)$ the source function, see Eq.~\eqref{myxclass:LocalOverlap}, and $\tau_{\rm total}^{m,c}(\nu)$ the total optical depth of each molecule $m$ and component $c$. Furthermore, $I_{\rm bg}$ represents the background intensity and $J_\mathrm{CMB}$ the intensity of the cosmic microwave background. Under the LTE assumption, the kinetic temperature of the gas can be estimated from the rotation temperature: $T_{\rm rot} \approx T_{\rm kin}$. As these high-mass star-forming regions have high H$_2$ densities $n_{H_2} > 10^5$~cm$^{-3}$, LTE conditions can be assumed \citepads{2015PASP..127..266M}. Line profiles are assumed to be Gaussian, but finite source size, dust attenuation, and optical depth effects are taken into account as well. Additionally, molecular parameters (e.g.\, transition frequencies, Einstein A coefficients, partition functions) are taken from an embedded SQLite database containing entries from the Cologne Database for Molecular Spectroscopy \citepads[CDMS, ][]{2001A&A...370L..49M, 2005JMoSt.742..215M} and Jet Propulsion Laboratory database \citepads[JPL, ][]{1998JQSRT..60..883P} using the Virtual Atomic and Molecular Data Center \citepads[VAMDC, ][]{2016JMoSp.327...95E}. In contrast to the standard CDMS, the database used by XCLASS describes partition functions for more than 1000 molecules between 1.07~and 1000~K, so extrapolation is no longer necessary for most molecules. The contribution of each molecule is described by multiple emission and absorption components, where each component is specified by the source size $\theta_{\rm source}$, the rotation temperature $T_{\rm rot}$, the column density $N_{\rm tot}$, the line width $\Delta v$, and the velocity offset from the systemic velocity $v_{\rm off}$. Additionally, each component can be located at a certain distance $l$ along the line of sight. Moreover, XCLASS offers the possibility to fit these model parameters to observational data by using different optimization algorithms. The modeling can be done simultaneously with corresponding isotopologues and vibrationally excited states. The ratio with respect to the main species can be either fixed or used as an additional fit parameter.

In line-crowded sources like Sgr~B2(M) line intensities from two neighbouring lines, which have central frequencies with (partly) overlapping width regions, do not simply add up if at least one line is optically thick. Here, local line overlap has to be taken into account, where photons emitted from one line can be absorbed by the other line. The extended XCLASS package provides the possibility to take local line overlap \citepads[described by][]{1991A&A...241..537C} from different components into account, where we define an average source function $S_l$ at frequency $\nu$ and distance $l$
\begin{equation}\label{myxclass:LocalOverlap}
  S_l (\nu) = \frac{\varepsilon_l (\nu)}{\alpha_l (\nu)} = \frac{\sum_{c, t} \tau_t^c (\nu) \, S_\nu (T_{{\rm rot}}^c)}{\sum_t \tau_t^c (\nu)}.
\end{equation}
Here, $\varepsilon_l$ represents the emission and $\alpha_l$ the absorption function, $T_{\rm rot}^c$ the excitation temperature, and $\tau_t^c$ the optical depth of transition $t$ and component $c$, respectively. For each frequency channel $\nu$, we take all transitions $t$ into account which belong to the current distance $l$ and whose Doppler-shifted transitions frequencies are located within a range of 5~$\Delta v_{\rm max}$. Here, $\Delta v_{\rm max}$ describes the largest line width of all components located at the current distance. Additionally, the optical depths of the individual lines included in Eq.~\eqref{myXCLASS:modelFirstDist} are replaced by their arithmetic mean at distance $l$, i.e.\
\begin{equation}\label{myxclass:LocalOverlapTau}
  \tau_{\rm total}^{l}(\nu) = \sum_c \left[\left[ \sum_t \tau_t^c (\nu) \right] + \tau_d^{c} (\nu)\right],
\end{equation}
where the sums run over both components $c$ and transitions $t$. Additionally, the dust opacity $\tau_d^{c}(\nu)$ is added as well. The iterative treatment of components at different distances, takes non-local effects into account as well. In the optically thin limit, Eq.~\eqref{myxclass:LocalOverlap} is equal to the traditional approach of describing a line as a sum of several Gaussians.

Extinction from dust is very important in the submillimeter and THz regime, particularly for a massive source like Sgr~B2(M). In XCLASS, dust opacity $\tau_d^{c}(\nu)$ of a component $c$ is described using the equation
\begin{align}\label{myxclass:dustOpacity}
  \tau_d^{c}(\nu) &= \tau_{d, {\rm ref}}^{c} \cdot \left(\frac{\nu}{\nu_{\rm ref}} \right)^{\beta^{c}} \nonumber\\
                  &= \left(N_H^{c} \cdot \kappa^{c}_{\nu_{\rm ref}} \cdot m_{H_2} \cdot \frac{1}{\chi_{\rm gas-dust}}\right) \cdot \left(\frac{\nu}{\nu_{\rm ref}} \right)^{\beta^{c}}.
\end{align}
Here, $N_H^{c}$ describes the hydrogen column density (in cm$^{-2}$), $\kappa^{c}_{\nu_{\rm ref}}$ the dust mass opacity for a certain type of dust \citepads[in cm$^{2}$~g$^{-1}$, ][]{1994A&A...291..943O}, and $\beta^{c}$ the spectral index. Additionally, $\nu_{\rm ref}$ = 230~GHz indicates the reference frequency for $\kappa^{c}_{\nu_{\rm ref}}$, $m_{H_2}$ the mass of a hydrogen molecule, and $1 / \chi_{\rm gas-dust}$ the dust to gas ratio which is set here to $1/100$ \citepads{1983QJRAS..24..267H}. The equation is valid for dust and gas well mixed.

For Sgr~B2(M), we assume a two layer model where the first layer (hereafter called core-layer) describes contributions from the 27~continuum sources identified by \citetads{2017A&A...604A...6S} and the second layer (envelope-layer) features located in the envelope. Here, we assume that all components belonging to a layer have the same distance to the observer. The source size used for the core layer is given by the angular size of the sum of solid angles of each core. For the envelope layer we assume beam filling, i.e.\ all molecules located in the envelope cover the full beam. For both layers we assume a dust mass opacity of $\kappa_{1300 \, \mu{\rm m}} = 0.511$~cm$^{2}$ g$^{-1}$ \citepads[agglomerated grains with thin ice mantles in cores;][]{1994A&A...291..943O} and a spectral index of 1.6. The dust temperature and the hydrogen column density used for the core layer is derived from the 3D radiative transfer model presented in \citetads{2016A&A...588A.143S} and is set to $T_{\rm dust}^{\rm core} = 95$~K and $N_{H_2}^{\rm core} = 1.1 \times 10^{24}$~cm$^{-2}$, respectively. In order to obtain the remaining parameters for the envelope-layer, we fit the dust temperature and the column density, $N_{H_2}$ to the HIFI continuum, while we keep the other dust parameters fixed. We obtain a dust temperature of $T_{\rm dust}^{\rm env} = 57$~K and a column density of $N_{H_2}^{\rm env} = 1.8 \times 10^{24}$~cm$^{-2}$. In order to get a smooth continuum we subtracted the continuum from the data and add a synthetic dust continuum using the aforementioned dust model.

In the first steps of the analysis of the line survey, we estimate initial parameter sets for each molecule showing at least one transition within the HIFI survey using parameters from previous surveys of Sgr~B2(M) and N \citepads{2013A&A...559A..47B, 2014ApJ...789....8N} and the new XCLASS GUI included in the extended XCLASS package. Here, we describe the continuum by the aforementioned two layer dust model. Furthermore, we assume that emission lines are modeled by components within the core layer, because these lines require in general excitation temperatures above the continuum level, which occur only in the hot cores. In contrast to that, absorption features are described by components located in the envelope layer, because here low excitation temperatures below the continuum level are needed. Additionally, some absorption lines have velocities far away from the source velocity of Sgr~B2 and cannot originate from the cores. For species with complex line shapes like H$_2$O or NH$_3$, this assumption is no longer valid, because these molecules have to be described in Non-LTE, which is beyond the scope of this paper. After adjusting the parameters by eye we use the Levenberg-Marquardt algorithm to improve the description further. A molecule is claimed as identified if all lines in the survey range which have, based on the fit, an intensity $>3\sigma$ at the line frequency, are not blended, and are detected with a line strength commensurate with the fit prediction. Additionally, we take \ion{N}{ii} (singly-ionized nitrogen), CH$_3$OCH$_3$ (dimethyl ether), NH$_2$CHO (formamide), and N$_2$O (dinitrogen monoxide) into account, although only weak lines below the 3$\sigma$ level are observed. In order to model contributions of other species, we compute for each molecule the total spectrum caused by all other identified molecules and store the numerator $\varepsilon_l$ and denominator $\alpha_l$ of Eq.~\eqref{myxclass:LocalOverlap} representing the emission and absorption functions, respectively, as function of frequency $\nu$ and layer $l$. Afterwards, we repeat fitting each identified molecule, where we add the emission $\varepsilon_l^{\rm other}$ and absorption $\alpha_l^{\rm other}$ functions describing the contributions from the other molecules to Eq.~\eqref{myxclass:LocalOverlap}, i.e.\
\begin{equation}\label{myxclass:LocalOverlapExt}
  S_l (\nu) = \frac{\sum_t \tau_t^c \, S_\nu (T_{{\rm kin}, t}^c) + \varepsilon_l^{\rm other} (\nu)}{\sum_t \tau_t^c  (\nu) + \alpha_l^{\rm other} (\nu)},
\end{equation}
to describe all other species as well. This procedure is repeated until the fit results show only minor changes.

    \subsection{Isotope ratios}\label{subsec:IsotopeRatios}

In order to simplify the comparison of our results with those derived by \citetads{2014ApJ...789....8N}, we used the same isotope ratios, i.e.\ set the $^{12}$C/$^{13}$C ratio to 20, the $^{16}$O/$^{18}$O to 250 and $^{32}$S/$^{34}$S to 13. Additionally, we take the ratios described by \citetads{2014ApJ...789....8N} for $^{16}$O/$^{17}$O to be 800, for $^{32}$S/$^{33}$S to be 75, for $^{14}$N/$^{15}$N to be 182, and for $^{35}$Cl/$^{37}$Cl to be 3. If deuterated isotopologues are identified, we leave the ratio as an additional fit parameter. Although some isotopic ratios are known to vary across the Galaxy, we kept the isotope ratios constant for all components along the line of sight because our main emphasis was a discussion of the lines originating in Sgr~B2 itself.

    \subsection{Statistics of the HIFI Spectrum}\label{subsec:Statistics}

A spectral line is considered to be detected when the peak intensity exceeds the 3$\sigma$ level of the random noise, which is calculated by first subtracting the continuum and applying astropy's \citepads{2018AJ....156..123T} sigma-clipping algorithm to each band, respectively. Additionally, we only take into account those lines with line widths comparable to that of all other detected lines (5 -- 40~km~s$^{-1}$), see Fig.~\ref{fig:KDEParam}. In this way one excludes single channel spikes, noise features, and other instrumental effects.

\begin{figure}[!tb]
   \centering
   \includegraphics[width=1.0\columnwidth]{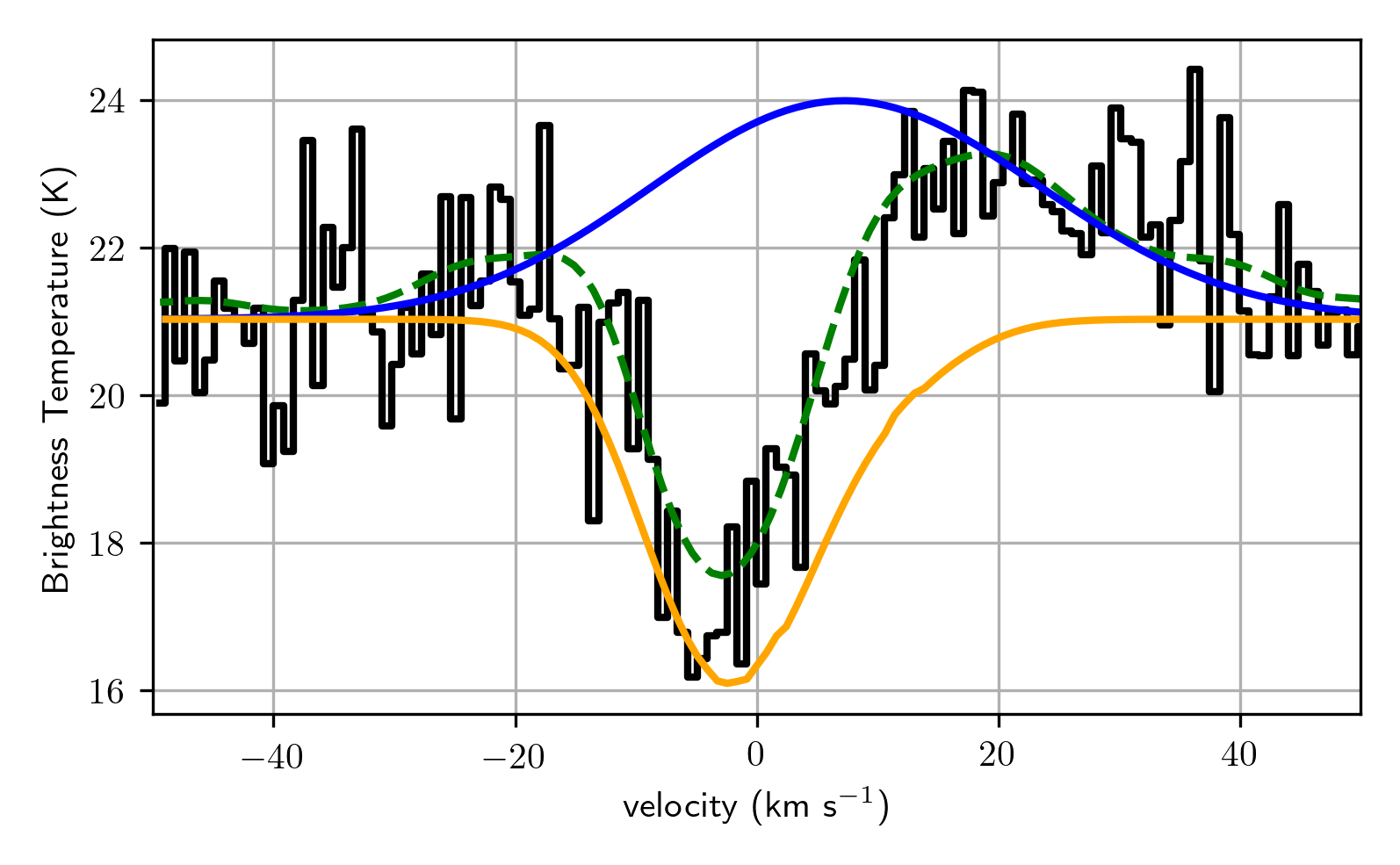}\\
   \caption{Spectrum of the 1837816.8~MHz transition of OH. The full LTE model of OH (green dashed line) is superposed on the HIFI spectrum (black histogram). The contribution of the core and the envelope are shown as blue and orange lines, respectively. In order to determine the integrated intensities of different components and velocity ranges we take only those components into account, which belong to a specific range. By doing so we neglect the contributions from other components as shown here. The sum of integrated intensities of core and envelope contributions is about a factor two higher than the integrated intensity of the full LTE model.}
   \label{fig:SummedInt}
\end{figure}

\begin{figure*}[!htb]
   \centering
   \includegraphics[width=1.0\textwidth]{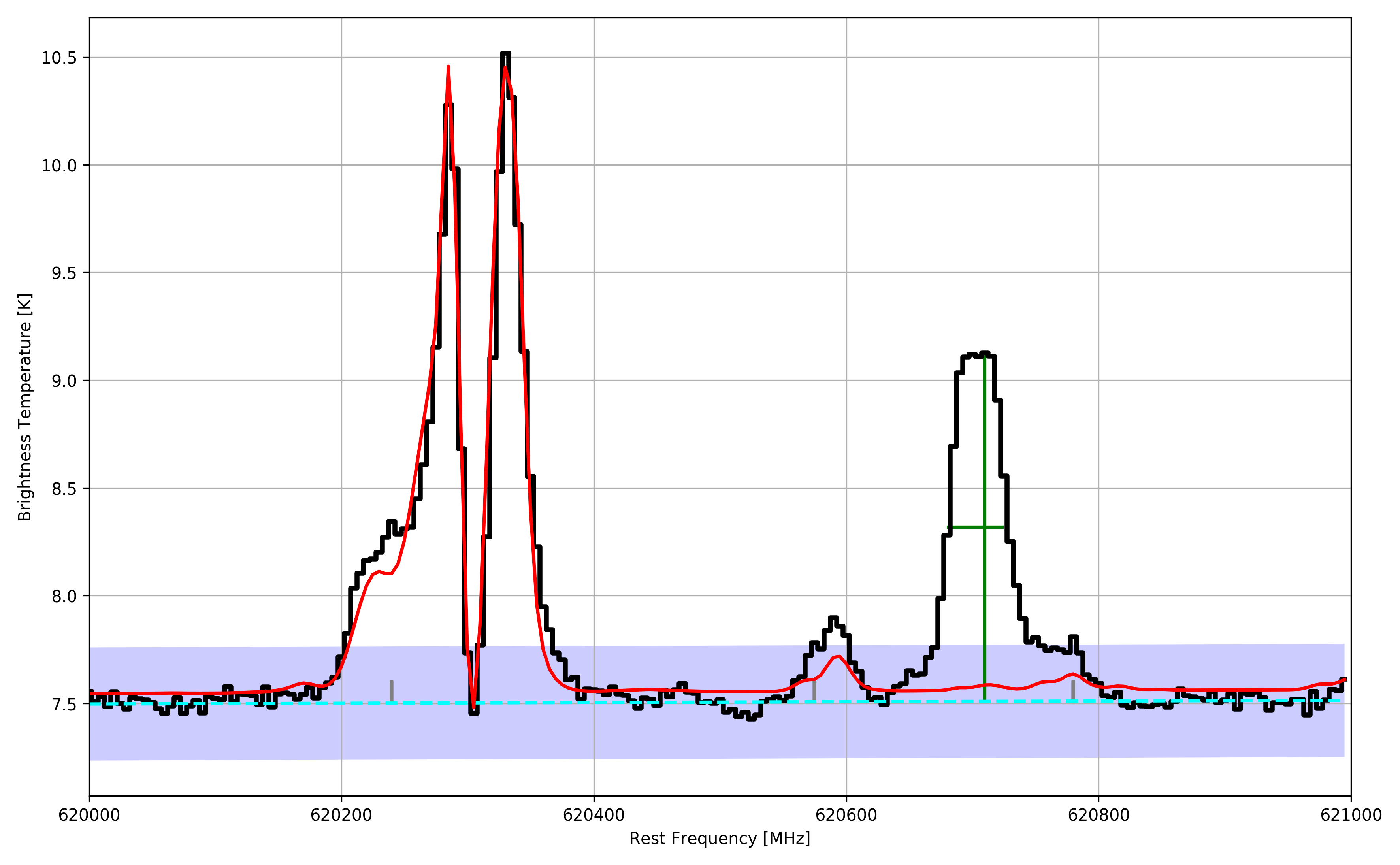}\\
   \caption{Example of unidentified peaks in the HIFI survey of Sgr~B2(M) in band 1b. The observed spectrum is shown in black, the full model in red, and the cyan dashed line marks the continuum level. The blue area represents the $+/-$3$\sigma$ level, i.e.\ all peaks that do not protrude from this area are ignored. The height and width of the unidentified feature near 620.7~GHZ is described by the horizontal and vertical green line, respectively. The peaks indicated by the small grey vertical lines are ignored, because their line widths are below the lower limit of the allowed range of line widths.}
   \label{fig:UPeakExample}
\end{figure*}

In order to determine the number of unidentified peaks, we use scipy's \citepads{2020SciPy-NMeth} peak finding algorithm {\sc scipy.signal.find\_peaks} to identify peaks in the observational data. Additionally, we calculate for each peak the corresponding width. Some features show a highly asymmetric line shape caused by contributions from neighbouring lines or other species. In order to derive a reliable width, we determine the left and right line width, i.e.\ we use the interpolated positions of left and right intersection points of a horizontal line at the respective evaluation height of the peak. If both parts differ by less than a factor two, i.e.\ less asymmetric line shape, we take the sum of both parts as the final line width. But, if both parts differ by more than a factor of two, we use twice the smaller part as final line width. By calculating the line width in the aforementioned way, we reduce the number of erroneous line widths caused by line blending, see Fig.~\ref{fig:UPeakExample}.

\begin{figure}[!b]
   \centering
   \includegraphics[width=1.0\columnwidth]{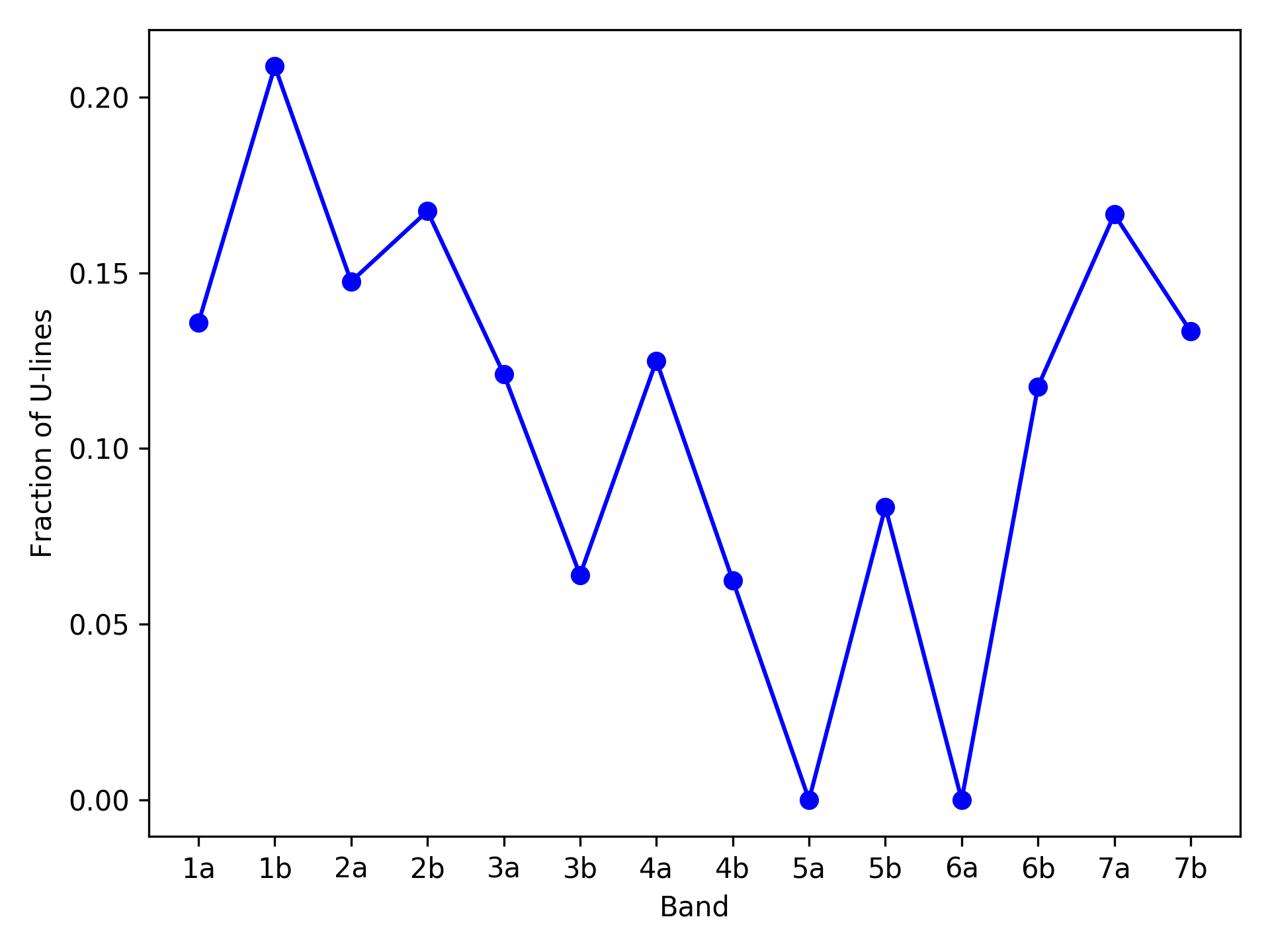}\\
   \caption{Fraction of unidentified peaks as function of band.}
   \label{fig:UPeakBand}
\end{figure}

All features showing heights above the 3$\sigma$ level and line widths which are comparable to that of all other detected lines, are compared with our model. Similar to \citepads{2014ApJ...789....8N} for Sgr~B2(N), we take a peak as unidentified, if the model for the corresponding feature is at least a factor of five weaker than the observed line intensity. The aforementioned method for emission lines is applied to absorption features as well, but here we have to start the identification process with the inverted spectrum to apply the scipy algorithm. As shown in Fig.~\ref{fig:UPeakBand}, between 5 -- 21~\% of all peaks in a single band are unidentified except for band 6a, where all peaks are identified. From this analysis, we estimate 1276 features (including both assigned and unassigned) in the Sgr~B2(M) survey, where 14~\% are not identified, which is similar to the fraction of unidentified lines ($\approx$ 8\%) in Sgr~B2(N) reported by \citetads{2014ApJ...789....8N}.\\

\begin{figure}[!b]
   \centering
   \includegraphics[width=1.0\columnwidth]{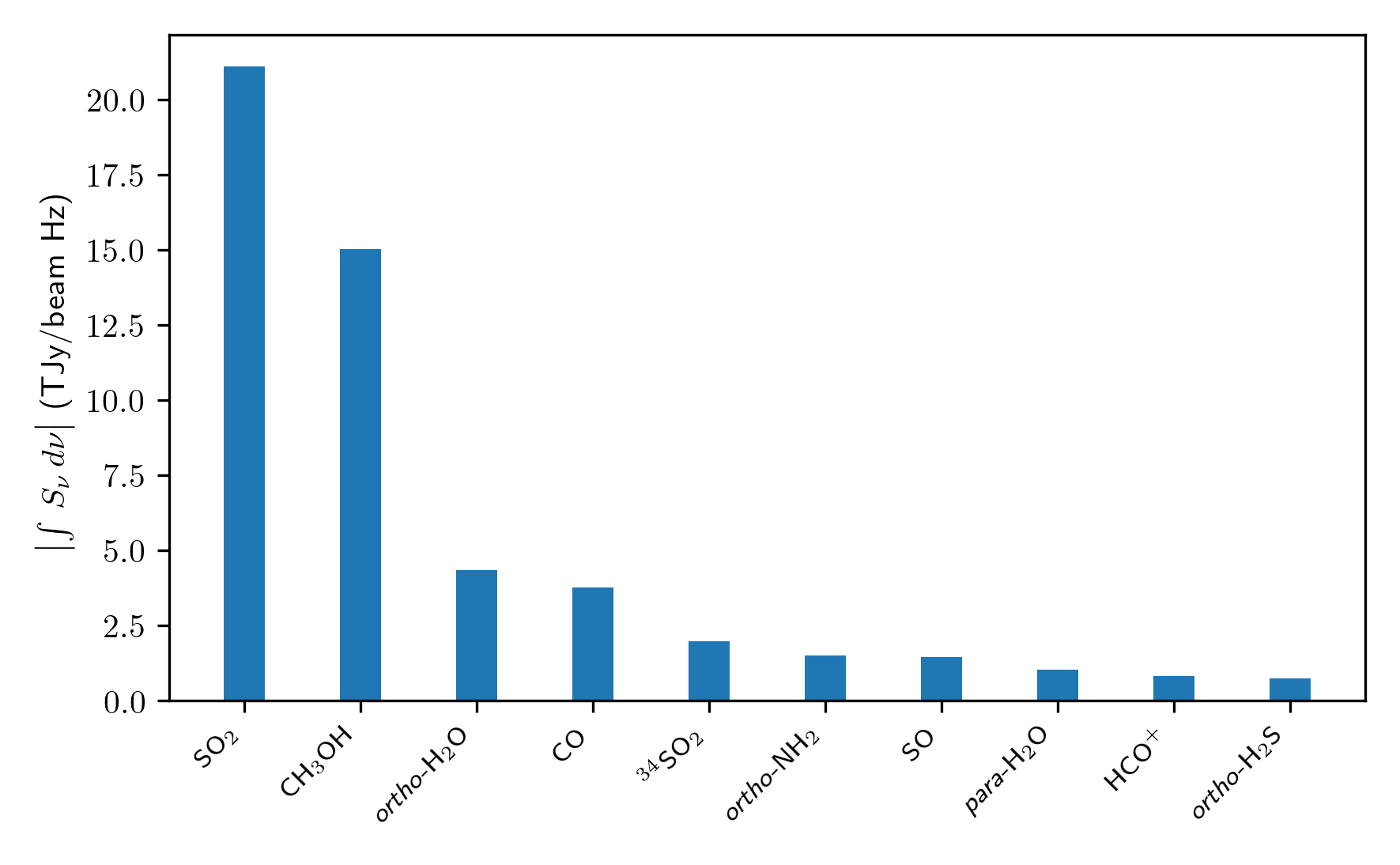}\\
   \caption{The ten strongest cooling molecules in Sgr~B2(M) in the frequency ranges covered with HIFI.}
   \label{fig:CoolingMolecules}
\end{figure}

\begin{table}[!tb]
    \centering
    \caption{Velocity ranges corresponding to different kinematic components, taken from \citetads{2014ApJ...785..135L}. Velocity offsets $v_{\rm off}$ are computed relative to the source velocity of 64~(km~s$^{-1}$).}
    \begin{tabular}{lcc}
        \hline
        \hline
        Component            & $v_{\rm LSR}$~(km~s$^{-1}$) & $v_{\rm off}$~(km~s$^{-1}$)\\
        \hline
        Galactic center (GC) &              $-$92 to $-$69 &            $-$156 to $-$133\\
        Norma arm            &              $-$47 to $-$13 &            $-$111 to  $-$77\\
        Galactic center (GC) &               $-$9 to 8     &             $-$73 to  $-$56\\
        Scutum arm           &              12 to 22       &             $-$52 to  $-$42\\
        Sagittarius B2       &              47 to 89       &             $-$17 to     25\\
        \hline
    \end{tabular}
    \label{Tab:VelStruc}
\end{table}

Some molecules show a superposition of different absorption components with systemic velocities distributed over a wide velocity range. Due to the differential rotation of the Milky Way, spectral features from clouds at different galactocentric radii can be associated with different velocity ranges, see \citepads{1979MNRAS.188..445W, 1994A&A...290..259G, 2006PASJ...58..335S, 2008AJ....135.1301V}, which are described in Table~\ref{Tab:VelStruc}. Streaming motions in the arms complicate the determination of the distance and hence the assignment to a specific spiral arm.\\

The integrated line intensities for each species are shown in Table~\ref{Tab:IntComp}. The values are derived from the best-fit LTE models, where we subtract the continuum before we integrate the spectra. Additionally, we compute the integrated line intensities for both core and envelope components as well as for different velocity ranges associated with different parts along the line of sight as described in Table~\ref{Tab:VelStruc}. In order to compute the integrated intensity for each spectral feature, we use the LTE parameters of the full model and take only those components into account which belong to the corresponding range. In doing so, we neglect contribution from other components, which may cause discrepancies between the sum of different parts and the total values, see Fig.~\ref{fig:SummedInt}.


\begin{center}
\begin{table*}[!htb]
\caption{Integrated line intensities for each molecule in the model to the Sgr~B2(M) spectral survey. The last column indicates the number of peaks with intensities above the 3$\sigma$ level.}
\label{Tab:IntComp}
\centering
\tiny
\begin{tabular}{lrrrrrrrr}
    \hline
    \hline
    Molecule:                                & \multicolumn{7}{c}{$\left| \int \, S_\nu \, d \nu \right|$~(GJy/beam~Hz)} & Number \\
    \cline{2-8}
            & $-$92 to $-$69 & $-$47 to $-$13 & $-$9 to 8 & 12 to 22 & 47 to 89 & Core & Total & Peaks \\
             & (km~s$^{-1}$) & (km~s$^{-1}$) & (km~s$^{-1}$) & (km~s$^{-1}$) & (km~s$^{-1}$)  &   &   \\
             & (GC) & (Norma arm) & (GC) & (Scutum arm) & (Env. Sgr~B2)  & (Sgr~B2) &   \\
    \hline
    \hline
    $^{12}$C                                 &  30.0 &  25.9 &  11.1 &  45.5 &     0 & 114.3 & 259.1 & 11 \\
    $^{13}$C                                 &   1.7 &   1.6 &   0.6 &   2.6 &     0 &  16.0 &  23.1 & 1 \\
    $^{12}$C$^+$                             & 386.8 & 418.5 & 902.3 & 263.9 & 404.4 &   1.3 & 2678.1 & 6 \\
    $^{14}$N$^+$                             &     0 &     0 & 132.7 &     0 &     0 &     0 & 132.7 & 0 \\
    $^{36}$ArH$^+$                           &  11.7 &   7.6 &   8.8 &   2.4 &     0 &     0 &  46.4 & 3 \\
    CH                                       &     0 &  93.5 &  12.5 &     0 & 448.3 &     0 & 839.9 & 9 \\
    $^{13}$CH                                &     0 &   5.8 &   0.7 &     0 &  27.4 &     0 &  47.3 & 0 \\
    CH$^+$                                   & 545.4 & 488.4 &     0 &     0 & 525.6 &   1.3 & 2001.3 & 7 \\
    $^{13}$CH$^+$                            &  49.2 &  23.1 &     0 &     0 &  63.4 &     0 & 266.2 & 6 \\
    CD$^+$                                   &     0 &     0 &   1.7 &  37.1 &  29.6 &     0 & 109.3 & 1 \\
    CO                                       &     0 &  38.2 &     0 & 453.8 & 6583.7 & 3764.8 & 9091.7 & 23 \\
    $^{13}$CO                                &     0 &     0 &     0 &     0 & 2994.3 & 133.0 & 3114.1 & 7 \\
    C$^{17}$O                                &     0 &     0 &     0 &     0 & 397.5 &  49.1 & 442.0 & 8 \\
    C$^{18}$O                                &     0 &     0 &     0 &     0 & 984.5 &  89.5 & 1071.1 & 7 \\
    $^{13}$C$^{17}$O                         &     0 &     0 &     0 &     0 &  15.5 &  77.0 &  92.2 & 2 \\
    $^{13}$C$^{18}$O                         &     0 &     0 &     0 &     0 &  27.7 &  16.8 &  44.5 & 2 \\
    C$_3$                                    &     0 &     0 &     0 &     0 & 1229.2 &     0 & 1229.1 & 6 \\
    CCH                                      &     0 &     0 &     0 &     0 &     0 &  38.2 &  38.2 & 4 \\
    CH$_2$NH                                 &     0 &     0 &     0 &     0 &  45.9 &     0 &  45.9 & 4 \\
    CH$_3$CN                                 &     0 &     0 &     0 &     0 &     0 & 109.5 & 109.5 & 4 \\
    CH$_3$NH$_2$                             &     0 &     0 &     0 &     0 &     0 &  34.8 &  34.8 & 1 \\
    CH$_3$OCH$_3$                            &     0 &     0 &     0 &     0 &     0 &  92.1 &  92.1 & 0 \\
    CH$_3$OH                                 &     0 &     0 &     0 &     0 & 589.7 & 15008.6 & 14423.9 & 345 \\
    $^{13}$CH$_3$OH                          &     0 &     0 &     0 &     0 &  58.0 & 354.5 & 351.2 & 9 \\
    CN                                       &     0 &     0 &     0 &     0 &     0 &  95.0 &  95.0 & 6 \\
    $^{13}$CN                                &     0 &     0 &     0 &     0 &     0 &  12.0 &  12.0 & 0 \\
    CS                                       &     0 &     0 &     0 &     0 &     0 & 420.7 & 420.7 & 10 \\
    $^{13}$CS                                &     0 &     0 &     0 &     0 &     0 &  28.4 &  28.4 & 3 \\
    C$^{33}$S                                &     0 &     0 &     0 &     0 &     0 &  15.2 &  15.2 & 0 \\
    C$^{34}$S                                &     0 &     0 &     0 &     0 &     0 &  44.0 &  44.0 & 5 \\
    \emph{ortho}-H$_2$CO                     &     0 &     0 &     0 &     0 &     0 & 314.7 & 314.7 & 18 \\
    \emph{para}-H$_2$CO                      &     0 &     0 &     0 &     0 &     0 & 181.1 & 181.1 & 14 \\
    H$_2$CS                                  &     0 &     0 &     0 &     0 &     0 &  57.6 &  57.6 & 1 \\
    \emph{ortho}-H$_2$Cl$^+$                 &     0 &     0 &  31.8 &   9.2 &  48.2 &     0 & 149.0 & 4 \\
    \emph{para}-H$_2$Cl$^+$                  &   1.5 &   2.6 &   4.6 &   0.5 &     0 &  10.0 &  19.2 & 3 \\
    \emph{ortho}-H$_2$O$^+$                  &     0 & 441.0 & 907.3 &     0 & 738.2 &  70.0 & 3706.3 & 9 \\
    \emph{para}-H$_2$O$^+$                   & 276.5 & 419.8 & 272.1 &  47.8 & 355.2 & 107.9 & 1710.4 & 13 \\
    \emph{ortho}-H$_2$O                      & 2797.7 & 3030.3 &     0 & 2065.3 & 3430.7 & 4332.4 & 11663.4 & 12 \\
    \emph{para}-H$_2$O                       &     0 &     0 &     0 &     0 & 2918.4 & 1041.7 & 3300.9 & 16 \\
    HDO                                      &     0 &     0 &     0 &     0 &  78.4 &  67.4 & 112.1 & 3 \\
    \emph{ortho}-H$_2 \! ^{17}$O             &     0 &     0 &     0 &     0 & 226.2 & 233.2 & 349.3 & 5 \\
    \emph{para}-H$_2 \! ^{17}$O              &     0 &     0 &     0 &     0 &  95.0 & 184.0 & 237.5 & 3 \\
    \emph{ortho}-H$_2 \! ^{18}$O             &     0 &     0 &     0 &     0 & 587.2 & 663.5 & 1046.6 & 7 \\
    \emph{para}-H$_2 \! ^{18}$O              &     0 &     0 &     0 &     0 & 233.7 & 540.8 & 695.6 & 5 \\
    \emph{ortho}-H$_2$S                      &  32.9 &  46.3 &  37.6 &     0 & 625.0 & 740.6 & 1410.6 & 19 \\
    \emph{ortho}-H$_2 \! ^{33}$S             &     0 &     0 &     0 &     0 &  16.6 &  33.7 &  37.3 & 2 \\
    \emph{ortho}-H$_2 \! ^{34}$S             &   2.5 &   3.7 &   3.2 &     0 & 100.8 & 299.9 & 348.1 & 10 \\
    \emph{para}-H$_2$S                       &     0 &     0 &     0 &     0 & 351.3 & 635.4 & 466.3 & 7 \\
    \emph{para}-H$_2 \! ^{33}$S              &     0 &     0 &     0 &     0 &   9.3 &  23.4 &  27.8 & 1 \\
    \emph{para}-H$_2 \! ^{34}$S              &     0 &     0 &     0 &     0 &  48.2 &  88.5 & 105.7 & 2 \\
    \emph{ortho}-H$_3$O$^+$                  & 164.9 &  76.2 &     0 &     0 & 478.8 & 345.4 & 1230.9 & 3 \\
    \emph{para}-H$_3$O$^+$                   &     0 &     0 &     0 &     0 & 887.5 &     0 & 887.5 & 2 \\
    HCCCN                                    &     0 &     0 &     0 &     0 &     0 &  17.4 &  17.4 & 3 \\
    HCN                                      &     0 &     0 &     0 &     0 &  82.2 & 347.9 & 361.9 & 13 \\
    H$^{13}$CN                               &     0 &     0 &     0 &     0 &   8.8 &   5.8 &  13.7 & 1 \\
    HCN, v$_2$=1                             &     0 &     0 &     0 &     0 &     0 & 384.5 & 384.5 & 4 \\
    HCO$^+$                                  &     0 &     0 &     0 &     0 & 224.3 & 827.8 & 883.2 & 12 \\
    H$^{13}$CO$^+$                           &     0 &     0 &     0 &     0 &  28.6 &  60.5 &  87.3 & 3 \\
    HC$^{18}$O$^+$                           &     0 &     0 &     0 &     0 &   2.6 &  13.1 &  15.7 & 1 \\
    HCS$^+$                                  &     0 &     0 &     0 &     0 &     0 &  17.5 &  17.5 & 2 \\
    HC(O)NH$_2$                              &     0 &     0 &     0 &     0 &     0 & 424.0 & 424.0 & 0 \\
    HCl                                      &     0 &     0 &     0 &  26.1 & 178.8 &     0 & 200.3 & 3 \\
    H$^{37}$Cl                               &     0 &     0 &     0 &  14.5 & 105.5 &     0 & 117.5 & 4 \\
    HF                                       & 524.6 & 695.3 & 496.3 & 158.4 & 606.5 &     0 & 2689.2 & 3 \\
    HNCO                                     &     0 &     0 &     0 &     0 & 1211.6 & 518.6 & 1586.8 & 29 \\
    HN$^{13}$CO                              &     0 &     0 &     0 &     0 &  66.4 &  30.0 &  88.3 & 0 \\
    HNC                                      &     0 &     0 &     0 &     0 &     0 & 206.1 & 206.1 & 9 \\
    \hline
    \hline
\end{tabular}
\end{table*}
\end{center}

\begin{center}
\begin{table*}
\setcounter{table}{1}
\centering
\tiny
\caption{(Continue)}
\begin{tabular}{lrrrrrrrr}
    \hline
    \hline
    Molecule:                                & \multicolumn{7}{c}{$\left| \int \, S_\nu \, d \nu \right|$~(GJy/beam~Hz)} & Number \\
    \cline{2-8}
            & $-$92 to $-$69 & $-$47 to $-$13 & $-$9 to 8 & 12 to 22 & 47 to 89 & Core & Total & Peaks \\
             & (km~s$^{-1}$) & (km~s$^{-1}$) & (km~s$^{-1}$) & (km~s$^{-1}$) & (km~s$^{-1}$)  &   &   \\
             & (GC) & (Norma arm) & (GC) & (Scutum arm) & (Env. Sgr~B2)  & (Sgr~B2) &   \\
    \hline
    \hline
    HN$^{13}$C                               &     0 &     0 &     0 &     0 &     0 &  28.4 &  28.4 & 2 \\
    HNC, v$_2$=1                             &     0 &     0 &     0 &     0 &     0 & 503.4 & 503.4 & 12 \\
    HNO                                      &     0 &     0 &     0 &     0 &  13.8 &     0 &  13.8 & 3 \\
    HOCO$^+$                                 &     0 &     0 &     0 &     0 &  61.5 &     0 &  61.5 & 5 \\
    N$_2$H$^+$                               &     0 &     0 &     0 &     0 &     0 &  68.0 &  68.0 & 4 \\
    N$_2$O                                   &     0 &     0 &     0 &     0 &     0 &  23.2 &  23.2 & 0 \\
    \emph{ortho}-NH$_2$                      &     0 & 422.6 & 553.5 &     0 & 2365.2 & 1505.9 & 4130.7 & 22 \\
    \emph{para}-NH$_2$                       &     0 &  29.5 &  71.5 &  23.5 & 1321.5 & 132.2 & 1718.4 & 4 \\
    \emph{ortho}-NH$_3$                      &   1.7 &  90.6 &  27.1 &     0 &  88.1 & 387.2 & 701.0 & 9 \\
    \emph{para}-NH$_3$                       &     0 & 124.9 & 313.1 & 103.7 & 3509.2 &     0 & 4271.6 & 13 \\
    NH                                       &     0 & 538.7 & 480.1 &     0 & 2198.5 &   3.2 & 4456.0 & 14 \\
    NO                                       &     0 &     0 &     0 &     0 &     0 & 379.7 & 379.7 & 13 \\
    NS                                       &     0 &     0 &     0 &     0 &     0 &  40.8 &  40.8 & 8 \\
    OCS                                      &     0 &     0 &     0 &     0 &     0 &  36.7 &  36.7 & 6 \\
    OH$^+$                                   & 1507.1 & 1719.2 & 633.3 & 322.4 & 732.2 &     0 & 5892.2 & 15 \\
    OH                                       &     0 &     0 &     0 &     0 & 433.2 & 563.7 & 476.2 & 2 \\
    SH$^+$                                   &  31.5 &  29.4 &  10.9 &  79.3 &  92.3 &     0 & 256.6 & 8 \\
    $^{34}$SH$^+$                            &     0 &     0 &     0 &     0 &   4.5 &     0 &   4.5 & 1 \\
    SO$^+$                                   &     0 &     0 &     0 &     0 &     0 &  57.3 &  57.3 & 8 \\
    SO$_2$                                   &     0 &     0 &     0 &     0 &     0 & 21089.4 & 21089.4 & 545 \\
    $^{33}$SO$_2$                            &     0 &     0 &     0 &     0 &     0 & 361.1 & 361.1 & 4 \\
    $^{34}$SO$_2$                            &     0 &     0 &     0 &     0 &     0 & 1976.1 & 1976.1 & 27 \\
    SO                                       &     0 &     0 &     0 &     0 &     0 & 1460.0 & 1460.0 & 43 \\
    $^{34}$SO                                &     0 &     0 &     0 &     0 &     0 & 587.0 & 587.0 & 37 \\
    SiO                                      &     0 &     0 &     0 &     0 &     0 &  31.4 &  31.4 & 5 \\
    \hline
    \hline
\end{tabular}
\end{table*}
\end{center}

Similar to Sgr~B2(N) \citepads{2014ApJ...789....8N}, methanol is the main contributor to the HIFI spectrum besides SO$_2$. These two molecules together with CO also contribute the most (64.61~\%) to the core of Sgr~B2(M). The envelope layer describing the envelope of Sgr~B2(M) and other kinetic features along the line of sight, see Table~\ref{Tab:VelStruc} are dominated by \emph{ortho}-H$_2$O, CO and OH$^{+}$. In general, \emph{ortho}-H$_2$O is the molecule with the strongest integrated
intensity over all its transitions in any velocity range, except the envelope around Sgr~B2(M), which consists mainly of CO, $^{13}$CO, and \emph{para}-H$_2$O. Additionally, \emph{ortho}-H$_2$O is not detected within the range between $-9$ and 8~km~s$^{-1}$ associated with the Galactic center, which is dominated by \emph{ortho}-H$_2$O$^{+}$ and OH$^{+}$. The part of the Norma arm ($-$47 to $-$13~km~s$^{-1}$), which is covered by our observation, contains besides \emph{ortho}-H$_2$O mainly OH$^{+}$, CH$^{+}$, HF and NH. Similarly OH$^{+}$ and CO are the molecules with the strongest integrated intensities across all their transitions, apart from \emph{ortho}-H$_2$O, in the Scutum arm (12 to 22~km~s$^{-1}$).

In general, complex molecules like NH$_2$CHO (HC(O)NH$_2$), CH$_3$NH$_2$ or CH$_3$OCH$_3$ exist only in the core of Sgr~B2(M), while some simpler molecules / ions like ArH$^{+}$ or $^{14}$N$^{+}$, are clearly not associated with Sgr~B2(M) but with the clouds and diffuse gas located throughout the Galactic arms. Additionally, all sulfur bearing molecules and their isotopologues contribute only to the cores, except SH$^{+}$ and H$_2$S. SH$^{+}$ is mainly found in the envelope, but it is also distributed along the line of sight. In addition, all cyanide molecules except HCN, which is observed in the envelope as well, are identified only in the cores. In contrast, the detected halogen molecules HF, HCl, and H$_2$Cl$^{+}$ and hydrocarbons CH (except CH$^{+}$ and CCH) appear only in the envelope. \emph{Para}-H$_2$Cl$^{+}$ shows a small contribution to the core layer as well. Most of the detected simple O-bearing and NH-bearing molecules are seen both in the core and in the envelope.

Finally, Table~\ref{Tab:IntComp} can be used to identify the major molecular coolants in Sgr~B2(M) within the frequency ranges covered with HIFI, which are shown in Fig.~\ref{fig:CoolingMolecules}. Due to the fact that the gas can cool through emission of the photons, we take only the emission components into account. Additionally, contributions from absorption lines are ignored, because they only heat the gas when the energy of absorption is released via collisions, otherwise it is radiated (isotropically), i.e.\ in a certain sense it is only a scattering. Since the densities in the absorption components are expected to be much lower than the critical densities of the absorbing molecules, collisional deexcitation and therefore heating of the gas seems unlikely.

    \section{Results}\label{sec:Results}

    \begin{figure*}[!htb]
       \centering
       \includegraphics[width=1.0\textwidth]{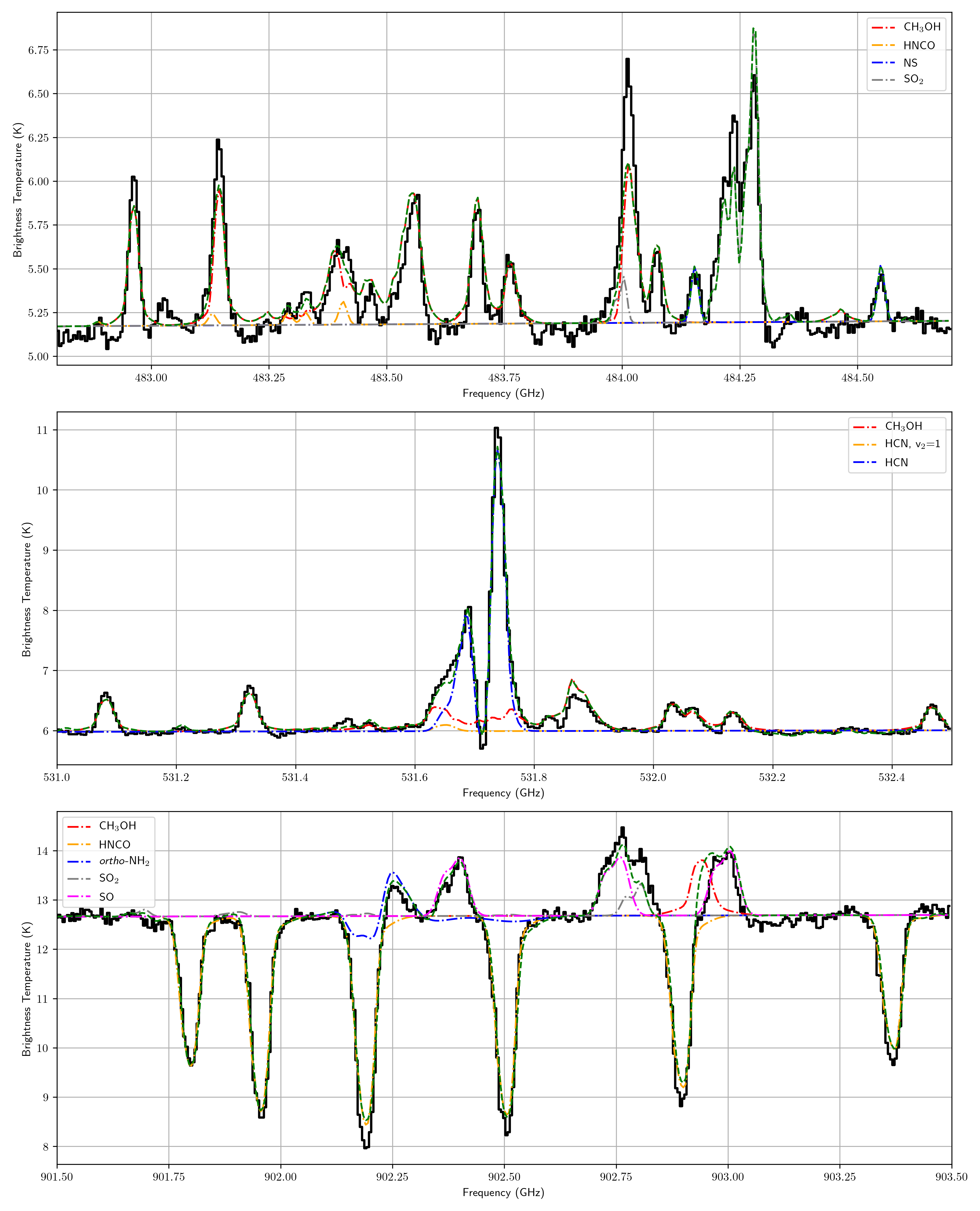}\\
       \caption{Excerpts of the HIFI survey of Sgr~B2(M) in bands 1a and 3b (solid black line), with the full model (dashed green line) and those of the molecules contributing to the spectral range of each panel.}
       \label{fig:PlotExcerpt}
    \end{figure*}

In this section we describe the molecules and their isotopologues and vibrational excited states detected in the Sgr~B2(M) HIFI survey, see Fig.~\ref{fig:PlotExcerpt}. The derived LTE parameters are described in Tabs.~\ref{CoreLTE:parameters} -- \ref{EnvLTE:parameters}. The detected molecules are divided into nine \emph{families} (simple O-bearing molecules (Sect.~\ref{subsec:SimpleOBearingMol}), complex O-bearing molecules (Sect.~\ref{subsec:ComplexOBearingMol}), NH-bearing molecules (Sect.~\ref{subsec:NHBearingMol}), N- and O-bearing molecules (Sect.~\ref{subsec:NOBearingMol}), cyanide molecules (Sect.~\ref{subsec:CyanideMol}), S-bearing molecules (Sect.~\ref{subsec:SBearingMol}), carbon and hydrocarbons (Sect.~\ref{subsec:CHMol}), halogen molecules (Sect.~\ref{subsec:HalogenMol}), other molecules (Sect.~\ref{subsec:OtherMol})), which are defined by chemical relationships; e.g.\, having the same heavy-atom backbone or the same functional group, and thus possibly related chemistry.

Since $^{12}$CO and \emph{ortho}/\emph{para}-H$_2$O have extremely complex line shapes and their source geometry is too complex as well as their optical depths are too high to be described by our modeling approach, these two molecules are only considered by purely effective line shape fits to the spectrum.

In the following sections, the quantum number $J = N + S + L$ refers to the total angular momentum of the molecule, where $N$, $S$, and $L$ describe the rotation, electron spin, and electron orbital angular momentum quantum numbers, respectively. For species without an electronic angular momentum, i.e\ $J = N$, the rotation transitions are indicated by $\Delta J$, while for species with electronic angular momentum the rotation levels are labeled as $N_J$. Symmetric rotor energy levels are described by $J_K$, where $K$ refers to the angular momentum along the symmetry axis. Energy levels for asymmetric rotors, are labeled as $J_{K_a,K_c}$, where $K_a$ and $K_c$ represent the angular momentum along the axis of symmetry in the oblate and prolate symmetric top limits, respectively.


A figure showing spectra of each detected molecule can be found in appendix\footnote{The fitted spectra are published as online material.}. For molecules with a large number of transitions, a representative sample of the detected transitions is shown.\\

    \begin{figure*}[!htb]
       \centering
       \includegraphics[width=1.0\textwidth]{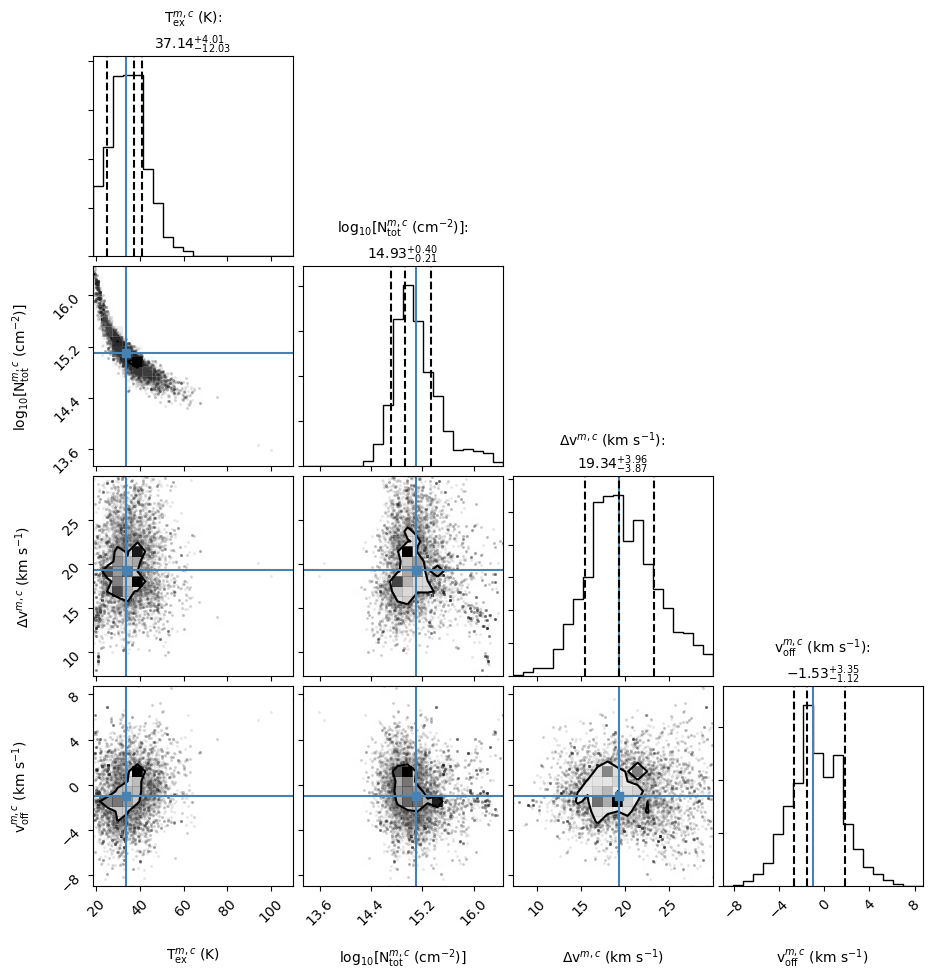}\\
       \caption{A corner plot \citepads{corner} showing the one and two dimensional projections of the posterior probability distributions of the model parameters of CN. On top of each column the probability distribution for each free parameter is shown together with the value of the best fit and the corresponding left and right errors. The left and right dashed lines indicates the lower and upper limit of the corresponding HPD interval, respectively. The dashed line in the middle indicates the mode of the distribution. The blue lines indicate the parameter values of the best fit. The plots in the lower left corner describe the projected 2D histograms of two parameters and the contours the HPD regions, respectively. In order to get a better estimation of the errors, we determine the error of the column density on log scale and use the velocity offset (v$_{\rm off}$) related to the source velocity of v$_{\rm LSR}$ = 64~km~s$^{-1}$.}
       \label{fig:ErrorEstim}
    \end{figure*}

A reliable estimation of the errors of the model parameters described in Tabs.~\ref{CoreLTE:parameters} -- \ref{EnvLTE:parameters} is not feasible within an acceptable time. We have therefore determined the errors of the four model parameters for CN as a proxy for the other parameters to give the reader some guidance on the reliability of the model parameters. Here, we use again Eq.~\eqref{myxclass:LocalOverlapExt} to take the contributions of all other species into account. The errors were estimated using the \texttt{emcee}\footnote{\url{https://emcee.readthedocs.io/en/stable/}} package \citepads{2013PASP..125..306F}, which implements the affine-invariant ensemble sampler of \citetads{2010CAMCS...5...65G}, to perform a Markov chain Monte Carlo (MCMC) algorithm approximating the posterior distribution of the model parameters by random sampling in a probabilistic space.

The MCMC algorithm starts at the estimated maximum of the likelihood function, i.e.\ the model parameters of CN described in Tabs.~\ref{CoreLTE:parameters} -- \ref{EnvLTE:parameters}, and draws 46 samples (walkers) of model parameters from the likelihood function in a small ball around the a priori preferred position. In the following, 100 burn-in steps are performed to let the walkers explore the parameter space. Afterwards, we used further 200 steps to sample the posterior.

After finishing the algorithm the probability distribution and the corresponding highest posterior density (HPD) interval of each free parameter is calculated. A HPD interval is basically the shortest interval on a posterior density for some given confidence level, i.e.\ 68~\% for 1$\sigma$, 95.4~\% for 2$\sigma$ etc. For the error estimation of CN, we are considering a 1$\sigma$ (68~\%) confidence interval, i.e.\ the HPD interval is the shortest interval that contains 68~\% of the probability of the posterior. In order to compute a HPD interval we rank-order the MCMC trace. We know that the number of samples included in the HPD is 0.68 (or another confidence level) times the total number of MCMC samples. Therefore, we consider all intervals containing this many samples and find the shortest interval. In the case of a normal distribution, an HPD interval coincides with the usual probability range, which is symmetric about the mean and includes the quantiles $\frac{n \, \sigma}{2}$ and $1 - \frac{n \, \sigma}{2}$. So, the error estimation algorithm reports the (first) mode and then the bounds on the HPD intervals. Here, the mode must not coincidence with the best fit result of the previously applied algorithms.

The posterior distributions of the individual parameters are shown in Fig.~\ref{fig:ErrorEstim}. For all histograms we find a unimodal distribution, i.e.\ there is only one best fit within the given parameter ranges.

    \subsection{Simple O-bearing Molecules}\label{subsec:SimpleOBearingMol}

\textbf{CO}. Carbon monoxide (CO) and its five isotopologues $^{13}$CO, C$^{17}$O, C$^{18}$O, $^{13}$C$^{17}$O, and $^{13}$C$^{18}$O are clearly identified, see Figs.~\ref{fig:co}, \ref{fig:c-13-o}, \ref{fig:co-17}, \ref{fig:co-18}, \ref{fig:c-13-o-17}, \ref{fig:c-13-o-18}. The vibrational excited states CO,v=1 and CO,v=2 are not detected. The energetically low lying transitions of CO up to $J_{\rm up}$~=~7 have a P-Cygni line shape with additional strong red-shifted absorptions at 21 and 78~km~s$^{-1}$. Additionally, transitions up to $J_{\rm up}$~=~13 show an inverse P-Cygni line shape, whereas transitions with higher $J$ as well as all transitions of the five isotopologues are seen in pure emission.


\textbf{CH$_3$OH}. Methanol (CH$_3$OH) is detected through a large number (351 features with intensities above the 3$\sigma$ level) of transitions ranging from the ground state to energies of about 1000~K, see Fig.~\ref{fig:ch3oh}. Methanol contributes strongly to the HIFI spectrum and the interaction of torsional and rotational motions leads to multiple energy states and a complex spectrum, which is described by a cold ($T_{\rm rot}$~=~39~K) and three warm emission ($T_{\rm rot}$~=~89 -- 308~K) in conjunction with two cold absorption ($T_{\rm rot}$~=~2 -- 10~K) components. Compared to Sgr~B2(N) \citepads{2014ApJ...789....8N}, we detect fewer absorption components but use a cold ($T_{\rm rot}$~=~39~K) and warm ($T_{\rm rot}$~=~308~K) emission component, which might be caused by the fact that the entry for methanol in the database has been extended considerably in $J$, $K$, $v_t$, and in frequency since 2014 and is now based on \citetads{2008JMoSp.251..305X}. The $^{13}$CH$_3$OH isotopologue is also fit by the same model, see Fig.~\ref{fig:c13h3oh}.


\textbf{H$_2$CO}. Transitions of \emph{ortho}- and \emph{para}-formaldehyde (H$_2$CO) have been detected in emission, see Figs.~\ref{fig:oh2co}, \ref{fig:ph2co}. The contribution of \emph{ortho}-H$_2$CO is fitted by two warm emission components ($T_{\rm rot}$~=~44 -- 68~K), whereas \emph{para}-H$_2$CO requires slightly higher temperatures ($T_{\rm rot}$~=~45 -- 74~K).




\textbf{HCO$^+$}. In good agreement to Sgr~B2(N) \citepads{2014ApJ...789....8N}, the energetically low lying transitions of formyl radical (HCO$^+$), ranging from $J_{\rm up}$~=~6 -- 14, are clearly observed, see Fig.~\ref{fig:hco+}. Transitions up to $E_{\rm low}$~=~282.4~K show self-absorption. The isotopologues H$^{13}$CO$^+$ (Fig.~\ref{fig:hc13o+}) and HC$^{18}$O$^+$ (Fig.~\ref{fig:hco18+}) are clearly identified, whereas neither HC$^{17}$O$^+$ , nor DCO$^+$ nor vibrational excited states are seen in the HIFI survey.


\textbf{HOCO$^+$}. Protonated carbon dioxide (HOCO$^+$) is identified through absorption transitions, see Fig.~\ref{fig:hoco+}, which are, in agreement with Sgr~B2(N) \citepads{2014ApJ...789....8N}, well described by two cold ($T_{\rm rot}$~=~6 -- 7~K) components. The isotopologue HO$^{13}$CO$^+$ is not clearly observed.


\textbf{OH}. Congruent with \citetads{2002ApJ...576L..77G}, we also clearly observe intra-ladder rotational transitions ($^2\Pi_{1/2} J = 3/2^+ -- 1/2^-$) of hydroxyl (OH) near 1835 and 1838~GHz, see Fig.~\ref{fig:oh}. But in contrast to \citetads{2002ApJ...576L..77G} and opposed to observations of Sgr~B2(N) \citepads{2014ApJ...789....8N}, OH shows self-absorption around these frequencies, which is well described by a warm ($T_{\rm rot}$~=~241~K) emission and a cold ($T_{\rm rot}$~=~30~K) absorption component. Since OH shows only one emission and absorption feature within the HIFI survey, respectively, the determined temperatures are poorly constrained and should only be considered as an upper limit.

\textbf{OH$^+$}. The ground-state transitions ($N$~=~1 -- 0) of the hydroxyl radical (OH$^+$) are seen in broad and deep absorptions ranging from $-$116 up to 66~km~s$^{-1}$, see Fig.~\ref{fig:oh+}. Neither higher excited lines nor its isotopologues OD$^{+}$ and $^{18}$OH$^{+}$ are detected.

\textbf{H$_2$O}. Observations of fundamental rotational transitions of \emph{ortho}- and \emph{para}-H$_2^{16}$O and H$_2^{18}$O in absorption are partly described in \citetads{2010A&A...521L..26L}, see Figs.~\ref{fig:oh2o}, \ref{fig:ph2o}, \ref{fig:oh2o-18}, \ref{fig:ph2o-18}. Additionally, as already described by \citetads{2010A&A...521L..38C}, the ground state and the 2$_{1,2}$ -- 1$_{0,1}$ transition of HDO was observed in absorption, see Fig.~\ref{fig:hdo}, whereas higher excited lines up to $E_{\rm low}$~=~66.4~K are seen in emission. Furthermore, we detect the ground-state transitions of \emph{ortho}- and \emph{para}-H$_2^{17}$O in absorption around 64~km~s$^{-1}$, Figs.~\ref{fig:oh2o-17}, \ref{fig:ph2o-17}. For \emph{ortho}-H$_2^{17}$O, we do not see higher excited lines except the 3$_{1,2}$ -- 3$_{0,3}$, and 3$_{2,1}$ -- 3$_{1,2}$-transitions, which are observed in emission. For \emph{para}-H$_2^{17}$O, we detect the next two excited states, which are closest to the ground state in emission. The ground-state transition of \emph{ortho}-H$_2^{17}$O is not clearly observed, because it is strongly blended by the \emph{para}-H$_2$O$^{+}$ transition at 607~GHz. Neither D$_2$O nor vibrationally excited states are observed except for H$_2$O,v$_2$=1, where we see a line at the correct frequency (658~GHz), but it cannot fit together with the other water lines because it probably originates from a very small, very hot region.


\textbf{H$_2$O$^+$}. The detection of ground-state transitions ($N_{K_a, K_c}$~=~1$_{10}$ -- 1$_{01}$) of \emph{ortho}- and \emph{para}-oxidaniumyl (H$_2$O$^+$) in the HIFI survey was previously described by \citetads{2010A&A...521L..11S}, see Figs.~\ref{fig:oh2o+}, \ref{fig:ph2o+}, with slightly different parameters caused by our extended dust model.


\textbf{H$_3$O$^+$}. As already described in \citetads{2014ApJ...785..135L}, metastable inversion transitions of hydronium (H$_3$O$^+$) are observed in the HIFI survey, see Figs.~\ref{fig:oh3o+}, \ref{fig:ph3o+}. We clearly identified the ground-state, the 3$_3^+$ -- 3$_3^-$, and 4$_3^-$ -- 3$_3^-$ transitions of \emph{ortho}-H$_3$O$^+$, where the energetically lowest states are seen in absorption while the transition at 1031~GHz is observed in emission. The contribution of \emph{ortho}-H$_3$O$^+$ is well described by one cold emission ($T_{\rm rot}$~=~39~K) and seven cold absorption components ($T_{\rm rot}$~=~12 -- 28~K). Due to the fact that \emph{ortho}-H$_3$O$^+$ shows only one emission feature within the HIFI range, the temperature of the emission component is not well constrained and should only be considered as an upper limit. Additionally, we detect the ground-state, the 2$_2^+$ -- 2$_2^+$, and 2$_1^-$ -- 2$_1^-$ transitions of \emph{para}-H$_3$O$^+$ in absorption and use three cold components with excitation temperatures between 5 and 31~K to model the contribution. \citetads{2014ApJ...785..135L} derived temperatures of $T_{\rm rot}$~=~177~$\pm$~54~K and $T_{\rm rot}$~=~546~$\pm$~80~K for H$_3$O$^+$ using a combination of Gaussian fits and rotational diagrams but without distinguishing between \emph{ortho} and \emph{para} states. Moreover, \citetads{2014ApJ...785..135L} consider only metastable levels of H$_3$O$^+$, that can be well described in LTE. However, we take all transitions into account, including those lines which are very far from LTE.

\textbf{SiO}. In contrast to Sgr~B2(N) \citepads{2014ApJ...789....8N}, the energetically low lying transitions ($J_{\rm up}$~=~12 -- 18) of silicon monoxide (SiO) are clearly observed in emission, see Fig.~\ref{fig:sio}, and modeled with a cold ($T_{\rm rot}$~=~25~K) and a warm ($T_{\rm rot}$~=~104~K) component. Our temperatures agree quite well with those derived by \citetads{2013A&A...559A..47B} for their emission components. Other isotopologues or vibrationally excited states could not be identified.


    \subsection{Complex O-bearing Molecules}\label{subsec:ComplexOBearingMol}

\textbf{CH$_3$OCH$_3$}. As in previous line surveys \citetads{1976A&A....48..159W, 1986ApJS...60..819C, 1998ApJS..117..427N, 2005A&A...444..521F, 2013A&A...559A..47B, 2014ApJ...789....8N} CH$_3$OCH$_3$ (dimethyl ether) is detected in weak lines, see Fig.~\ref{fig:ch3och3}, although a large number of transitions is included in the survey. Similar to \citetads{2014ApJ...789....8N} we used a single warm ($T_{\rm rot}$~=~100~K) component to describe CH$_3$OCH$_3$ in Sgr~B2(M).

\textbf{Others}. In contrast to other line surveys \citepads{1991ApJS...76..617T, 1998ApJS..117..427N, 2004ApJ...600..234F, 2013A&A...559A..47B} HCOOH (formic acid), H$_2$CCO (ketene), C$_2$H$_5$OH (ethanol), vinyl cyanide (C$_2$H$_3$CN), ethyl cyanide (C$_2$H$_5$CN), and methyl formate (CH$_3$OCHO) are not conclusively detected.

    \subsection{NH-bearing Molecules}\label{subsec:NHBearingMol}

\textbf{CH$_3$NH$_2$}. Unlike in Sgr~B2(N) \citepads{2014ApJ...789....8N}, where CH$_3$NH$_2$ (methylamine) was clearly detected, we found only a weak detection from the cores, see Fig.~\ref{fig:ch3nh2}. Most of the lines are well described by a single component with an excitation temperature of $T_{\rm rot} = 14$~K in contrast to \citetads{2013A&A...559A..47B} which used an emission ($T_{\rm rot} = 50$~K) and a cold absorption component ($T_{\rm rot} = 2.7$~K).

\textbf{CH$_2$NH}. In contrast to \citetads{1991ApJS...76..617T} and \citetads{2013A&A...559A..47B}, we see CH$_2$NH (methylene imine) only in absorption in the envelope with a quite low excitation temperature of $T_{\rm rot} = 11$~K, see Fig.~\ref{fig:ch2nh}. For Sgr~B2(N), \citetads{2014ApJ...789....8N} found CH$_2$NH within the cores ($T_{\rm rot} = 150$~K) and within the envelope ($T_{\rm rot}$~=~6 -- 11~K). The isotopologues $^{13}$CH$_2$NH and CH$_3^{15}$NH$_2$ could not be detected.

\textbf{NH}. Similar to Sgr~B2(N) \citepads{2014ApJ...789....8N} and in good agreement with \citetads{2007MNRAS.377.1122P} and \citetads{2013A&A...556A.137E}, three broad absorption features around 946, 974, and 1000~GHz have been clearly identified as hyperfine transitions (($J$~=~0,N~=~1) $\leftarrow$ ($J$~=~1,N~=~0), ($J$~=~2,N~=~1) $\leftarrow$ ($J$~=~1,N~=~0), ($J$~=~1,N~=~1) $\leftarrow$ ($J$~=~1,N~=~0)) of NH (Nitrogen monohydride) with excitation temperatures between 5 and 14~K, see Fig.~\ref{fig:nh}. 

\textbf{NH$_2$}. The Amino radical (NH$_2$) is clearly identified within the core and the envelope, see Figs.~\ref{fig:onh2}, \ref{fig:pnh2}. As in Sgr~B2(N) \citepads{2014ApJ...789....8N}, we see the $1_{11}$ -- $0_{00}$, $2_{02}$ -- $1_{11}$ and $3_{13}$ -- $2_{02}$ transitions of \emph{ortho}-NH$_2$ as well as the $2_{12}$ -- $1_{01}$ transitions of \emph{para}-NH$_2$ in absorption, while higher-energy transitions are seen in emission. Especially, the $2_{11}$ -- $2_{02}$ \emph{ortho}-transitions show strong emission features.

\textbf{NH$_3$}. Ammonia (NH$_3$) is found in absorption in the envelope and in line of sight clouds, see Figs.~\ref{fig:onh3}, \ref{fig:pnh3}. In contrast to Sgr~B2(N) \citepads{2014ApJ...789....8N}, the ground-state transition ($J$~=~0, K~=~1) $\leftarrow$ ($J$~=~0, K~=~0) near 572~GHz of \emph{ortho}-NH$_3$ shows self-absorption around 64~km~s$^{-1}$. The higher-excited transitions ($J$~=~2, K~=~0) $\leftarrow$ ($J$~=~1, K~=~0) and ($J$~=~3, K~=~0) $\leftarrow$ ($J$~=~2, K~=~0) at $E_{\rm low}$~=~28 and 86~K are seen in absorption only. All three observed transitions show multiple absorptions from $-$149 to 65~km~s$^{-1}$. Additionally, we detect ground state and energetically low lying transitions of \emph{para}-NH$_3$ up to $E_{\rm low}$~=~58~K in absorption, where the three energetically lowest transitions show multiple absorptions between $-$103 and 64~km~s$^{-1}$. Neither the \emph{ortho} nor the \emph{para} form of $^{15}$NH$_3$, NH$_2$D, NH$_3$D$^{+}$, ND$_3$, and NH$_3$,v$_2$=1 could be detected.


\textbf{N$_2$H$^{+}$}. In contrast to \citetads{2013A&A...559A..47B} and in good agreement with Sgr~B2(N) \citepads{2014ApJ...789....8N}, we see the dyazenilium ion (N$_2$H$^{+}$) only in emission through four transitions, from $J_{\rm up}$~=~6 to 9, see Fig.~\ref{fig:n2h+}. Its isotopologues N$_2$D$^{+}$, $^{15}$NNH$^{+}$, N$^{15}$NH$^{+}$, $^{15}$NND$^{+}$, N$^{15}$ND$^{+}$ and the vibrational state N$_2$H$^{+}$,v$_2$=1 could not be identified.

\textbf{N$^{+}$}. As in Sgr~B2(N) \citepads{2014ApJ...789....8N}, the fine-structure transition ($^2$P$_{1}$ -- $^2$P$_{0}$) of the single ionized atomic nitrogen [\ion{N}{ii}] at 1461.1~GHz is identified in absorption around 64~km~s$^{-1}$, see Fig.~\ref{fig:14n+}. In contrast to \citetads{2013A&A...556A.137E}, we can not find [\ion{N}{ii}] in emission.

    \subsection{N- and O-bearing Molecules}\label{subsec:NOBearingMol}

\textbf{HNCO}. As in previous line surveys \citepads{1991ApJS...77..255S, 1998ApJS..117..427N, 2013A&A...559A..47B} we also see isocyanic acid (HNCO) in Sgr~B2(M), see Fig.~\ref{fig:hnco}. All b-type transitions between the $K_a$~=~0 and 1 ladders, in the $P$ ($\Delta J$~=~-1), $Q$ ($\Delta J$~=~0), and $R$ ($\Delta J$~=~+1) branches, are observed in this survey up to $J$~=~17. Caused by the strong continuum in the far-infrared, these lines are all observed in absorption. Similar to Sgr~B2(N) \citepads{2014ApJ...789....8N}, we use three cold ($T_{\rm rot}$~=~13 -- 17~K) absorption components with slightly different velocities, which is consistent with \citetads{1986ApJ...305..405C}, who found a rotational temperature of 10~K for transitions with $E_{\rm low} <$ 40~K in the millimeter ($K_a$~=~0 a-type transitions). Like in Sgr~B2(N), we also detect high-energy ($J_{\rm up}$~=~22) a-type transitions in emission with lower-state energies ranging from 300 -- 900~K, originating from the cores. In agreement with \citetads{2014ApJ...789....8N} and in contrast to \citetads{2013A&A...559A..47B}, we use a single core component with a rotational temperature of 300~K to describe these emissions.

Additionally, we also identified the $K_a$~=~1 -- 0 $Q$ branch of HN$^{13}$CO, see Fig.~\ref{fig:hnc13o}. Due to the fact that the carbon nucleus is located very near the center of mass of the molecule, the change of the $B$ and $C$ rotational constants from HNCO to HN$^{13}$CO ($\delta B / B$ and $\delta C / C \sim 3 \times 10^{-5}$) is very small. Therefore, a-type transitions of HN$^{13}$CO are blended with their HN$^{12}$CO counterparts. However, the $A$ rotational constant changes by $\sim$2~GHz between isotopologues, which shifts the b-type transitions in frequency and thus allows their detection. Other isotopologues like H$^{15}$NCO, HNC$^{17}$O, or HNC$^{18}$O are not detected.

\textbf{NH$_2$CHO}. A-type ($\mu_a$~=~3.61~debye) and b-type ($\mu_b$~=~0.85~debye) transitions of formamide (NH$_2$CHO or HC(O)NH$_2$) are only weakly detected in emission, in contrast to Sgr~B2(N) \citepads{2014ApJ...789....8N}. Although the HIFI bandwidth covers an enormous number of transitions up to $J_{\rm up}$~=~92 we see only the 23$_{2,21}$ -- 22$_{2,20}$, 24$_{2,22}$ -- 23$_{2,21}$,26$_{2,24}$ -- 25$_{2,23}$, and 27$_{2,25}$ -- 26$_{2,24}$ transitions, which are well described with a single hot component ($T_{\rm rot}$~=~300~K), see Fig.~\ref{fig:nh2cho}.


\textbf{NO}. Similar to Sgr~B2(N) \citepads{2014ApJ...789....8N}, Nitric oxide (NO) is clearly identified through a large number of transitions in emission, see Fig.~\ref{fig:no}. The lines are well described by a cooler ($T_{\rm rot}$~=~25~K), warmer (85~K), and hotter (325~K) components which are associated with the hot core. Its isotopologues $^{15}$NO, N$^{17}$O, N$^{17}$O, N$^{18}$O, and $^{15}$N$^{17}$O are not seen in the HIFI survey.

\textbf{HNO}. The three low-energy transitions 1$_{11}$ -- 0$_{00}$, 2$_{12}$ -- 1$_{01}$, and 1$_{10}$ -- 1$_{01}$ of HNO (nitroxyl) are detected in absorption from the Sgr~B2(M) envelope, see Fig.~\ref{fig:hno}.

\textbf{N$_2$O}. In agreement with \citetads{1998ApJS..117..427N}, N$_2$O (dinitrogen monoxide) is identified in emission through nine transitions, from $J_{\rm up}$~=~20 to 29, see Fig.~\ref{fig:n2o}.

    \subsection{Cyanide Molecules}\label{subsec:CyanideMol}

\textbf{HCN}. Hydrogen cyanide (HCN), its isotopologues (H$^{13}$CN), and the vibrational excited state HCN,$v_2$=1 are clearly observed both in emission and absorption, see Figs.~\ref{fig:hcn}, \ref{fig:hc13n} \ref{fig:hcnv21}, and already described in \citetads{2010A&A...521L..46R}.


\textbf{HNC}. Like in Sgr~B2(N) \citepads{2014ApJ...789....8N}, the energetically low lying transitions of isocyanide (HNC) up to $J_{\rm up}$~=~12 are observed, see Fig.~\ref{fig:hnc}. Its isotopologue HN$^{13}$C is also detected, see Fig.~\ref{fig:hnc13}. Similar to HCN, we also observe the vibrationally excited state HNC,$v_2$=1, see Fig.~\ref{fig:hncv21}.


\textbf{CH$_3$CN}. Methyl cyanide (CH$_3$CN) is observed in many transitions with lower energies up to $E_{\rm low} \sim$~1200~K, see Fig.~\ref{fig:ch3cn}. In contrast to other observations \citepads{1997A&A...320..957D, 2013A&A...559A..47B, 2020MNRAS.497.1521A}, we see CH$_3$CN only in emission, which can be fit well with a single hot component. An additional second component, as used for Sgr~B2(N) \citepads{2014ApJ...789....8N}, is not necessary. The derived rotational temperature of $T_{\rm rot}$~=~187~K, corresponds quite well to the temperature of the emission component ($T_{\rm rot}$~=~200~K) described by \citetads{2013A&A...559A..47B}. The absence of absorption features in the HIFI survey is caused by the fact, that the energetically lowest transition of CH$_3$CN has a lower energy of $E_{\rm low}$~=~310~K well above the continuum level. Neither the $^{13}$C, nor the $^{15}$N isotopologues, nor the rotational transitions in the $\nu_8$~=~1 state are detected.




\textbf{CN}. The energetically lowest hyperfine transitions ($E_{\rm low} \le$~152~K) near 566, 680, 793, and 907~GHz of the cyano radical (CN) are clearly observed in emission, see Fig.~\ref{fig:cn}. In agreement with \citetads{2013A&A...559A..47B} and \citetads{2014ApJ...789....8N} (for Sgr~B2(N)), we use a single cold component with a slightly lower temperature of $T_{\rm rot}$~=~34~K. Absorptions, as described by \citetads{2013A&A...559A..47B}, are not seen. The isotope $^{13}$CN is also detected in emission, see Fig.\ref{fig:c13n}.


\textbf{HCCCN}. In opposite to Sgr~B2(N) \citepads{2014ApJ...789....8N}, we clearly observe the energetically low lying transitions ($E_{\rm low} \le$~672.2~K) of cyanoacetylene (HCCCN), see Fig.~\ref{fig:hc3n}. We use,  \citepads[in good agreement with][]{1987ApJ...313L...5G, 1991ApJS...77..255S, 2013A&A...559A..47B}, a cold ($T_{\rm rot}$~=~31~K) and a warm component ($T_{\rm rot}$~=~90~K) to describe the contribution of HCCCN to the HIFI spectrum. Neither its isotopologues (H$^{13}$CCCN, HC$^{13}$CCN, HCC$^{13}$CN, HCCC$^{15}$N, DCCCN, D$^{13}$CCCN, DC$^{13}$CCN, DCC$^{13}$CN, DCCC$^{15}$N, H$^{13}$C$^{13}$CCN, H$^{13}$CC$^{13}$CN, HC$^{13}$C$^{13}$CN, HC$^{13}$CC$^{15}$N) nor vibrationally excited states are detected.

    \subsection{S-bearing Molecules}\label{subsec:SBearingMol}

\textbf{H$_2$S}. In good agreement to \citetads{1994A&A...289..579T}, we clearly detected hydrogen sulfide (H$_2$S) and its isotopologues H$_2^{33}$S and H$_2^{34}$S in emission and absorption, see Figs.~\ref{fig:oh2s},~\ref{fig:ph2s},~\ref{fig:oh2s33},~\ref{fig:ph2s33},~\ref{fig:oh2s34},~\ref{fig:ph2s34}. In contrast to Sgr~B2(N), the ground-state transition (2$_{12}$ -- 1$_{01}$) of \emph{ortho}-H$_2$S shows self-absorption and a series of cold, red-shifted absorptions with velocities down to $-$104~km~s$^{-1}$. Even the ground-state transitions of both isotopologues show a combination of red-shifted emission and blue-shifted absorption features. Higher energy transitions are observed only in emission, except the (2$_{21}$ -- 1$_{10}$) transition ($E_{\rm low}$~=~8.1~K) of \emph{ortho}-H$_2$S, which shows self-absorption. \emph{Para}-H$_2$S and its isotopologues \emph{para}-H$_2^{33}$S and \emph{para}-H$_2^{34}$S are identified as well, whereas \emph{para}-H$_2^{33}$S is only weakly detected. All transitions are seen in emission, except the (2$_{02}$ -- 1$_{11}$) and (3$_{13}$ -- 2$_{02}$) transitions of \emph{para}-H$_2^{33}$S which show self-absorption as well. (Opposed to \emph{ortho}-H$_2$S, the ground state transition of \emph{para}-H$_2$S is not included in the HIFI survey).


\textbf{SO}. Sulfur monoxide (SO) is identified in emission, see Fig.~\ref{fig:so}. Similar to Sgr~B2(N) \citepads{2014ApJ...789....8N}, we use four components to describe the contribution of SO, where the two hot components have comparable temperatures ($T_{\rm rot}$~=~169 -- 174~K). But unlike Sgr~B2(N) and in good agreement with \citetads{2013A&A...559A..47B}, we use two cold ($T_{\rm rot}$~=~17 -- 28~K) instead of two warm components. The weaker $^{34}$SO lines are modeled with two warm components, see Fig.~\ref{fig:s-34-o}. The other isotopologues $^{33}$SO, $^{36}$SO, S$^{17}$O, S$^{18}$O, are not detected.


\textbf{SO$_2$}. Sulfur dioxide (SO$_2$) is clearly detected in emission, see Fig.~\ref{fig:so2}, and well described by three cold ($T_{\rm rot}$~=~15 -- 34~K), one warm ($T_{\rm rot}$~=~69~K), and two hot components ($T_{\rm rot}$~=~500~K). Its isotopologues $^{33}$SO$_2$ and $^{34}$SO$_2$ are also observed, see Figs.~\ref{fig:s33o2}, \ref{fig:s34o2}, but SO$^{17}$O and SO$^{18}$O and the vibrational excited state SO$_2$,v$_2$=1 are not conclusively detected.


\textbf{CS}. Rotational transitions of carbon monosulfide (CS) up to $J_{\rm up}$~=~22 ($E_{\rm low}$~=~542.7~K) are detected, see Fig.~\ref{fig:cs}, but in contrast to other line surveys \citepads{2013A&A...559A..47B, 2014ApJ...789....8N} only in emission. We use two warm components ($T_{\rm rot}$~=~41 -- 99~K) to model the contribution of CS. The isotopologues $^{13}$CS, C$^{33}$S, C$^{34}$S are also clearly observed, see Figs.~\ref{fig:c13s}, \ref{fig:cs33}, \ref{fig:cs34}, whereas the vibrational excited states CS,v=1, CS,v=2, CS,v=3, and CS,v=4 are not seen.


\textbf{H$_2$CS}. In agreement with Sgr~B2(N) \citepads{2014ApJ...789....8N}, thioformaldehyde (H$_2$CS) is weakly observed in emission, see Fig.~\ref{fig:h2cs}, and modeled with a single warm component ($T_{\rm rot}$~=~120~K). A low temperature contribution as described by \citetads{2013A&A...559A..47B} is not seen.


\textbf{OCS}. Comparable to Sgr~B2(N) \citepads{2014ApJ...789....8N}, all transitions of carbonyl sulfide (OCS) ranging from $J_{\rm up}$~=~40 to 60 ($E_{\rm low}$~=~455 -- 1032~K) are observed, see Fig.~\ref{fig:ocs} in emission and well described by two warm components ($T_{\rm rot}$~=~135 -- 173~K) with slightly different velocity offsets ($v_{\rm off}$~=~57.5 -- 66.5~km~s$^{-1}$). The isotopologues $^{17}$OCS, $^{18}$OCS, O$^{13}$CS, OC$^{33}$S, OC$^{34}$S, OC$^{36}$S, O$^{13}$C$^{33}$S, O$^{13}$C$^{34}$S, $^{18}$O$^{13}$CS, and $^{18}$O$^{13}$C$^{34}$S are not reliably detected.


\textbf{HCS$^{+}$}. In contrast to Sgr~B2(N) \citepads{2014ApJ...789....8N}, transitions of thiomethylium (HCS$^{+}$) are weakly observed in emission and modeled by a single warm ($T_{\rm rot}$~=~51~K) component, see Fig.~\ref{fig:hcs+}.

\textbf{NS}. Similar to Sgr~B2(N) \citepads{2014ApJ...789....8N}, nitrogen sulfide (NS) is clearly detected in emission, see Fig.~\ref{fig:ns}. But in contrast to \citetads{2013A&A...559A..47B} we use a colder emission component ($T_{\rm rot}$~=~70~K) describing the lower-energy ($\Omega$~=~(1/2)) ladder, and do not see any absorption lines of NS, which is caused by the fact that the energetically lowest transition within the HIFI survey has $E_{\rm low}$~=~110~K. Neither the $^{15}$NS nor the N$^{33}$S, nor the N$^{34}$S isotopologues are seen.


\textbf{SO$^{+}$}. Similar to \citetads{1998ApJS..117..427N}, we clearly identify the sulfoxide ion (SO$^{+}$) in emission, see Fig.~\ref{fig:so+}. The detected transitions (20/2 -- 18/2(f) to 39/2 -- 37/2(e)) have lower energies from 110 to 400~K and are well described by two warm components ($T_{\rm rot}$~=~50 -- 90~K).

\textbf{SH$^{+}$}. Consistent with \citetads{2011A&A...525A..77M}, we also observed sulfoniumylidene (or sulfanylium) (SH$^{+}$) in absorption, see Fig.~\ref{fig:sh+}. The ground-state transitions (1$_2$ -- 0$_1$ and 1$_1$ -- 0$_1$) at 526 and 683~GHz are seen with multiple velocity components between $-$144 and 66~km~s$^{-1}$, whereas the higher-excited lines (2$_3$ -- 1$_2$ and 2$_1$ -- 1$_0$) at 1083 and 1231~GHz are described by only a few velocity components around 64~km~s$^{-1}$. Additionally, we observed the ground-state transitions of $^{34}$SH$^{+}$ in absorption, see Fig.~\ref{fig:s-34-h+}.


    \subsection{Carbon and Hydrocarbons}\label{subsec:CHMol}

\textbf{CCH}. We observe ethynyl (CCH) emission toward the hot cores and detect transitions from $N$ = 6 -- 5 to 21 -- 20 ($E_{\rm up}$ = 63 -- 879~K), see Fig.~\ref{fig:cch}. The doublets, caused by the unpaired electron, are clearly separated in the survey and well described by a single component without a red-shifted wing as described by \citetads{2013A&A...559A..47B}. In contrast to \citetads{2014ApJ...789....8N} we use a lower excitation temperature of T$_{\rm rot}$~=~27~K with an order of magnitude higher column density of N$_{\rm tot}$~=~$6.4 \cdot 10^{15}$~cm$^{-2}$ . Neither the isotopologues C$^{13}$CH, CC$^{13}$H, and CCD nor the vibrational excited states CCH,v$_2$=1, CCH,v$_2$=2, and CCH,v$_3$=1 could be detected reliably.

\textbf{CH}. In addition to the six hyperfine components of CH (methylylidene radical) in the ground electronic state ($X \, ^2\Pi$), ($J$~=~3/2,N~=~1) $\leftarrow$ ($J$~=~1/2,N~=~1) near 532.8~and 536.8~GHz described by \citetads{2010A&A...521L..14Q}, we observe the lowest transitions in the $F_1$ ladder ($J$~=~5/2,N~=~2) $\leftarrow$ ($J$~=~3/2,N~=~1) at 1657.0 and 1661.1~GHz, see Fig.~\ref{fig:ch}. But in contrast to the ground-state transitions which are seen with multiple velocity components between $-$93 and 65~km~s$^{-1}$, the higher-energy transitions are described by a few velocity components around 64~km~s$^{-1}$. In contrast to the isotopologue $^{13}$CH, which is clearly identified, see Fig.\ref{fig:c13h}, the vibrational excited state CH,v=1 is not observed.

    \begin{figure*}[!htb]
        \centering
        \begin{subfigure}[t]{1.0\columnwidth}
           \includegraphics[width=1.0\columnwidth]{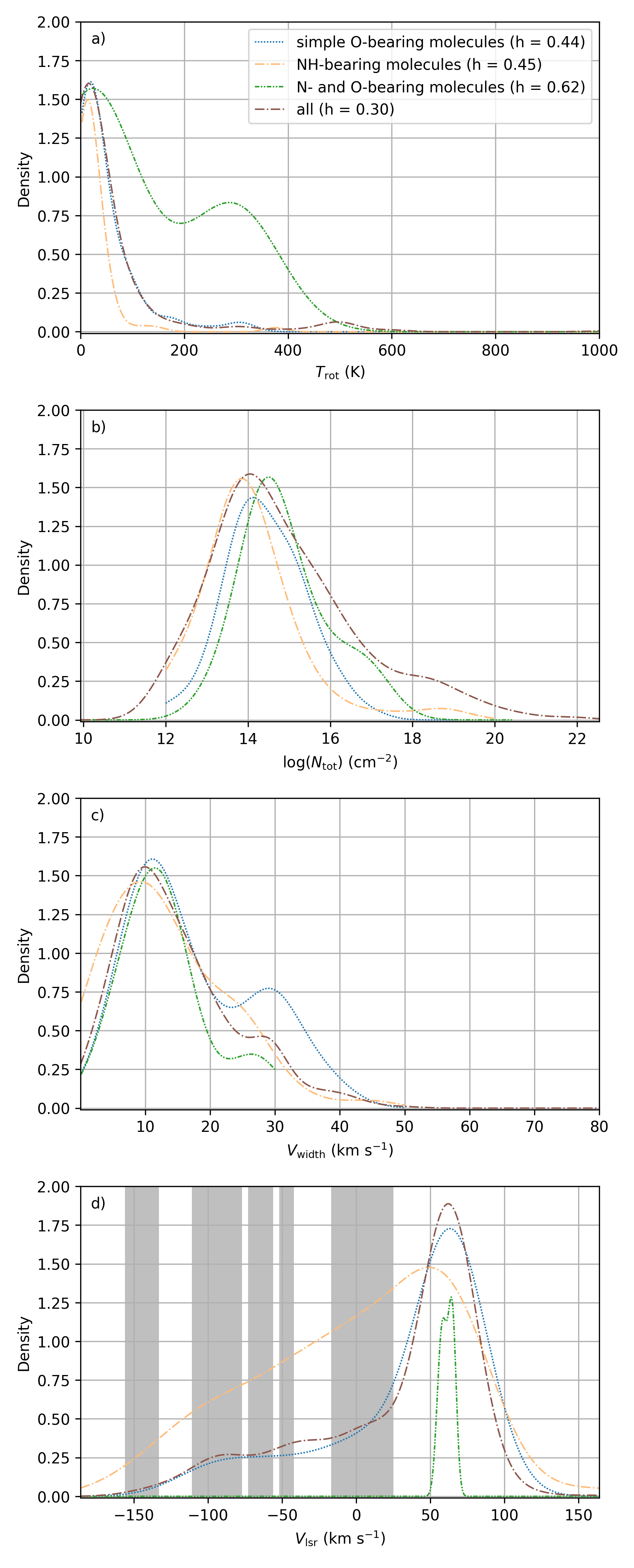}\\
        \end{subfigure}
	\quad
        \begin{subfigure}[t]{1.0\columnwidth}
           \includegraphics[width=1.0\columnwidth]{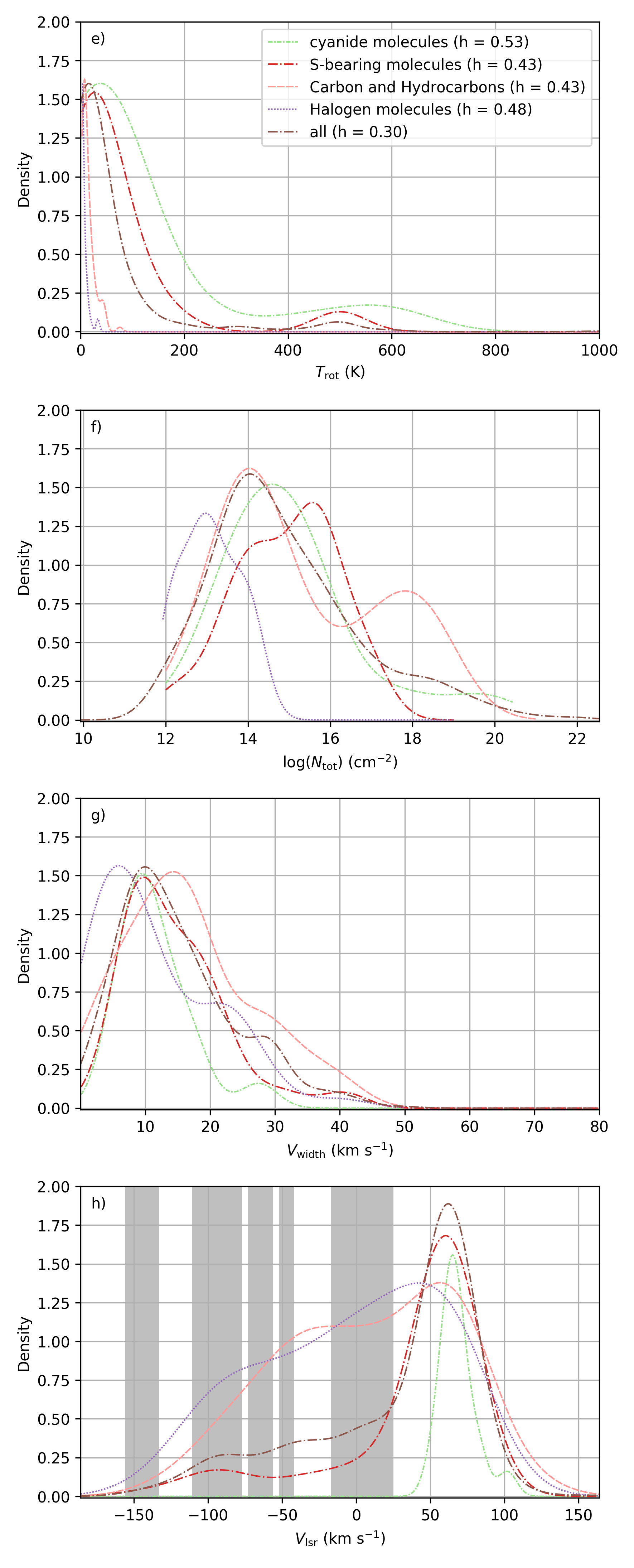}\\
       \end{subfigure}
       \caption{Normalized KDE for different model parameters and molecule families defined in Sect.~\ref{sec:Results}. Additionally, the bandwidth $h$ of all KDEs used for each molecule family is given as well}. The grey areas in d) indicate the velocity ranges described in Table~\protect\ref{Tab:VelStruc}. The families "complex O-bearing molecules" (Sect.~\ref{subsec:ComplexOBearingMol}) and "other molecules" (Sect.~\ref{subsec:OtherMol})) are not shown, because they contain only one molecule, respectively.
       \ContinuedFloat
       \label{fig:KDEParam}
    \end{figure*}

    \begin{figure*}[p]
        \centering
        \begin{subfigure}[t]{0.97\columnwidth}
           \includegraphics[width=1.0\columnwidth]{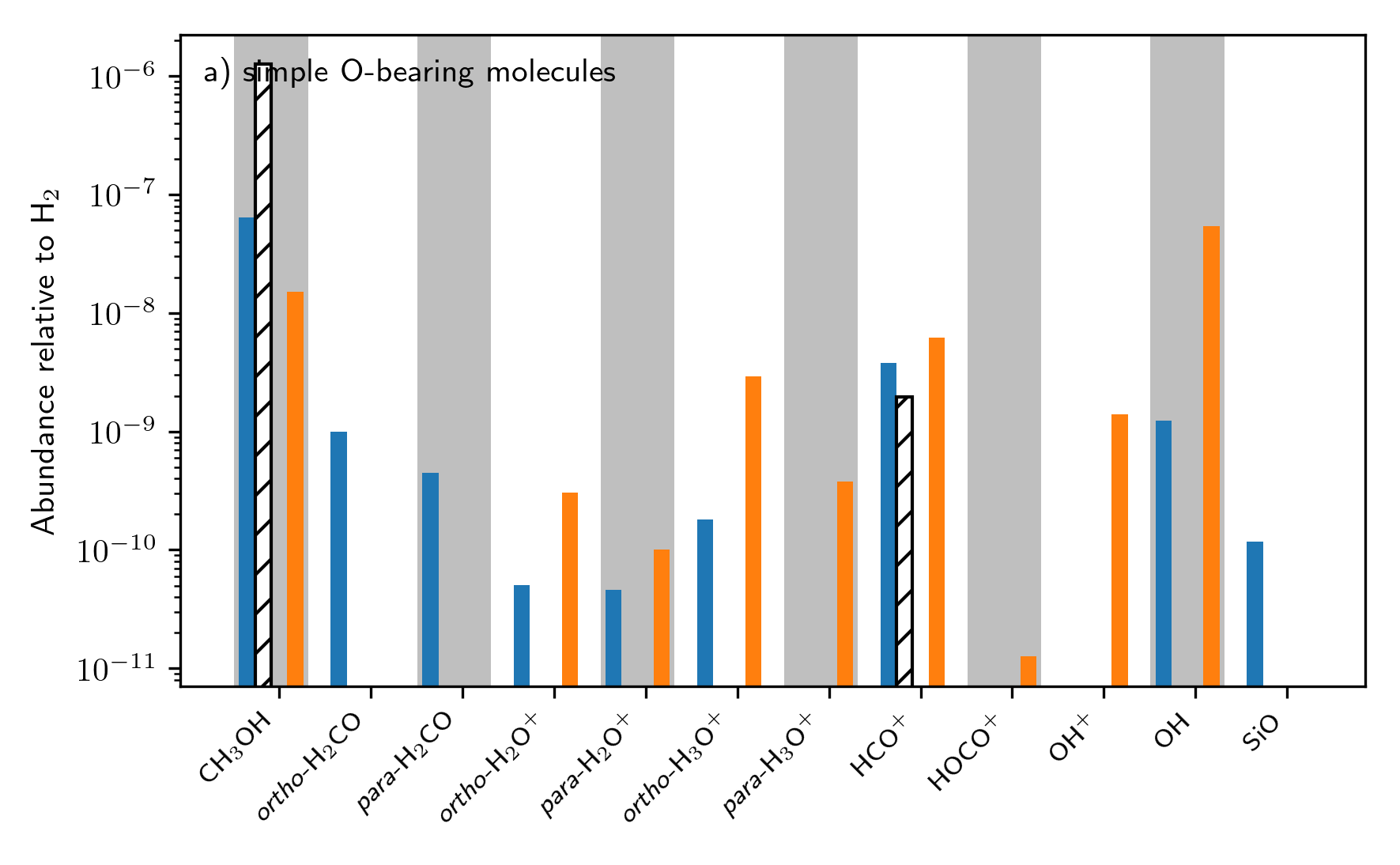}\\
        \end{subfigure}
        \begin{subfigure}[t]{0.97\columnwidth}
           \includegraphics[width=1.0\columnwidth]{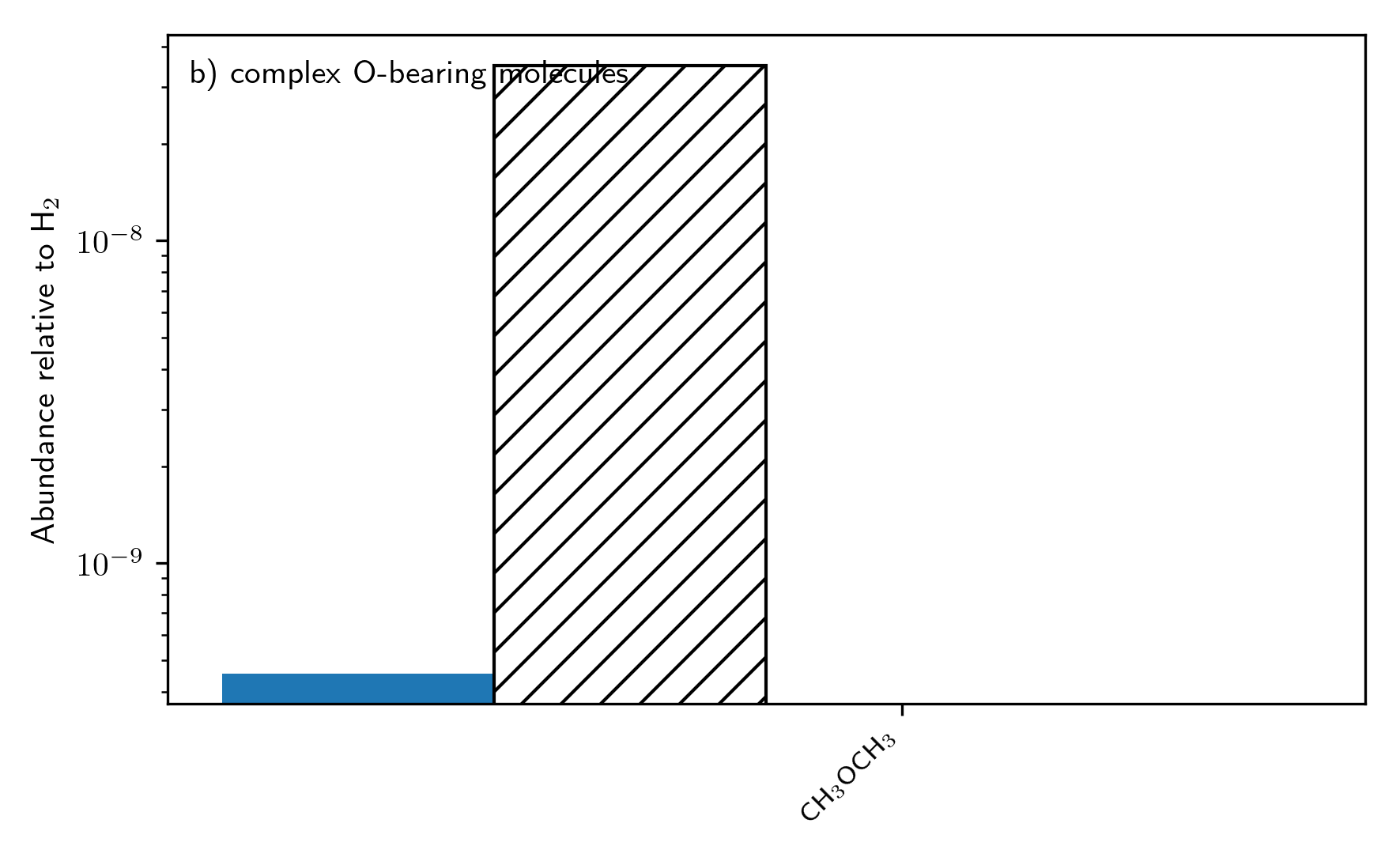}\\
        \end{subfigure}
        \begin{subfigure}[t]{0.97\columnwidth}
           \includegraphics[width=1.0\columnwidth]{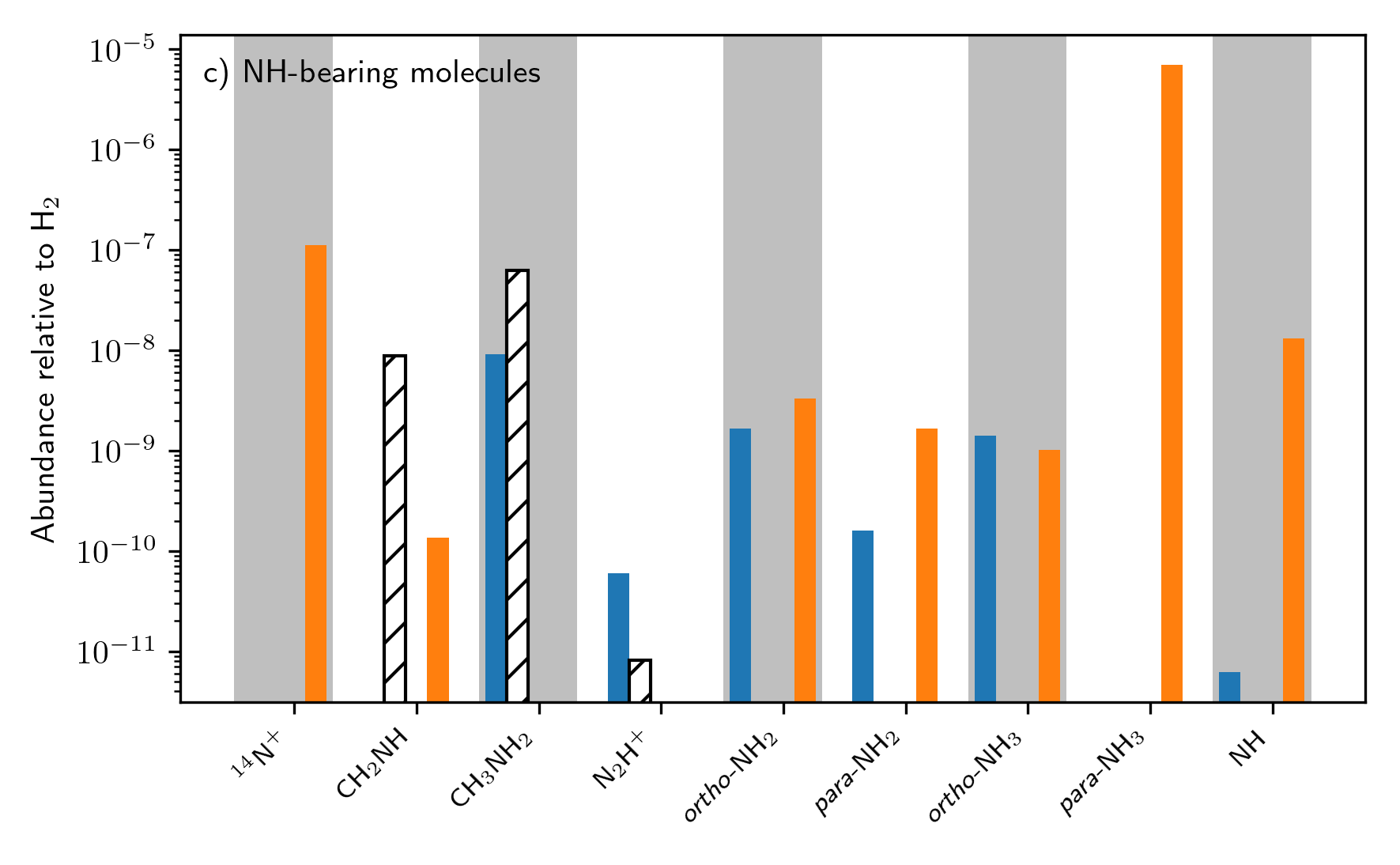}\\
        \end{subfigure}
        \begin{subfigure}[t]{0.97\columnwidth}
           \includegraphics[width=1.0\columnwidth]{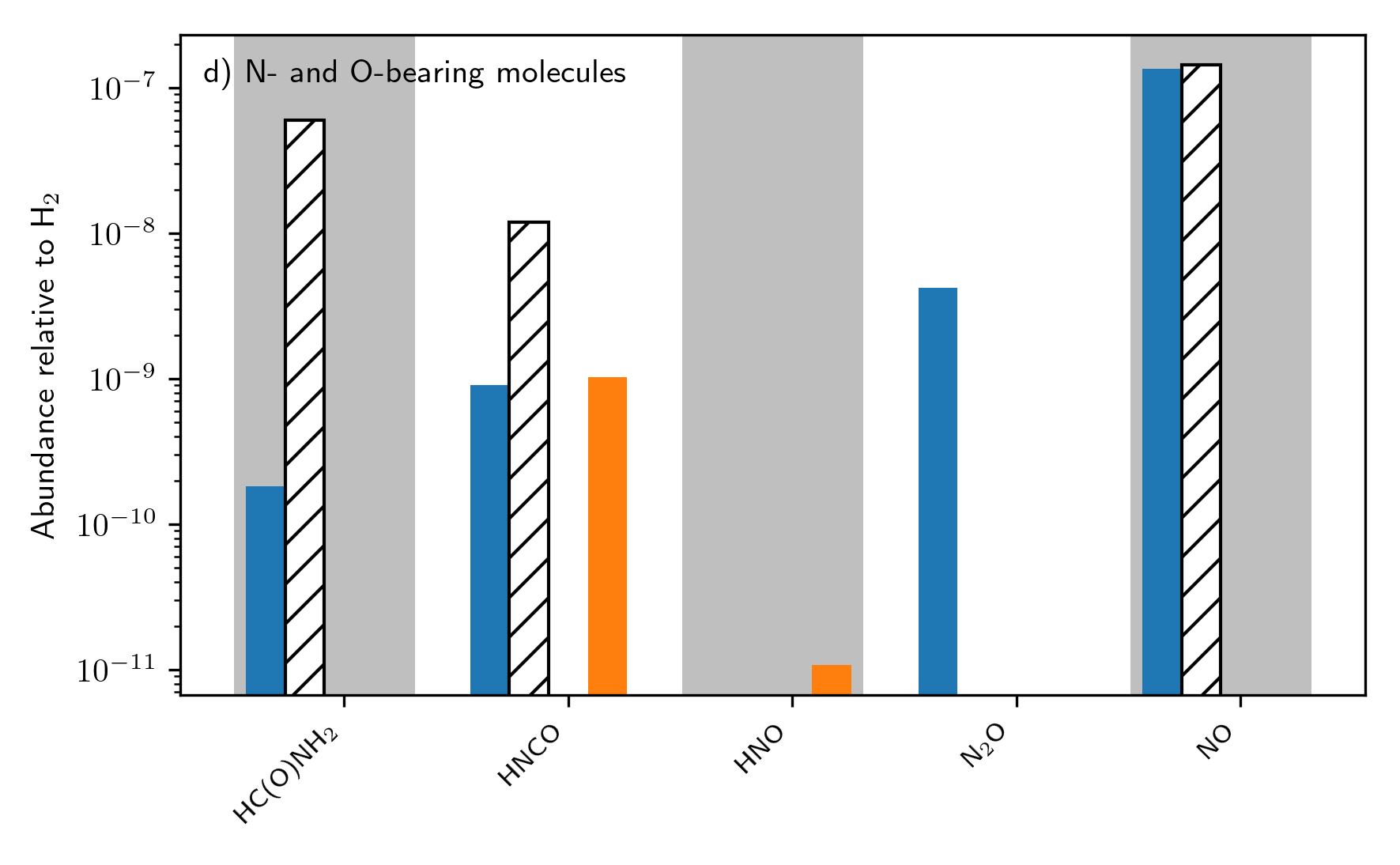}\\
        \end{subfigure}
	\quad
        \begin{subfigure}[t]{0.97\columnwidth}
           \includegraphics[width=1.0\columnwidth]{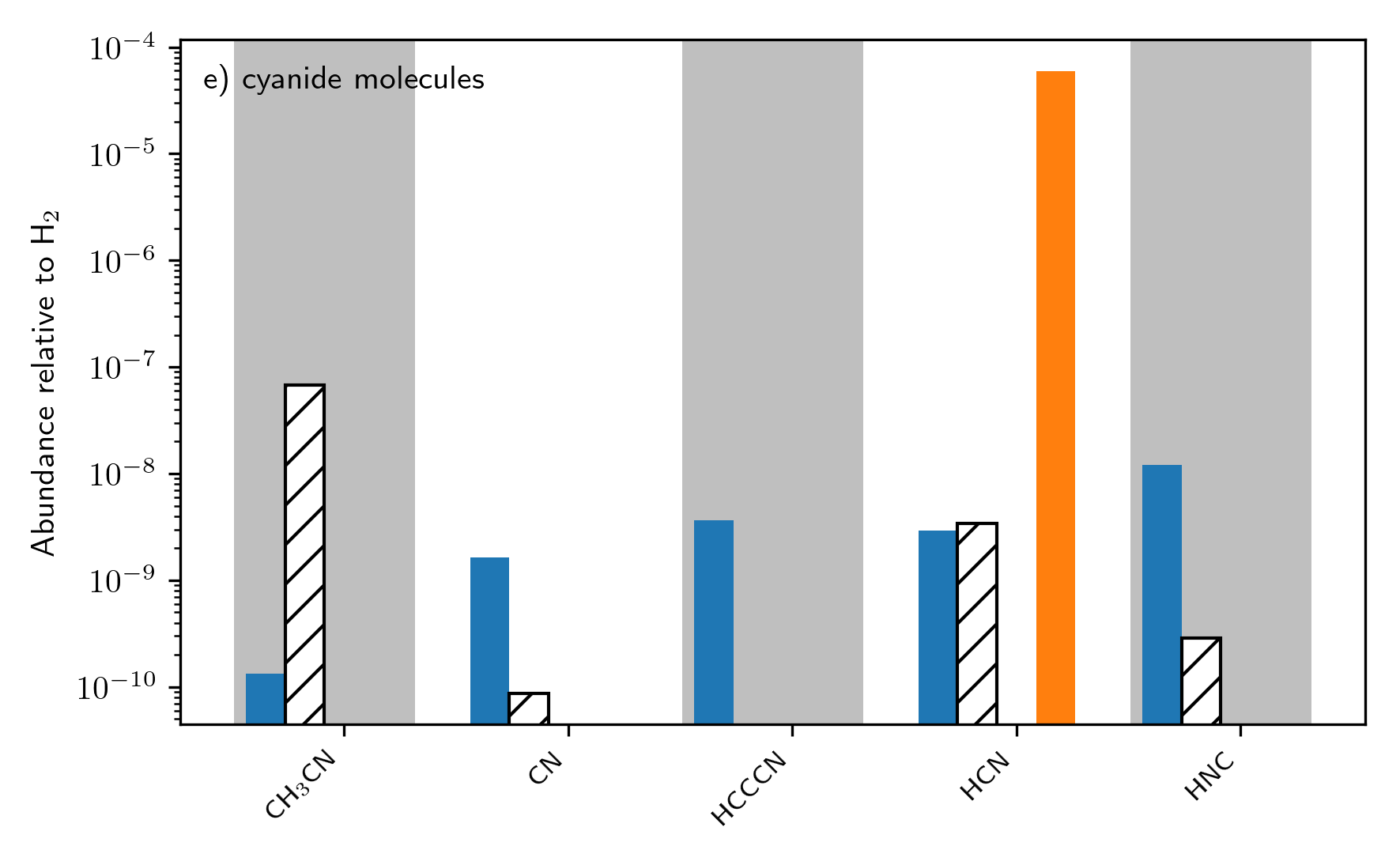}\\
       \end{subfigure}
        \begin{subfigure}[t]{0.97\columnwidth}
           \includegraphics[width=1.0\columnwidth]{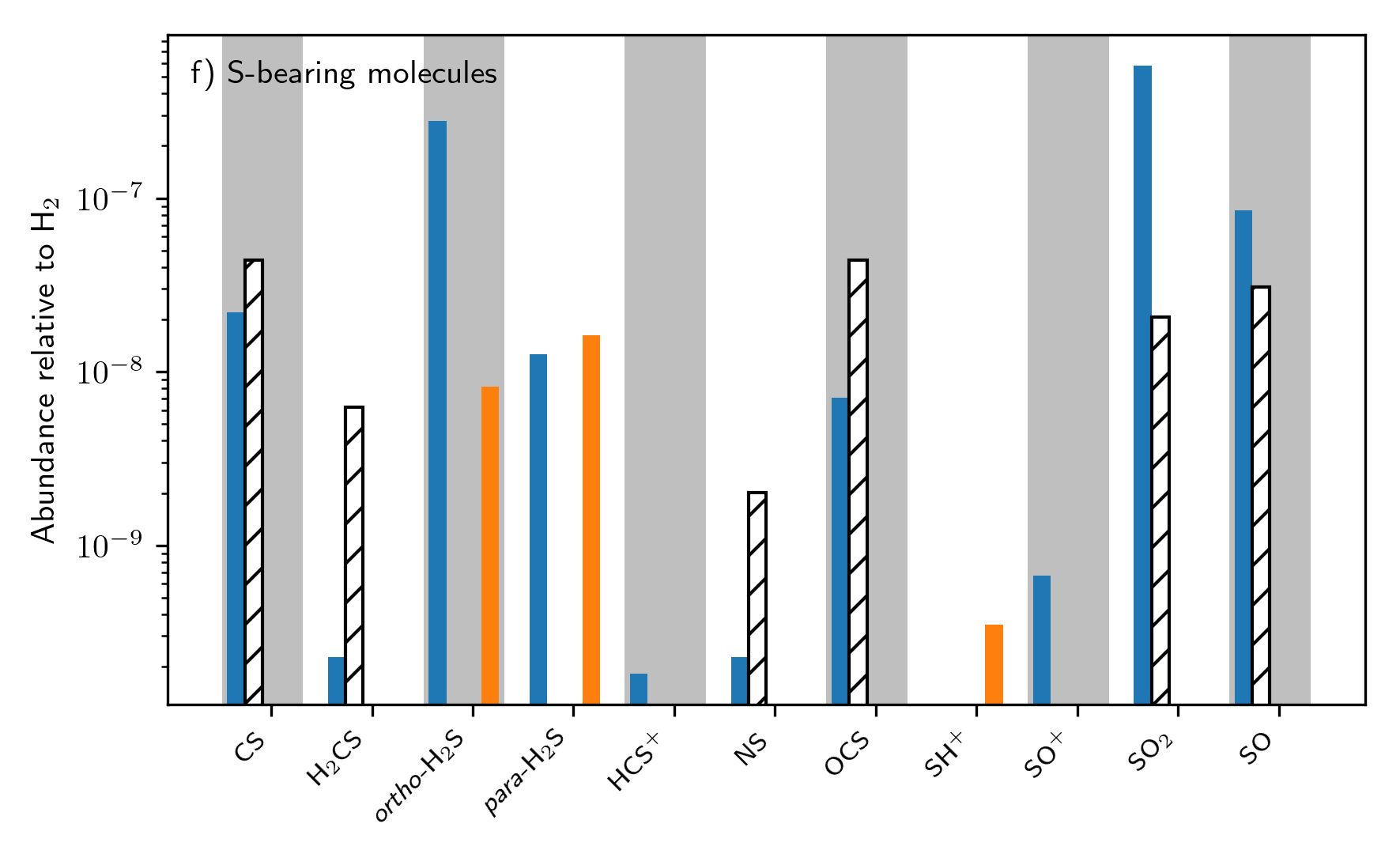}\\
       \end{subfigure}
        \begin{subfigure}[t]{0.97\columnwidth}
           \includegraphics[width=1.0\columnwidth]{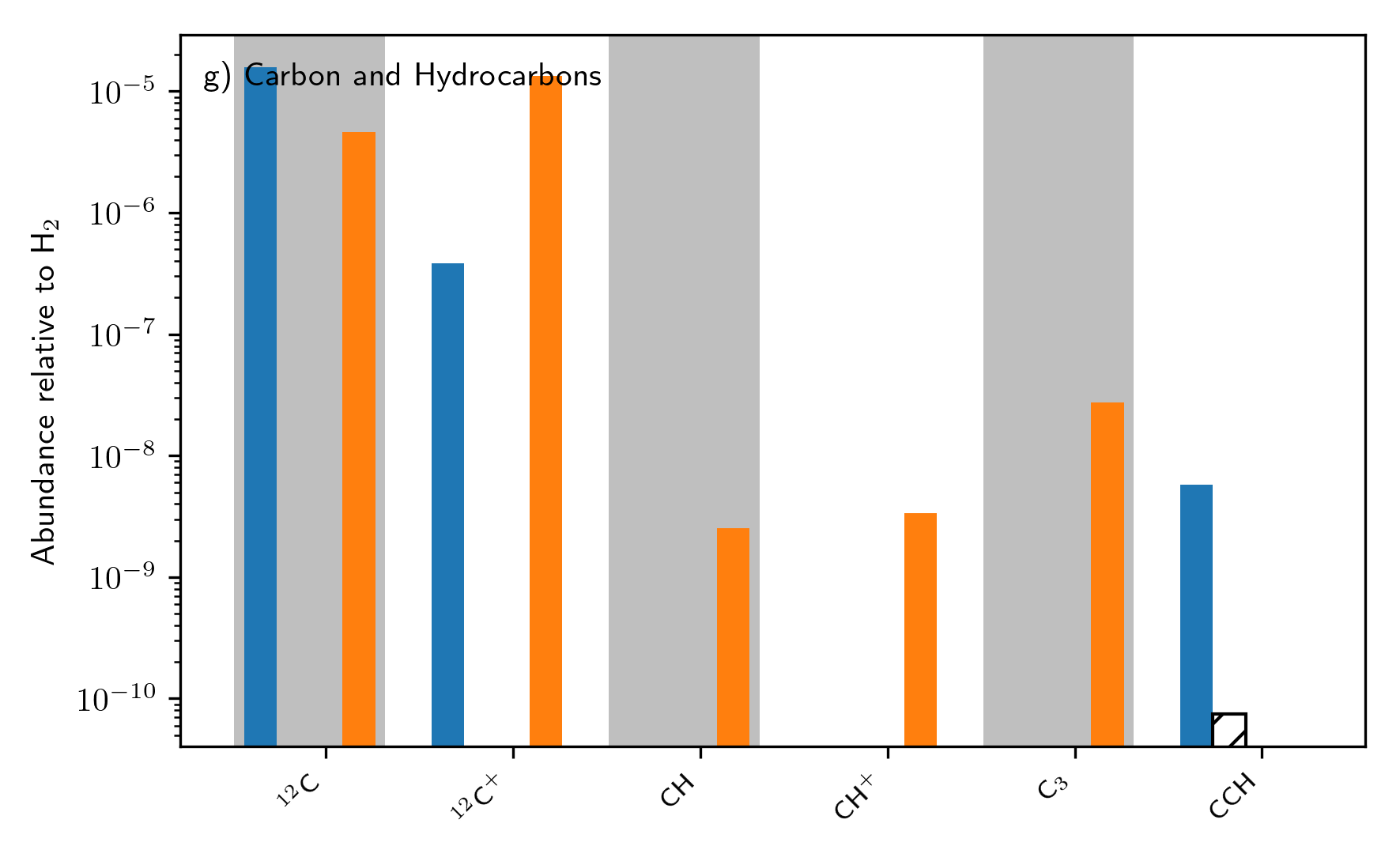}\\
       \end{subfigure}
        \begin{subfigure}[t]{0.97\columnwidth}
           \includegraphics[width=1.0\columnwidth]{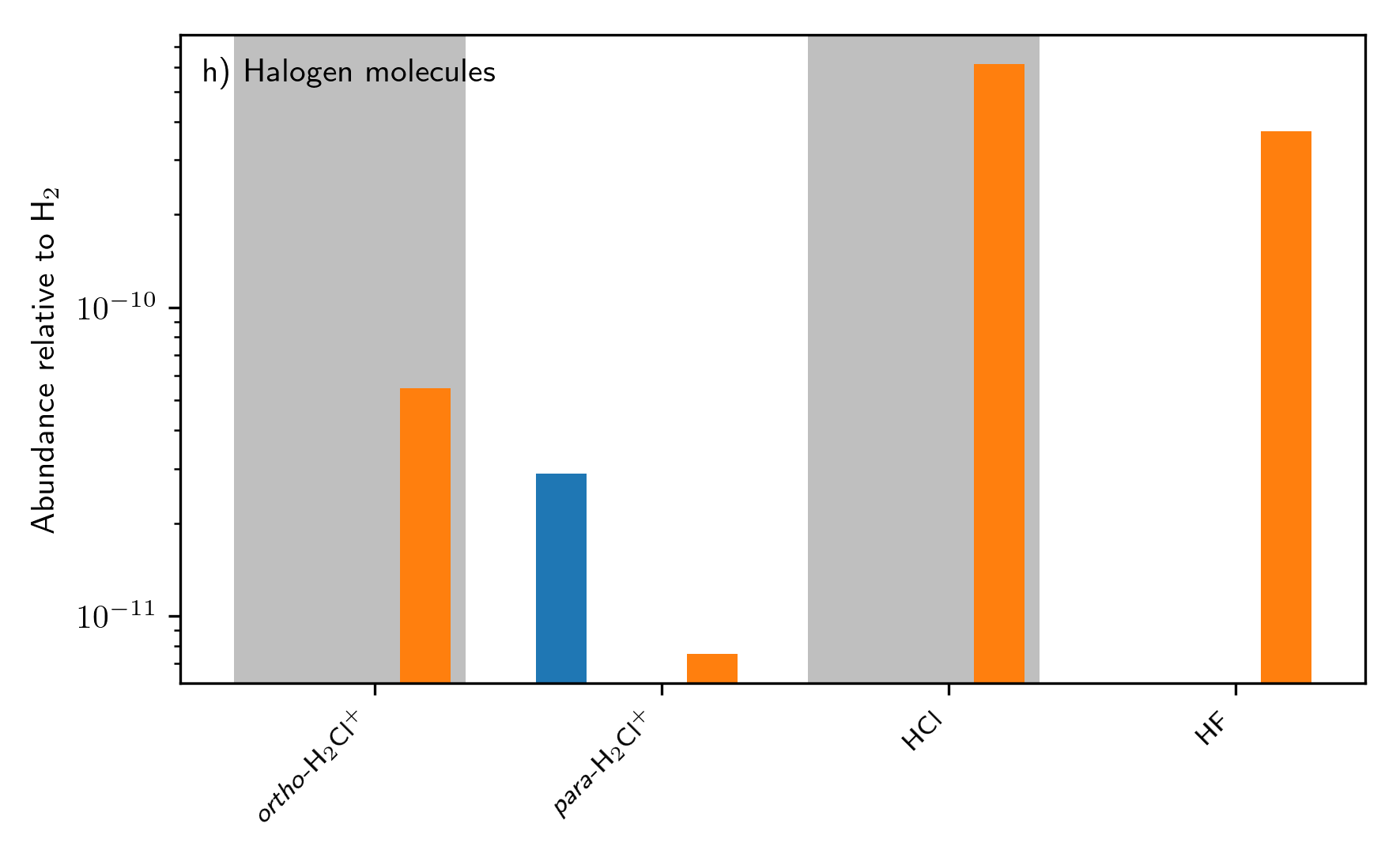}\\
       \end{subfigure}
       \caption{Abundances for different molecule families a) -- g) as described in Sect.~\ref{sec:Results}. The blue and orange bars indicate the abundances of core and envelope components, respectively. The hatched bars indicate abundances in the hot core of Sgr~B2(N) with respect to H$_2$, derived by \citetads{2014ApJ...789....8N}. Since the distinction between \emph{ortho} and \emph{para} was not done by \citetads{2014ApJ...789....8N}, we compare the abundances of the \emph{ortho} and \emph{para} molecules with the unresolved results of Sgr~B2(N).}
       \label{fig:Abun}
    \end{figure*}

\textbf{CH$^{+}$}. As in Sgr~B2(N) \citep[described by ][]{2012A&A...540A..87G, 2014ApJ...789....8N} the methylylidene ion (CH$^+$) and its isotopologue $^{13}$CH$^+$ are clearly detected, see Figs.~\ref{fig:ch+}, \ref{fig:c-13-h+}. Similar to CH, the energetically low lying ($J$~=~1 -- 0) transition is seen with multiple velocity components between $-$128 and 73~km~s$^{-1}$, whereas the next rotational transition ($J$~=~2 -- 1) at 1669.2~GHz is centered around 64~km~s$^{-1}$. Additionally, we observe CD$^{+}$ but only two transitions are located within the HIFI survey, so that we are not able to derive reliable physical parameters, see Fig.~\ref{fig:cd+}.

\textbf{C}. Similar to Sgr~B2(N) \citepads{2013A&A...556A.137E, 2014ApJ...789....8N} the neutral atomic carbon [{\sc C i}] and its isotopologue $^{13}$C are clearly identified as well, see Figs.~\ref{fig:12c}, \ref{fig:13c}. The $^3$P$_1$ -- $^3$P$_0$ transition at 492~GHz and the $^3$P$_2$ -- $^3$P$_1$ transition at 809~GHz are modeled by multiple components within the core and the envelope layer.

\textbf{C$^{+}$}. Due the fact that only one transition, the $^2$P$_{3/2}$ -- $^2$P$_{1/2}$ fine structure transition at 1901~GHz, is included in the HIFI survey, a reliable quantitative description of the single ionized atomic carbon was not possible, see Fig.~\ref{fig:12c+}. As reported by \citepads{2014ApJ...789....8N} for Sgr~B2(N), we also see [\ion{C}{ii}] in multiple absorption features around 64~km~s$^{-1}$.

\textbf{C$_3$}. As reported by \citetads{2000ESASP.456...99G, 2000ApJ...534L.199C, 2007MNRAS.377.1122P, 2014ApJ...789....8N} we also see vibrational excited transitions of the nonpolar linear carbon trimer, C$_3$ in absorption around 64~km~s$^{-1}$, see Fig.~\ref{fig:c3}. The rovibrational $Q(12)$, $Q(10)$, $Q(8)$, $Q(6)$, $Q(4)$, $Q(2)$, $P(4)$, and $P(2)$ transitions are well described with two cold components with excitation temperatures of $T_{\rm rot}$~=~8 -- 10~K, in good agreement with \citetads{2000ApJ...534L.199C, 2020A&A...633A.120G}. But we need an additional warm component with $T_{\rm rot}$~=~46~K to describe absorptions with lower-state energies ranging from 26 -- 68~K. Energetically lower lying transitions of the $Q$ band were not detected.

    \subsection{Halogen Molecules}\label{subsec:HalogenMol}

\textbf{HF}.The detection of ($J$~=~1) $\leftarrow$ ($J$~=~0) transition of HF (ethynyl radical) at 1232~GHz in the HIFI survey was previously described by \citetads{2011ApJ...734L..23M}, see Figs.~\ref{fig:hf}, with slightly different parameters caused by our extended dust model. Its isotopologues DF is not detected.

\begin{table*}[t]
    \centering
    \tiny
    \caption{\emph{ortho}/\emph{para} ratios for different molecules and velocity ranges. Results derived for molecules marked with '*' might be affected by the underlying LTE assumption. $^{(a)}$ \citetads{1999ApJ...518..733D},  $^{(b)}$ \citetads{2013JPCA..11710018G}, $^{(c)}$ \citetads{2010A&A...521L..11S}, $^{(d)}$ \citetads{2019A&A...632A...8P}, $^{(e)}$ \citetads{2014ApJ...781..114C}, $^{(f)}$ \citetads{2019MNRAS.487.3392F}, $^{(g)}$ \citetads{2016A&A...596A..35L}, $^{(h)}$ \citetads{2013ApJ...770L...2F}.}
    \begin{tabular}{lcccccccc}
        \hline
        \hline
        Molecule         & $-$92 to $-$69 & $-$47 to $-$13 & $-$9 to 8 & 12 to 22 & 47 to 89 & Core  & Total & high temperature \\
        \emph{ortho / para}  & (km~s$^{-1}$) & (km~s$^{-1}$) & (km~s$^{-1}$) & (km~s$^{-1}$) & (km~s$^{-1}$) &      &     & limit \\
 & (GC) & (Norma arm) & (GC) & (Scutum arm) & (Env. Sgr~B2) &      &     &     \\
        \hline
        \hline
        H$_2$CO          &  -   &  -   &  -   &  -   &     -   &    2.2 &     2.2 & 3$^{(a)}$  \\
        H$_2$Cl$^+$      &  -   &  -   &  3.0 &  4.6 &     -   &    -   &     2.2 & 3$^{(b)}$  \\
        H$_2$O$^+$       &  -   &  1.2 &  4.6 &  -   &     2.4 &    1.1 &     2.6 & 3$^{(c)}$  \\
        $^*$H$_2$S       &  -   &  -   &  -   &  -   &     0.4 &   31.2 &    10.4 & 3$^{(e)}$  \\
        $^*$H$_3$O$^+$   &  -   &  -   &  -   &  -   &     4.2 &    -   &     4.2 & 1$^{(f)}$  \\
        NH$_2$           &  -   &  5.4 &  3.4 &  -   &     1.7 &   10.3 &     2.4 & 3$^{(g)}$  \\
        $^*$NH$_3$       &  -   &  3.3 &  0.1 &  -   & $<$0.1  &    -   & $<$0.1  & 1$^{(h)}$  \\
        \hline
        \hline
    \end{tabular}
    \label{Tab:OrthoPara}
\end{table*}

\textbf{HCl}. As in Sgr~B2(N) \citep[described by ][]{2014ApJ...789....8N} we clearly detect the ($J$~=~1 -- 0) and ($J$~=~2) $\leftarrow$ ($J$~=~1) hyperfine transitions of H$^{35}$Cl (hydrogen chloride) and its isotopologue H$^{37}$Cl, see Figs.~\ref{fig:hcl} and \ref{fig:hcl37}. The transitions are modeled with five velocity components between 19 and 68~km~s$^{-1}$ and a mean column density of $N_{\rm tot} = 9.8 \times 10^{13}$~cm$^{-2}$, which is in good agreement with the column density of $N_{\rm tot} = 1.63 \times 10^{14}$~cm$^{-2}$ derived by \citetads{1995ApJ...447L.125Z}.

\textbf{H$_2$Cl$^{+}$}. In agreement to Sgr~B2(N) \citep[described by ][]{2014ApJ...789....8N}  H$_2$Cl$^+$ (chloronium) is clearly detected in Sgr~B2(M) as well, see Figs.~\ref{fig:oh2cl+}, \ref{fig:ph2cl+}. Here, we modeled the \emph{ortho} and \emph{para} form separately, where both states are described by $K_a + K_c$ being odd and even, respectively.

    \subsection{Other Molecules}\label{subsec:OtherMol}

\textbf{ArH$^{+}$}. The detection of ArH$^+$ (Argonium) in the HIFI survey was previously described by \citetads{2014A&A...566A..29S}. We derived slightly different parameters due to our different modeling method, see Fig.~\ref{fig:arh+}.

    \section{Discussion}\label{sec:Discussion}

\subsection{Physical parameters for molecular families}\label{subsec:PhysicalParameters}

In order to visualize the distribution of the derived physical parameters we computed normalized kernel density estimations \citepads[KDE][]{rosenblatt1956, parzen1962} for the fitted parameters of the different families introduced in Sect.~\ref{sec:Results}, which are presented in Fig.~\ref{fig:KDEParam}, where we take all detected molecules and their isotopologues into account. (In the following we exclude CH$_3$OCH$_3$ and ArH$^{+}$ because their respective families contain only one molecule). The KDE offers a visual representation of the individual parameters with the aim to see their spread and to search for general trends\footnote{As a kind of continuous replacement of a discrete histogram the KDE is a function ${\widehat {f}}_{h}(x)$ applied to a distribution $x_{i}$ of $n$ points
\begin{equation}
  {\widehat {f}}_{h}(x) = \frac{1}{n \, h} \sum _{i = 1}^{n} K \Big[ \frac{x - x_{i}}{h} \Big],
\end{equation}
where $K(x)$ is called the \emph{kernel function} that is mostly a Gaussian and $h > 0$ indicates the bandwidth that controls the amount of smoothing. Here, Silverman's rule \citepads{1986desd.book.....S} is used to compute the bandwidth $h$ of each KDE.}. This is usually not easy to accomplish by just looking at numbers.

For the excitation temperatures we find mostly bimodal KDEs, presented in panels Fig.~\ref{fig:KDEParam}~a) and e), for all molecular families and for the total distribution of all molecules. The majority of the KDEs have two maxima around 20 and 500~K. For halogen molecules both maxima are shifted to 3 and 33~K, respectively, because these molecules appear only in the envelope and in clouds along the line of sight, except \emph{para}-H$_2$Cl$^+$, which has one weak core component. Furthermore, the KDE for carbon and hydrocarbons shows even three local maxima at 8, 42, and 76~K, which might be caused by the fact that these molecules are mostly located in the envelope and in foreground clouds. The KDEs of the other families have their second maxima at higher temperatures, because these families contain molecules which are predominantly associated with the core of Sgr~B2(M). The strong maximum at low temperatures appearing in all KDEs is due to the fact that the hot cores cover only a small fraction of the large HIFI beam. Therefore, molecules contained in the cold envelope dominate the HIFI spectrum. Similar to that, most of the column density KDEs, see Fig.~\ref{fig:KDEParam}~b) and f), show unimodal distributions, with a maximum around $1 \cdot 10^{14}$~cm$^{-2}$ for each family of molecules. Only the KDE for Carbon and Hydrocarbons have a second pronounced maximum at around $1 \cdot 10^{18}$~cm$^{-2}$. Additionally, the line width KDEs, shown in panels Fig.~\ref{fig:KDEParam}~c) and g), of most molecular families are rather unimodal with minor side maxima. The most abundant line width is around 9~km~s$^{-1}$, with a side maximum around 29~km~s$^{-1}$. Finally, the distribution of the velocity offsets, described in panels Fig.~\ref{fig:KDEParam}~d) and h), nicely fits to the velocity ranges (grey areas) associated with structures along the line of sight, see Table~\ref{Tab:VelStruc}.\\

In Fig.~\ref{fig:Abun}, the abundances of all detected species associated with each family are shown in separate panels, using an H$_2$ column density of $1.1 \cdot 10^{24}$~cm$^{-2}$ for the core and $1.8 \cdot 10^{24}$~cm$^{-2}$ for envelope components, respectively. The abundances we quote are source-averaged abundances. It is possible that a given molecule exists only in a subvolume of the core as seen by Herschel --- either along the line of sight, or within the beam.  Therefore, the local abundances can deviate from the values determined here. This property is due to finite resolution, and is shared by all line surveys. The only way to improve this further is analyzing higher resolution data \citepads[][M\"{o}ller in prep.]{2017A&A...604A...6S}. This is outside the scope of the present paper, and of course even then there may be variations on a scale unresolved by the beam. Additionally, we compare the abundances with those derived by \citetads{2014ApJ...789....8N} for Sgr~B2(N), which were obtained, reduced and analyzed with similar methods, so that observation and modeling effects are minimized. Nevertheless, the two analyses differ fundamentally in some points, which is why the results cannot be compared without limitations. In contrast to \citetads{2014ApJ...789....8N}, we take dust attenuation within the envelope into account, which leads to different hydrogen column densities, i.e.\ our column density of the core layer is about seven times lower than the one used by \citetads{2014ApJ...789....8N} for Sgr~B2(N). However, the results cannot be directly compared, since we take local-overlap effects into account, which is important for line crowded sources like Sgr~B2(M) and N. Furthermore, for some molecules we distinguish between \emph{ortho} and \emph{para} states, see Sect.~\ref{subsec:OrthoParaRatios}, which makes a direct comparison difficult. In the following, we discuss similarities and differences between the molecular abundances in the cores of Sgr~B2(M) and Sgr~B2(N) for each molecular group.

\emph{O-bearing species}. The abundances of simple and complex oxygen-bearing species are shown in Fig.~\ref{fig:Abun}~a)~+~b), respectively, where the abundance of some molecules (NH$_3$ and H$_3$O$^+$) are influenced by the underlying LTE assumption. It is generally assumed that O- and C-bearing molecules have a common grain surface origin from the hydrogenation of CO on grain surfaces \citep[see e.g.,][]{2009A&A...505..629F, 2011ApJ...735...15G, 2013ApJ...778..158G}. Here, methanol is 19 times more abundant in Sgr~B2(N), while Sgr~B2(M) is richer in the formyl radical (HCO$^+$). The abundance of dimethyl ether (CH$_3$OCH$_3$) in our survey is more than 77 times lower than for Sgr~B2(N). Ignoring molecules, which have to be described in Non-LTE, we derive total abundances of $7.18 \cdot 10^{-8}$ and $2.22 \cdot 10^{-7}$ for all oxygen-bearing species within the core and envelope of Sgr~B2(M), respectively.

\emph{NH-bearing species}. Abundances of NH-bearing species are shown in Fig.~\ref{fig:Abun}~c). In contrast to Sgr~B2(N), methanamine (CH$_2$NH) is not detected in the core, which makes a direct comparison difficult. Methylamine (CH$_3$NH$_2$) is more abundant in Sgr~B2(N), while Sgr~B2(M) contains more dyazenilium (N$_2$H$^{+}$). The abundances for both species differ by a factor of 7. The description of ammonia might be complicated by Non-LTE effects. The total abundances for NH-bearing species within the cores of Sgr~B2(M) (excluding NH$_3$) is $1.10 \cdot 10^{-8}$. For the envelope of Sgr~B2(M) we get $1.30 \cdot 10^{-7}$, which is caused by the strong absorption lines of ammonia and the amino radical.

\emph{N- and O-bearing species}. In Fig.~\ref{fig:Abun}~d), the abundances of N- and O-bearing species are described. Here, the more complex molecules formamide (NH$_2$CHO or HC(O)NH$_2$) and isocyanic acid (HNCO) show a much higher abundance in Sgr~B2(N) than in M. The enhancement varies between 13 for HNCO and 133 for formamide. Opposed to that, the abundance of nitric oxide (NO) is nearly the same in both sources. The total abundance in the core of Sgr~B2(M) is $1.41 \cdot 10^{-7}$ and $1.04 \cdot 10^{-9}$ in the envelope, respectively.

\emph{CN-bearing species}. The abundances of CN-bearing species are presented in Fig.~\ref{fig:Abun}~e).
Similar to the N- and O-bearing species, the more complex CN-bearing species ethyl cyanide (CH$_3$CN) is more abundant in Sgr~B2(N) compared to M, while the lighter molecules isocyanide (HNC) and cyano radical (CN) have a higher abundance in Sgr~B2(M). The comparison of HCN must be considered with caution, because the underlying LTE assumption may no longer be valid. Furthermore, cyanoacetylene (HCCCN) was not detected in Sgr~B2(N)) but only an upper limit was described. The total abundance (without HCN) for the core is $1.76 \cdot 10^{-8}$ (Sgr~B2(M)). Except HCN, no CN-bearing molecule is contained in the envelope.

\emph{S-bearing species}. A comparison of the abundances of S-bearing species is displayed in Fig.~\ref{fig:Abun}~f). Compared to Sgr~B2(N), Sgr~B2(M) is richer in Sulfur monoxide (SO) and Sulfur dioxide (SO$_2$). The latter is the most abundant molecule in our survey after CO and 28 times more abundant than in Sgr~B2(N). In contrast to that we find less carbon monosulfide (CS), thioformaldehyde (H$_2$CS), nitrogen sulfide (NS), and carbonyl sulfide (OCS). The total abundances in the core and in the envelope of Sgr~B2(M) are $1.11 \cdot 10^{-6}$ and $2.61 \cdot 10^{-8}$, respectively.

\emph{Carbon and hydrocarbons}. Fig.~\ref{fig:Abun}~g) describes the abundances of carbon and hydrocarbons. Except CCH and CH$^+$ none of the species contained in this group is found in the core of Sgr~B2(N). The abundances of [\ion{C}{i}] and [\ion{C}{ii}] might be adversely affected by Non-LTE effects and non-clean off-positions. The total abundance (without [\ion{C}{i}] and [\ion{C}{ii}]) in the core of Sgr~B2(M) is $5.78 \cdot 10^{-9}$ and for the envelope $3.34 \cdot 10^{-8}$, respectively.

\emph{Halogen molecules}. Finally, the abundances of halogen molecules are shown in Fig.~\ref{fig:Abun}~h). Similar to the case described above, there is neither chloronium (H$_2$Cl$^+$) nor hydrogen chloride (H$^{35}$Cl) in the core of Sgr~B2(N). For Sgr~B2(M) we derived total abundances for the core and for the envelope of $2.90 \cdot 10^{-11}$ and for the envelope $1.05 \cdot 10^{-09}$, respectively.

\subsection{Ortho/para ratios}\label{subsec:OrthoParaRatios}

For some molecules for which \emph{ortho} and \emph{para} resolved molecular parameters are available, we derive the total and the velocity range dependent \emph{ortho}/\emph{para} ratios as well, see Table~\ref{Tab:OrthoPara}. For the latter, we sum up the column densities for all components with velocity offsets located within each velocity range for \emph{ortho} and \emph{para} molecules respectively and determine the corresponding ratios.

The derived ratios for some molecules may be incorrect because their excitation is not described well by the underlying LTE assumption. H$_2$S, H$_3$O$^{+}$,and NH$_3$ should be described in Non-LTE to get more reliable ratios. Additionally, the ground state transition of \emph{para}-H$_2$S is not included in the HIFI survey, which makes a comparison more difficult. For most other molecules and velocity ranges we found ratios between 1.2 and 5.4, except for NH$_2$, where we obtained a ratio of 10.3 for the core of Sgr~B2(M), which might be caused by the fact that only two emission features of \emph{para}-NH$_2$ are included in the HIFI survey leading to a less constrained LTE model. Additionally, we obtained a very low ratio of 0.1 in the range of $-$9 and 8~km~s$^{-1}$ for NH$_3$, caused by a very small contribution of \emph{ortho}-NH$_3$. However, the determination of the column density is not very reliable for such weak contributions. Furthermore, for H$_2$O$^{+}$, we get a ratio of 2.6. Although this is only half the mean ratio derived by \citetads{2010A&A...521L..11S}, both ratios agree quite well between $-$9 and 8~km~s$^{-1}$.\\

    \section{Conclusions}\label{sec:Conclusions}

We have presented a full HIFI chemical survey of the Sgr~B2(M) star-forming region, which spans a frequency range from 480 to 1907~GHz. A total of 92 isotopologues arising from 49 molecules are detected in this survey through transitions in emission and absorption. Taking local overlap effects into account and assuming a two-layer model representing the core and the envelope of Sgr~B2(M) with different dust concentrations, we performed LTE modeling using the XCLASS program, to the observed features of all molecules in the spectrum. About 14~\% of the data channels observed to have intensities over the 3$\sigma$ level are currently unidentified. Sulfur dioxide is the strongest contributor to the HIFI survey both in terms of total integrated intensity and number of detected lines. Based on the obtained column densities, we computed molecular abundances for each detected species and compared them with those derived by \citetads{2014ApJ...789....8N} for Sgr~B2(N). For all but one molecular families we find bimodal distributions of the excitation temperatures with maxima around 20 and 500~K. For carbon, hydrocarbons, and for halogen molecules both maxima are shifted to lower temperatures. Furthermore, we derive \emph{ortho} / \emph{para} ratios for different molecules and velocity regions ranging from 1.2 to 4.6. Ratios above five and below one might be falsified by Non-LTE effects.

\begin{acknowledgements}
HIFI has been designed and built by a consortium of institutes and university departments from across Europe, Canada and the United States under the leadership of SRON Netherlands Institute for Space Research, Groningen, The Netherlands and with major contributions from Germany, France and the US. Consortium members are: Canada: CSA, U. Waterloo; France: CESR, LAB, LERMA, IRAM; Germany: KOSMA, MPIfR, MPS; Ireland, NUI Maynooth; Italy: ASI, IFSI-INAF, Osservatorio Astrofisico di Arcetri- INAF; Netherlands: SRON, TUD; Poland: CAMK, CBK; Spain: Observatorio Astronómico Nacional (IGN), Centro de Astrobiologìa (CSIC-INTA). Sweden: Chalmers University of Technology – MC2, RSS \& GARD; Onsala Space Observatory; Swedish National Space Board, Stockholm University -- Stockholm Observatory; Switzerland: ETH Zurich, FHNW; USA: Caltech, JPL, NHSC. Support for this work was provided by NASA through
an award issued by JPL/Caltech. CSO is supported by the NSF, award AST-0540882. This work was supported by the Deutsche Forschungsgemeinschaft (DFG) through grant Collaborative Research Centre 956 (subproject A6 and C3, project ID 184018867) and from BMBF/Verbundforschung through the projects ALMA-ARC 05A14PK1 and ALMA-ARC 05A20PK1. Part of this research was carried out at the Jet Propulsion Laboratory, California Institute of Technology, under a contract with the National Aeronautics and Space Administration.
\end{acknowledgements}

    \bibliographystyle{aa}
    \bibliography{SgrB2ADS.bib}

    \newpage
    \appendix
    \clearpage
    \section{LTE parameters}

%
\tablefirsthead{%
\hline
\hline
Molecule      & $\theta^{m,c}$ & T$_{\rm ex}^{m,c}$ & N$_{\rm tot}^{m,c}$   & $\Delta$ v$^{m,c}$ & v$_{\rm LSR}^{m,c}$\\
              & ($\arcsec$)    & (K)                & (cm$^{-2}$)           & (km~s$^{-1}$)      & (km~s$^{-1}$)      \\
\hline
}

\tablehead{%
\multicolumn{6}{c}{(Continued)}\\
\hline
\hline
Molecule      & $\theta^{m,c}$ & T$_{\rm ex}^{m,c}$ & N$_{\rm tot}^{m,c}$   & $\Delta$ v$^{m,c}$ & v$_{\rm LSR}^{m,c}$\\
              & ($\arcsec$)    & (K)                & (cm$^{-2}$)           & (km~s$^{-1}$)      & (km~s$^{-1}$)      \\
\hline
}

\tabletail{%
\hline
\hline
}

\topcaption{LTE Parameters for the Sgr~B2(M) Full-band Model (Core Components)}
\tiny
\begin{supertabular}{lcccC{1cm}C{1cm}}\label{CoreLTE:parameters}
\centering
$^{12}$C      &   19.7         &                 44 &  6.7 $\times 10^{18}$ &                 19 &                  82\\
              &   19.7         &                 21 &  1.5 $\times 10^{16}$ &                 21 &                  78\\
              &   19.7         &                 34 &  5.4 $\times 10^{18}$ &                  9 &                  73\\
              &   19.7         &                 43 &  4.5 $\times 10^{18}$ &                 14 &                  60\\
              &   19.7         &                 19 &  1.2 $\times 10^{13}$ &                 20 &                  57\\
$^{13}$C      &   19.7         &                 44 &  3.3 $\times 10^{17}$ &                 19 &                  82\\
              &   19.7         &                 21 &  7.7 $\times 10^{14}$ &                 21 &                  78\\
              &   19.7         &                 34 &  2.7 $\times 10^{17}$ &                  9 &                  73\\
\end{supertabular}

%
\tablefirsthead{%
\hline
\hline
Molecule      & $\theta^{m,c}$ & T$_{\rm ex}^{m,c}$ & N$_{\rm tot}^{m,c}$   & $\Delta$ v$^{m,c}$ & v$_{\rm LSR}^{m,c}$\\
              & ($\arcsec$)    & (K)                & (cm$^{-2}$)           & (km~s$^{-1}$)      & (km~s$^{-1}$)      \\
\hline
}

\tablehead{%
\multicolumn{6}{c}{(Continued)}\\
\hline
\hline
Molecule      & $\theta^{m,c}$ & T$_{\rm ex}^{m,c}$ & N$_{\rm tot}^{m,c}$   & $\Delta$ v$^{m,c}$ & v$_{\rm LSR}^{m,c}$\\
              & ($\arcsec$)    & (K)                & (cm$^{-2}$)           & (km~s$^{-1}$)      & (km~s$^{-1}$)      \\
\hline
}

\tabletail{%
\hline
\hline
}
\newpage
\clearpage

\topcaption{LTE Parameters for the Sgr~B2(M) Full-band Model (Envelope Components)}
\tiny
\begin{supertabular}{lcccC{1cm}C{1cm}}\label{EnvLTE:parameters}
\centering
$^{12}$C      & ext.           &                  9 &  1.5 $\times 10^{18}$ &                 29 &                 115\\
              & ext.           &                  8 &  9.0 $\times 10^{17}$ &                 18 &                  43\\
              & ext.           &                 10 &  2.6 $\times 10^{18}$ &                 29 &                  20\\
              & ext.           &                  3 &  3.1 $\times 10^{17}$ &                 15 &                   0\\
              & ext.           &                  3 &  2.7 $\times 10^{13}$ &                 29 &                 -28\\
              & ext.           &                  7 &  3.4 $\times 10^{15}$ &                  3 &                 -38\\
              & ext.           &                  9 &  1.3 $\times 10^{18}$ &                 11 &                 -41\\
              & ext.           &                  9 &  1.5 $\times 10^{18}$ &                 28 &                 -84\\
\end{supertabular}
\newpage
\clearpage
\onecolumn

    \section{Plots}

In each figure presented in this section the HIFI spectrum (blue line) is shown together with the overall fit (green dashed line) and the contribution of the respective molecule (red line), respectively. The vertical gray dashed line indicates the source velocity of Sgr~B2(M).\\

\begin{figure*}[!htb]
    \centering
    \includegraphics[scale=0.80]{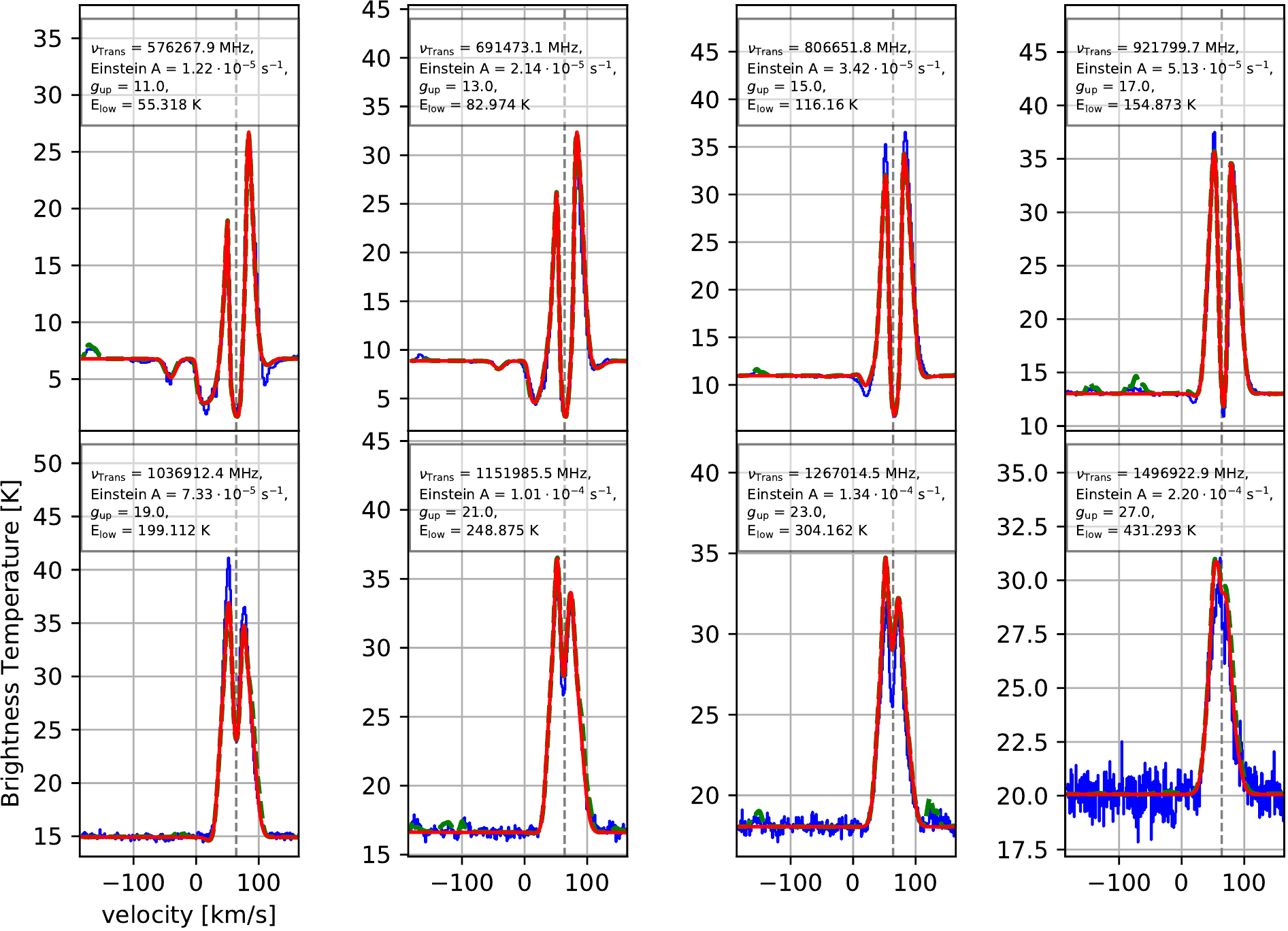}\\
    \caption{Selected transitions of CO (red line).}
    \label{fig:co}
\end{figure*}

\begin{figure*}[!htb]
    \centering
    \includegraphics[scale=0.80]{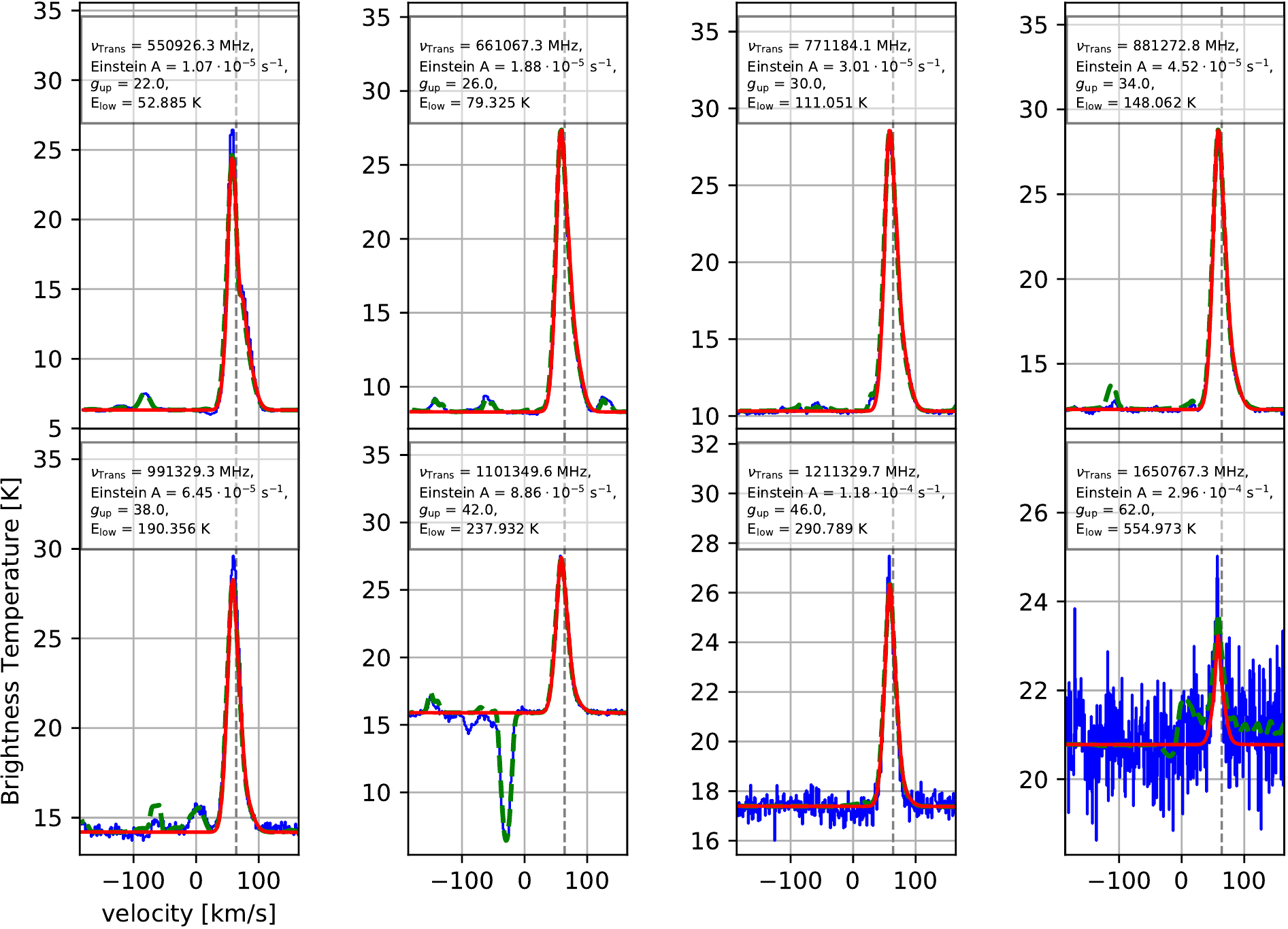}\\
    \caption{Selected transitions of $^{13}$CO (red line).}
    \label{fig:c-13-o}
\end{figure*}

\begin{figure*}[!htb]
    \centering
    \includegraphics[scale=0.80]{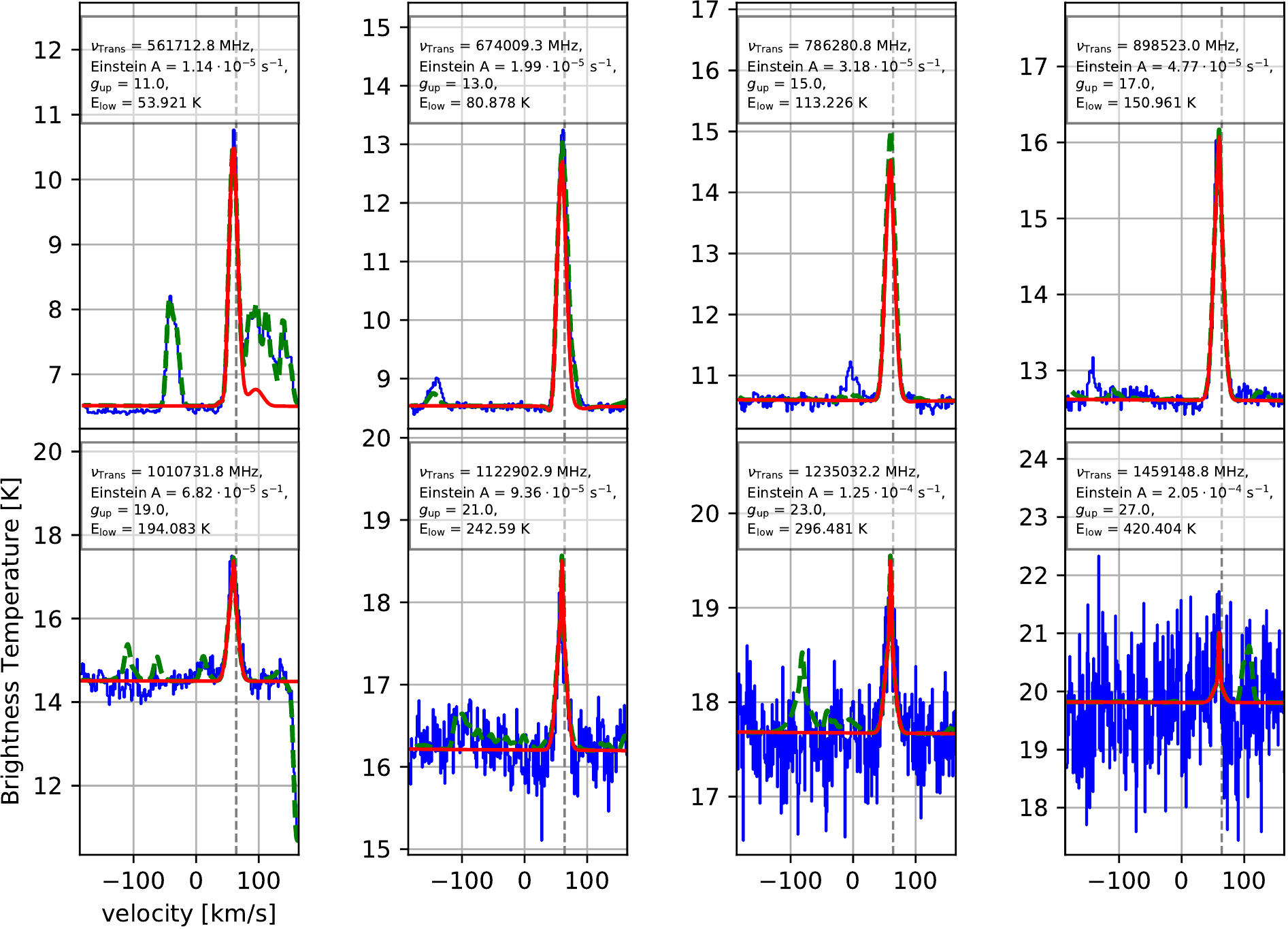}\\
    \caption{Selected transitions of C$^{17}$O (red line).}
    \label{fig:co-17}
\end{figure*}

\begin{figure*}[!htb]
    \centering
    \includegraphics[scale=0.80]{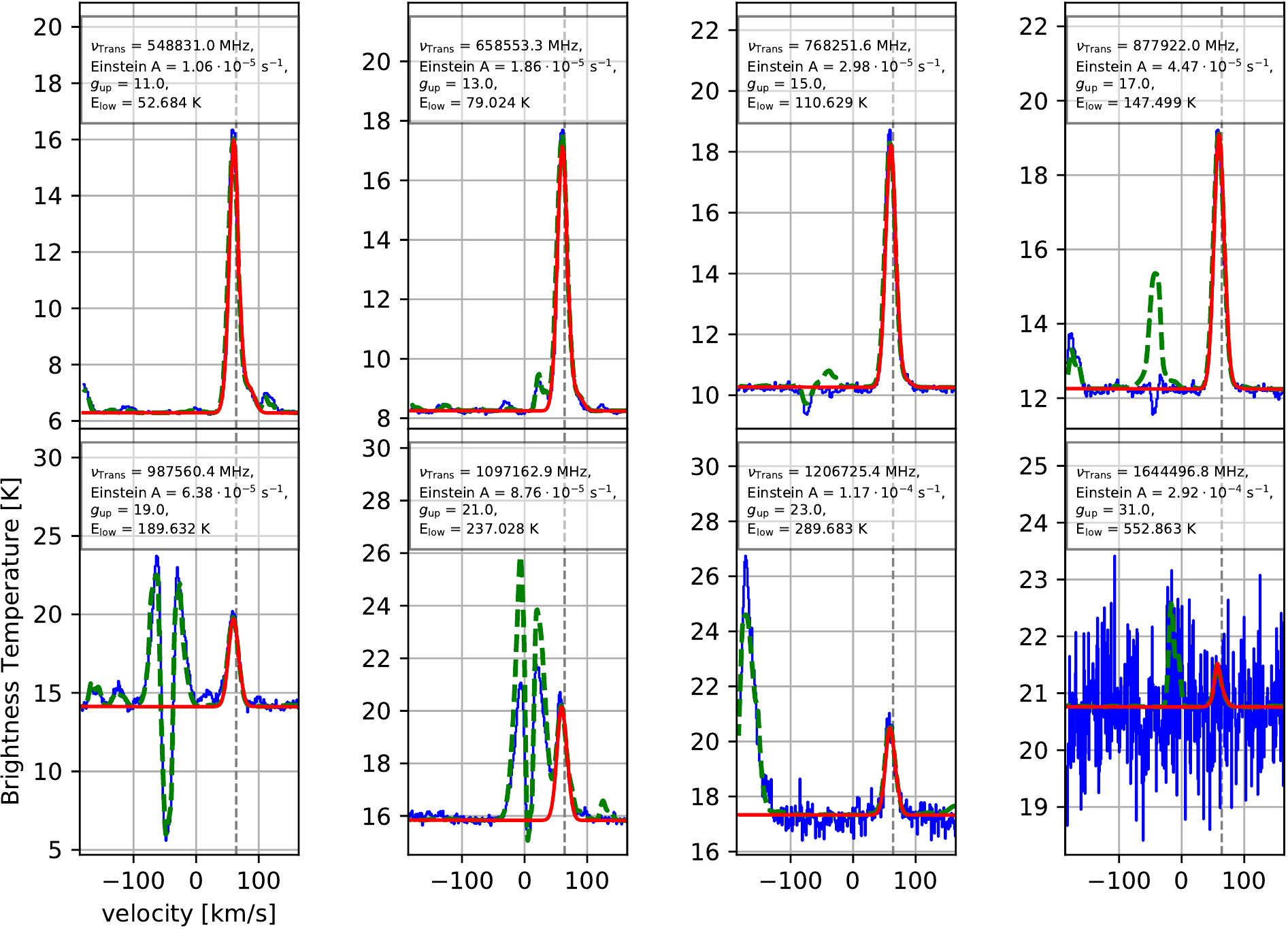}\\
    \caption{Selected transitions of C$^{18}$O (red line).}
    \label{fig:co-18}
\end{figure*}

\begin{figure*}[!htb]
    \centering
    \includegraphics[scale=0.80]{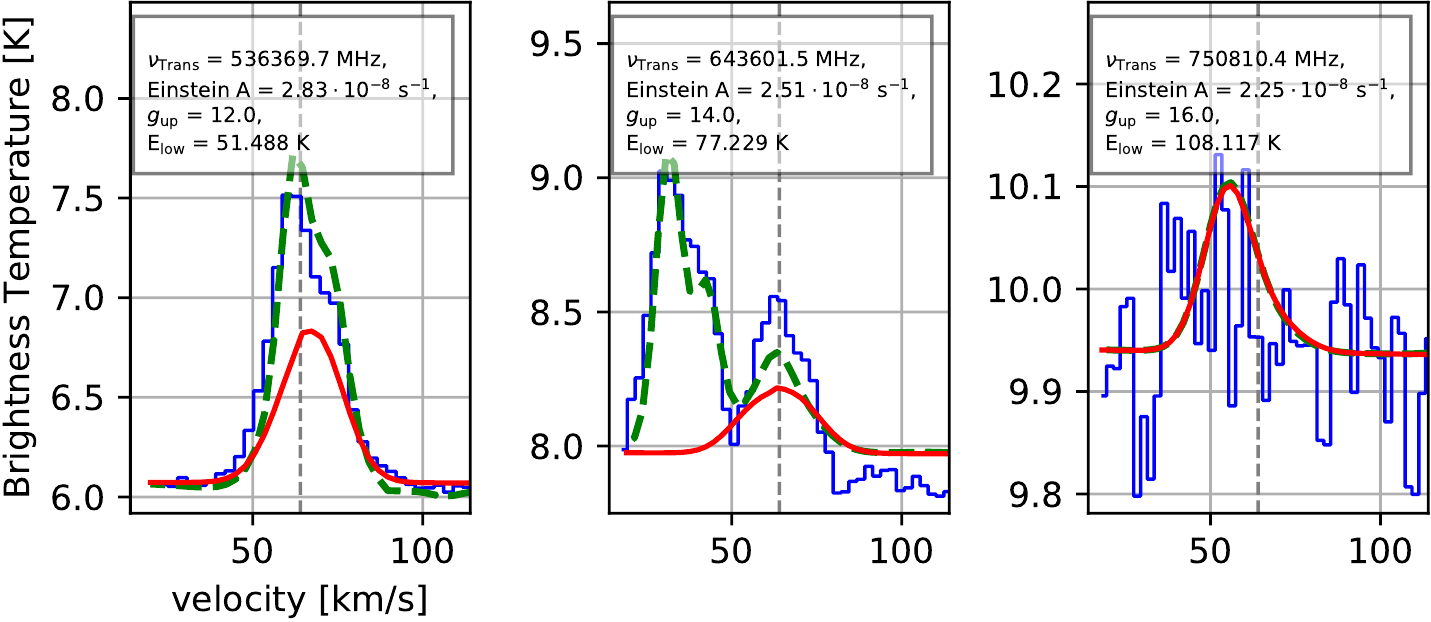}\\
    \caption{Selected transitions of $^{13}$C$^{17}$O (red line).}
    \label{fig:c-13-o-17}
\end{figure*}

\begin{figure*}[!htb]
    \centering
    \includegraphics[scale=0.80]{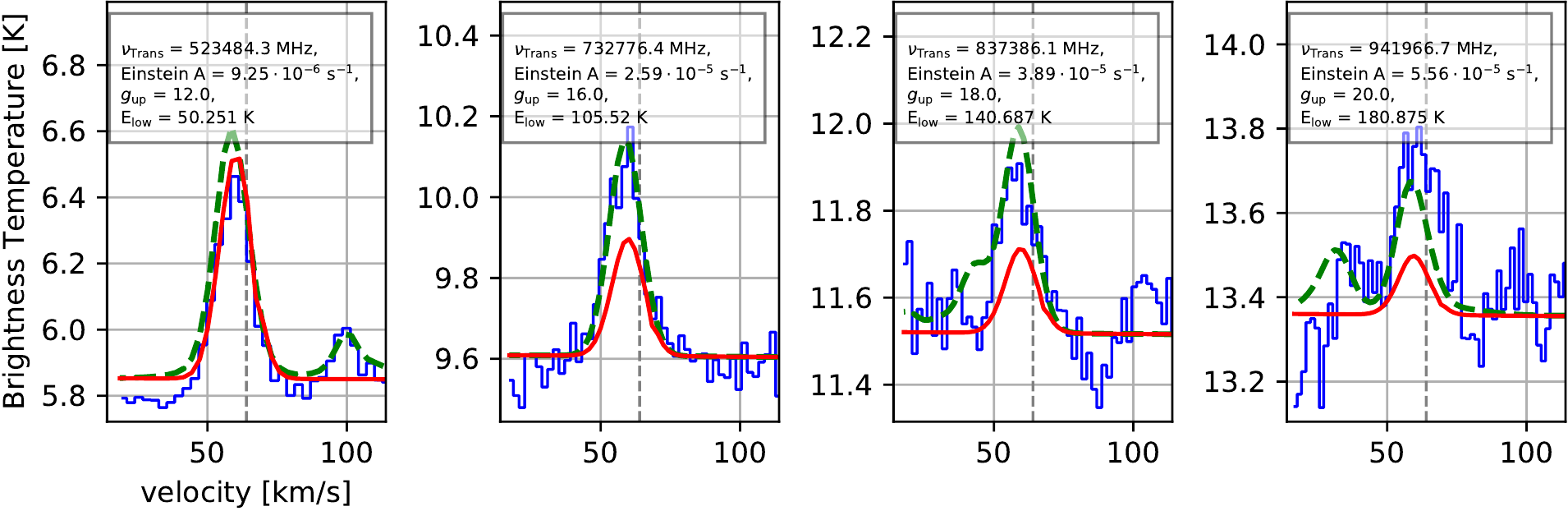}\\
    \caption{Selected transitions of $^{13}$C$^{18}$O (red line).}
    \label{fig:c-13-o-18}
\end{figure*}
\newpage

\clearpage

\begin{figure*}[!htb]
    \centering
    \includegraphics[scale=0.80]{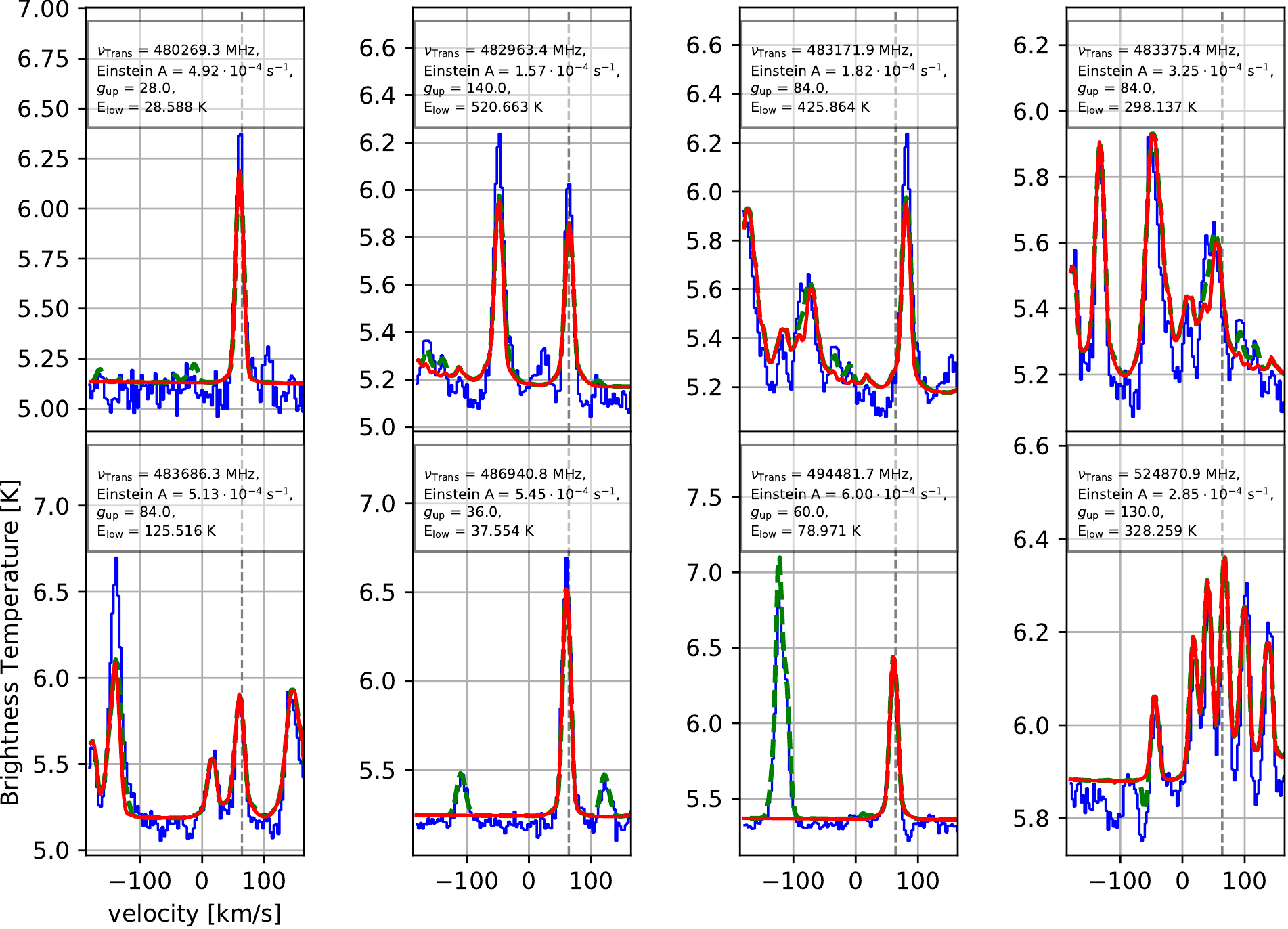}\\
    \caption{Selected transitions of CH$_3$OH (red line).}
    \label{fig:ch3oh}
\end{figure*}

\begin{figure*}[!htb]
    \centering
    \includegraphics[scale=0.80]{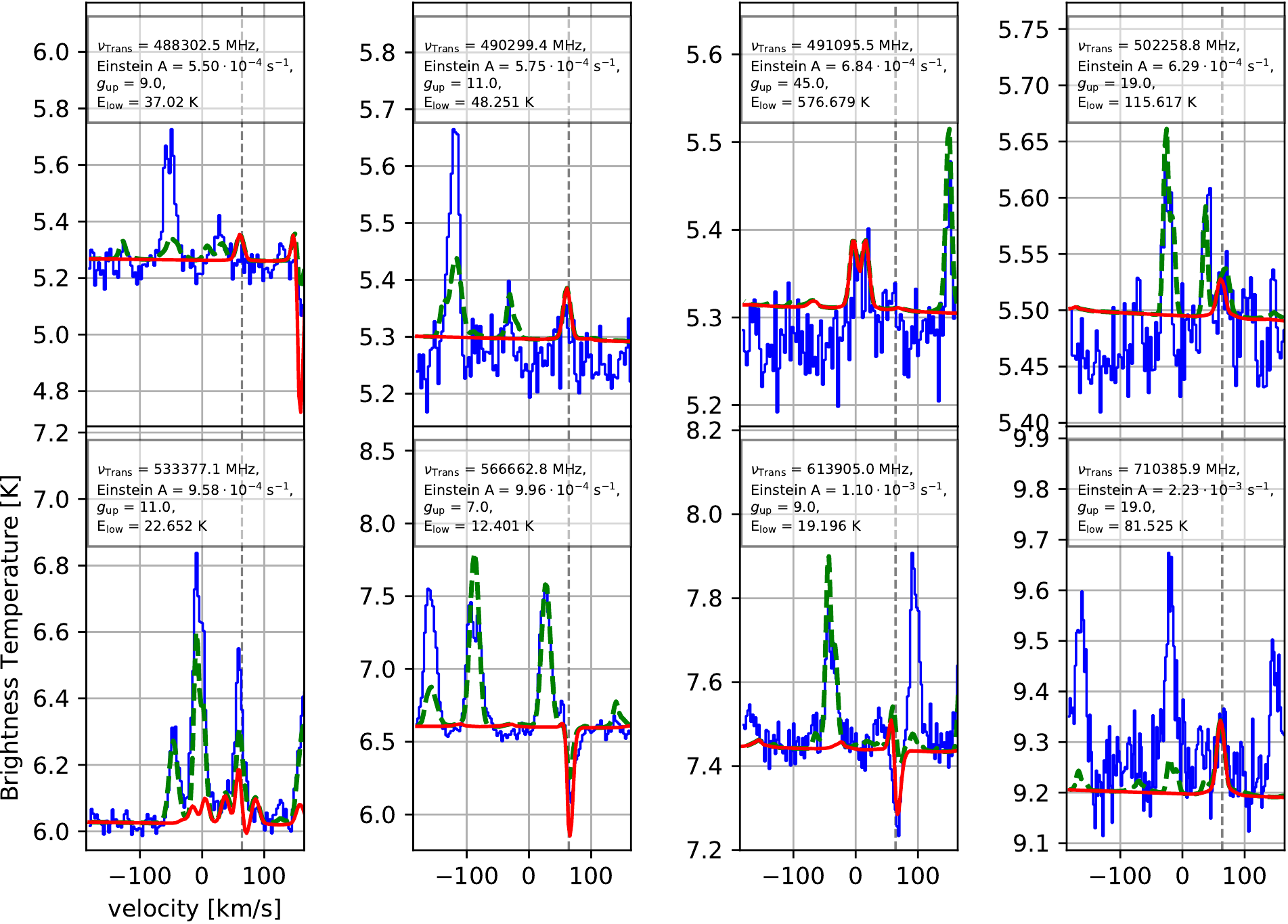}\\
    \caption{Selected transitions of $^{13}$CH$_3$OH (red line).}
    \label{fig:c13h3oh}
\end{figure*}
\newpage

\clearpage

\begin{figure*}[!htb]
    \centering
    \includegraphics[scale=0.80]{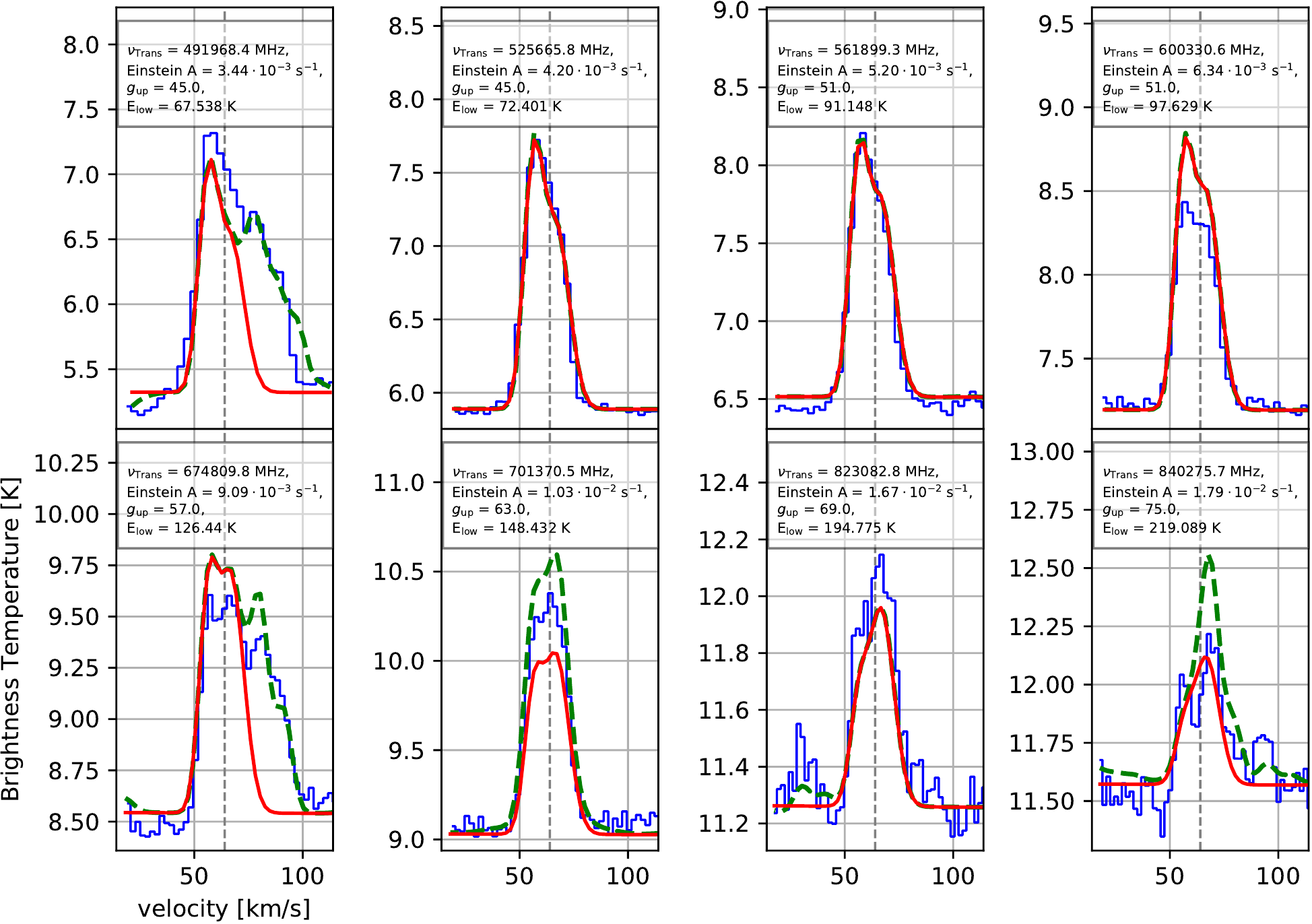}\\
    \caption{Selected transitions of \emph{ortho}-H$_2$CO (red line).}
    \label{fig:oh2co}
\end{figure*}

\begin{figure*}[!htb]
    \centering
    \includegraphics[scale=0.80]{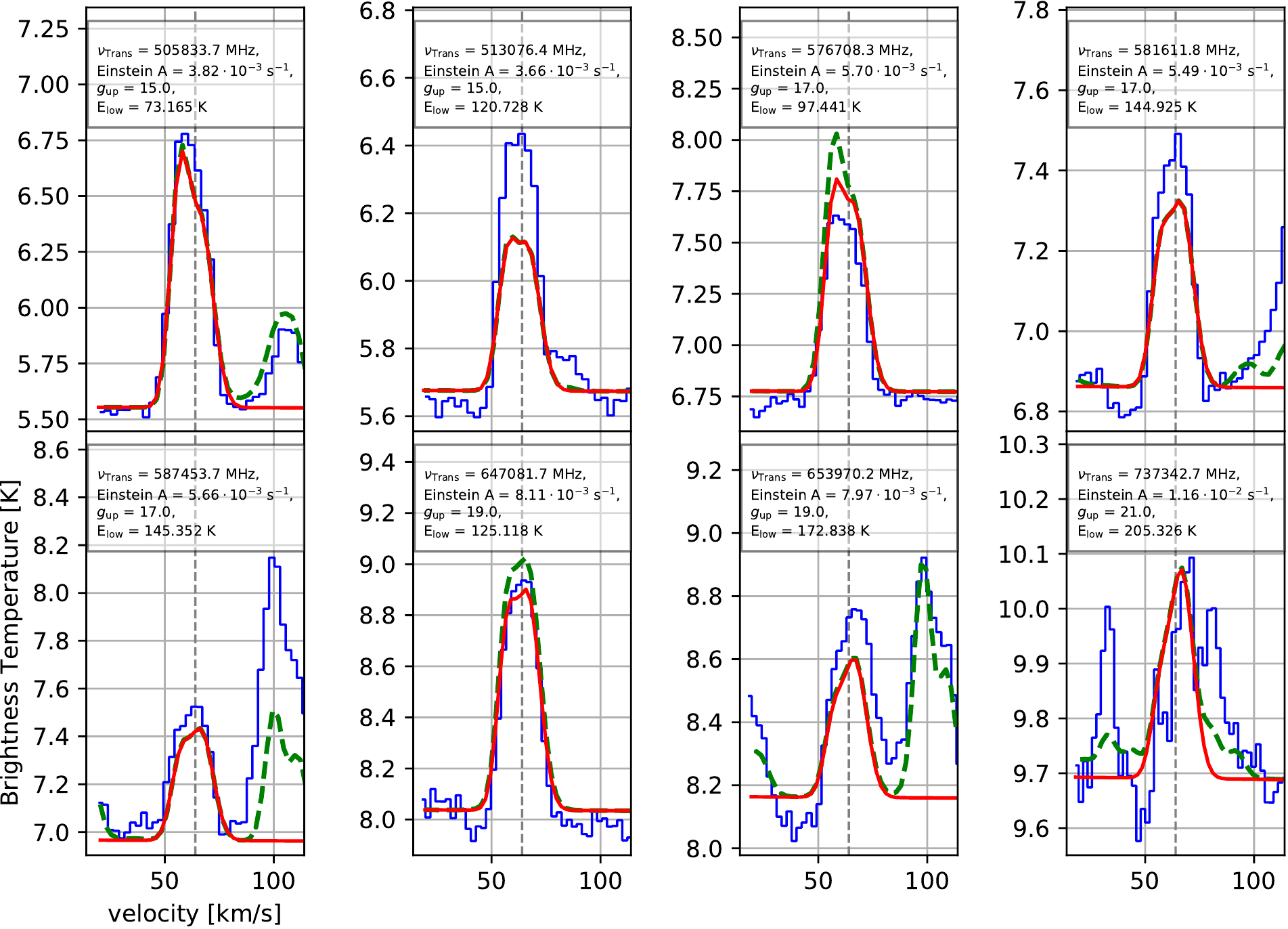}\\
    \caption{Selected transitions of \emph{para}-H$_2$CO (red line).}
    \label{fig:ph2co}
\end{figure*}
\newpage

\clearpage



\begin{figure*}[!htb]
    \centering
    \includegraphics[scale=0.80]{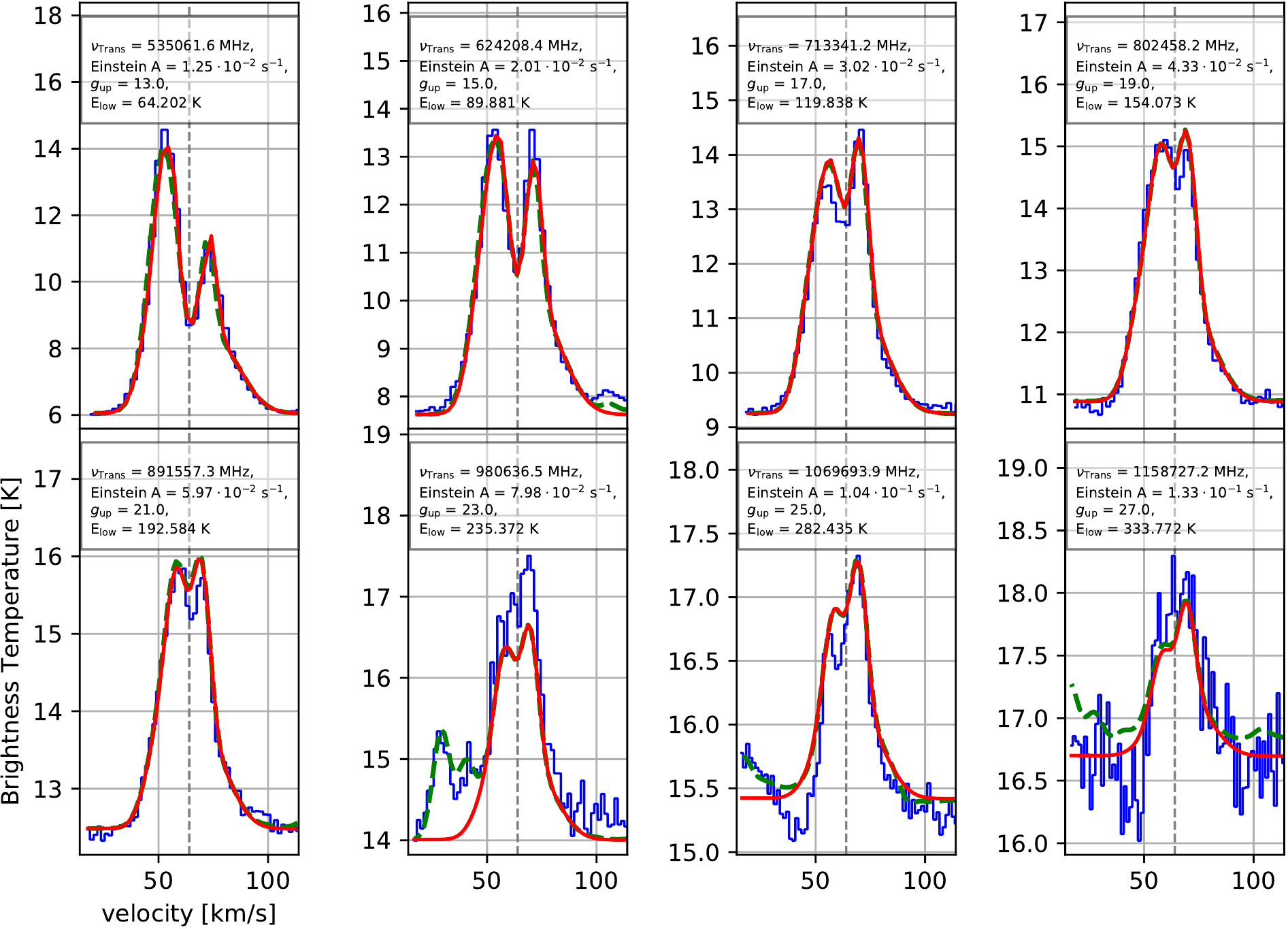}\\
    \caption{Selected transitions of HCO$^{+}$ (red line).}
    \label{fig:hco+}
\end{figure*}

\begin{figure*}[!htb]
    \centering
    \includegraphics[scale=0.80]{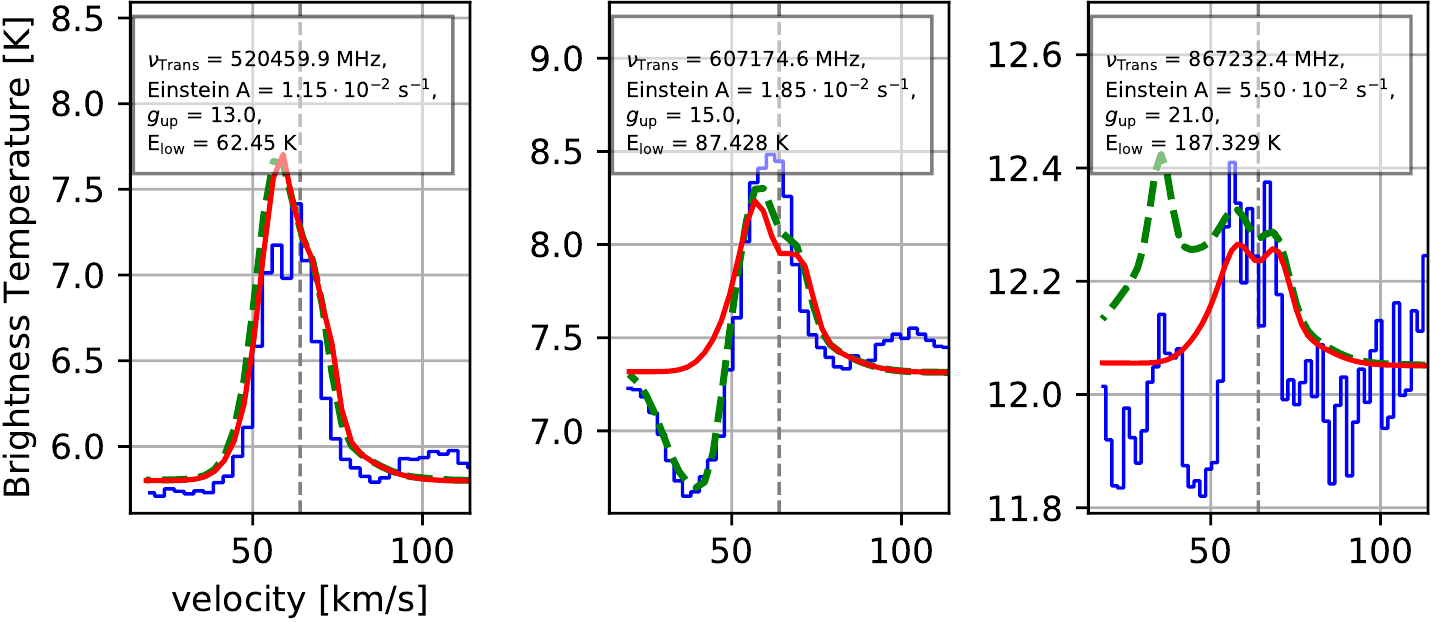}\\
    \caption{Selected transitions of H$^{13}$CO$^{+}$ (red line).}
    \label{fig:hc13o+}
\end{figure*}

\begin{figure*}[!htb]
    \centering
    \includegraphics[scale=0.80]{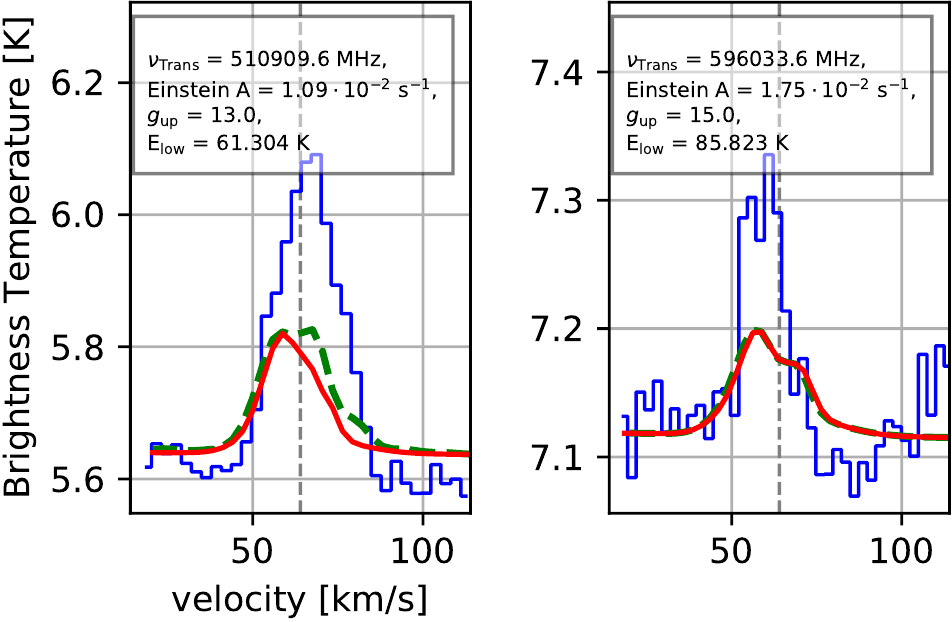}\\
    \caption{Selected transitions of HC$^{18}$O$^{+}$ (red line).}
    \label{fig:hco18+}
\end{figure*}
\newpage

\clearpage

\begin{figure*}[!htb]
    \centering
    \includegraphics[scale=0.80]{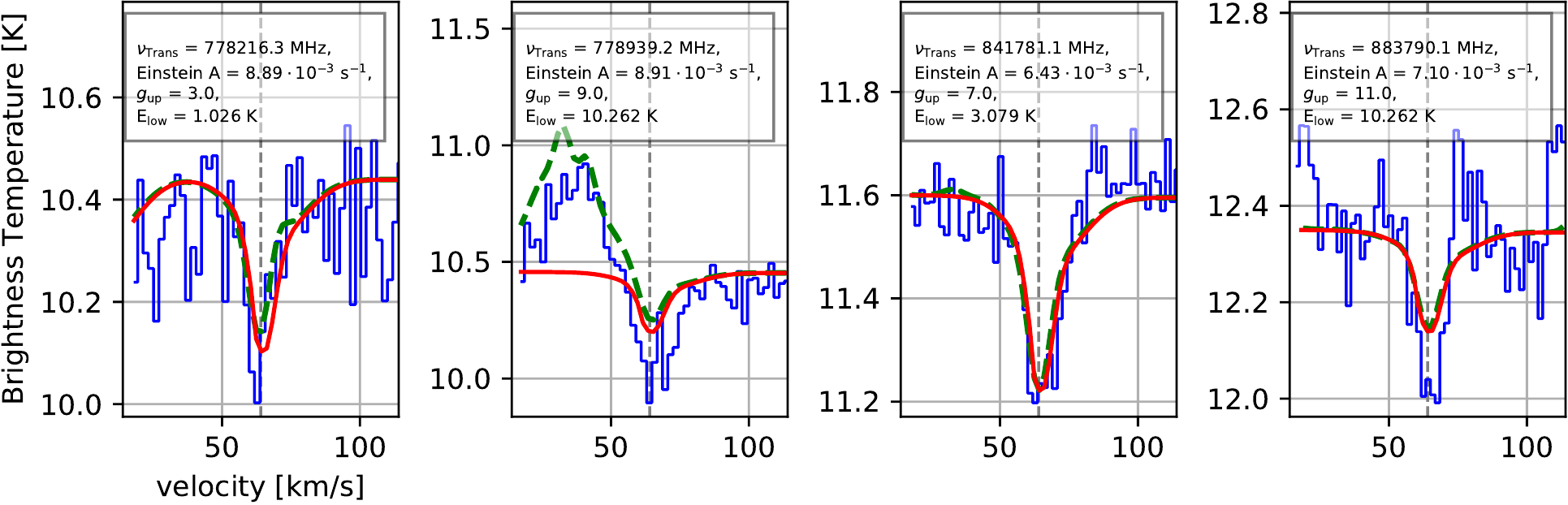}\\
    \caption{Selected transitions of HOCO$^{+}$ (red line).}
    \label{fig:hoco+}
\end{figure*}

\begin{figure*}[!htb]
    \centering
    \includegraphics[scale=0.80]{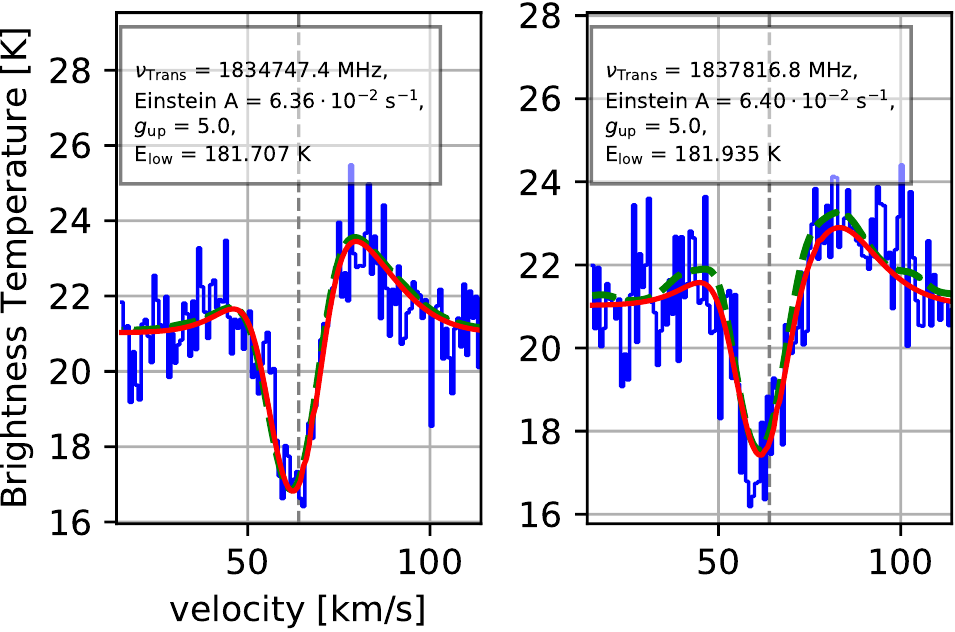}\\
    \caption{Selected transitions of OH (red line).}
    \label{fig:oh}
\end{figure*}

\begin{figure*}[!htb]
    \centering
    \includegraphics[scale=0.80]{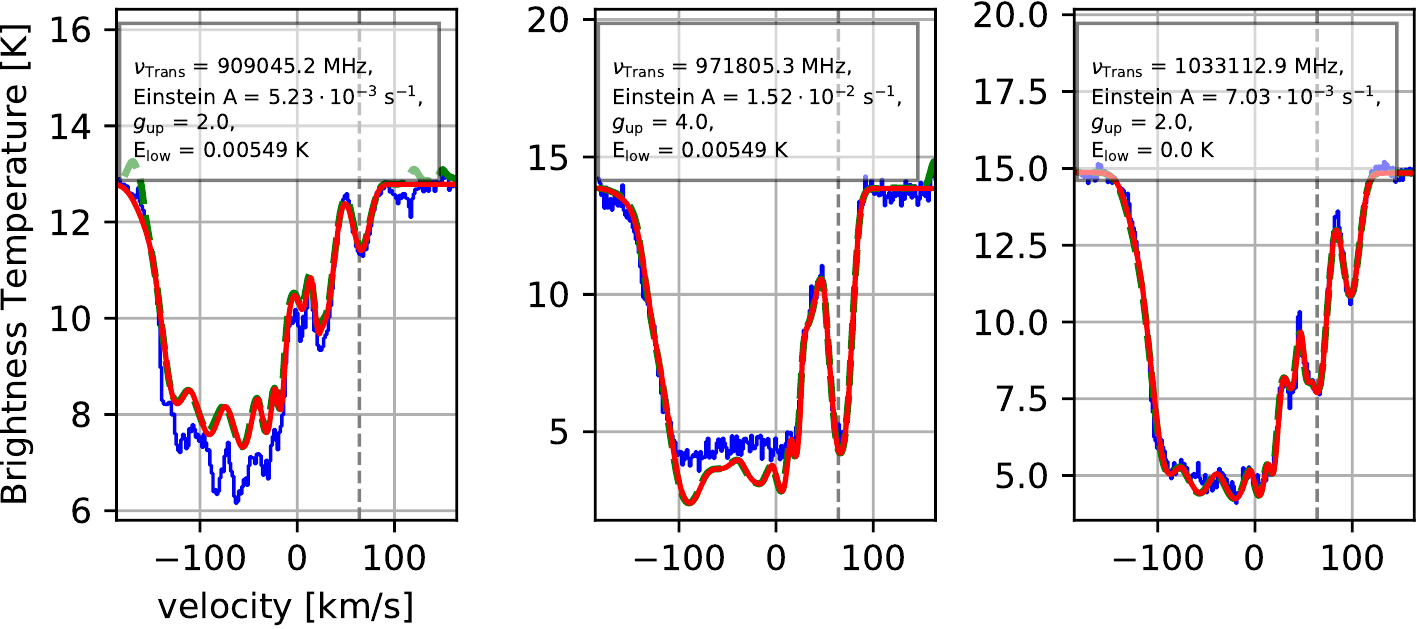}\\
    \caption{Selected transitions of OH$^{+}$ (red line).}
    \label{fig:oh+}
\end{figure*}
\newpage

\clearpage

\begin{figure*}[!htb]
    \centering
    \includegraphics[scale=0.80]{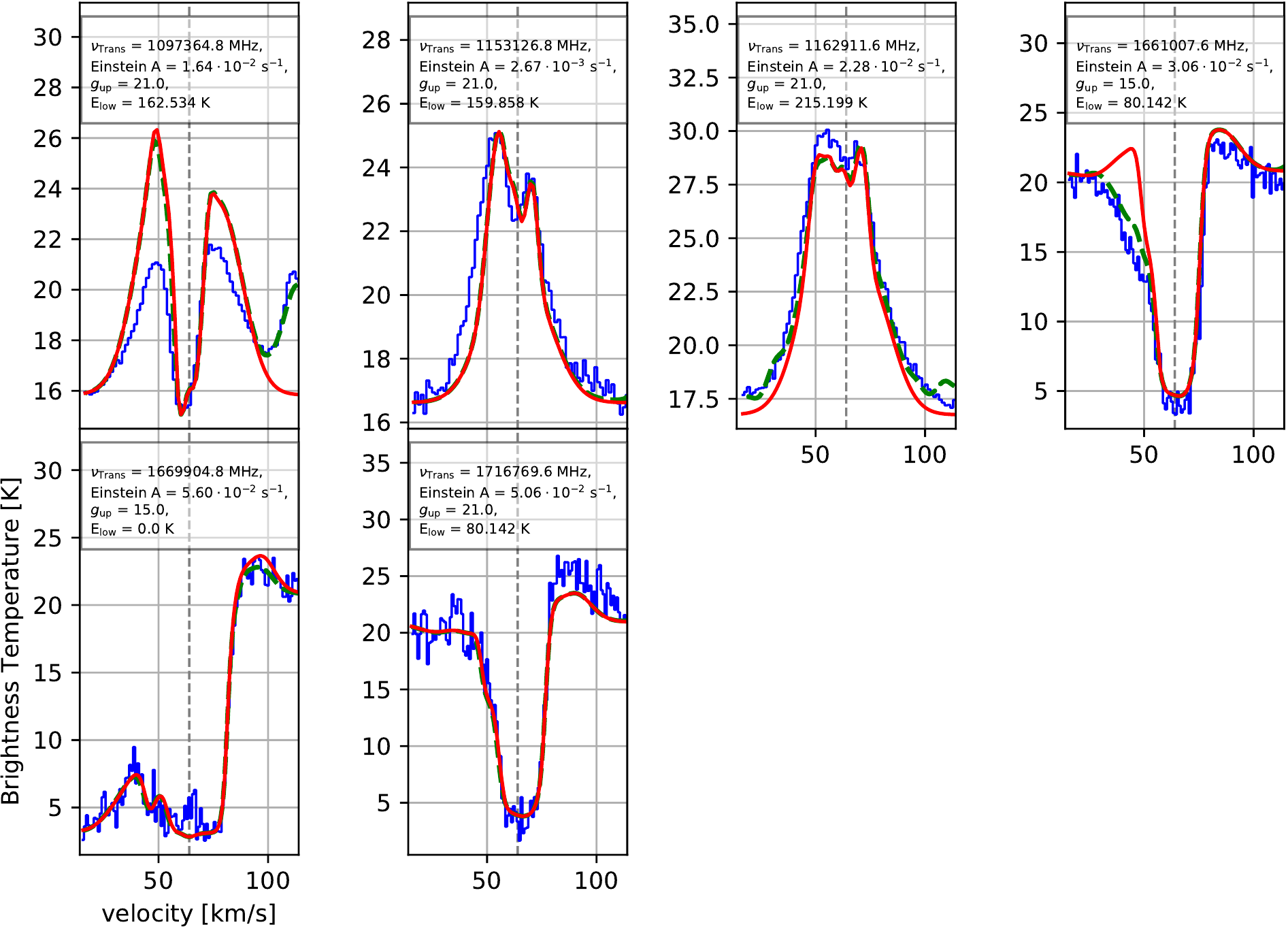}\\
    \caption{Selected transitions of \emph{ortho}-H$_2$O (red line).}
    \label{fig:oh2o}
\end{figure*}

\begin{figure*}[!htb]
    \centering
    \includegraphics[scale=0.80]{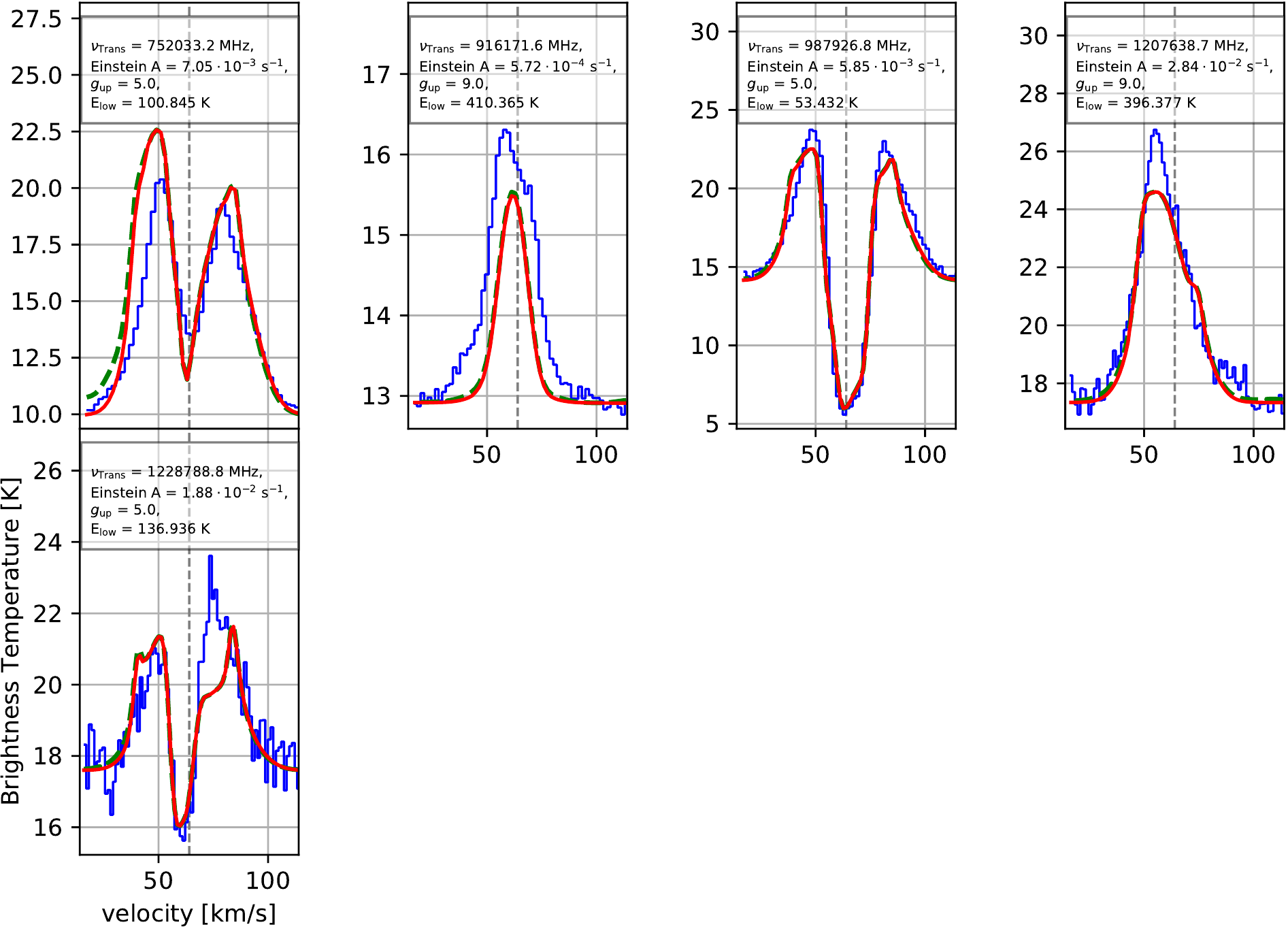}\\
    \caption{Selected transitions of \emph{para}-H$_2$O (red line).}
    \label{fig:ph2o}
\end{figure*}
\newpage

\clearpage

\begin{figure*}[!htb]
    \centering
    \includegraphics[scale=0.80]{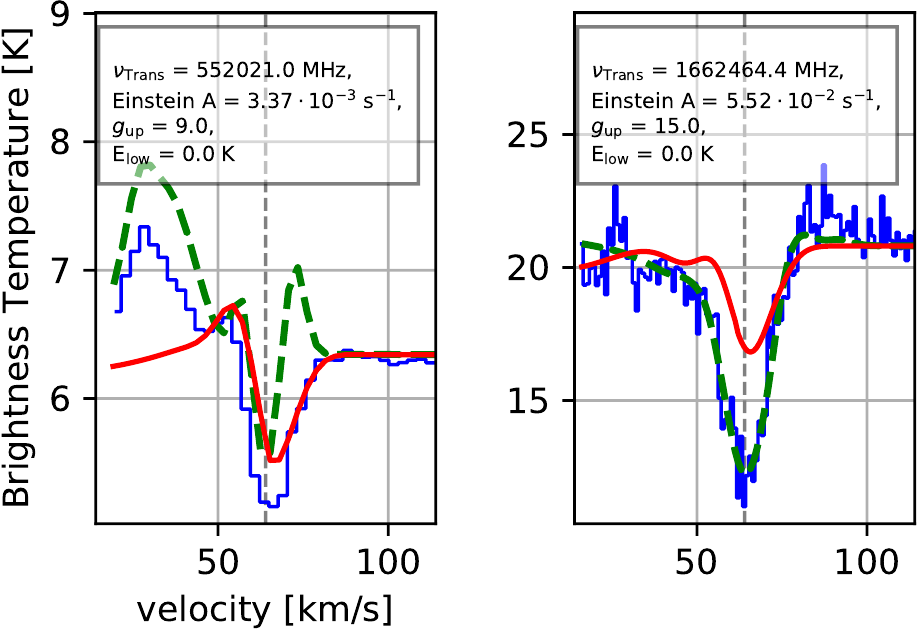}\\
    \caption{Selected transitions of \emph{ortho}-H$_2 \, ^{17}$O (red line).}
    \label{fig:oh2o-17}
\end{figure*}

\begin{figure*}[!htb]
    \centering
    \includegraphics[scale=0.80]{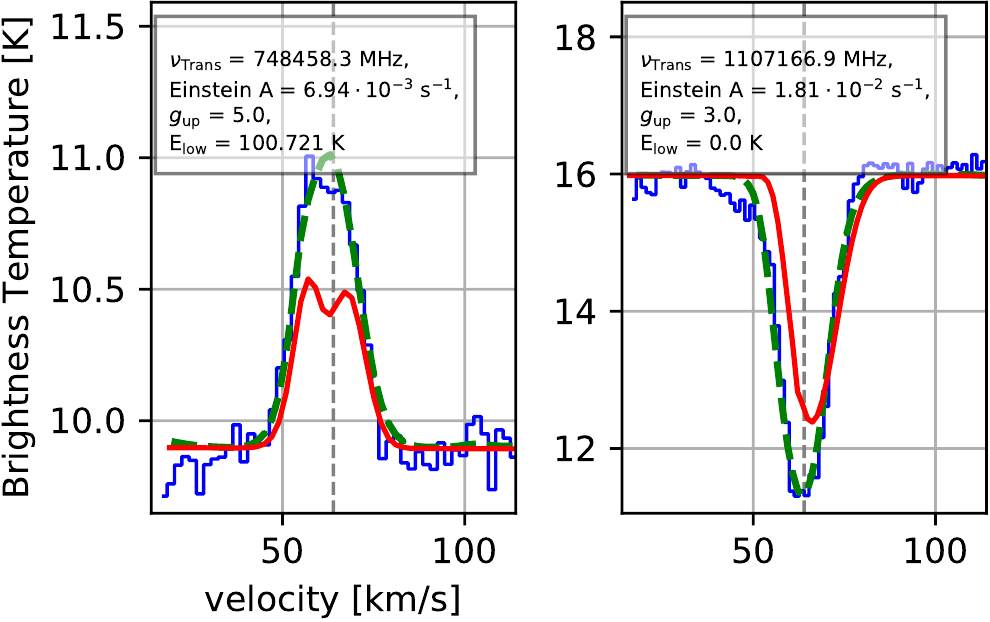}\\
    \caption{Selected transitions of \emph{para}-H$_2 \, ^{17}$O (red line).}
    \label{fig:ph2o-17}
\end{figure*}

\begin{figure*}[!htb]
    \centering
    \includegraphics[scale=0.80]{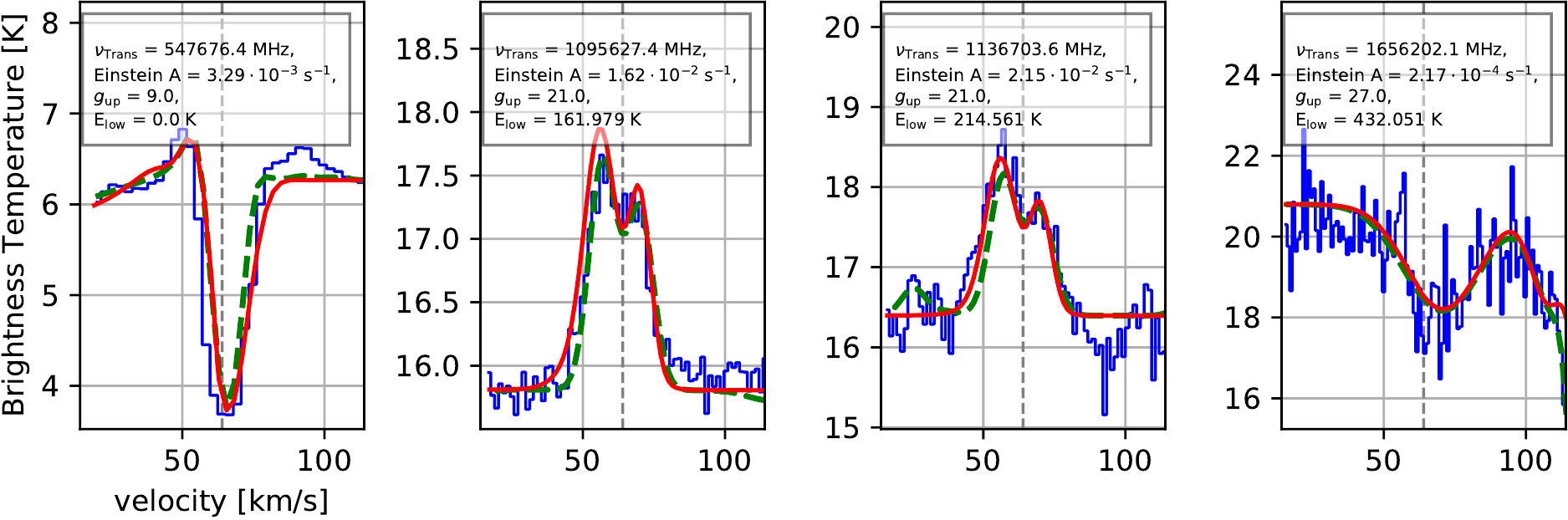}\\
    \caption{Selected transitions of \emph{ortho}-H$_2 \, ^{18}$O (red line).}
    \label{fig:oh2o-18}
\end{figure*}
\newpage

\clearpage

\begin{figure*}[!htb]
    \centering
    \includegraphics[scale=0.80]{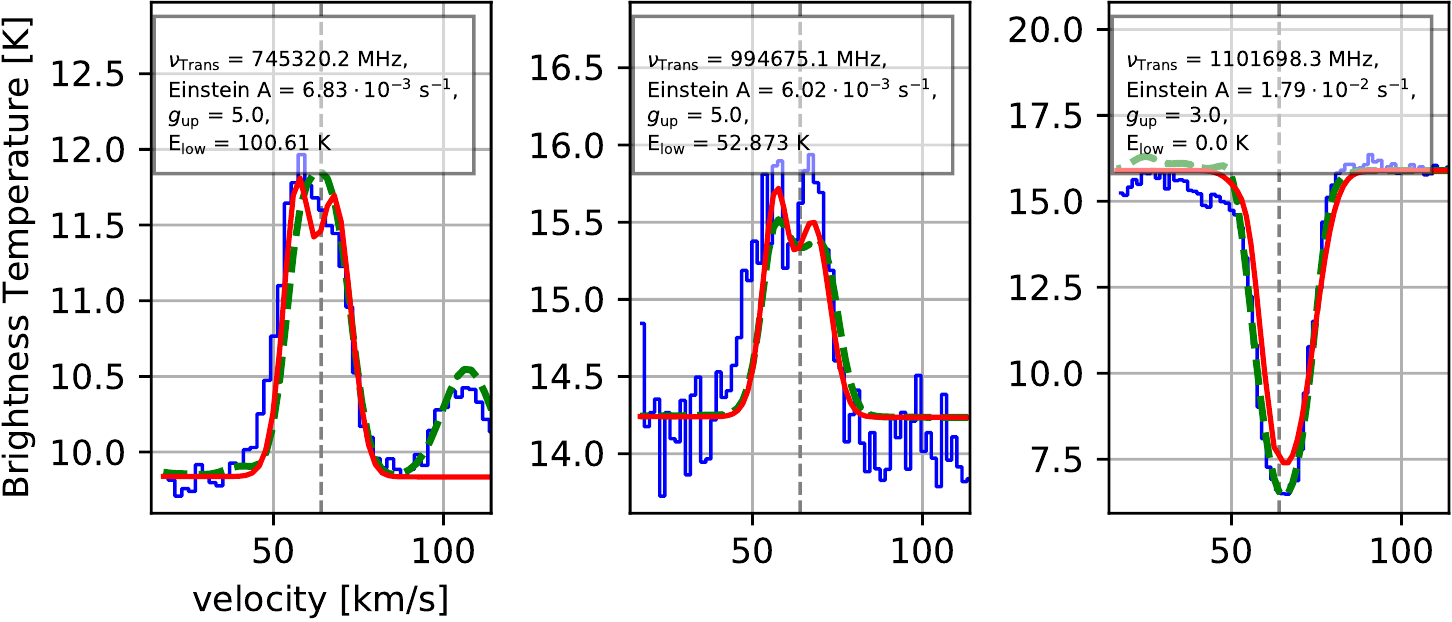}\\
    \caption{Selected transitions of \emph{para}-H$_2 \, ^{18}$O (red line).}
    \label{fig:ph2o-18}
\end{figure*}

\begin{figure*}[!htb]
    \centering
    \includegraphics[scale=0.80]{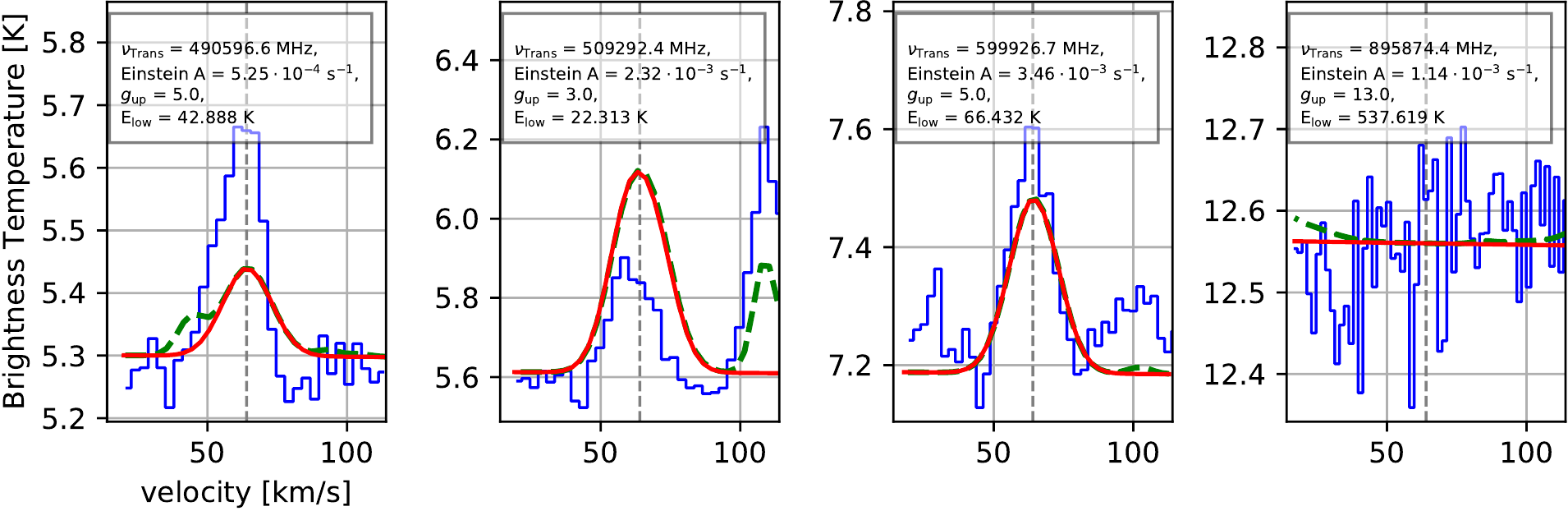}\\
    \caption{Selected transitions of HDO (red line).}
    \label{fig:hdo}
\end{figure*}

\begin{figure*}[!htb]
    \centering
    \includegraphics[scale=0.80]{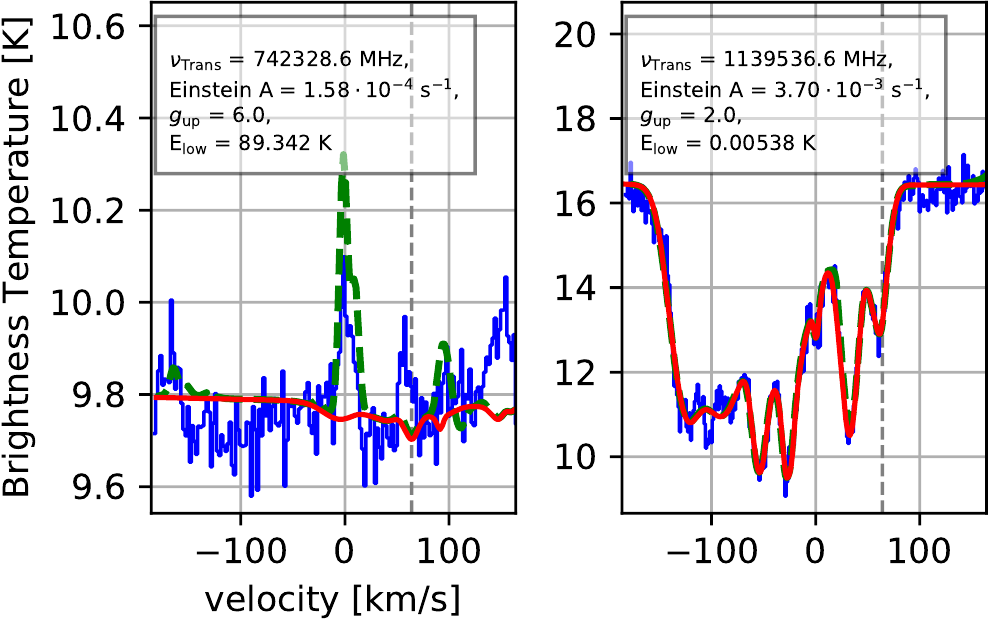}\\
    \caption{Selected transitions of \emph{ortho}-H$_2$O$^{+}$ (red line).}
    \label{fig:oh2o+}
\end{figure*}

\begin{figure*}[!htb]
    \centering
    \includegraphics[scale=0.80]{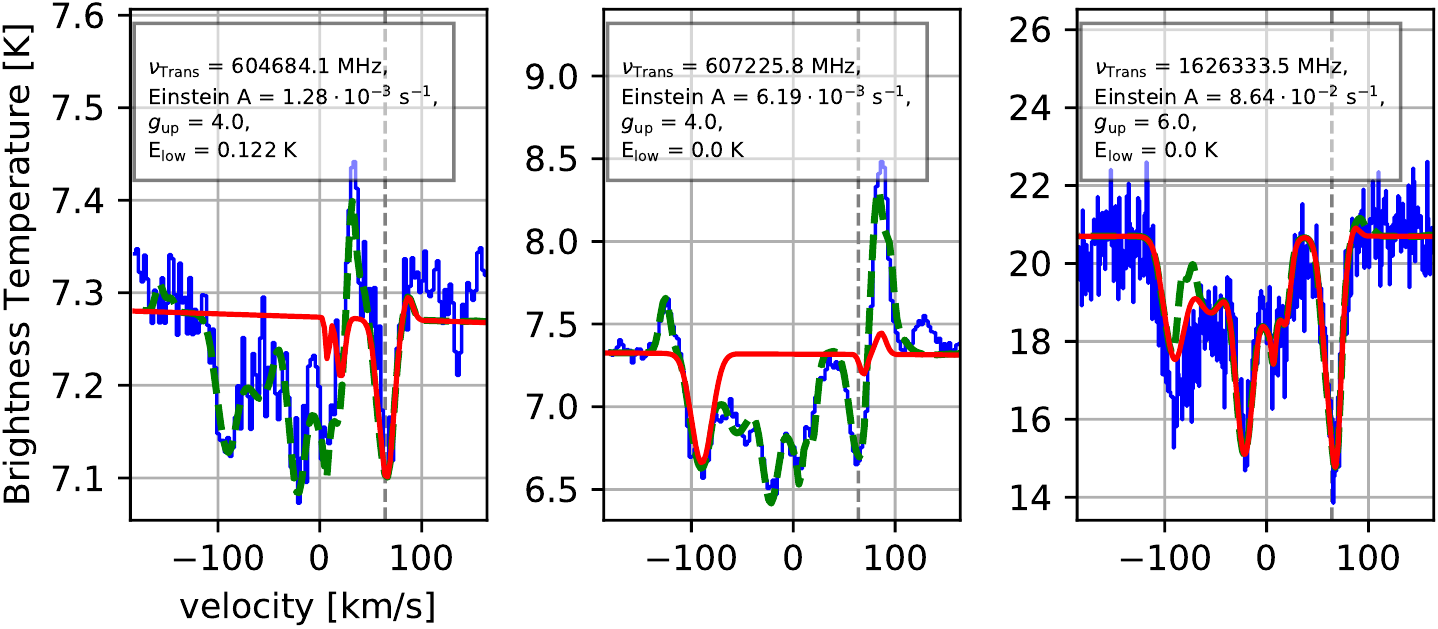}\\
    \caption{Selected transitions of \emph{para}-H$_2$O$^{+}$ (red line).}
    \label{fig:ph2o+}
\end{figure*}

\begin{figure*}[!htb]
    \centering
    \includegraphics[scale=0.80]{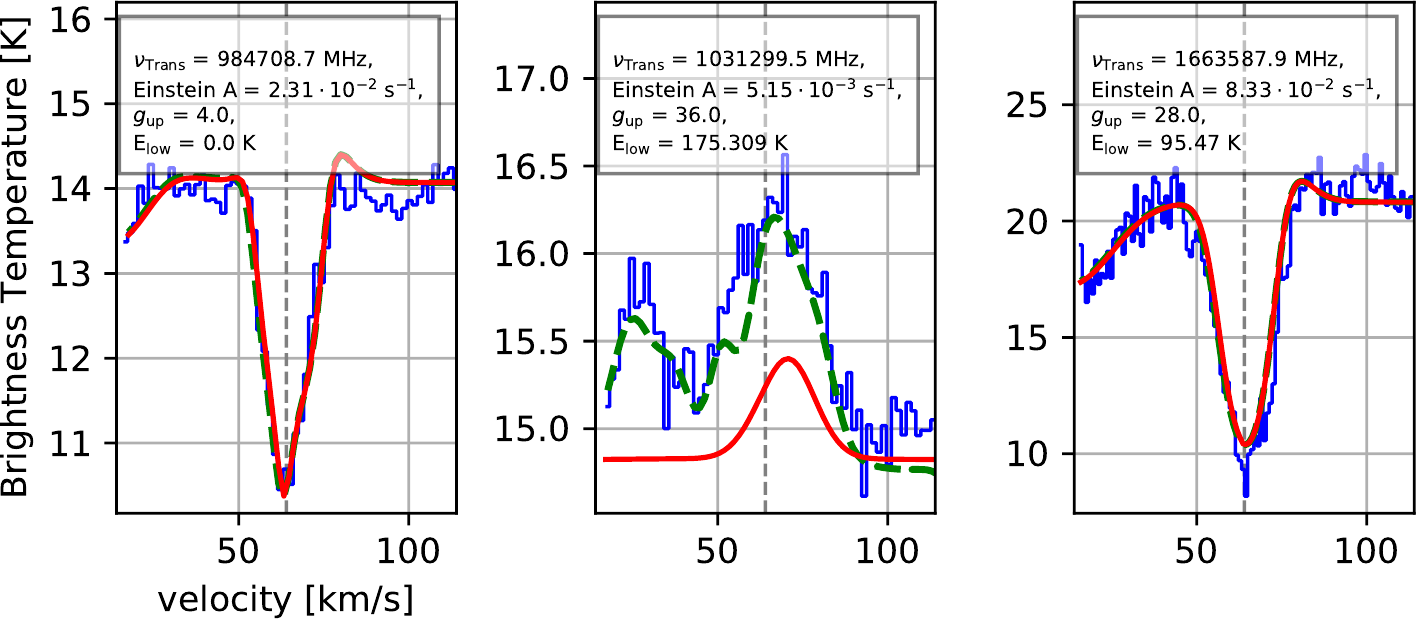}\\
    \caption{Selected transitions of \emph{ortho}-H$_3$O$^{+}$ (red line).}
    \label{fig:oh3o+}
\end{figure*}

\begin{figure*}[!htb]
    \centering
    \includegraphics[scale=0.80]{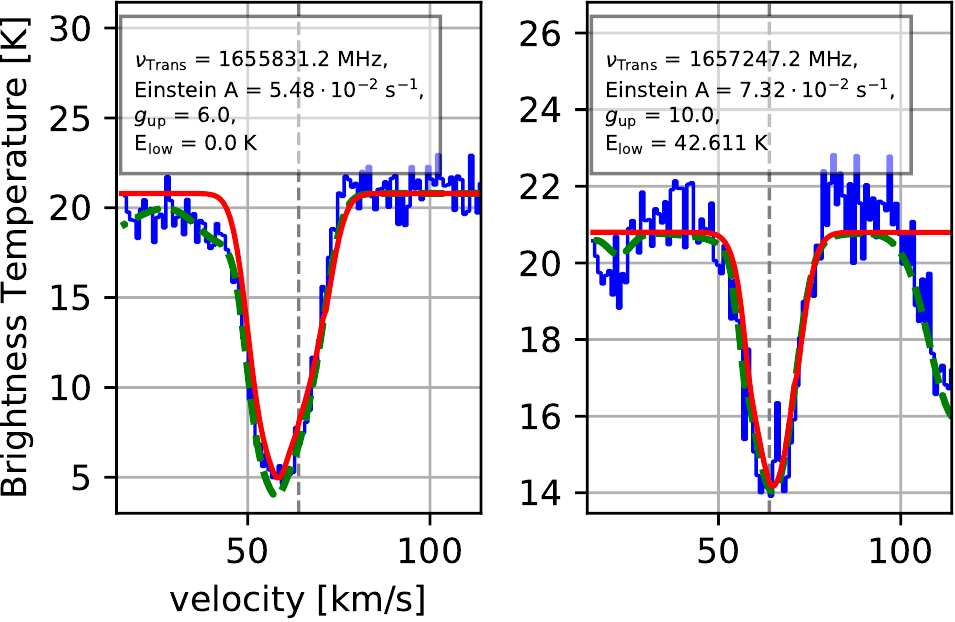}\\
    \caption{Selected transitions of \emph{para}-H$_3$O$^{+}$ (red line).}
    \label{fig:ph3o+}
\end{figure*}

\begin{figure*}[!htb]
    \centering
    \includegraphics[scale=0.80]{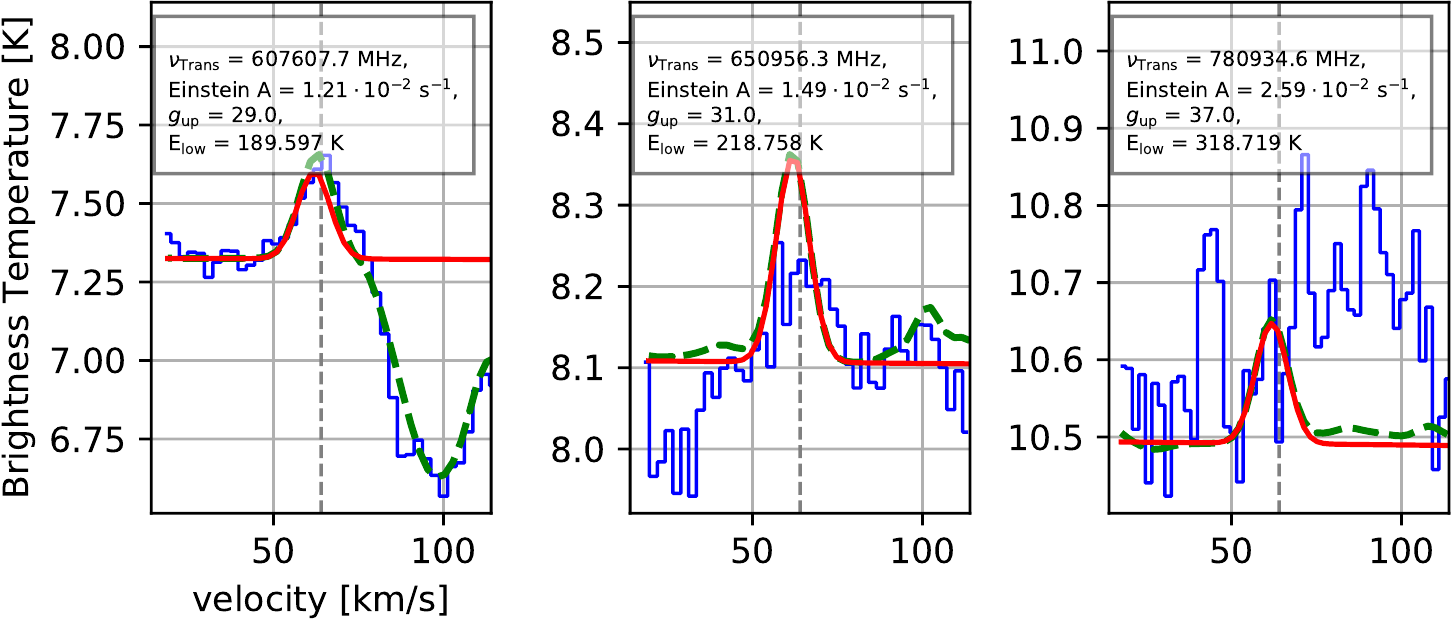}\\
    \caption{Selected transitions of SiO (red line).}
    \label{fig:sio}
\end{figure*}
\newpage

\clearpage

\begin{figure*}[!htb]
    \centering
    \includegraphics[scale=0.80]{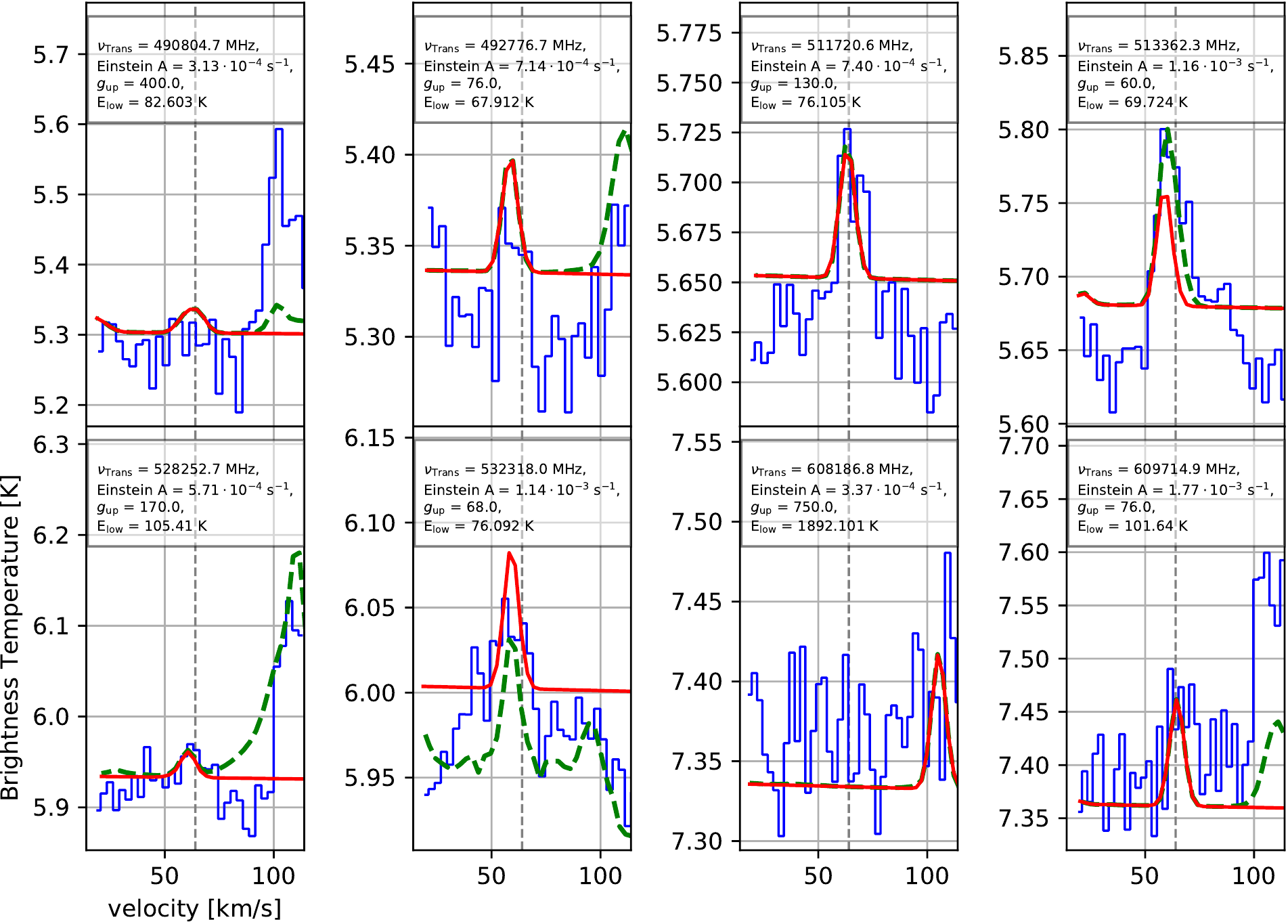}\\
    \caption{Selected transitions of CH$_3$OCH$_3$ (red line).}
    \label{fig:ch3och3}
\end{figure*}

\begin{figure*}[!htb]
    \centering
    \includegraphics[scale=0.80]{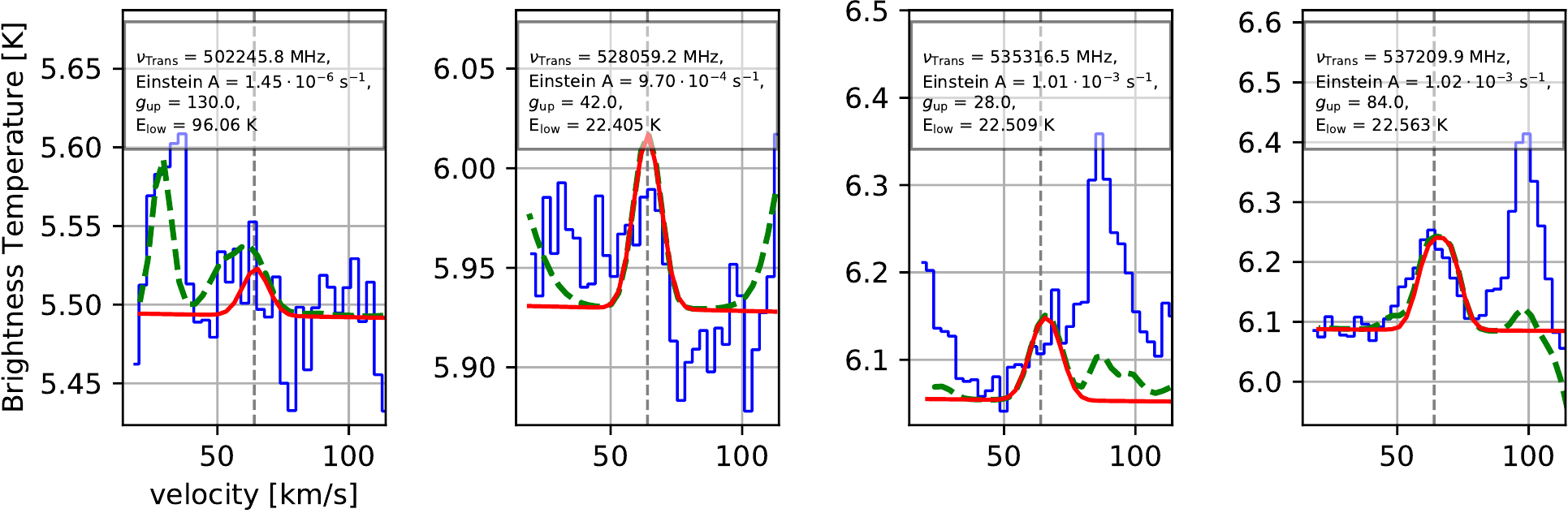}\\
    \caption{Selected transitions of CH$_3$NH$_2$ (red line).}
    \label{fig:ch3nh2}
\end{figure*}
\newpage

\clearpage

\begin{figure*}[!htb]
    \centering
    \includegraphics[scale=0.80]{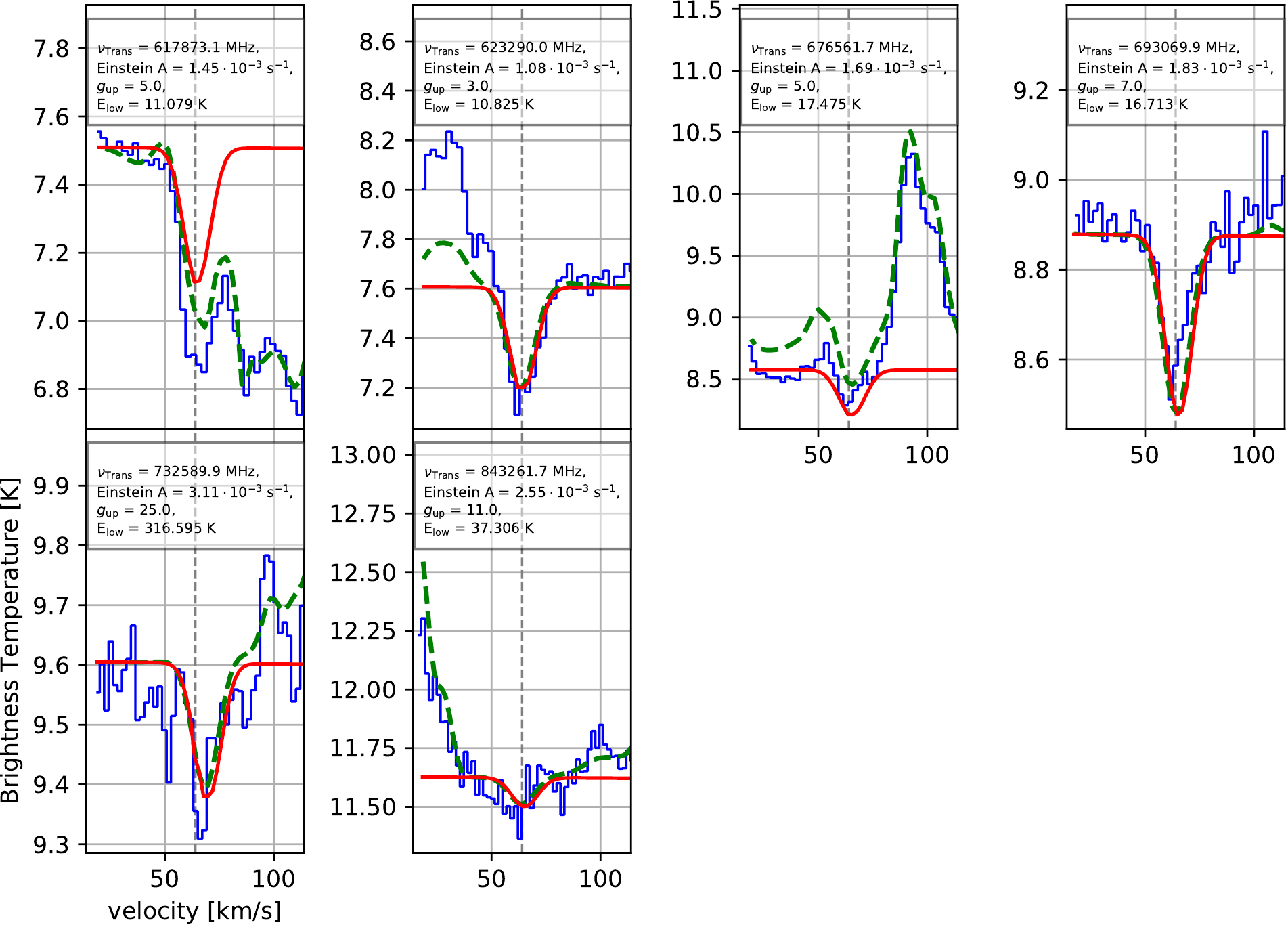}\\
    \caption{Selected transitions of CH$_2$NH (red line).}
    \label{fig:ch2nh}
\end{figure*}

\begin{figure*}[!htb]
    \centering
    \includegraphics[scale=0.80]{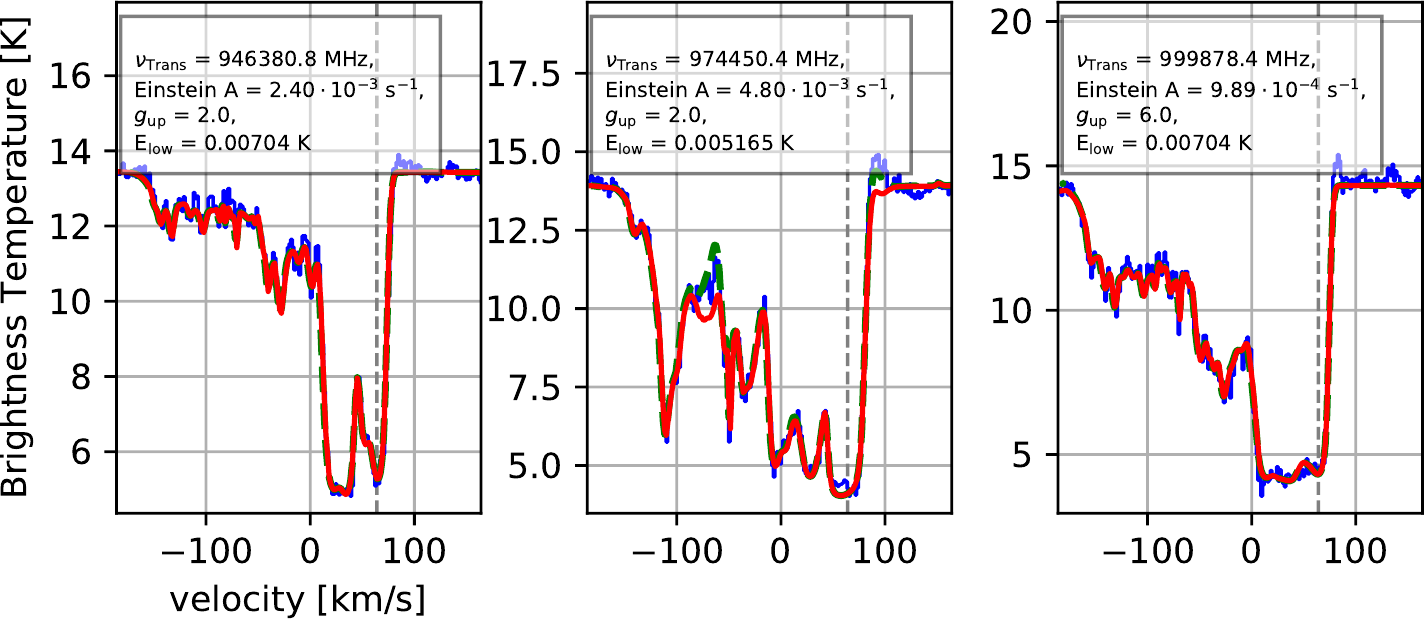}\\
    \caption{Selected transitions of NH (red line).}
    \label{fig:nh}
\end{figure*}

\begin{figure*}[!htb]
    \centering
    \includegraphics[scale=0.80]{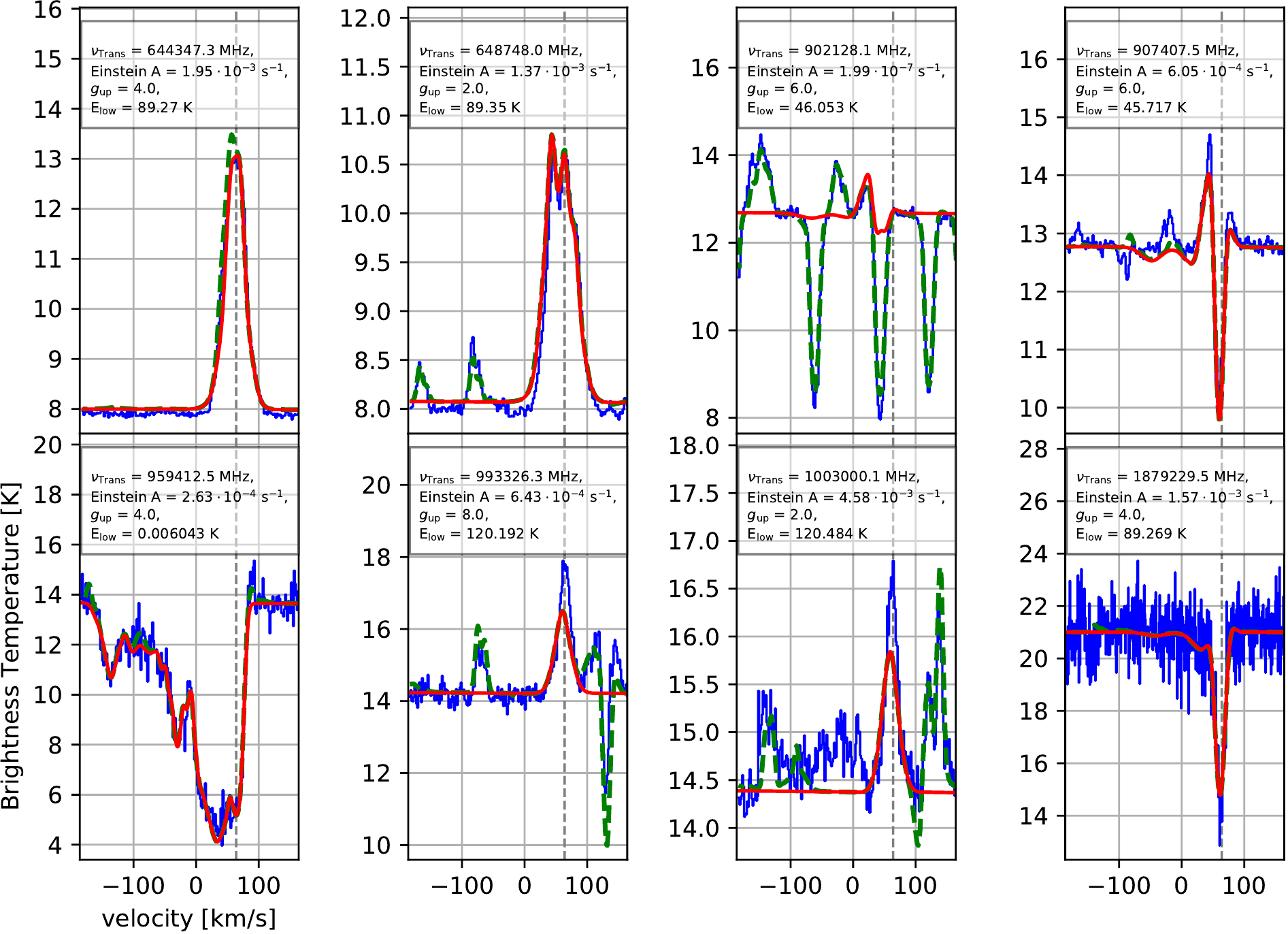}\\
    \caption{Selected transitions of \emph{ortho}-NH$_2$ (red line).}
    \label{fig:onh2}
\end{figure*}

\begin{figure*}[!htb]
    \centering
    \includegraphics[scale=0.80]{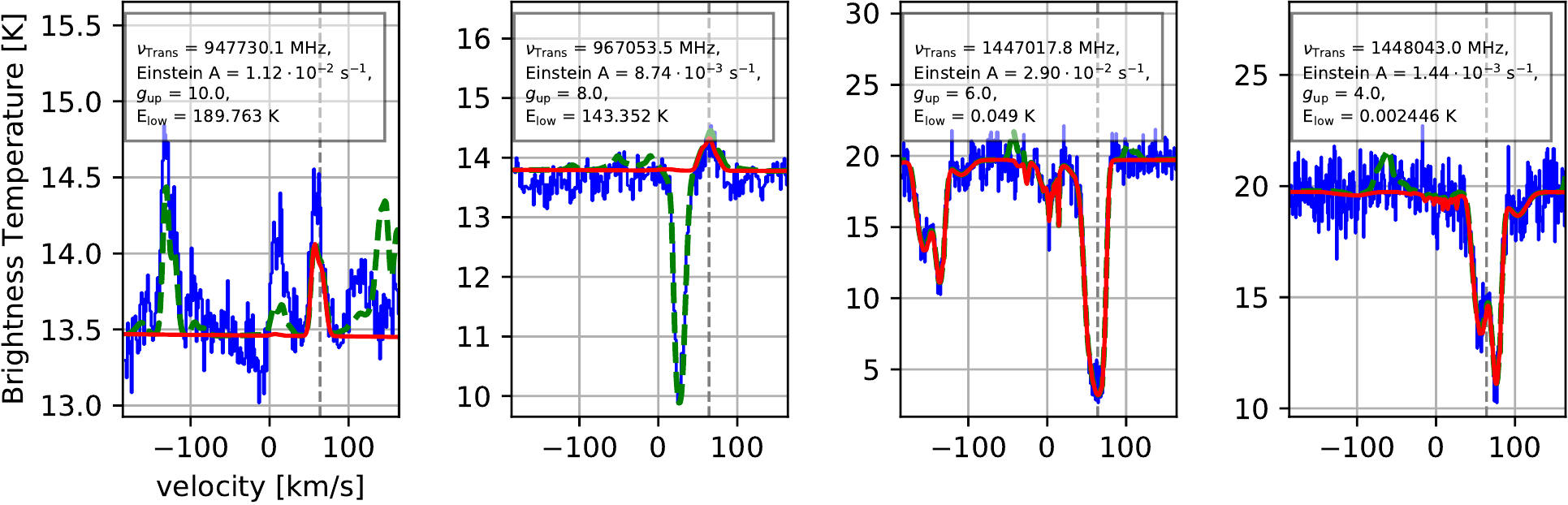}\\
    \caption{Selected transitions of \emph{para}-NH$_2$ (red line).}
    \label{fig:pnh2}
\end{figure*}

\begin{figure*}[!htb]
    \centering
    \includegraphics[scale=0.80]{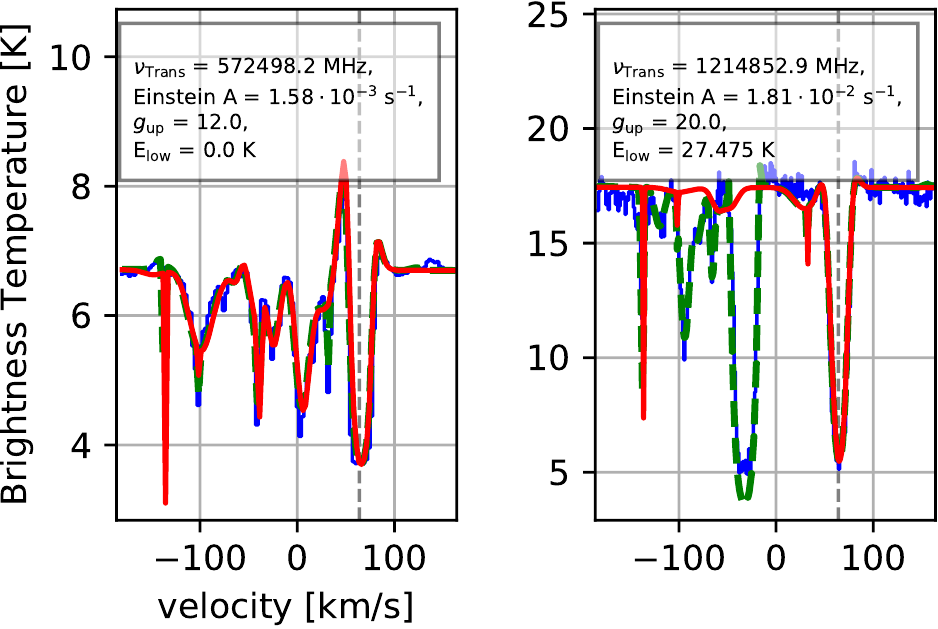}\\
    \caption{Selected transitions of \emph{ortho}-NH$_3$ (red line).}
    \label{fig:onh3}
\end{figure*}

\begin{figure*}[!htb]
    \centering
    \includegraphics[scale=0.80]{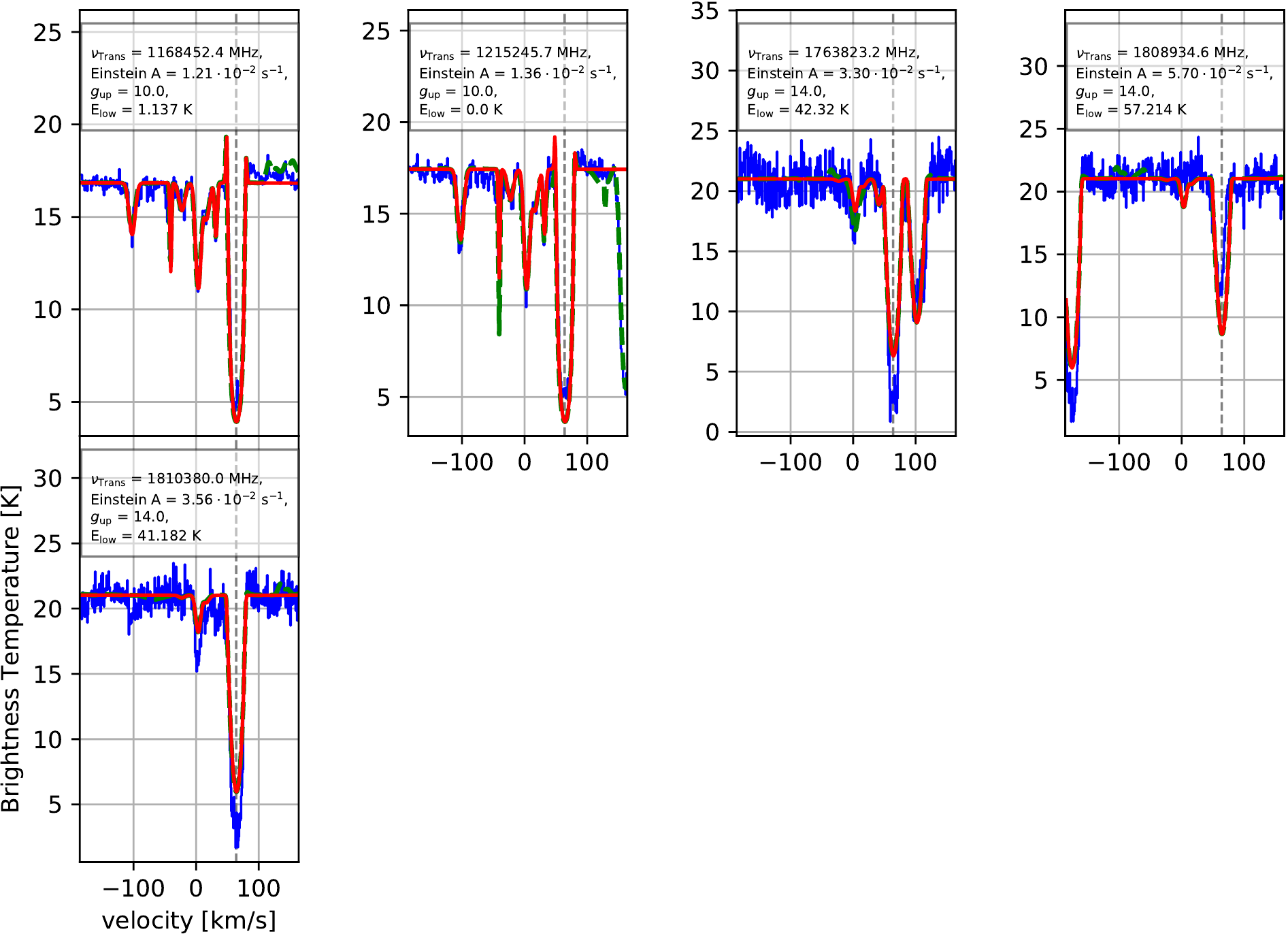}\\
    \caption{Selected transitions of \emph{para}-NH$_3$ (red line).}
    \label{fig:pnh3}
\end{figure*}

\begin{figure*}[!htb]
    \centering
    \includegraphics[scale=0.80]{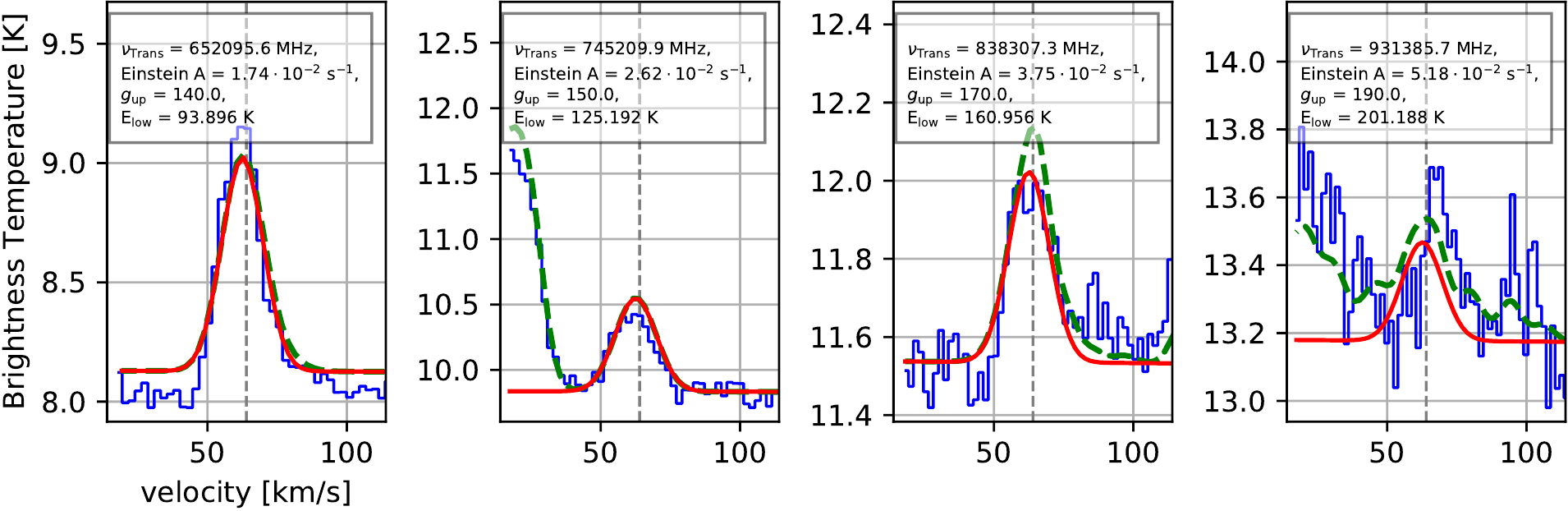}\\
    \caption{Selected transitions of N$_2$H$^{+}$ (red line).}
    \label{fig:n2h+}
\end{figure*}

\begin{figure*}[!htb]
    \centering
    \includegraphics[scale=0.80]{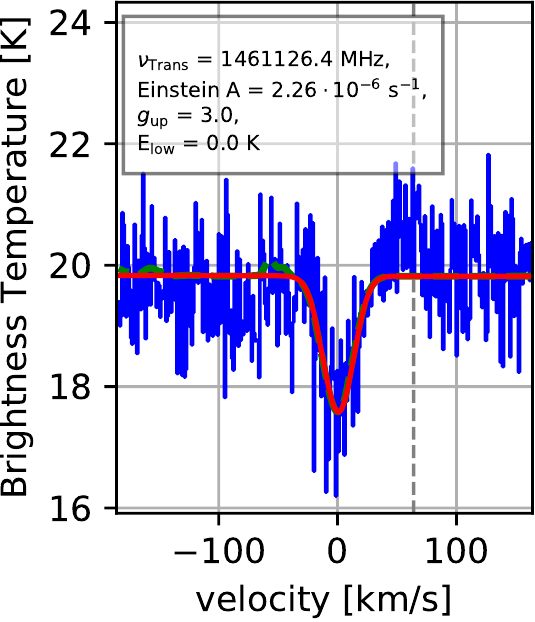}\\
    \caption{Selected transitions of $^{14}$N$^{+}$ (red line).}
    \label{fig:14n+}
\end{figure*}

\begin{figure*}[!htb]
    \centering
    \includegraphics[scale=0.80]{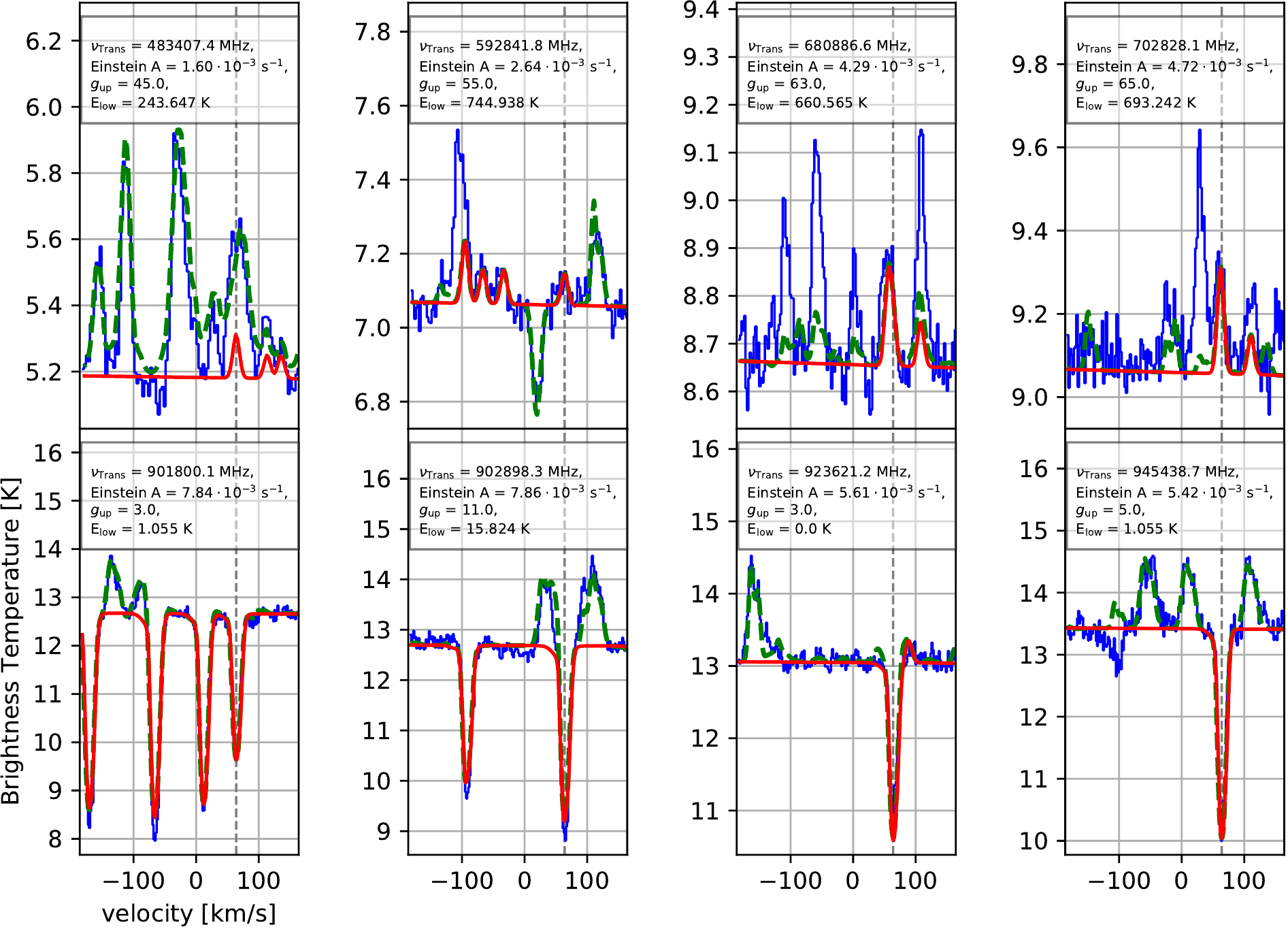}\\
    \caption{Selected transitions of HNCO (red line).}
    \label{fig:hnco}
\end{figure*}

\begin{figure*}[!htb]
    \centering
    \includegraphics[scale=0.80]{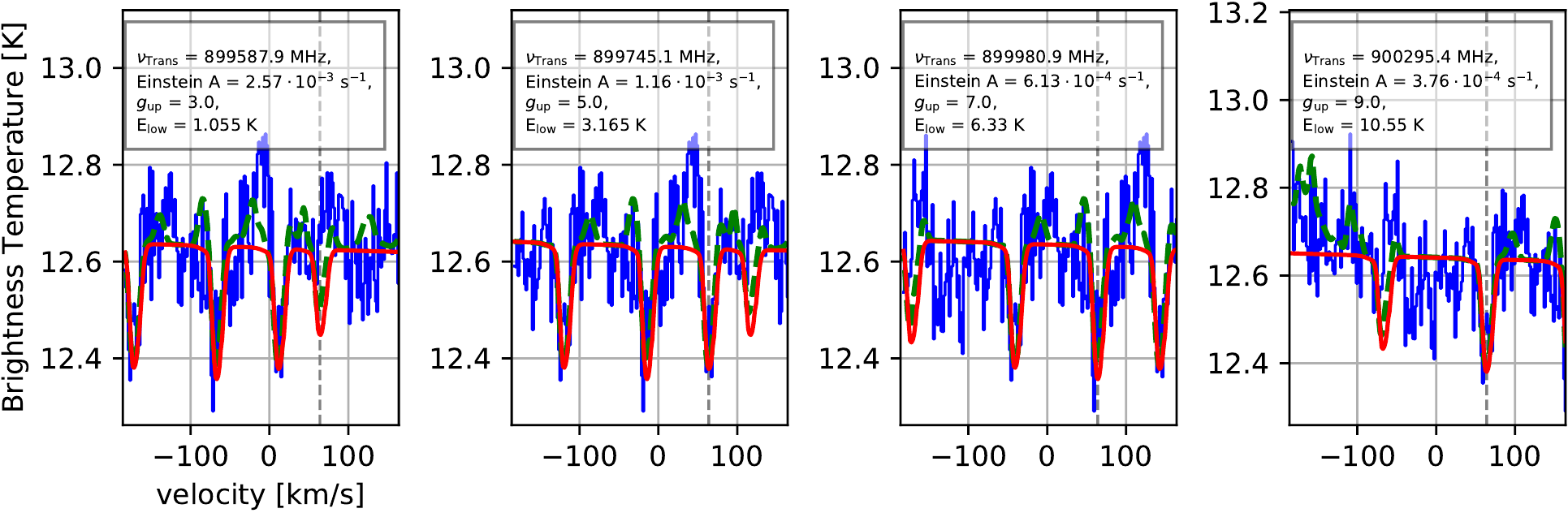}\\
    \caption{Selected transitions of HN$^{13}$CO (red line).}
    \label{fig:hnc13o}
\end{figure*}

\begin{figure*}[!htb]
    \centering
    \includegraphics[scale=0.80]{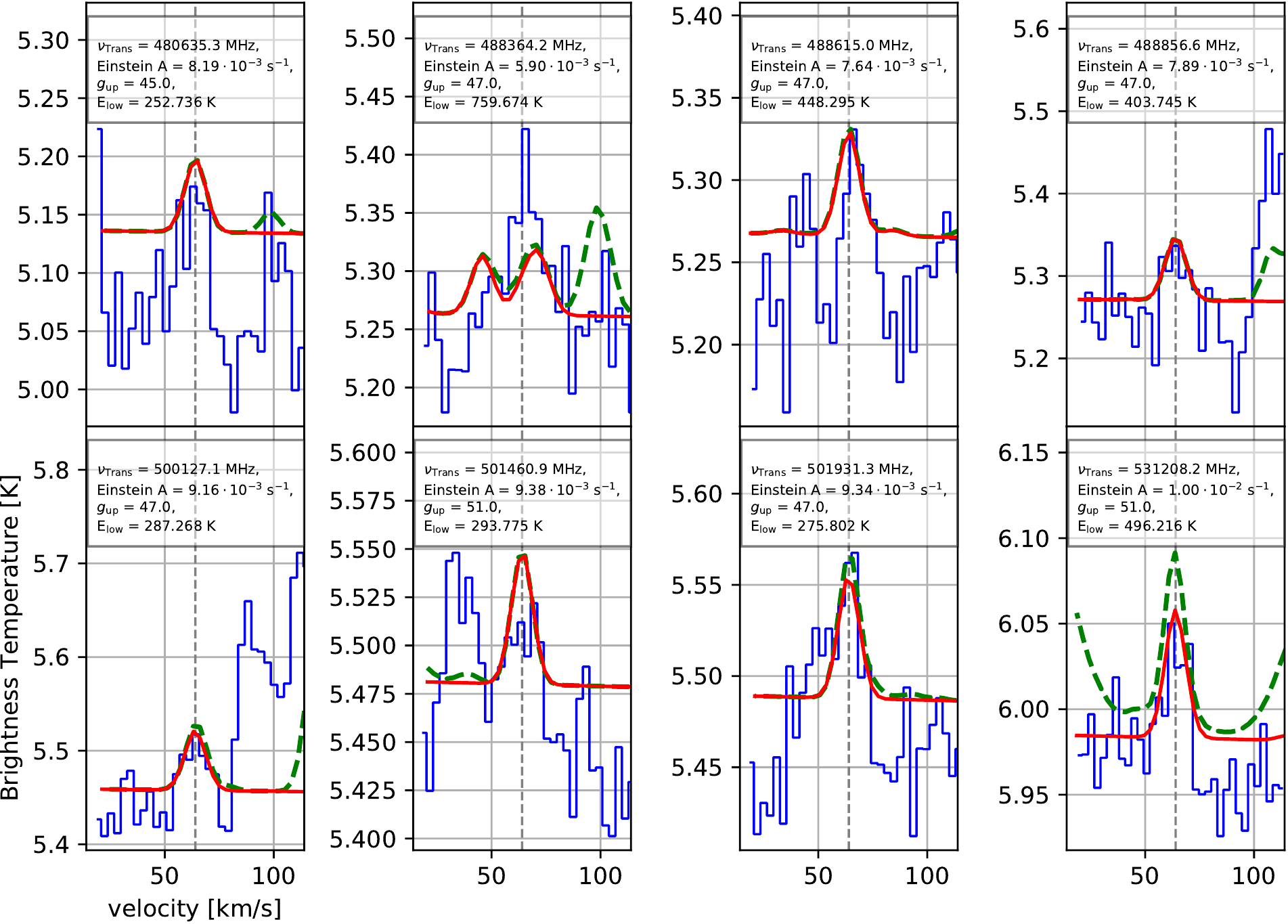}\\
    \caption{Selected transitions of NH$_2$CHO (red line).}
    \label{fig:nh2cho}
\end{figure*}
\newpage

\clearpage

\begin{figure*}[!htb]
    \centering
    \includegraphics[scale=0.80]{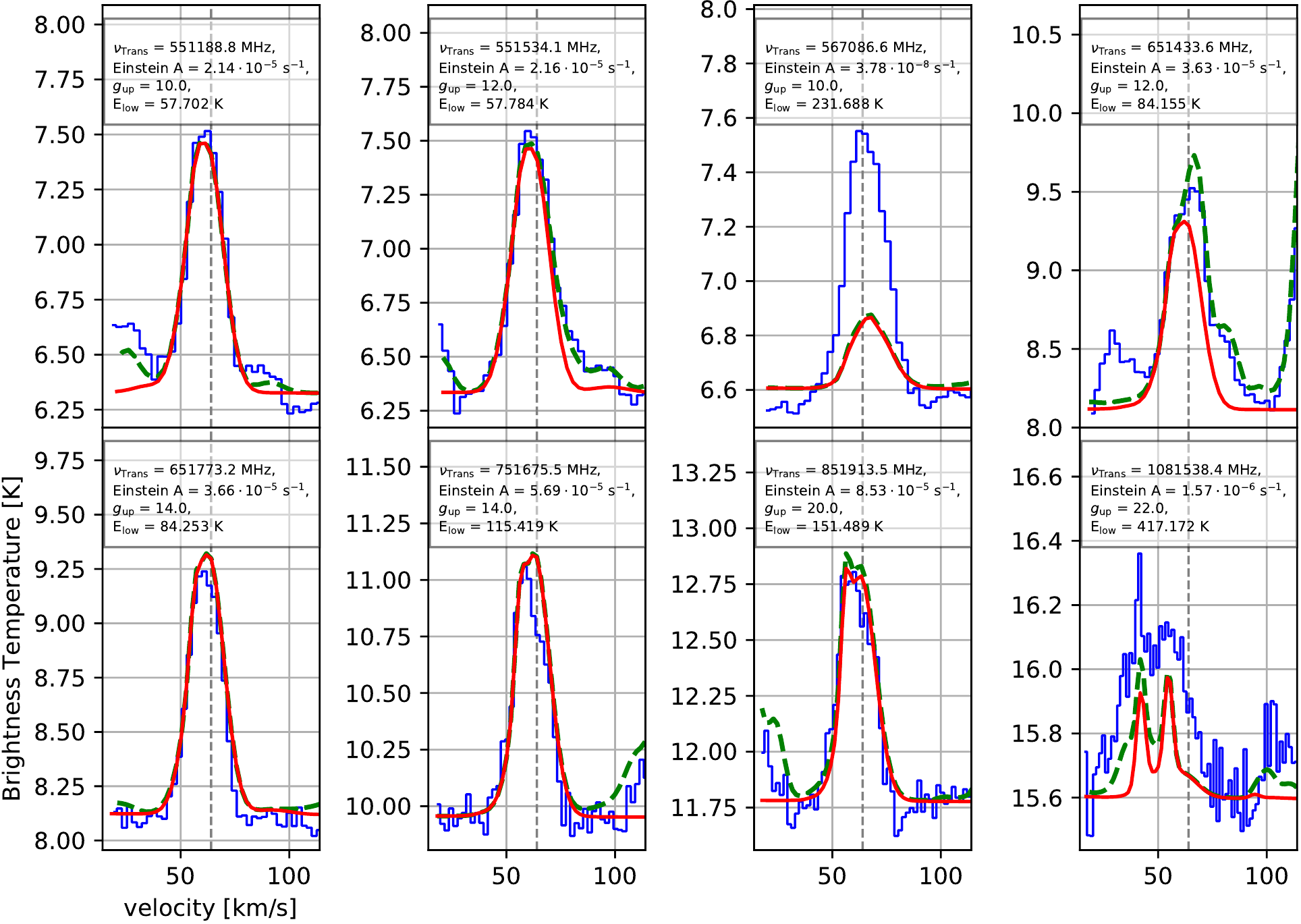}\\
    \caption{Selected transitions of NO (red line).}
    \label{fig:no}
\end{figure*}

\begin{figure*}[!htb]
    \centering
    \includegraphics[scale=0.80]{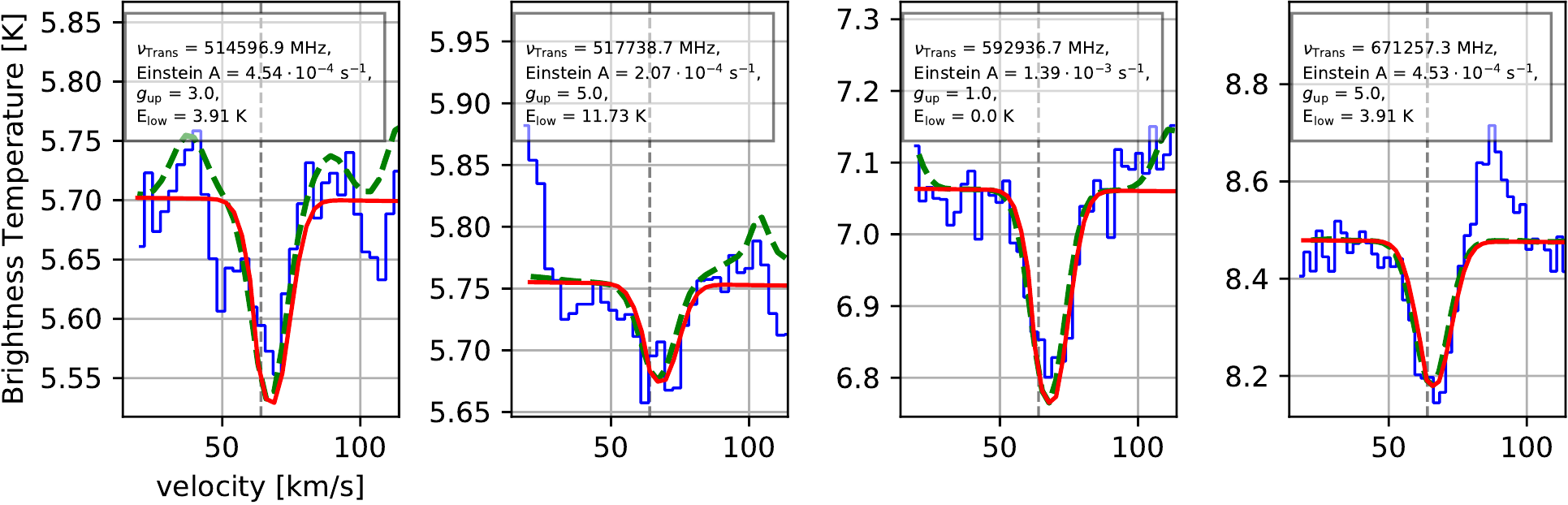}\\
    \caption{Selected transitions of HNO (red line).}
    \label{fig:hno}
\end{figure*}

\begin{figure*}[!htb]
    \centering
    \includegraphics[scale=0.80]{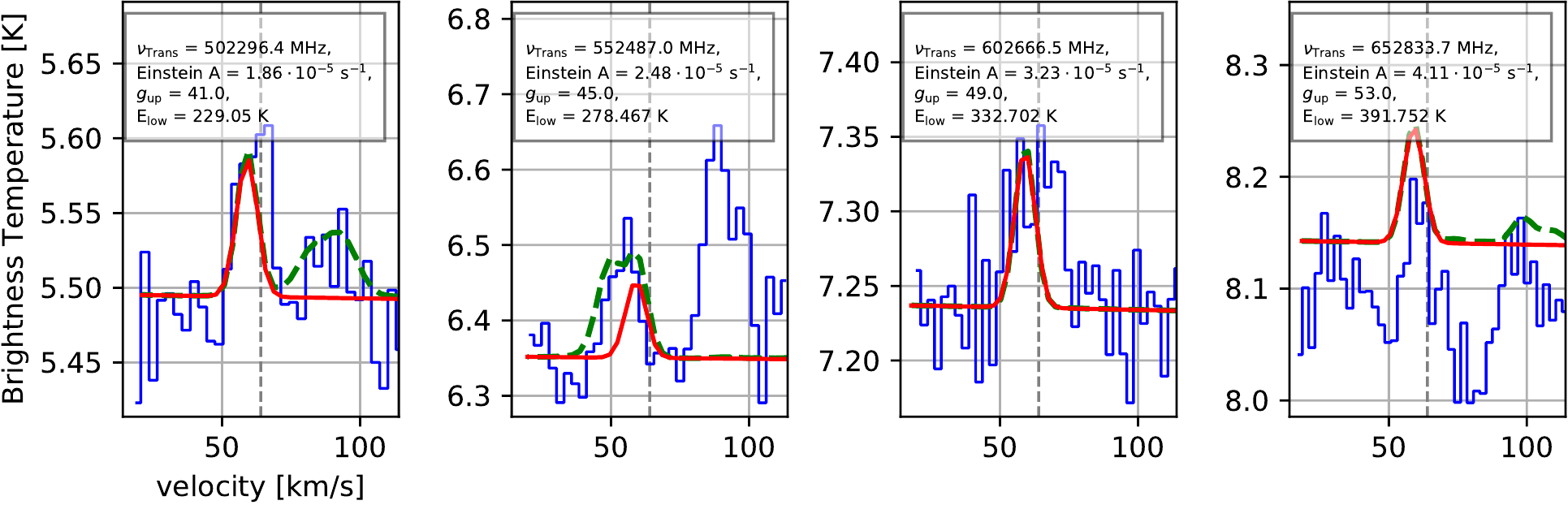}\\
    \caption{Selected transitions of N$_2$O (red line).}
    \label{fig:n2o}
\end{figure*}

\begin{figure*}[!htb]
    \centering
    \includegraphics[scale=0.80]{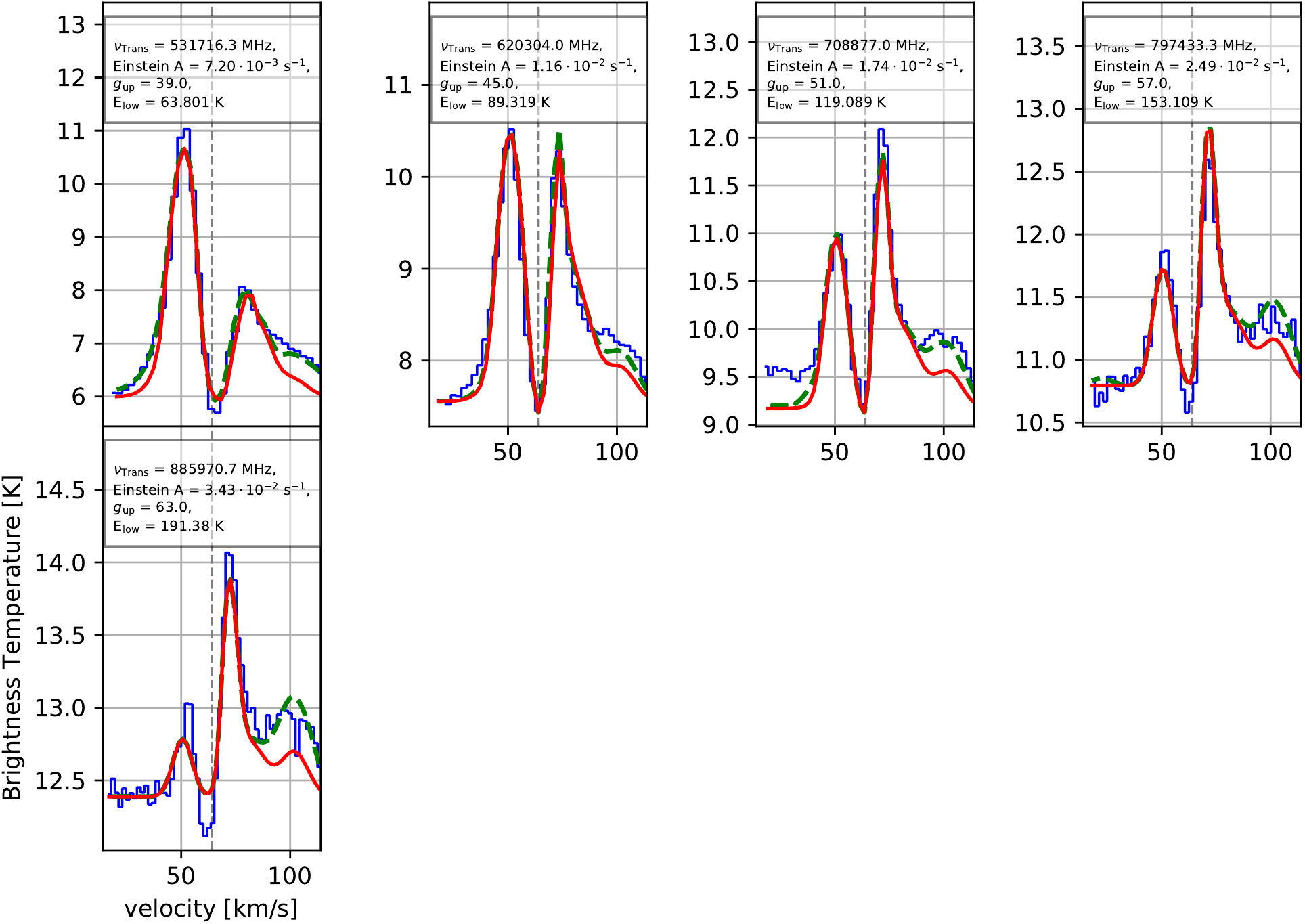}\\
    \caption{Selected transitions of HCN (red line).}
    \label{fig:hcn}
\end{figure*}

\begin{figure*}[!htb]
    \centering
    \includegraphics[scale=0.80]{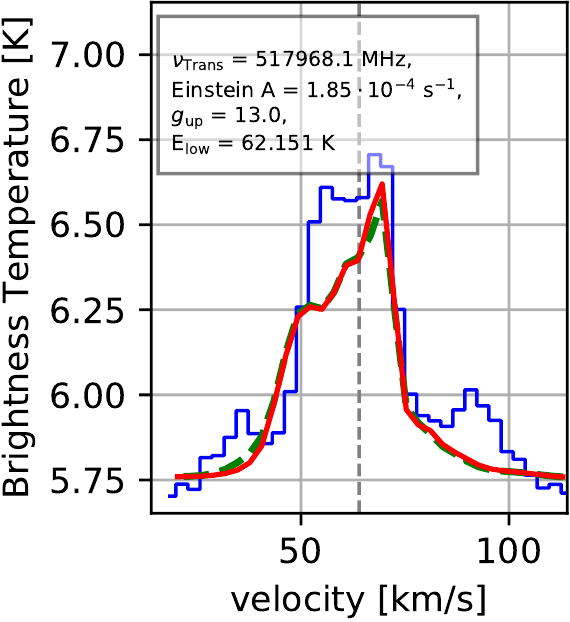}\\
    \caption{Selected transitions of H$^{13}$CN (red line).}
    \label{fig:hc13n}
\end{figure*}

\begin{figure*}[!htb]
    \centering
    \includegraphics[scale=0.80]{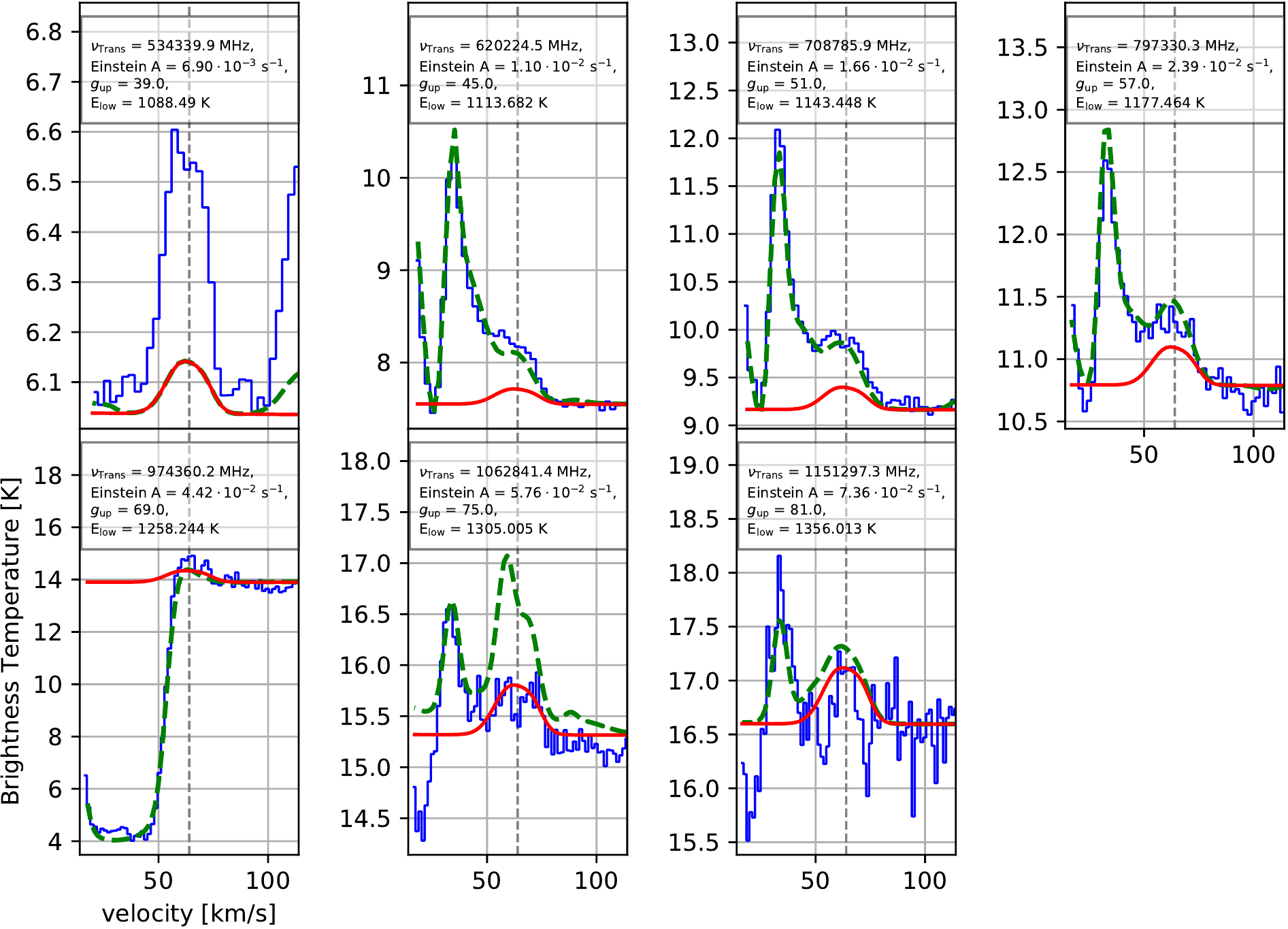}\\
    \caption{Selected transitions of HCN,v$_2$=1 (red line).}
    \label{fig:hcnv21}
\end{figure*}

\begin{figure*}[!htb]
    \centering
    \includegraphics[scale=0.80]{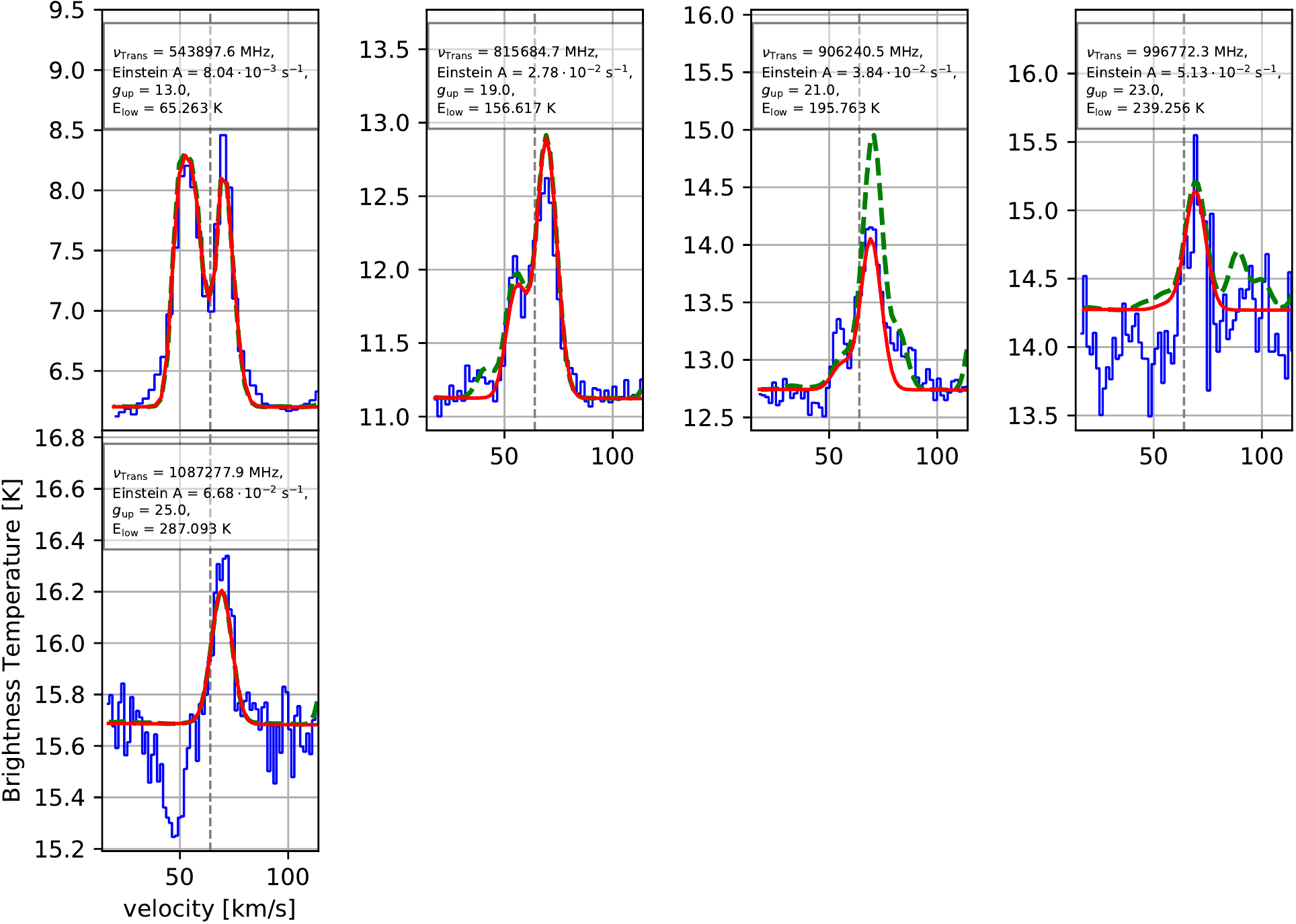}\\
    \caption{Selected transitions of HNC (red line).}
    \label{fig:hnc}
\end{figure*}

\begin{figure*}[!htb]
    \centering
    \includegraphics[scale=0.80]{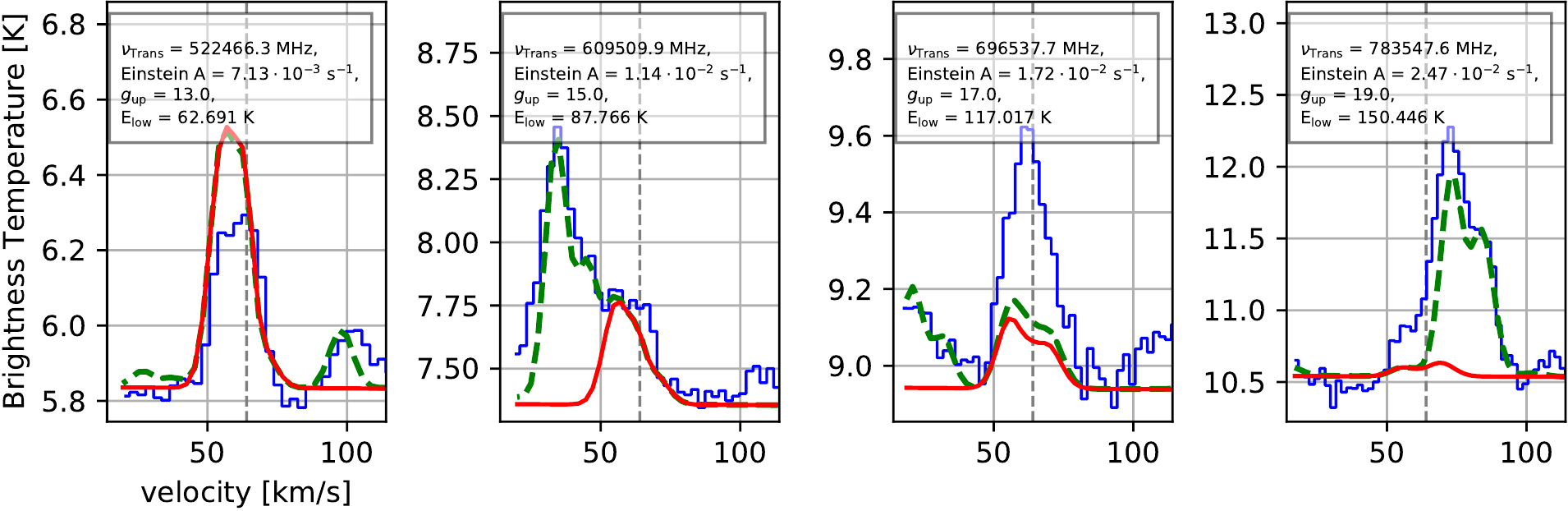}\\
    \caption{Selected transitions of HN$^{13}$C (red line).}
    \label{fig:hnc13}
\end{figure*}

\begin{figure*}[!htb]
    \centering
    \includegraphics[scale=0.80]{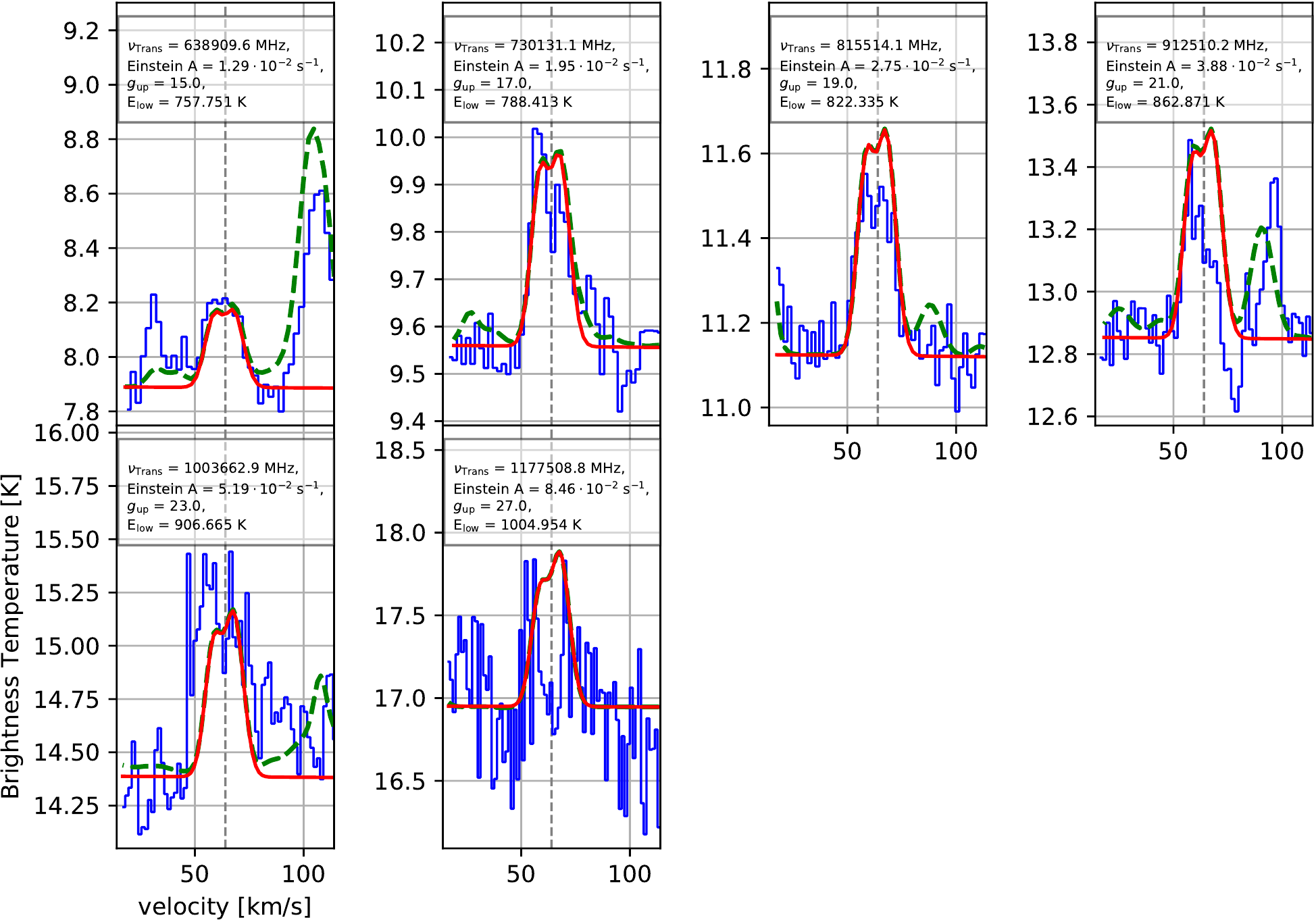}\\
    \caption{Selected transitions of HNC,v$_2$=1 (red line).}
    \label{fig:hncv21}
\end{figure*}

\begin{figure*}[!htb]
    \centering
    \includegraphics[scale=0.80]{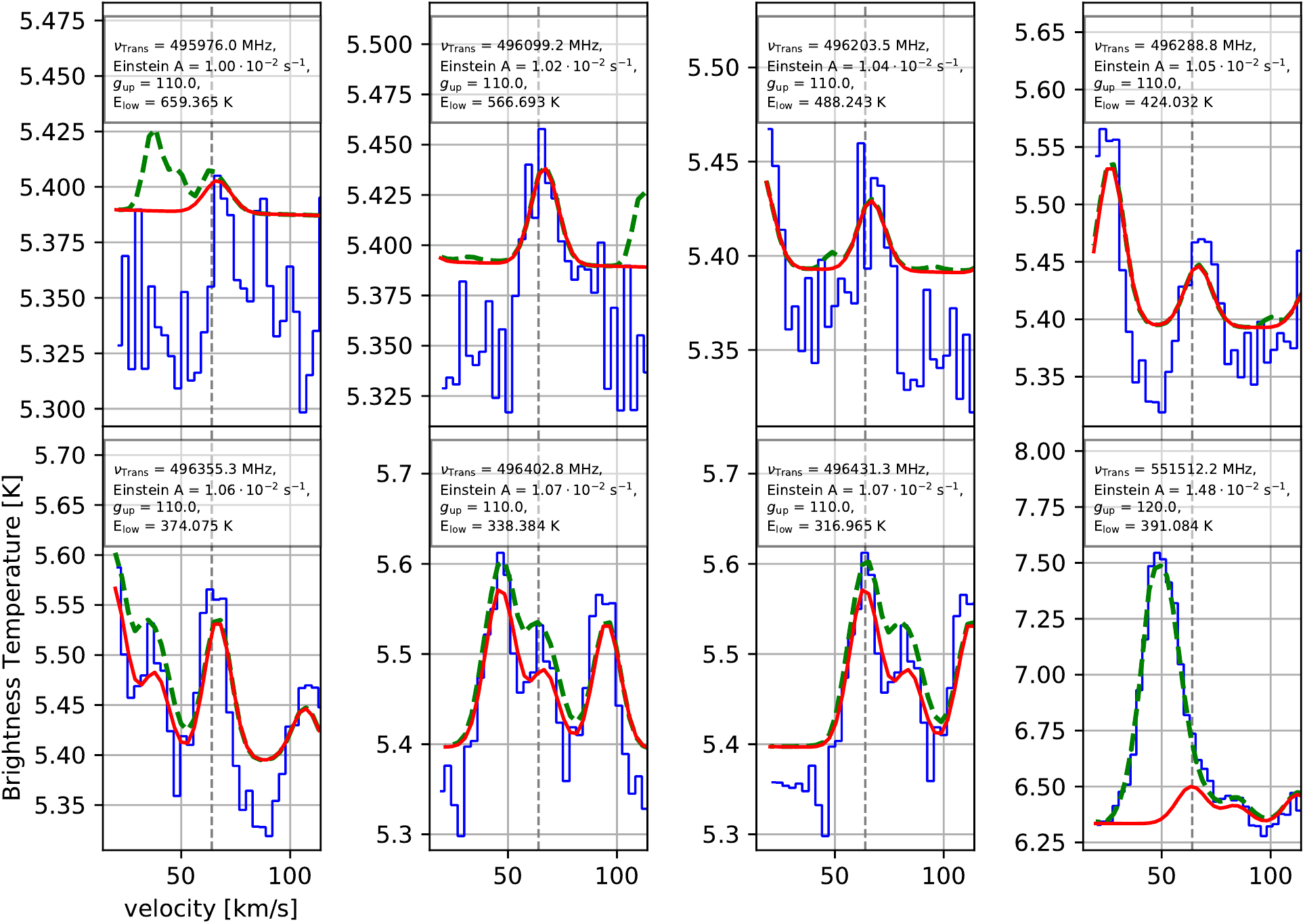}\\
    \caption{Selected transitions of CH$_3$CN (red line).}
    \label{fig:ch3cn}
\end{figure*}




\begin{figure*}[!htb]
    \centering
    \includegraphics[scale=0.80]{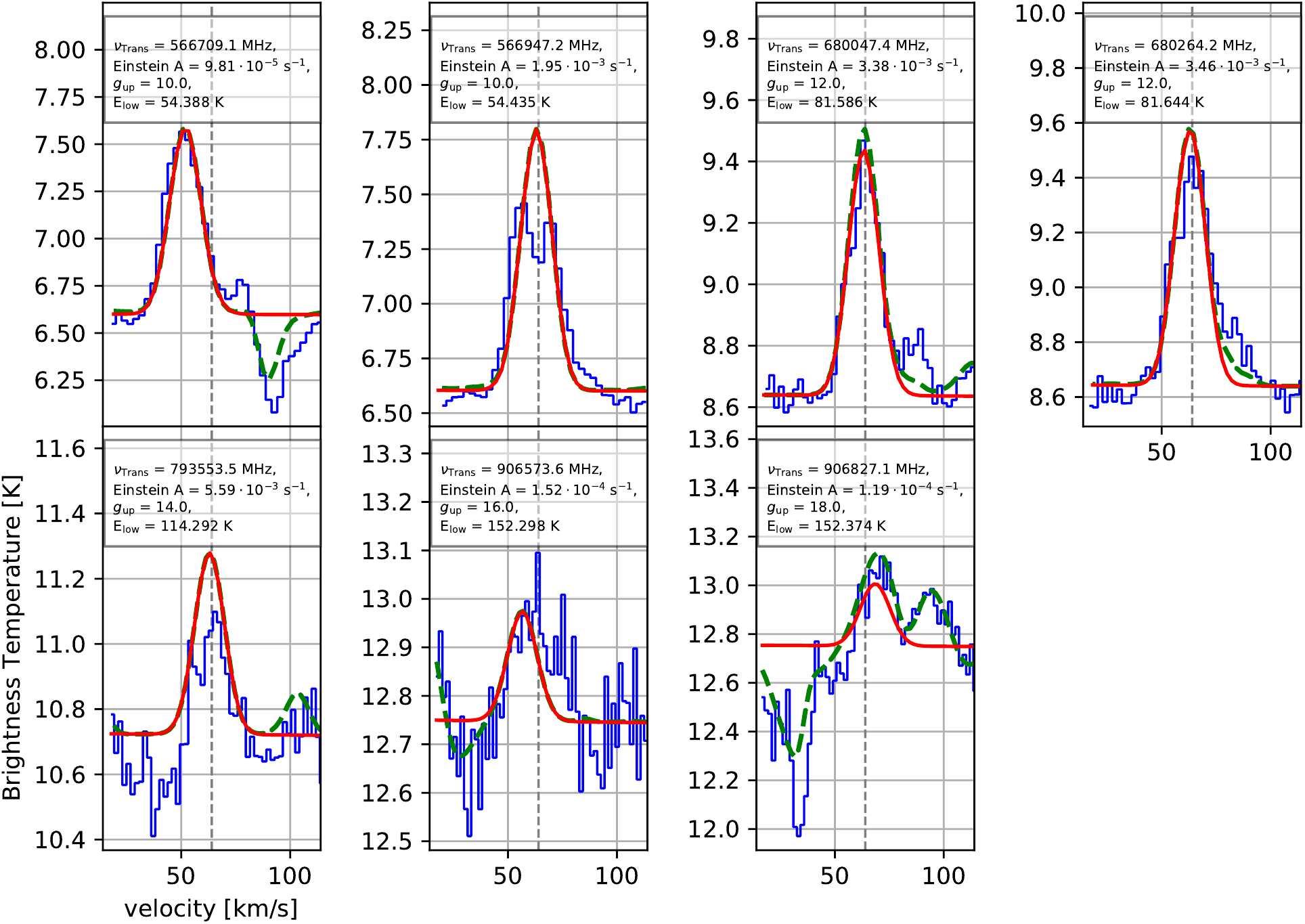}\\
    \caption{Selected transitions of CN (red line).}
    \label{fig:cn}
\end{figure*}

\begin{figure*}[!htb]
    \centering
    \includegraphics[scale=0.80]{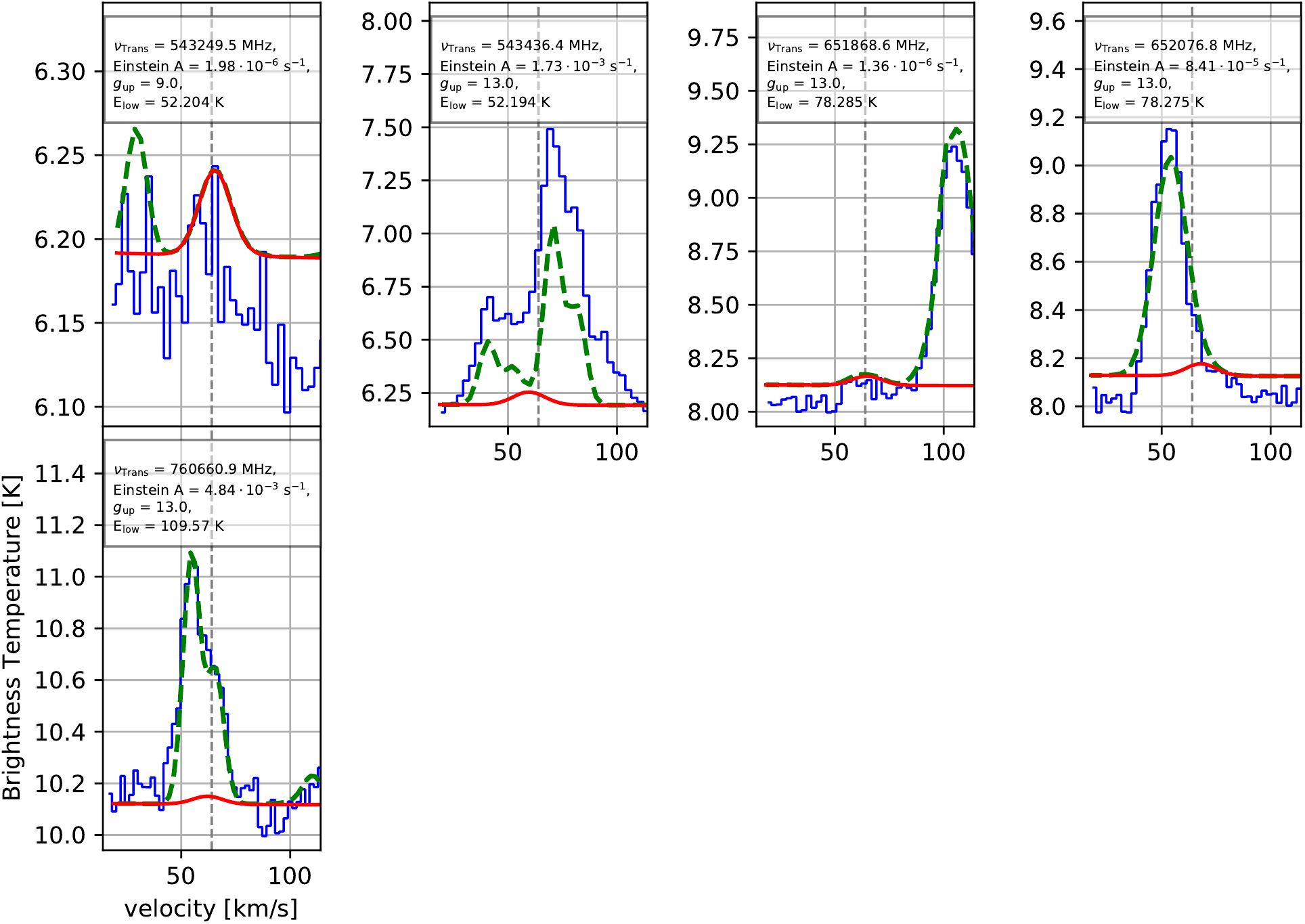}\\
    \caption{Selected transitions of $^{13}$CN (red line).}
    \label{fig:c13n}
\end{figure*}

\begin{figure*}[!htb]
    \centering
    \includegraphics[scale=0.80]{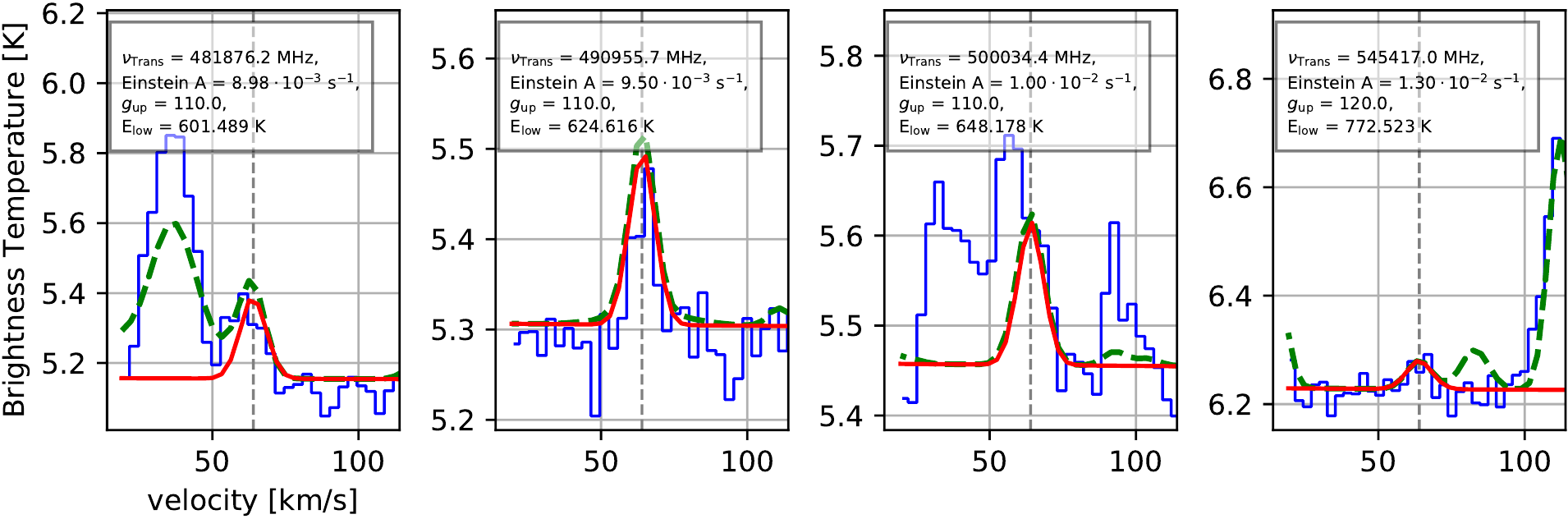}\\
    \caption{Selected transitions of HCCCN (red line).}
    \label{fig:hc3n}
\end{figure*}
\newpage

\clearpage

\begin{figure*}[!htb]
    \centering
    \includegraphics[scale=0.80]{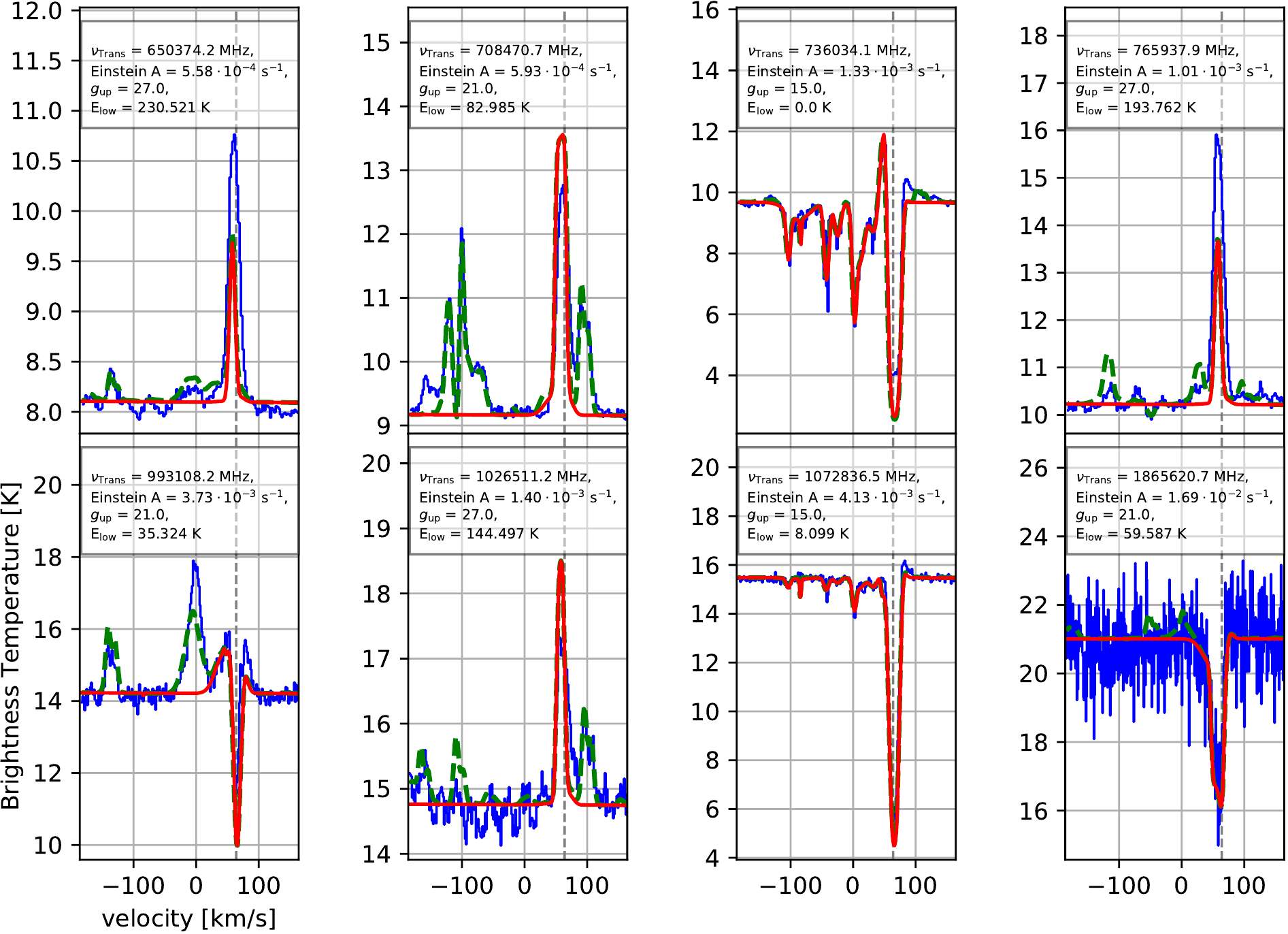}\\
    \caption{Selected transitions of \emph{ortho}-H$_2$S (red line).}
    \label{fig:oh2s}
\end{figure*}

\begin{figure*}[!htb]
    \centering
    \includegraphics[scale=0.80]{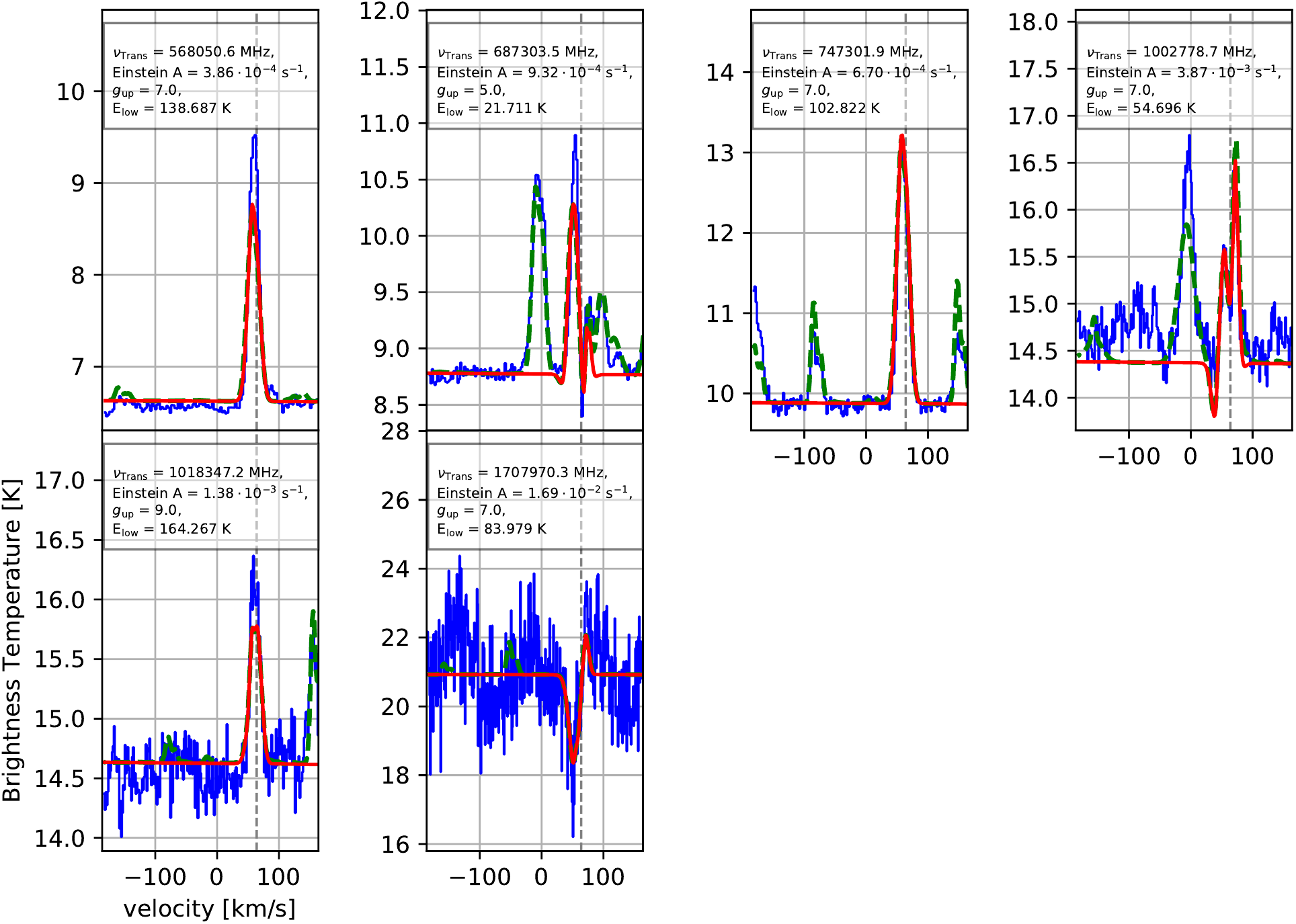}\\
    \caption{Selected transitions of \emph{para}-H$_2$S (red line).}
    \label{fig:ph2s}
\end{figure*}

\begin{figure*}[!htb]
    \centering
    \includegraphics[scale=0.80]{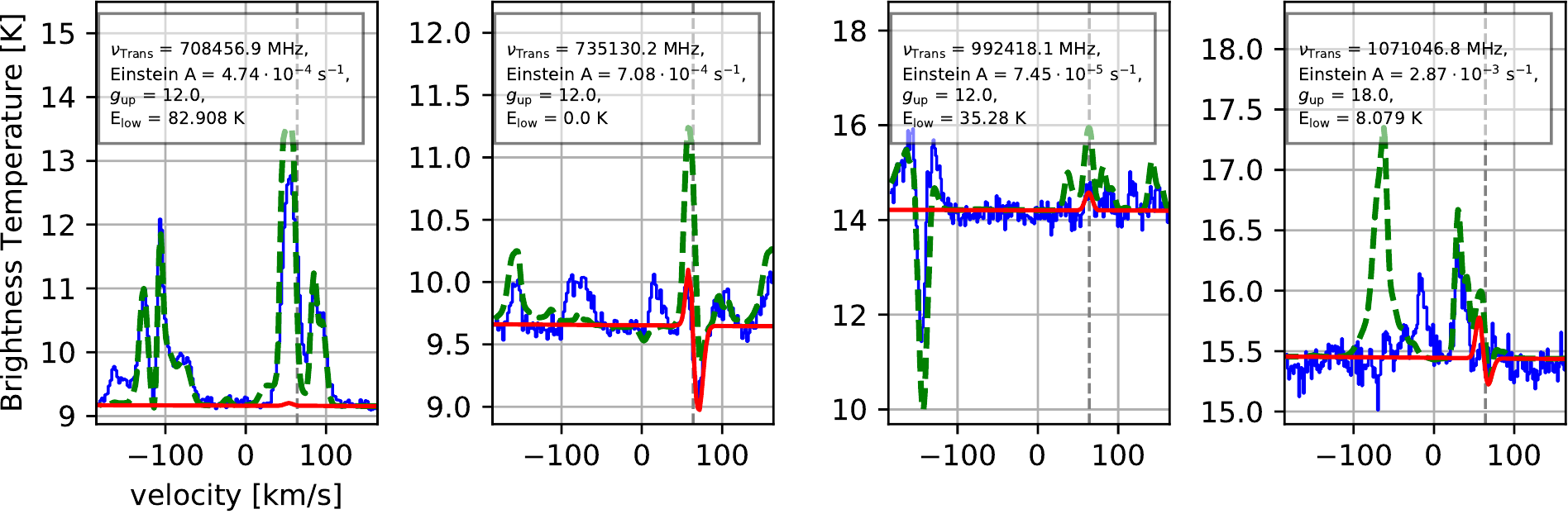}\\
    \caption{Selected transitions of \emph{ortho}-H$_2 \, ^{33}$S (red line).}
    \label{fig:oh2s33}
\end{figure*}

\begin{figure*}[!htb]
    \centering
    \includegraphics[scale=0.80]{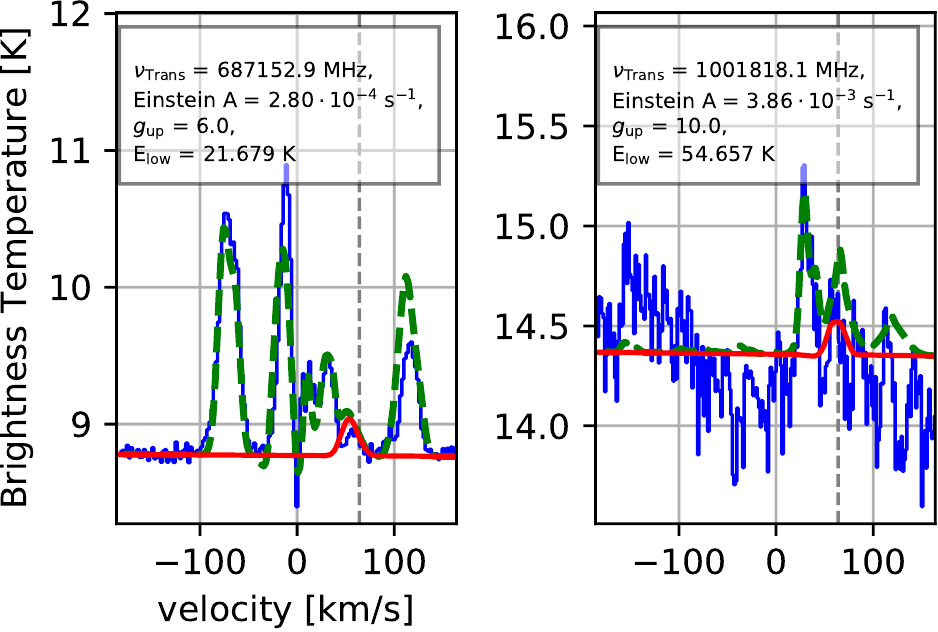}\\
    \caption{Selected transitions of \emph{para}-H$_2 \, ^{33}$S (red line).}
    \label{fig:ph2s33}
\end{figure*}

\begin{figure*}[!htb]
    \centering
    \includegraphics[scale=0.80]{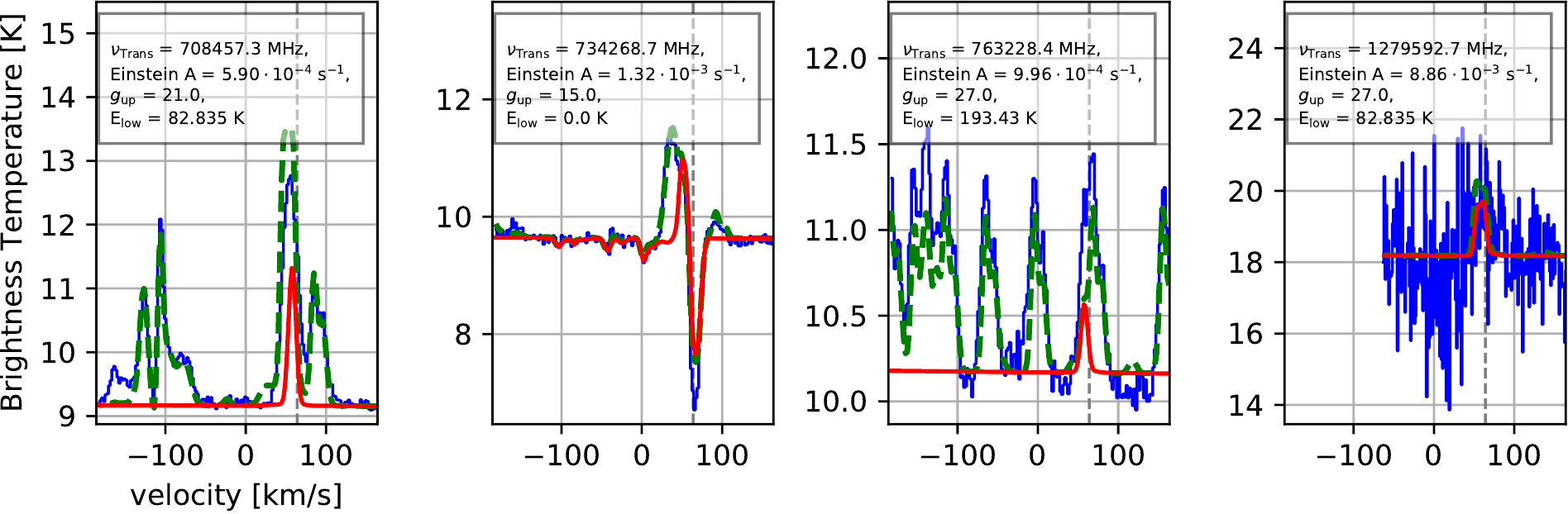}\\
    \caption{Selected transitions of \emph{ortho}-H$_2 \, ^{34}$S (red line).}
    \label{fig:oh2s34}
\end{figure*}

\begin{figure*}[!htb]
    \centering
    \includegraphics[scale=0.80]{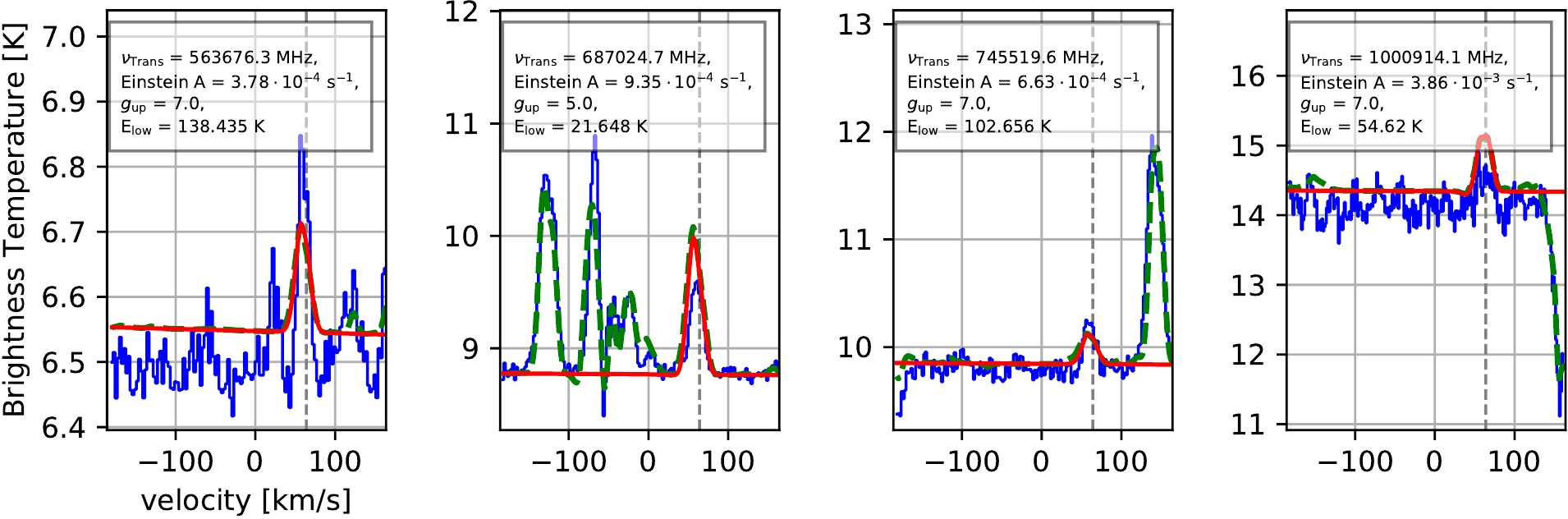}\\
    \caption{Selected transitions of \emph{para}-H$_2 \, ^{34}$S (red line).}
    \label{fig:ph2s34}
\end{figure*}

\begin{figure*}[!htb]
    \centering
    \includegraphics[scale=0.80]{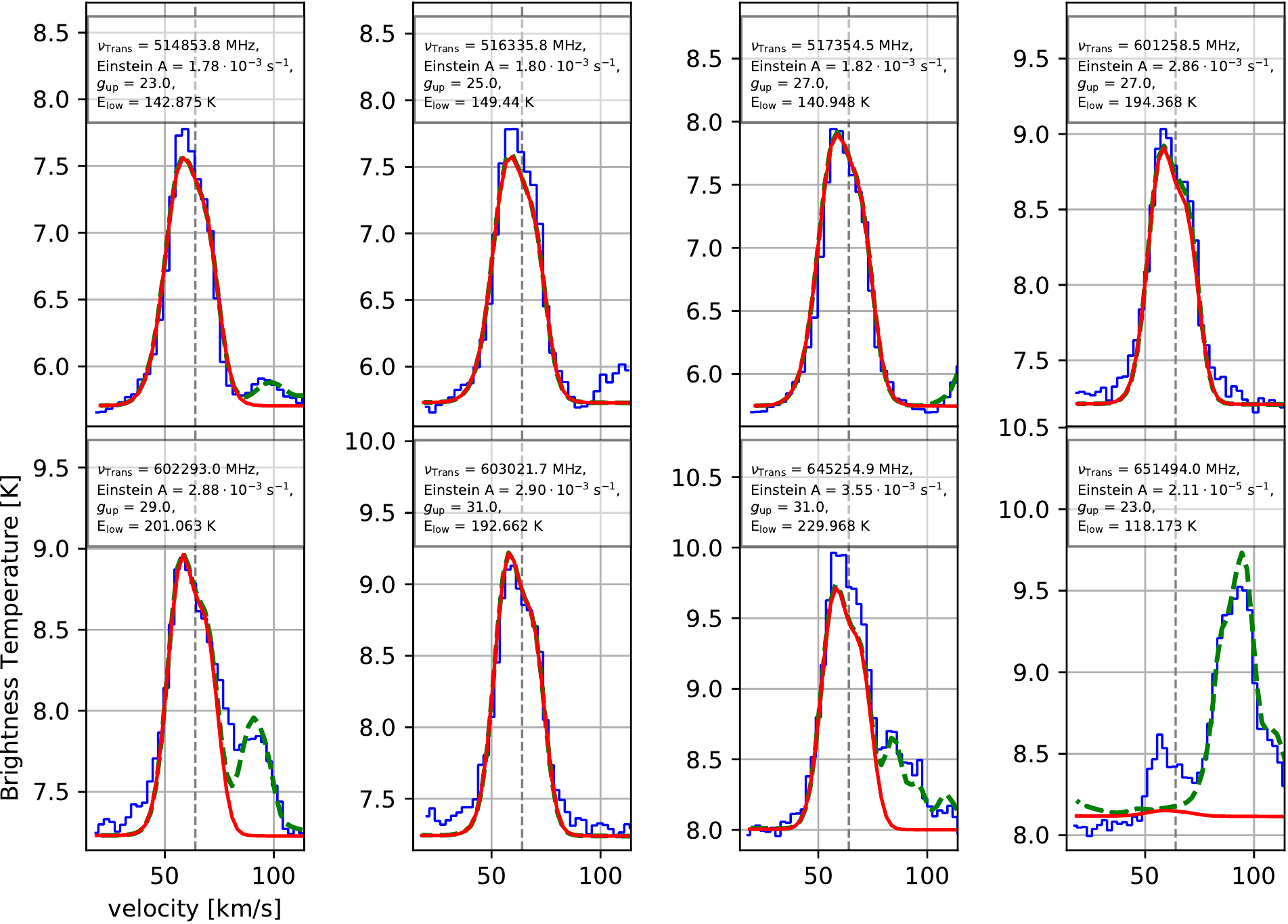}\\
    \caption{Selected transitions of SO (red line).}
    \label{fig:so}
\end{figure*}

\begin{figure*}[!htb]
    \centering
    \includegraphics[scale=0.80]{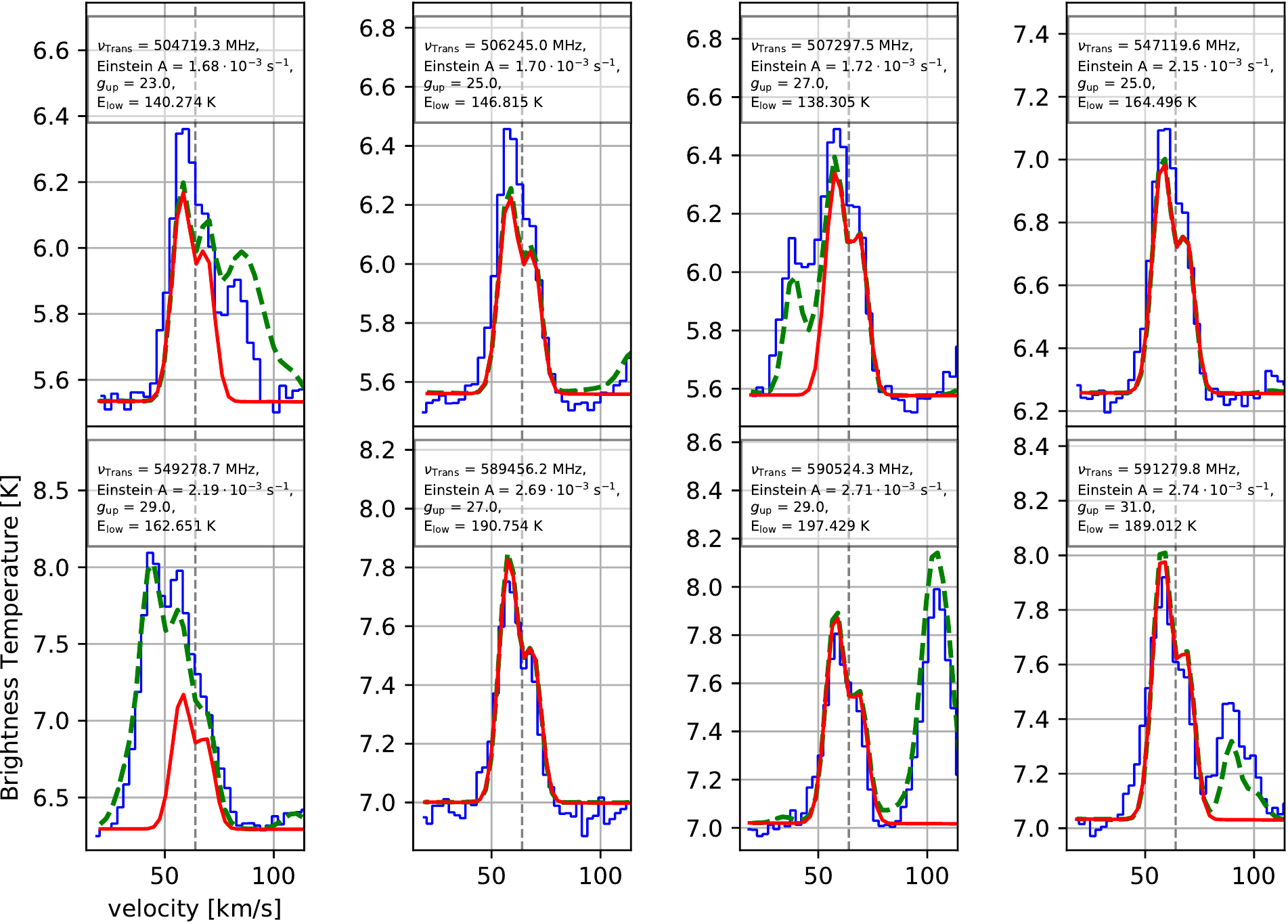}\\
    \caption{Selected transitions of $^{34}$SO (red line).}
    \label{fig:s-34-o}
\end{figure*}

\begin{figure*}[!htb]
    \centering
    \includegraphics[scale=0.80]{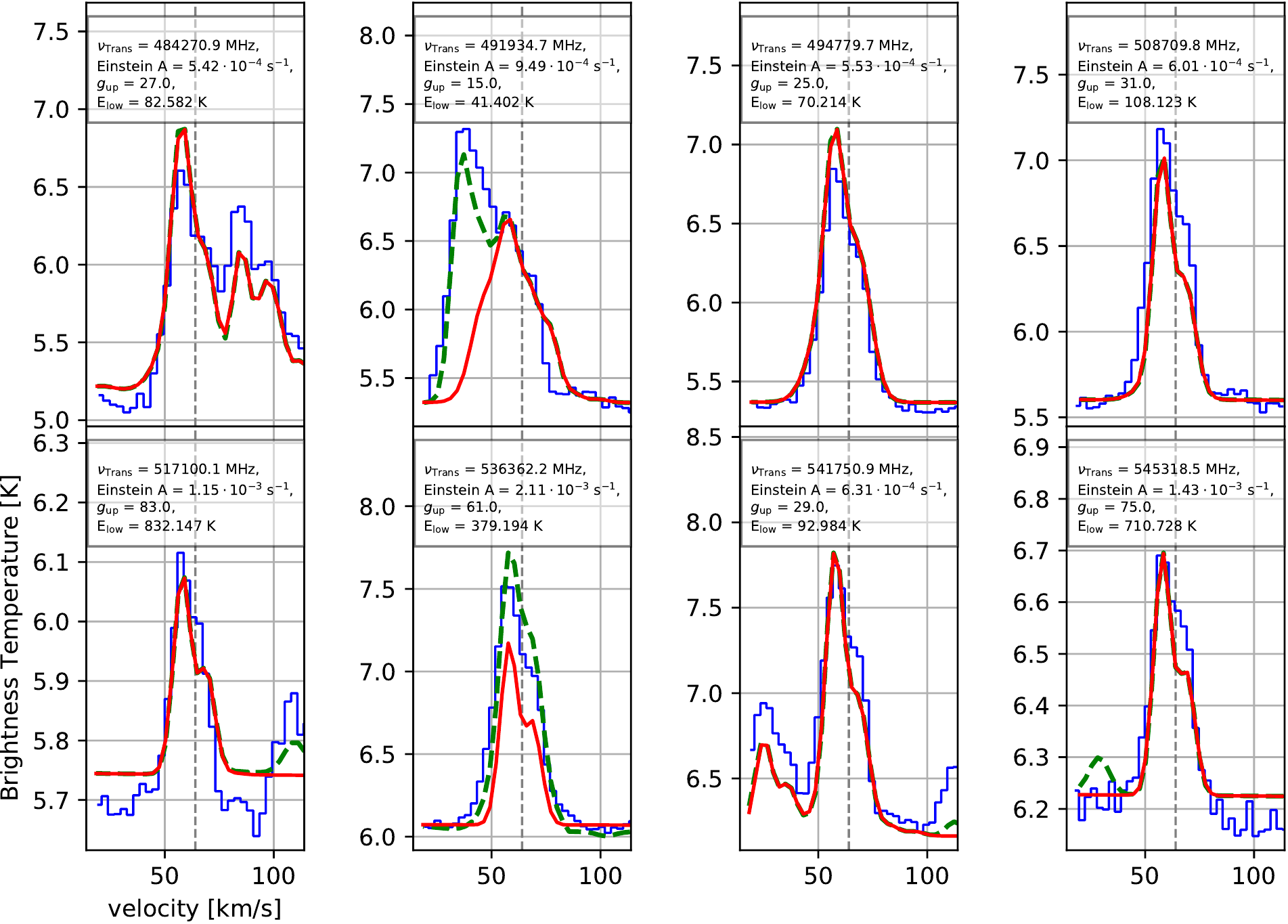}\\
    \caption{Selected transitions of SO$_2$ (red line).}
    \label{fig:so2}
\end{figure*}

\begin{figure*}[!htb]
    \centering
    \includegraphics[scale=0.80]{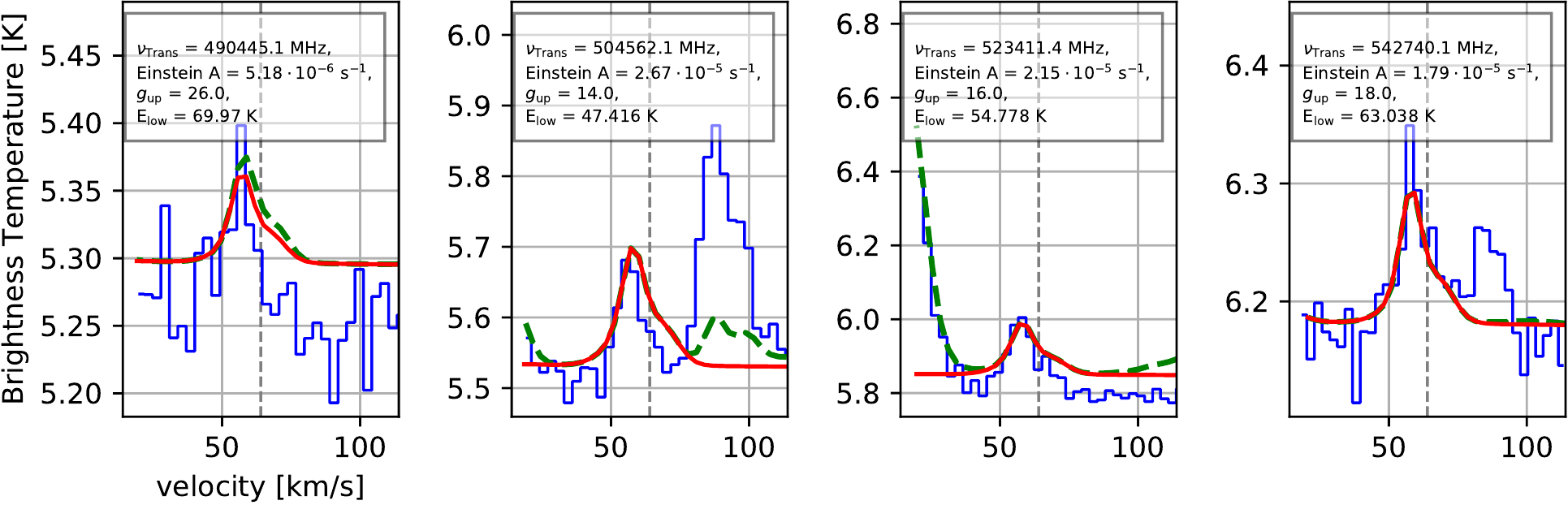}\\
    \caption{Selected transitions of $^{33}$SO$_2$ (red line).}
    \label{fig:s33o2}
\end{figure*}

\begin{figure*}[!htb]
    \centering
    \includegraphics[scale=0.80]{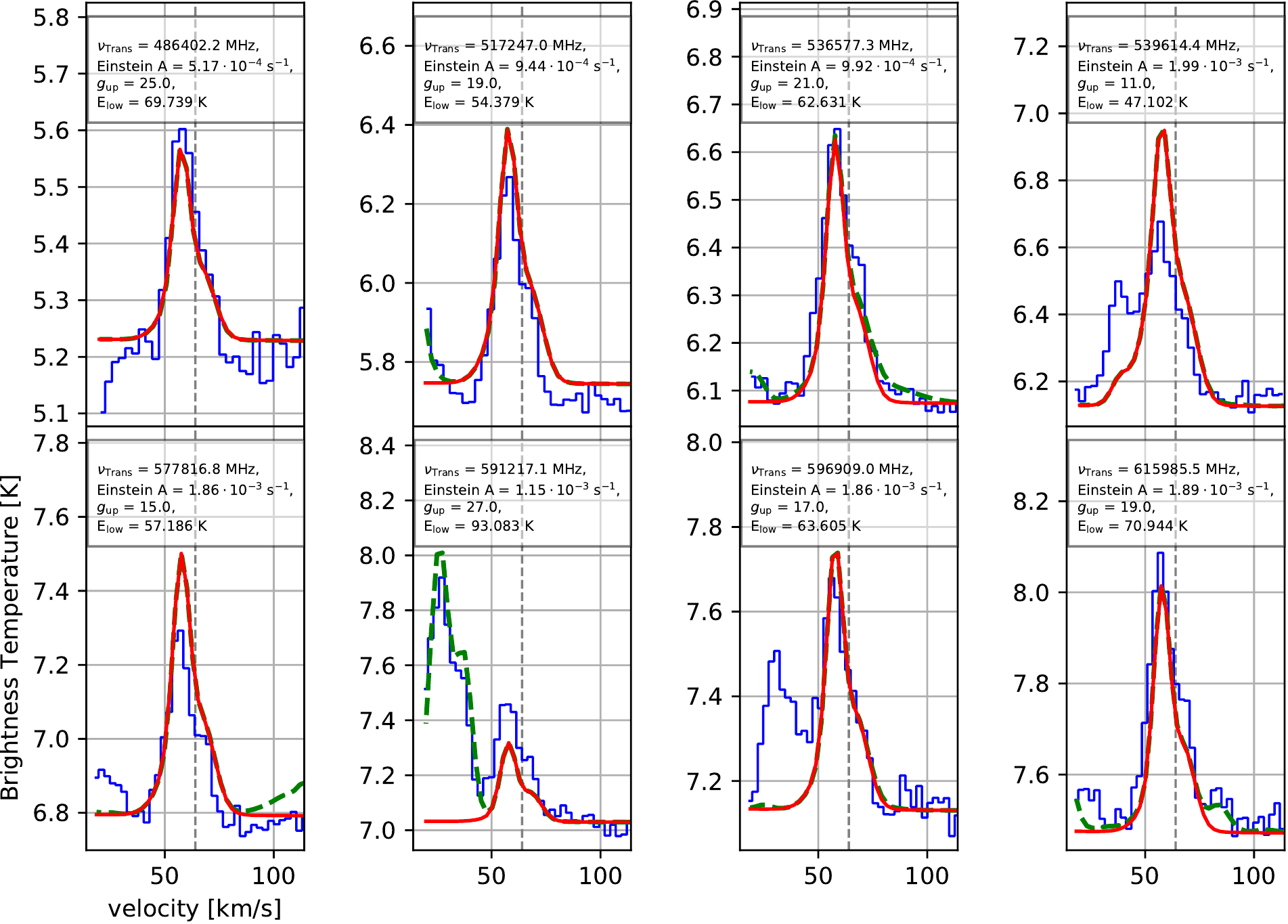}\\
    \caption{Selected transitions of $^{34}$SO$_2$ (red line).}
    \label{fig:s34o2}
\end{figure*}

\begin{figure*}[!htb]
    \centering
    \includegraphics[scale=0.80]{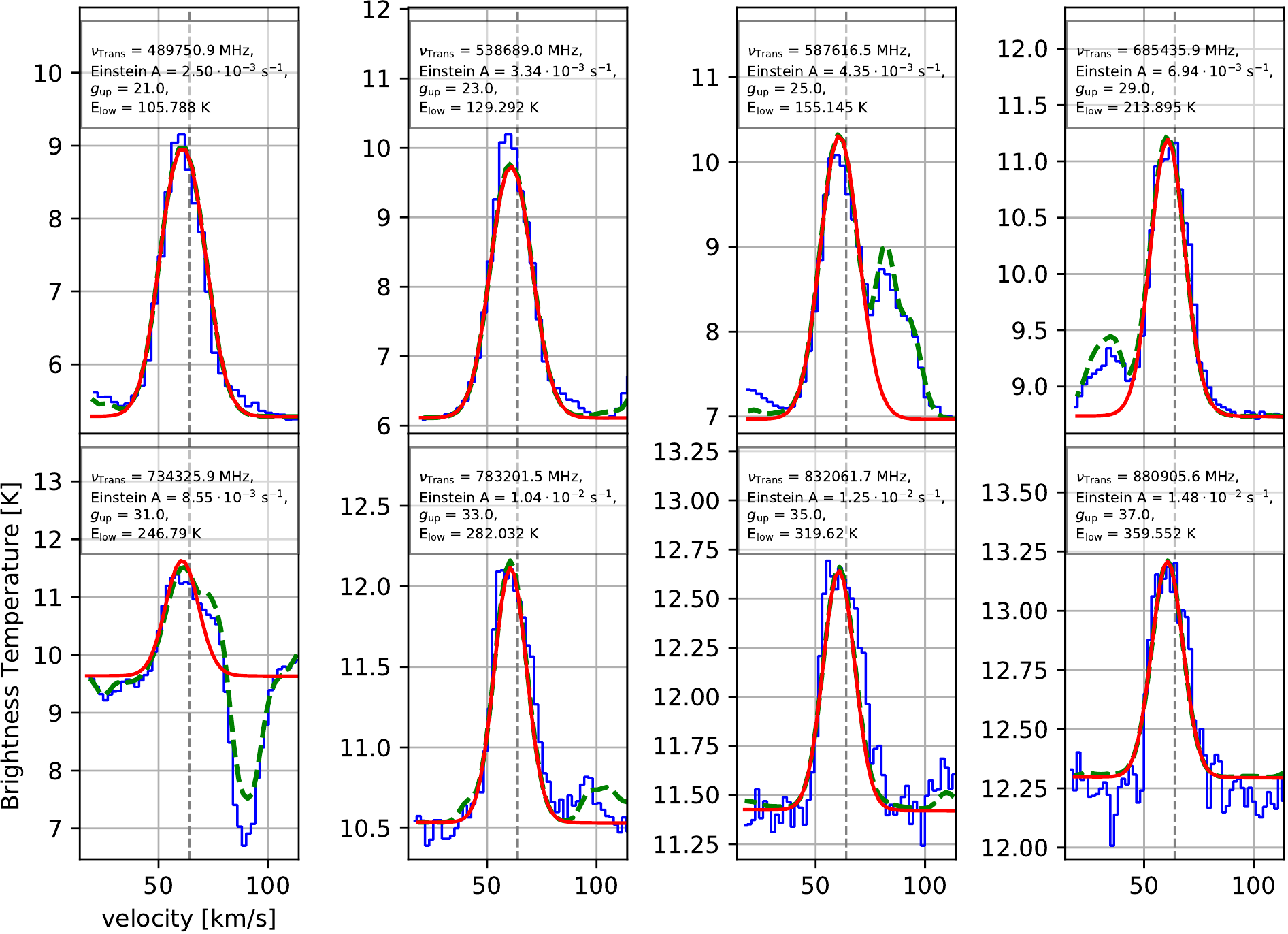}\\
    \caption{Selected transitions of CS (red line).}
    \label{fig:cs}
\end{figure*}

\begin{figure*}[!htb]
    \centering
    \includegraphics[scale=0.80]{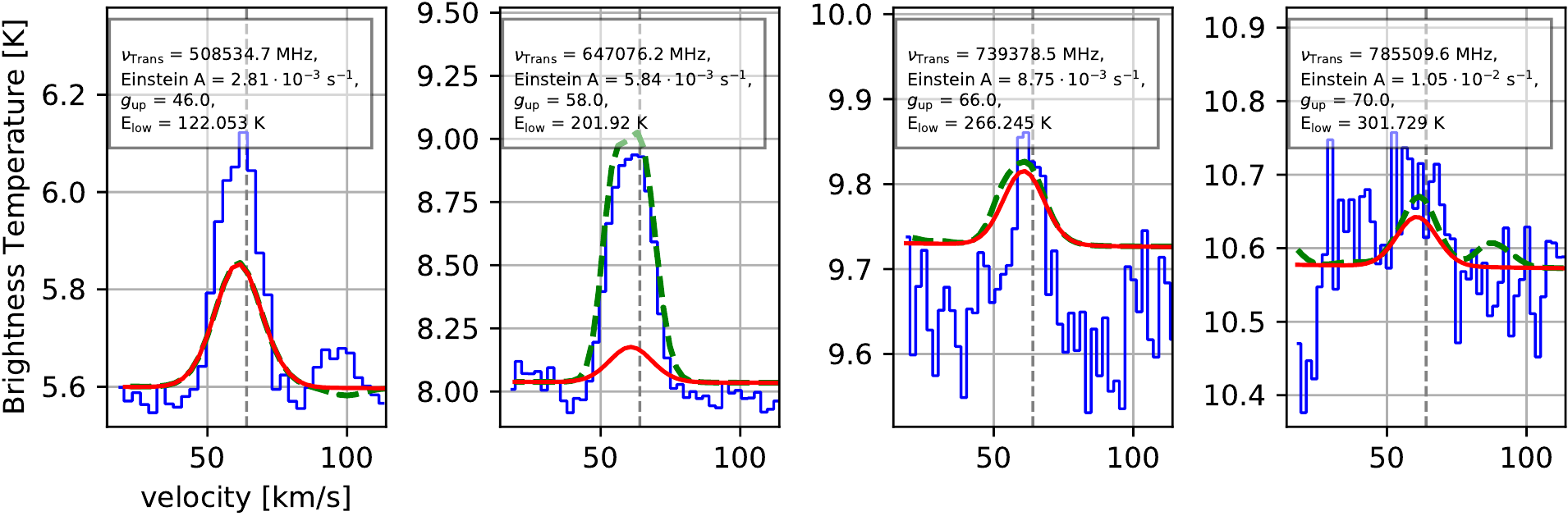}\\
    \caption{Selected transitions of $^{13}$CS (red line).}
    \label{fig:c13s}
\end{figure*}

\begin{figure*}[!htb]
    \centering
    \includegraphics[scale=0.80]{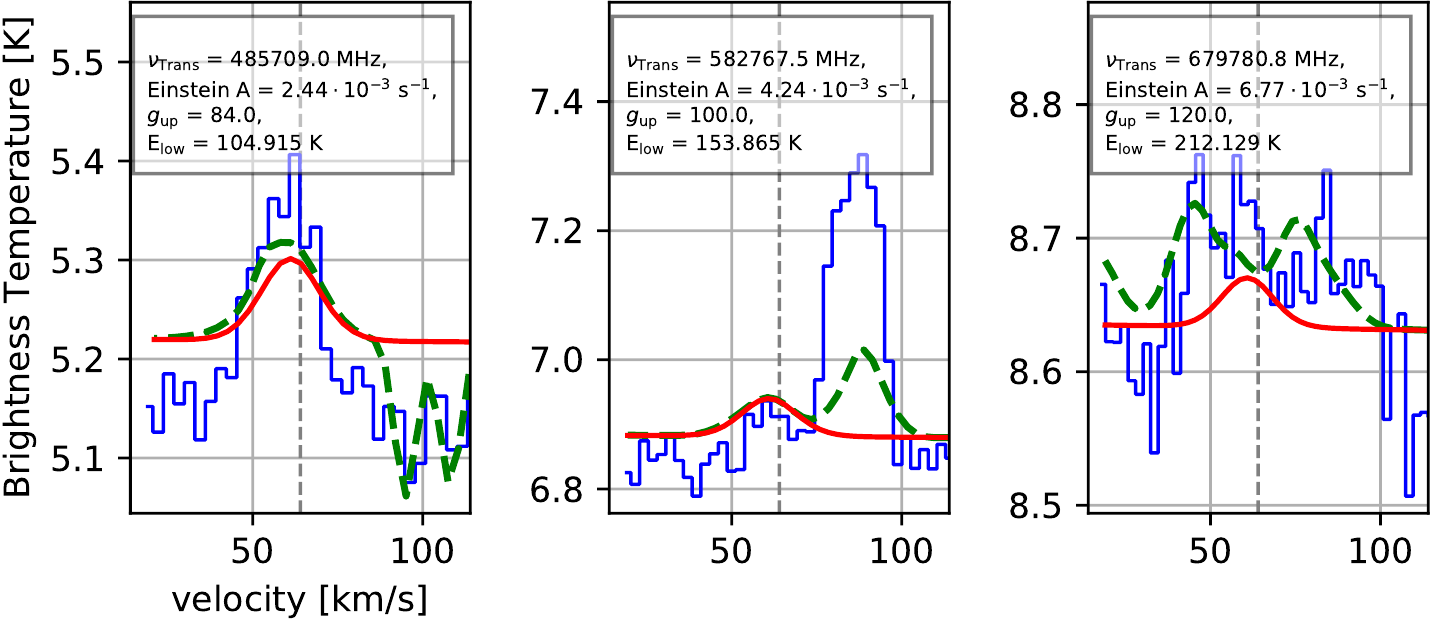}\\
    \caption{Selected transitions of C$^{33}$S (red line).}
    \label{fig:cs33}
\end{figure*}

\begin{figure*}[!htb]
    \centering
    \includegraphics[scale=0.80]{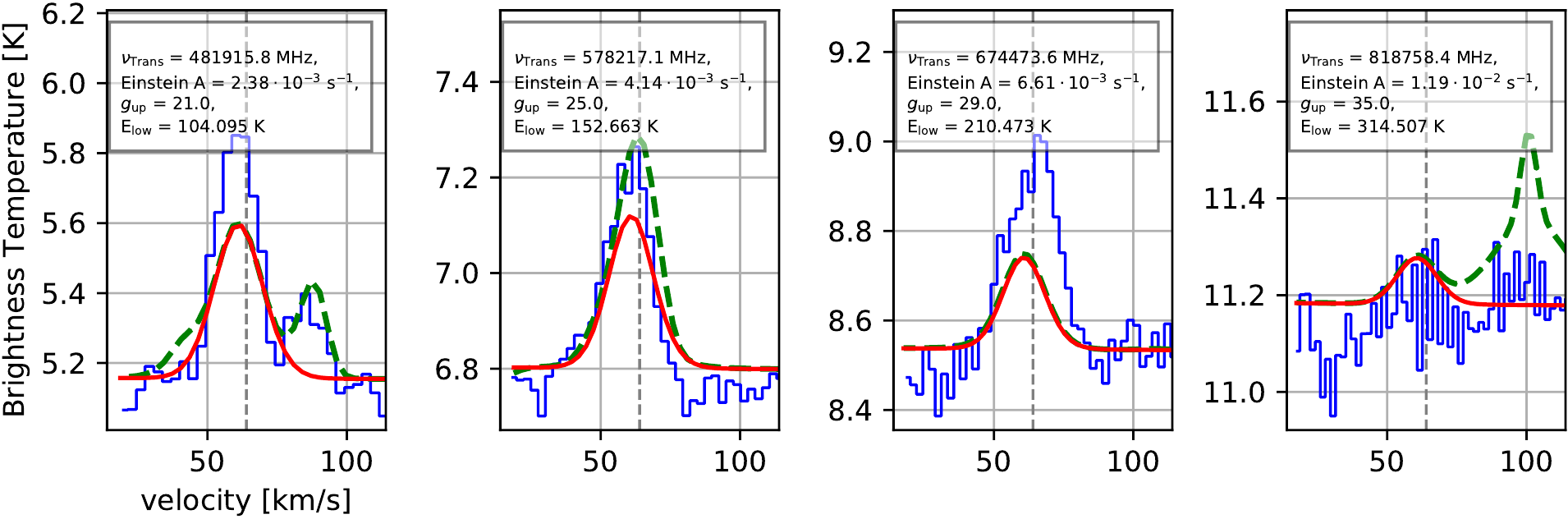}\\
    \caption{Selected transitions of C$^{34}$S (red line).}
    \label{fig:cs34}
\end{figure*}

\begin{figure*}[!htb]
    \centering
    \includegraphics[scale=0.80]{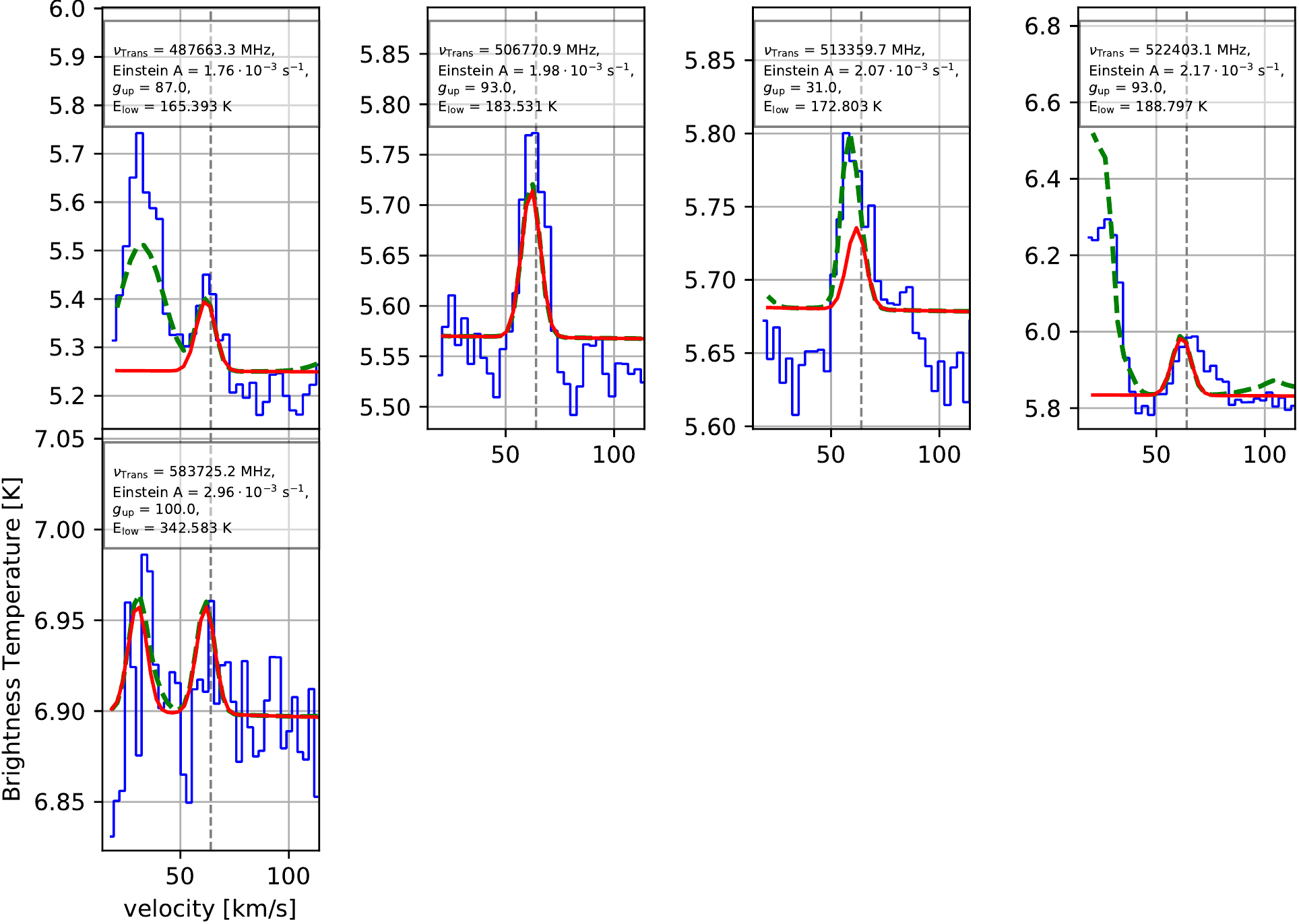}\\
    \caption{Selected transitions of H$_2$CS (red line).}
    \label{fig:h2cs}
\end{figure*}

\begin{figure*}[!htb]
    \centering
    \includegraphics[scale=0.80]{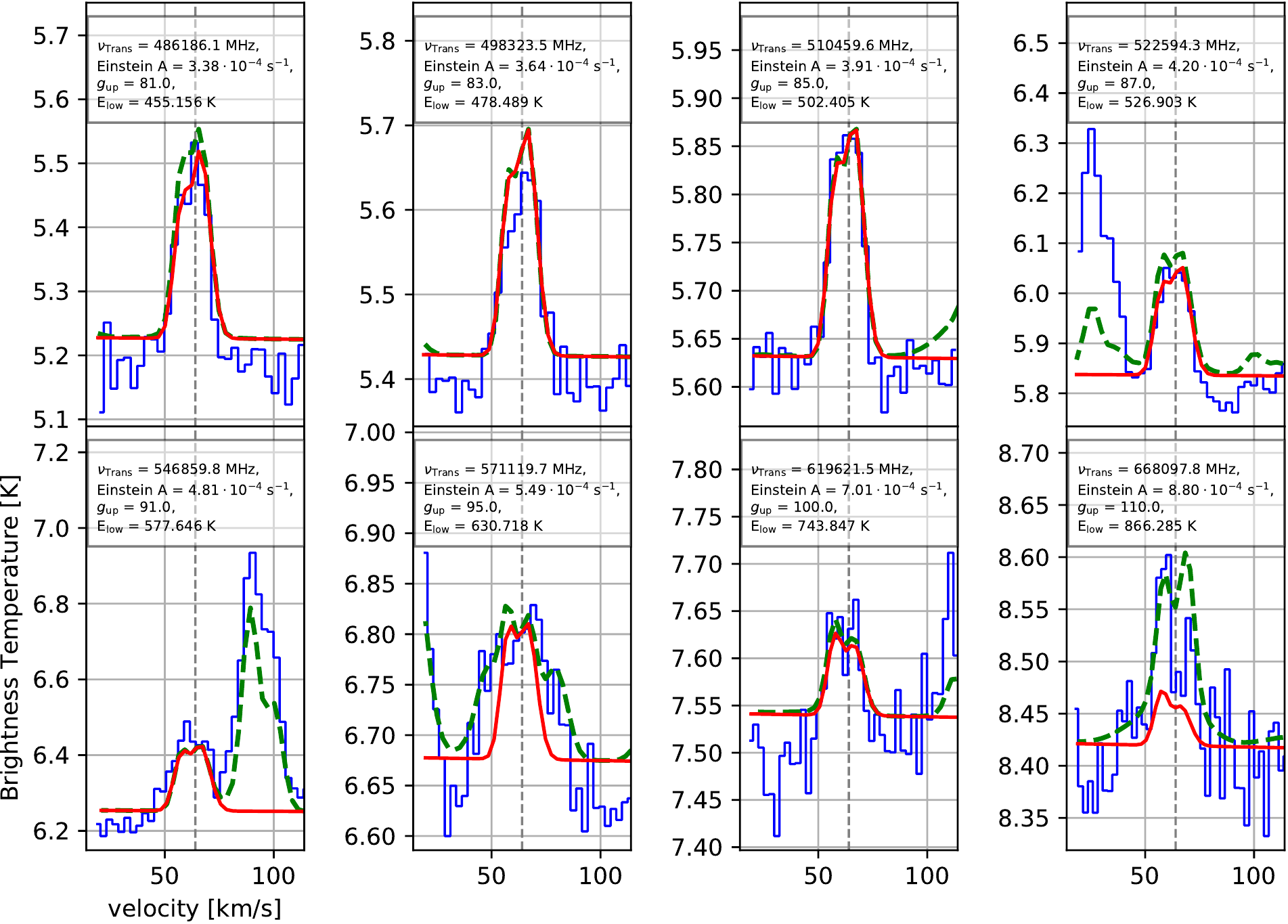}\\
    \caption{Selected transitions of OCS (red line).}
    \label{fig:ocs}
\end{figure*}

\begin{figure*}[!htb]
    \centering
    \includegraphics[scale=0.80]{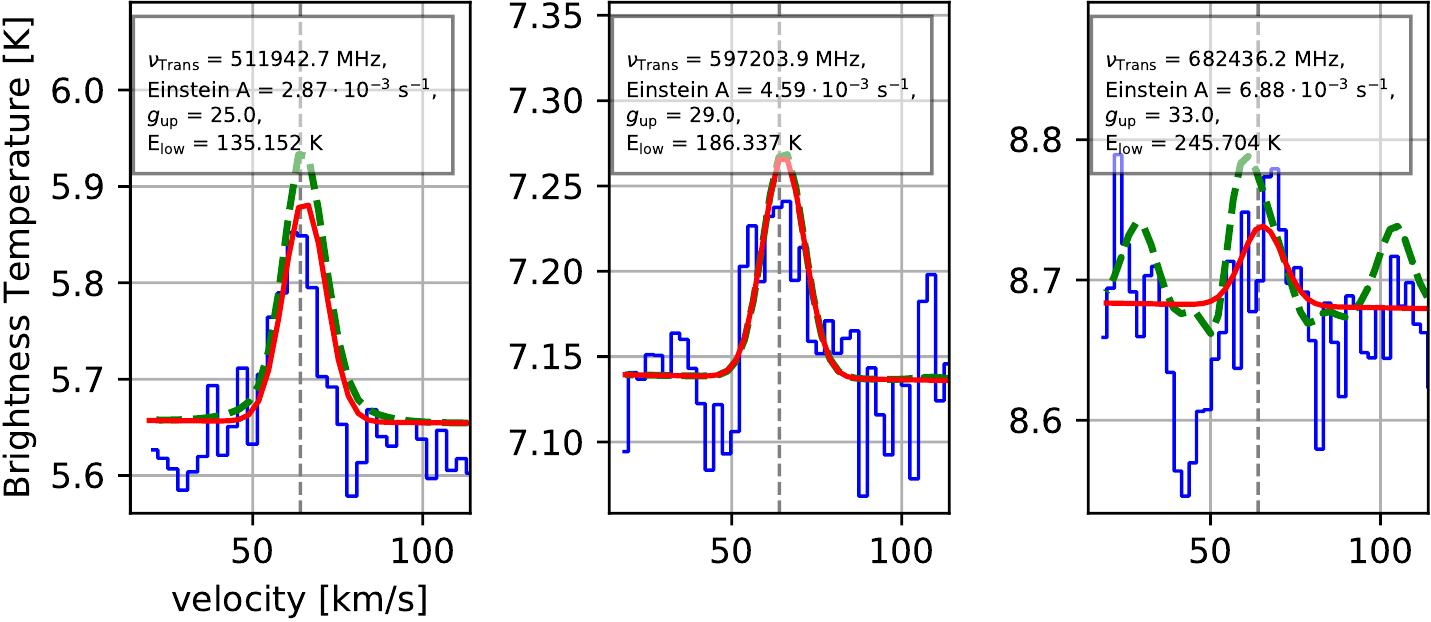}\\
    \caption{Selected transitions of HCS$^{+}$ (red line).}
    \label{fig:hcs+}
\end{figure*}

\begin{figure*}[!htb]
    \centering
    \includegraphics[scale=0.80]{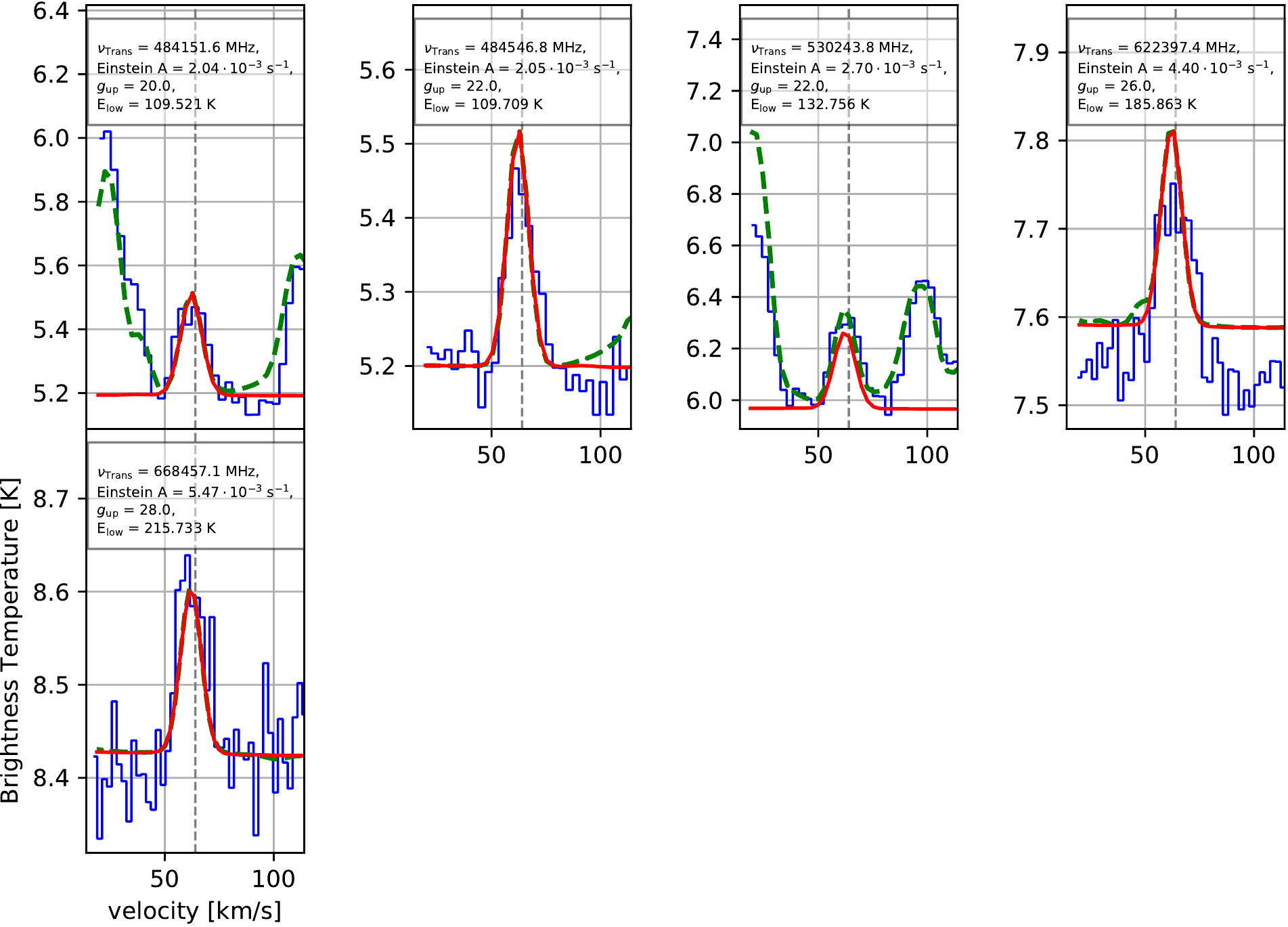}\\
    \caption{Selected transitions of NS (red line).}
    \label{fig:ns}
\end{figure*}

\begin{figure*}[!htb]
    \centering
    \includegraphics[scale=0.80]{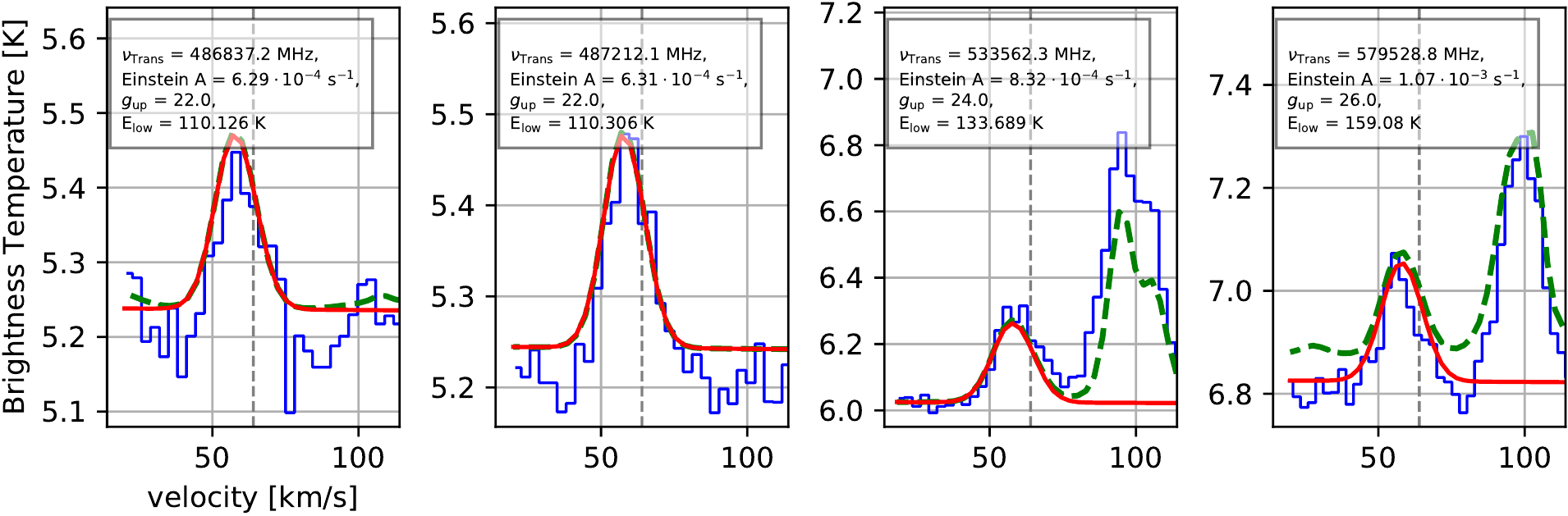}\\
    \caption{Selected transitions of SO$^{+}$ (red line).}
    \label{fig:so+}
\end{figure*}

\begin{figure*}[!htb]
    \centering
    \includegraphics[scale=0.80]{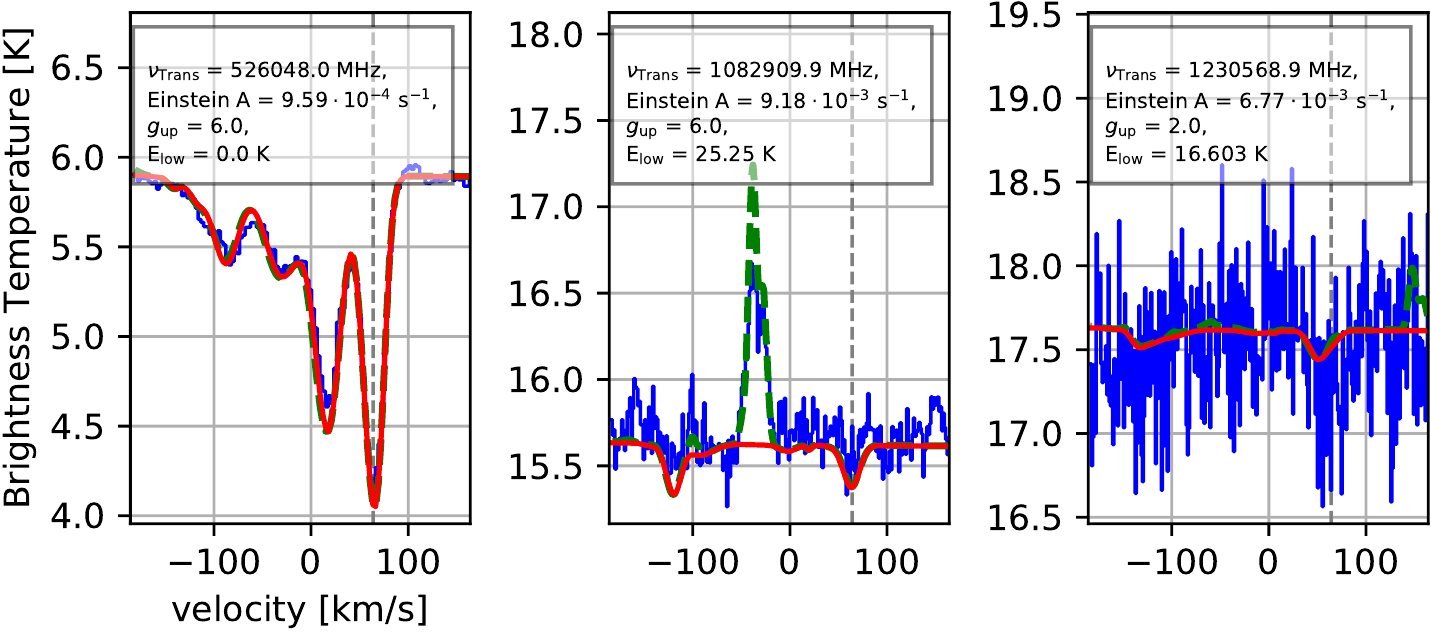}\\
    \caption{Selected transitions of SH$^{+}$ (red line).}
    \label{fig:sh+}
\end{figure*}

\begin{figure*}[!htb]
    \centering
    \includegraphics[scale=0.80]{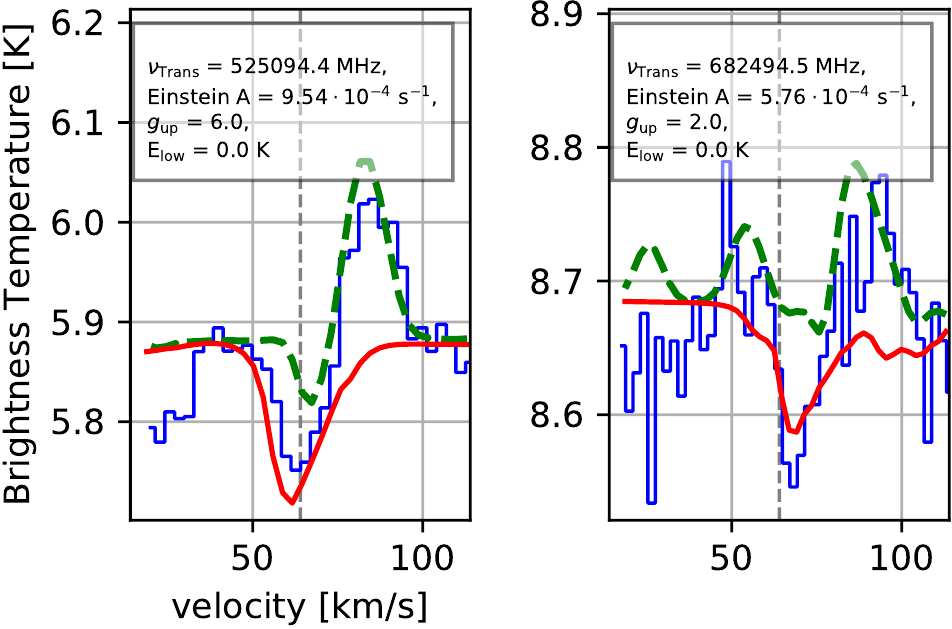}\\
    \caption{Selected transitions of $^{34}$SH$^{+}$ (red line).}
    \label{fig:s-34-h+}
\end{figure*}

\begin{figure*}[!htb]
    \centering
    \includegraphics[scale=0.80]{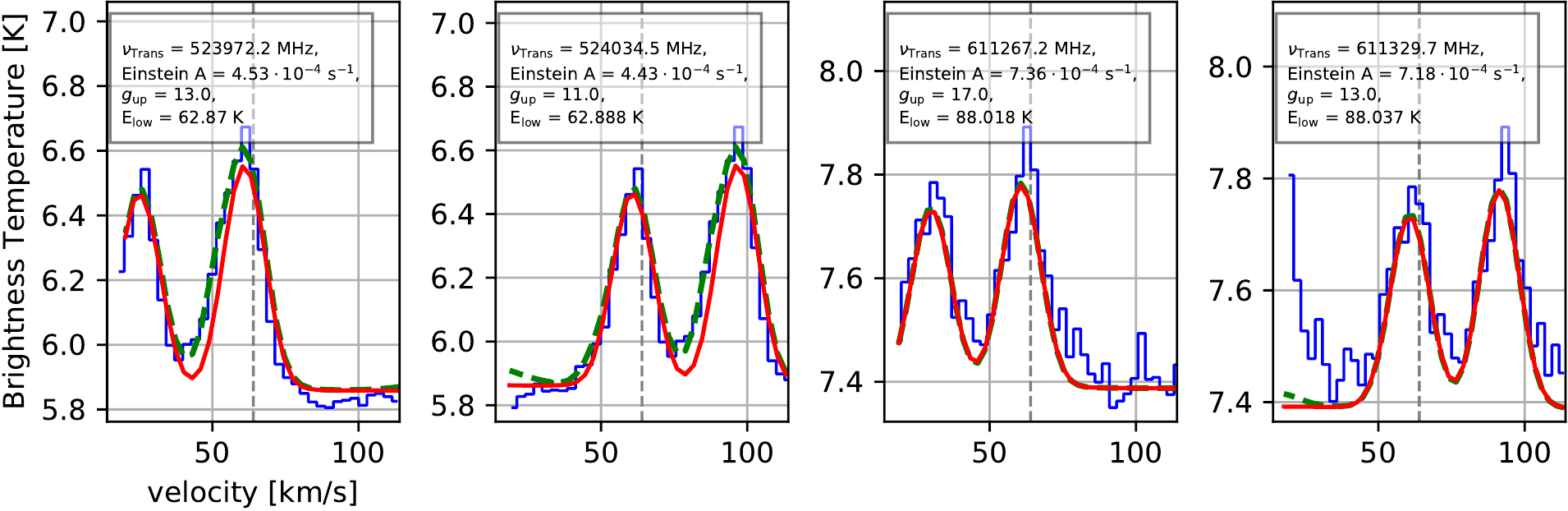}\\
    \caption{Selected transitions of CCH (red line).}
    \label{fig:cch}
\end{figure*}

\begin{figure*}[!htb]
    \centering
    \includegraphics[scale=0.80]{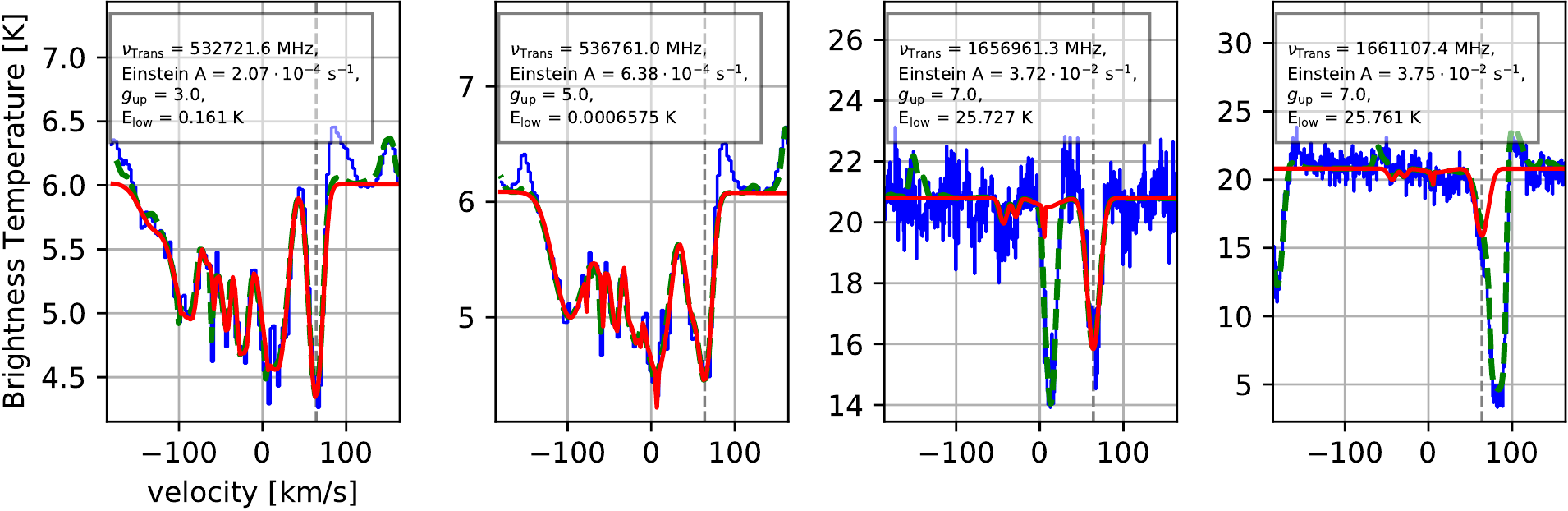}\\
    \caption{Selected transitions of CH (red line).}
    \label{fig:ch}
\end{figure*}

\begin{figure*}[!htb]
    \centering
    \includegraphics[scale=0.80]{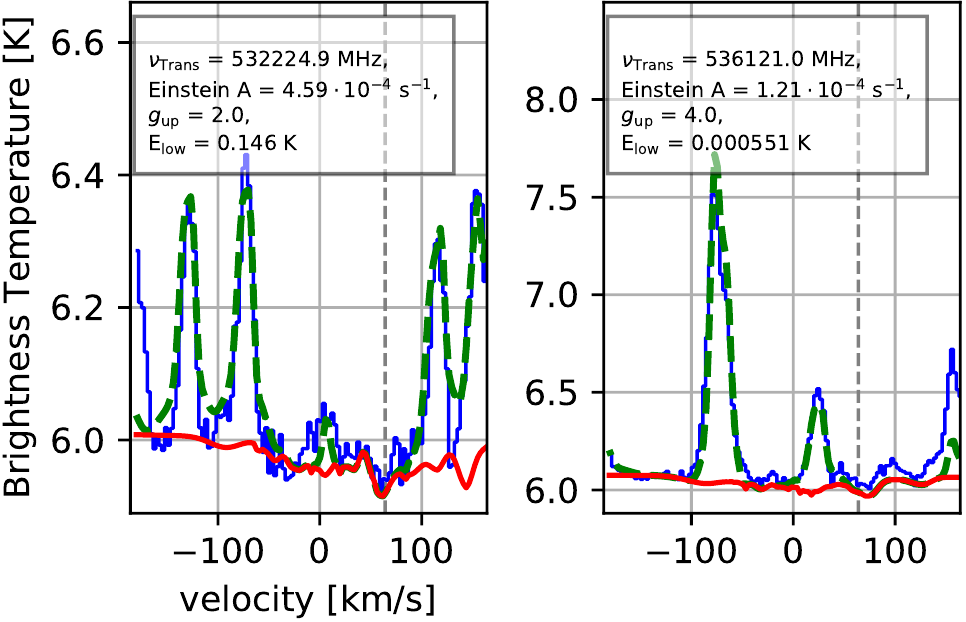}\\
    \caption{Selected transitions of $^{13}$CH (red line).}
    \label{fig:c13h}
\end{figure*}

\begin{figure*}[!htb]
    \centering
    \includegraphics[scale=0.80]{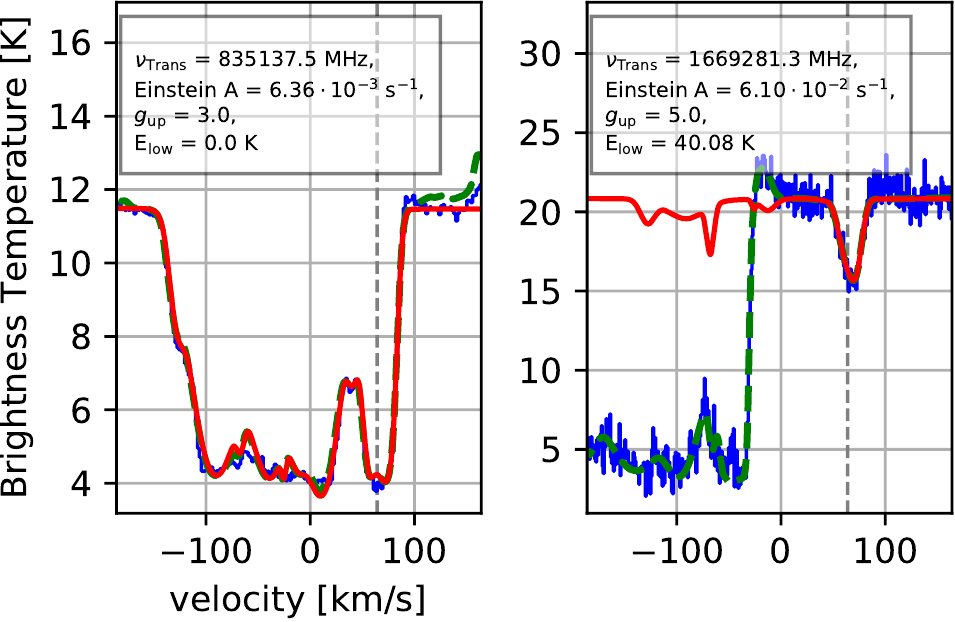}\\
    \caption{Selected transitions of CH$^{+}$ (red line).}
    \label{fig:ch+}
\end{figure*}

\begin{figure*}[!htb]
    \centering
    \includegraphics[scale=0.80]{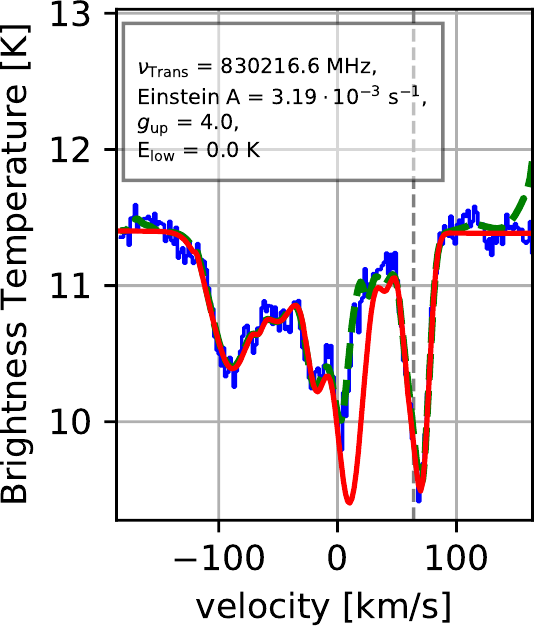}\\
    \caption{Selected transitions of $^{13}$CH$^{+}$ (red line).}
    \label{fig:c-13-h+}
\end{figure*}

\begin{figure*}[!htb]
    \centering
    \includegraphics[scale=0.80]{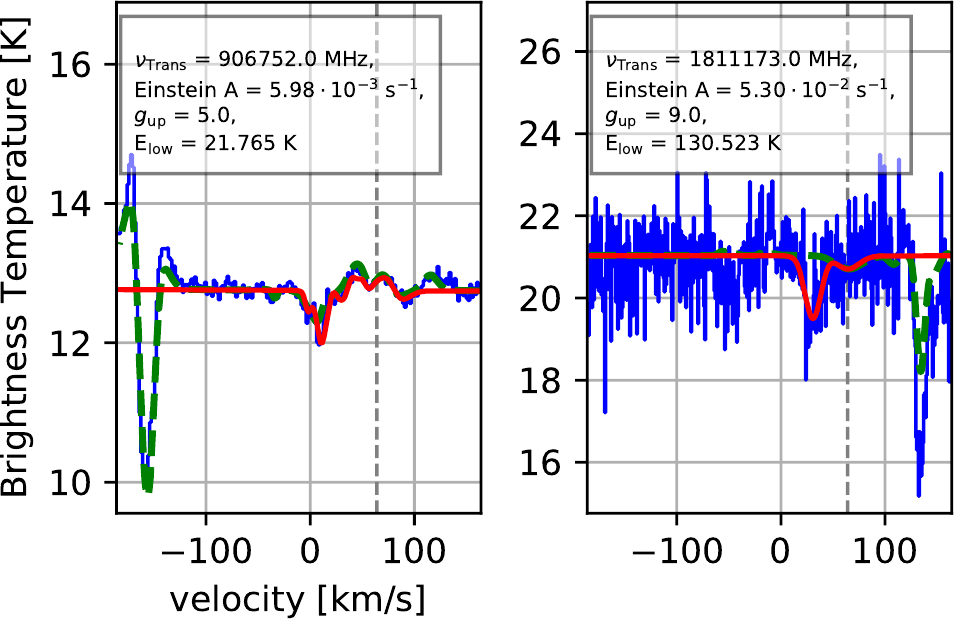}\\
    \caption{Selected transitions of CD$^{+}$ (red line).}
    \label{fig:cd+}
\end{figure*}
\newpage

\clearpage

\begin{figure*}[!htb]
    \centering
    \includegraphics[scale=0.80]{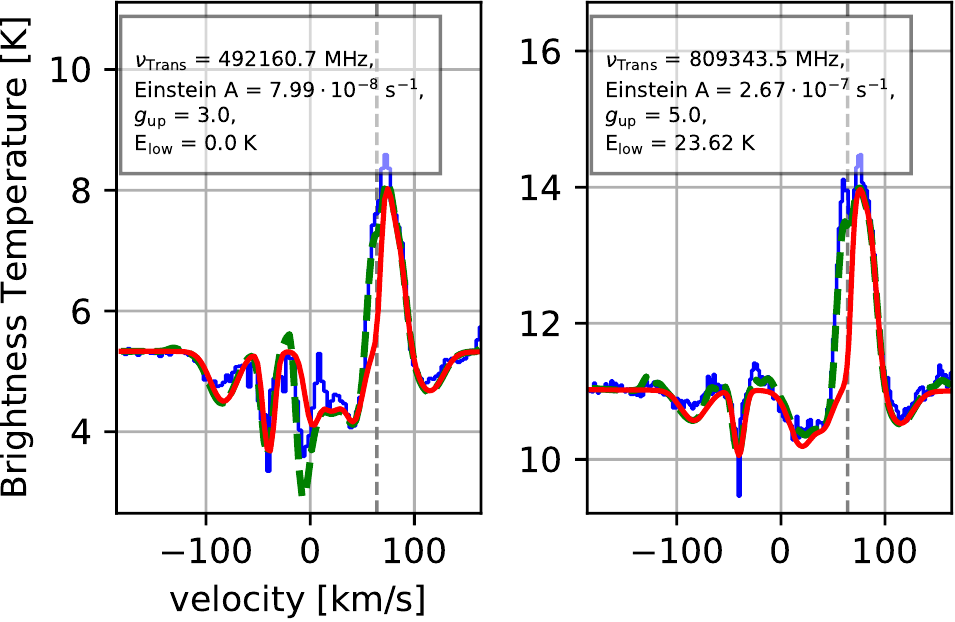}\\
    \caption{Selected transitions of $^{12}$C (red line).}
    \label{fig:12c}
\end{figure*}

\begin{figure*}[!htb]
    \centering
    \includegraphics[scale=0.80]{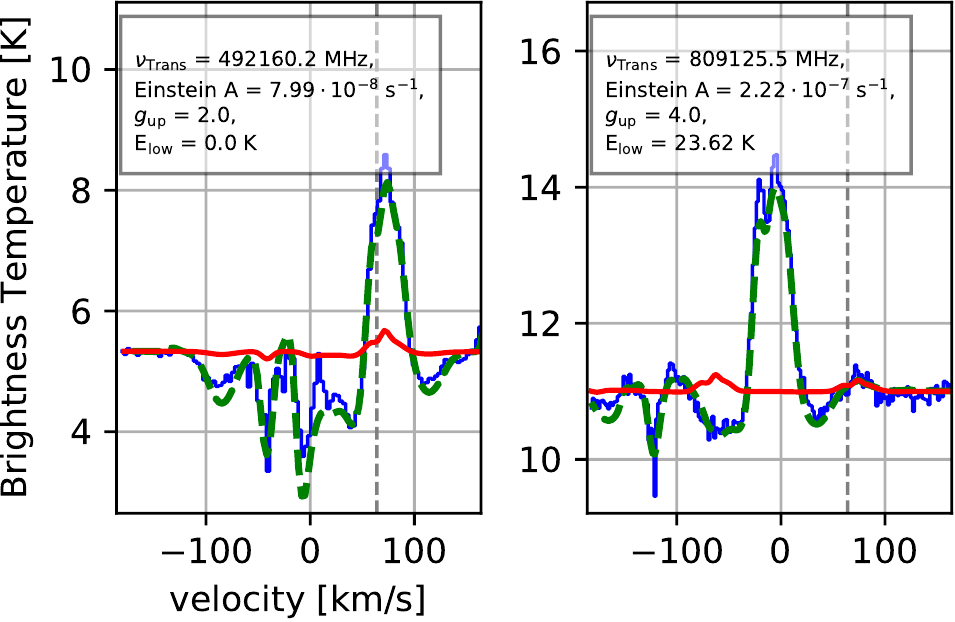}\\
    \caption{Selected transitions of $^{13}$C (red line).}
    \label{fig:13c}
\end{figure*}

\begin{figure*}[!htb]
    \centering
    \includegraphics[scale=0.80]{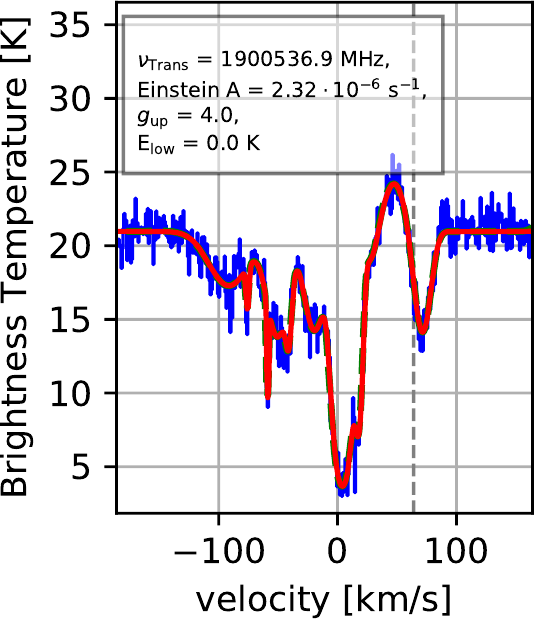}\\
    \caption{Selected transitions of $^{12}$C$^{+}$ (red line).}
    \label{fig:12c+}
\end{figure*}

\begin{figure*}[!htb]
    \centering
    \includegraphics[scale=0.80]{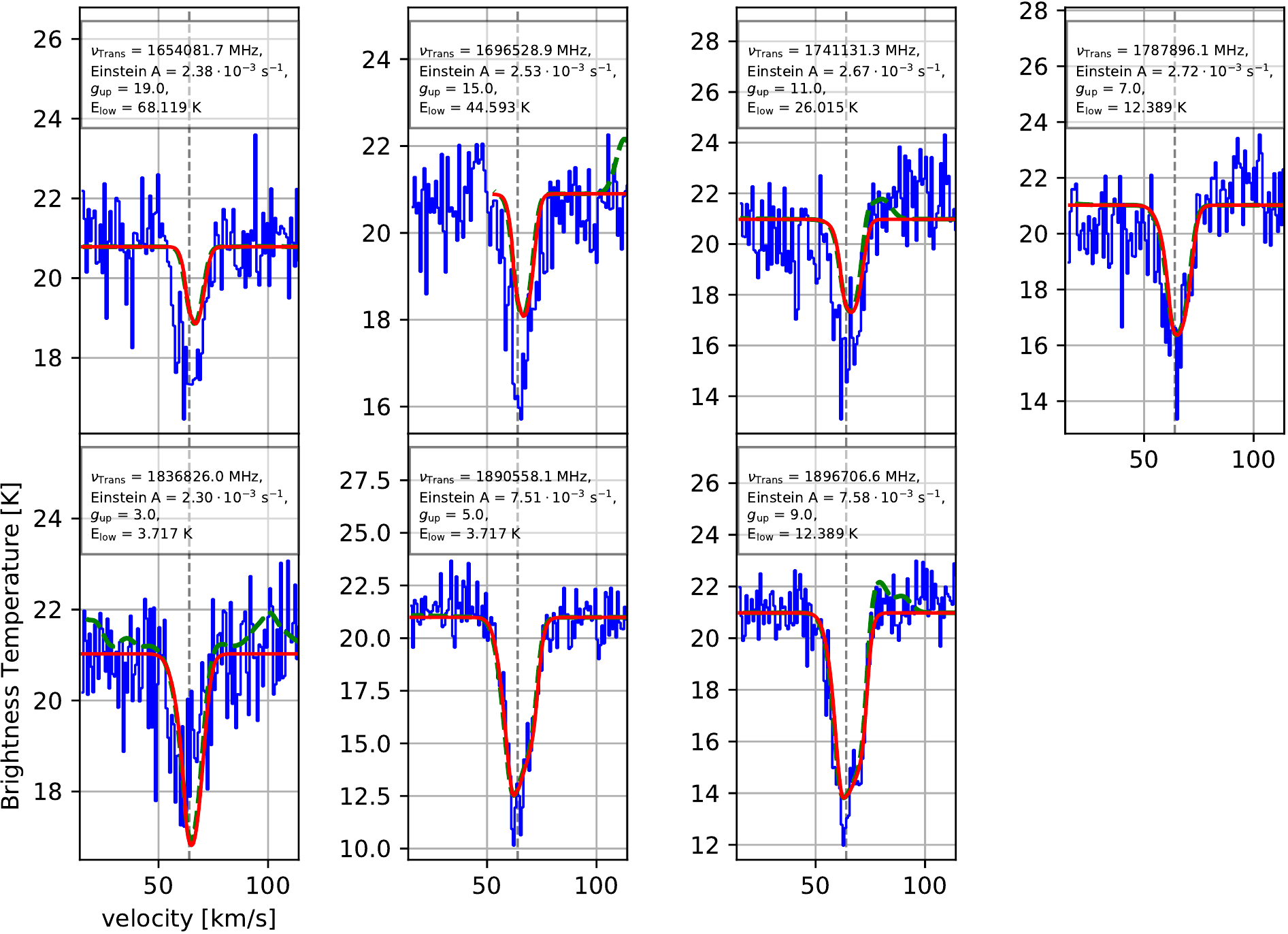}\\
    \caption{Selected transitions of C$_3$ (red line).}
    \label{fig:c3}
\end{figure*}
\newpage

\clearpage

\begin{figure*}[!htb]
    \centering
    \includegraphics[scale=0.80]{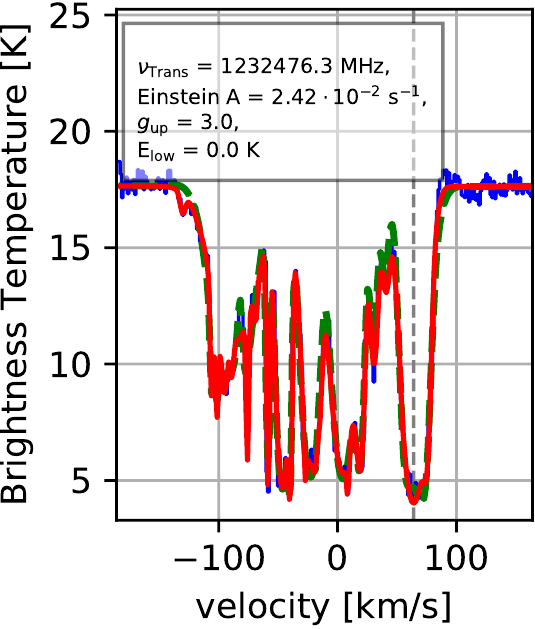}\\
    \caption{Selected transitions of HF (red line).}
    \label{fig:hf}
\end{figure*}

\begin{figure*}[!htb]
    \centering
    \includegraphics[scale=0.80]{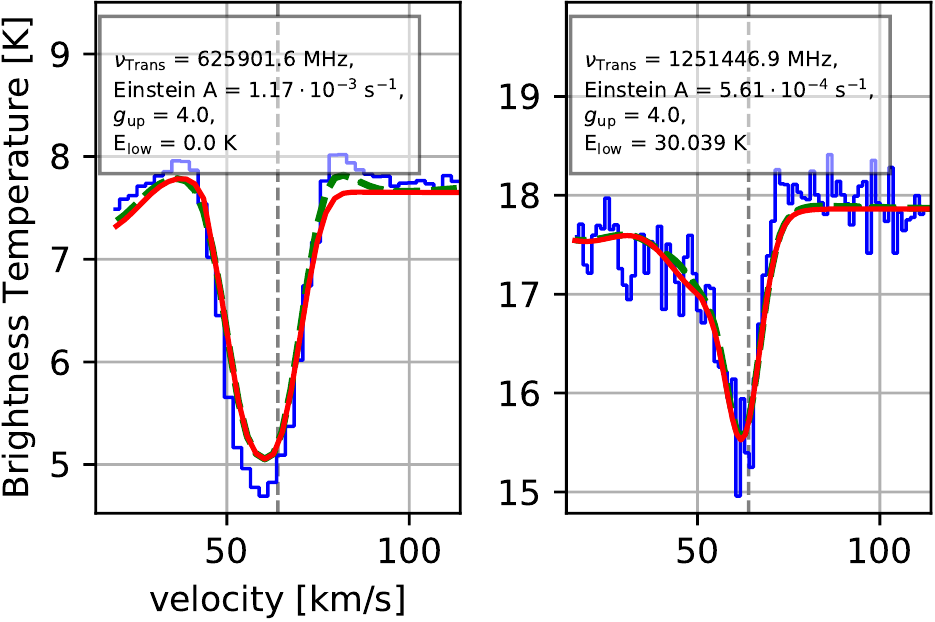}\\
    \caption{Selected transitions of HCl (red line).}
    \label{fig:hcl}
\end{figure*}

\begin{figure*}[!htb]
    \centering
    \includegraphics[scale=0.80]{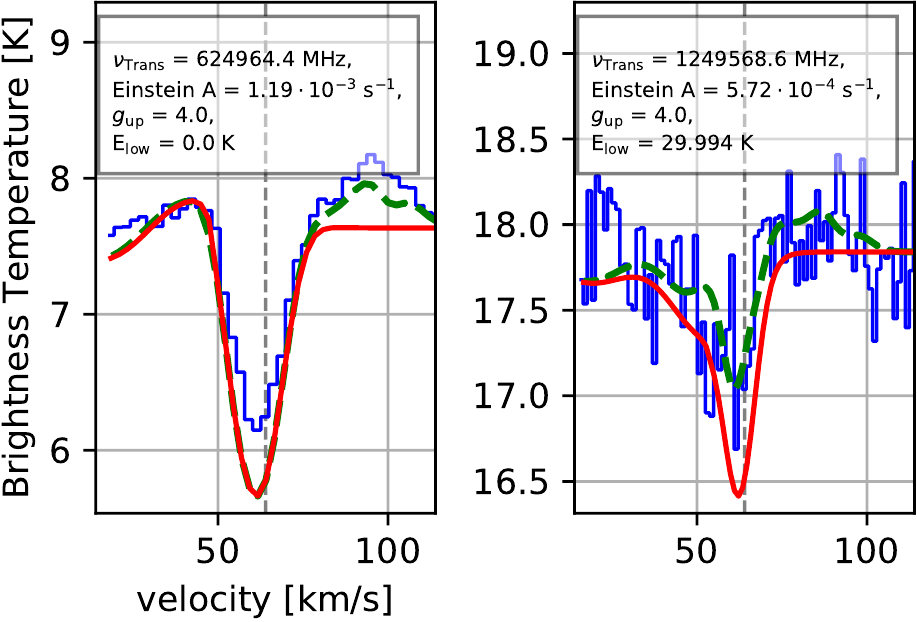}\\
    \caption{Selected transitions of H$^{37}$Cl (red line).}
    \label{fig:hcl37}
\end{figure*}

\begin{figure*}[!htb]
    \centering
    \includegraphics[scale=0.80]{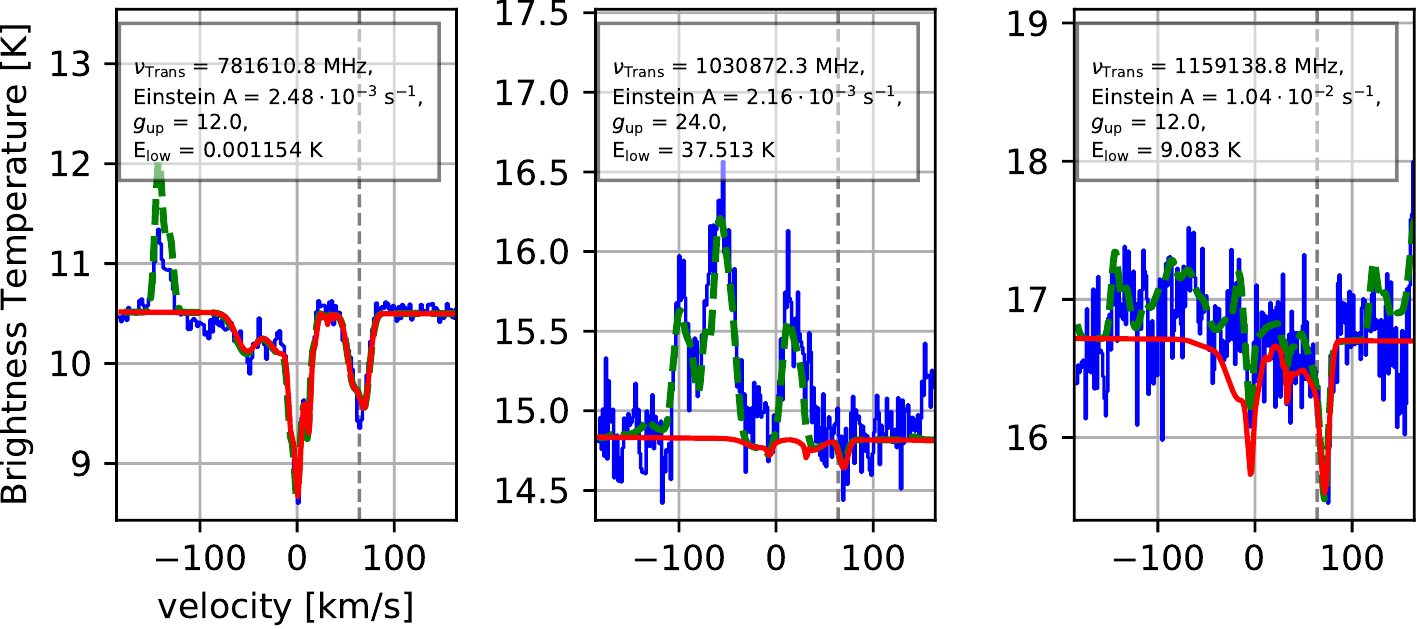}\\
    \caption{Selected transitions of \emph{ortho}-H$_2$Cl$^{+}$ (red line).}
    \label{fig:oh2cl+}
\end{figure*}

\begin{figure*}[!htb]
    \centering
    \includegraphics[scale=0.80]{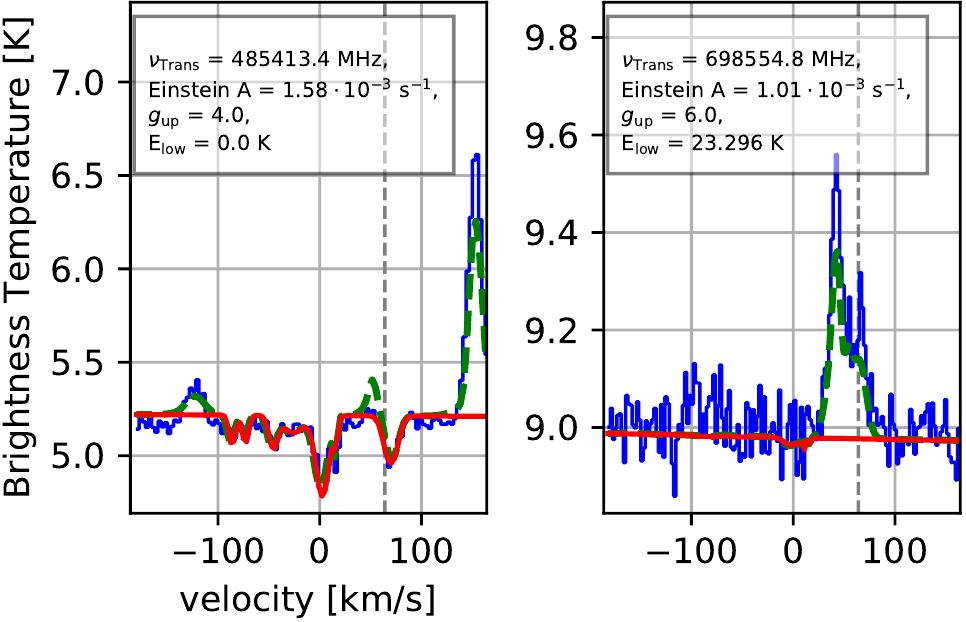}\\
    \caption{Selected transitions of \emph{para}-H$_2$Cl$^{+}$ (red line).}
    \label{fig:ph2cl+}
\end{figure*}

\begin{figure*}[!htb]
    \centering
    \includegraphics[scale=0.80]{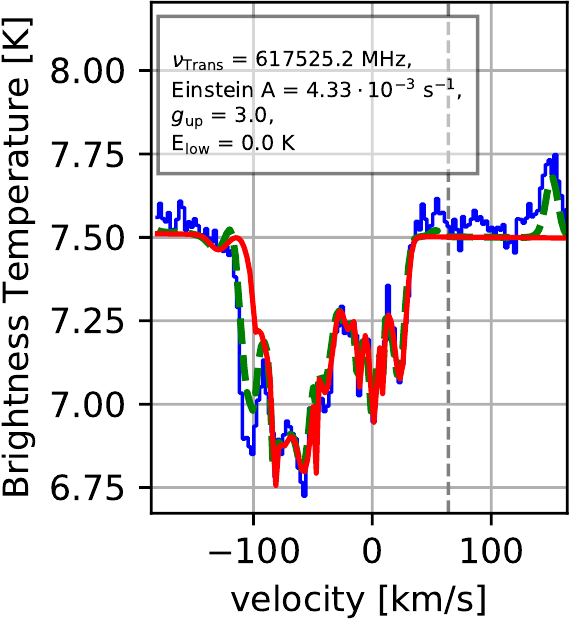}\\
    \caption{Selected transitions of $^{36}$ArH$^{+}$ (red line).}
    \label{fig:arh+}
\end{figure*}
\end{document}